\documentclass[apjl,iop]{emulateapj}
\usepackage{hyperref}
\usepackage{comment}
\usepackage{ifthen}

\newcommand{\forloop}[5][1]%
{%
\setcounter{#2}{#3}%
\ifthenelse{#4}%
	{%
	#5%
	\addtocounter{#2}{#1}%
	\forloop[#1]{#2}{\value{#2}}{#4}{#5}%
	}%
	{%
	}%
}%

\newcommand{\ctbd}[1]{}

\newcommand{\lc}{light curve}
\newcommand{\lcs}{light curves}
\newcommand{\Lc}{Light curve}

\newcommand{\masy}{\ensuremath{\rm mas\,yr^{-1}}}
\newcommand{\kms}{\ensuremath{\rm km\,s^{-1}}}
\newcommand{\ms}{\ensuremath{\rm m\,s^{-1}}}

\newcommand{\gcmc}{\ensuremath{\rm g\,cm^{-3}}}
\newcommand{\gmcc}{\ensuremath{\rm g\,cm^{-3}}}

\newcommand{\vsini}{\ensuremath{v \sin{i}}}
\newcommand{\feh}{\ensuremath{\rm [Fe/H]}}

\newcommand{\vmac}{\ensuremath{v_{\rm mac}}}
\newcommand{\vmic}{\ensuremath{v_{\rm mic}}}

\newcommand{\rsun}{\ensuremath{R_\sun}}
\newcommand{\msun}{\ensuremath{M_\sun}}
\newcommand{\lsun}{\ensuremath{L_\sun}}

\newcommand{\rstar}{\ensuremath{R_\star}}
\newcommand{\mstar}{\ensuremath{M_\star}}
\newcommand{\lstar}{\ensuremath{L_\star}}

\newcommand{\teffstar}{\ensuremath{T_{\rm eff\star}}}
\newcommand{\rhostar}{\ensuremath{\rho_\star}}
\newcommand{\loggstar}{\ensuremath{\log{g_{\star}}}}

\newcommand{\rpl}{\ensuremath{R_{p}}}
\newcommand{\mpl}{\ensuremath{M_{p}}}

\newcommand{\rhopl}{\ensuremath{\rho_{p}}}

\newcommand{\arstar}{\ensuremath{a/\rstar}}
\newcommand{\zrstar}{\ensuremath{\zeta/\rstar}}

\newcommand{\rjup}{\ensuremath{R_{\rm J}}}
\newcommand{\mjup}{\ensuremath{M_{\rm J}}}

\newcommand{\refsec}[1]{\mbox{\S\ \ref{sec:#1}}}

\newcommand{\reffigl}[1]{Figure~\ref{fig:#1}}
\newcommand{\refsecl}[1]{\mbox{Section \ref{sec:#1}}}

\newcommand{\reftabl}[1]{Table~\ref{tab:#1}}

\newcommand{\flwof}{\mbox{FLWO 1.2\,m}}

\newcommand{\loopand}{\ifnum\value{planetcounter}=4 and \else\fi}
\newcommand{\loopcomma}{\ifnum\value{planetcounter}<4 ,\else. \fi}
\newcommand{\loopcommanoperiod}{\ifnum\value{planetcounter}<4 ,\else \space\fi}
\newcommand{\loopcommanospace}{\ifnum\value{planetcounter}<4 ,\else \fi}

\newcommand{\hatcurhtrxxxxxA}{HTR316-006}                              
\newcommand{\hatcurfieldxxxxxA}{316}                                   
\newcommand{\hatcurCCraxxxxxA}{\ensuremath{07^{\mathrm h}52^{\mathrm m}15.20{\mathrm s}}}                            
\newcommand{\hatcurCCdecxxxxxA}{\ensuremath{+12{\arcdeg}08{\arcmin}21.9{\arcsec}}}                           
\newcommand{\hatcurCCmagxxxxxA}{11.762}                                
\newcommand{\hatcurCCtwomassxxxxxA}{2MASS~07521521+1208218}            
\newcommand{\hatcurCCgscxxxxxA}{GSC~0787-00340}                        
\newcommand{\hatcurCCtassmvxxxxxA}{\ensuremath{11.762\pm0.030}}        
\newcommand{\hatcurCCtassmvshortxxxxxA}{\ensuremath{11.8}}             
\newcommand{\hatcurCCtassmBxxxxxA}{\ensuremath{12.282\pm0.050}}        
\newcommand{\hatcurCCtassmBshortxxxxxA}{\ensuremath{12.3}}             
\newcommand{\hatcurCCtassmIxxxxxA}{\ensuremath{11.194\pm0.052}}        
\newcommand{\hatcurCCtassmIshortxxxxxA}{\ensuremath{11.2}}             
\newcommand{\hatcurCCtassmgxxxxxA}{\ensuremath{11.973\pm0.050}}        
\newcommand{\hatcurCCtassmgshortxxxxxA}{\ensuremath{12.0}}             
\newcommand{\hatcurCCtassmrxxxxxA}{\ensuremath{11.650\pm0.030}}        
\newcommand{\hatcurCCtassmrshortxxxxxA}{\ensuremath{11.7}}             
\newcommand{\hatcurCCtassmixxxxxA}{\ensuremath{11.550\pm0.020}}        
\newcommand{\hatcurCCtassmishortxxxxxA}{\ensuremath{11.6}}             
\newcommand{\hatcurCCtwomassJmagxxxxxA}{\ensuremath{10.816\pm0.021}}   
\newcommand{\hatcurCCtwomassHmagxxxxxA}{\ensuremath{10.545\pm0.020}}   
\newcommand{\hatcurCCtwomassKmagxxxxxA}{\ensuremath{10.500\pm0.018}}   
\newcommand{\hatcurCCcitJmagxxxxxA}{\ensuremath{10.836\pm0.021}}       
\newcommand{\hatcurCCcitHmagxxxxxA}{\ensuremath{10.541\pm0.021}}       
\newcommand{\hatcurCCcitKmagxxxxxA}{\ensuremath{10.524\pm0.018}}       
\newcommand{\hatcurCCbbJmagxxxxxA}{\ensuremath{10.880\pm0.023}}        
\newcommand{\hatcurCCbbHmagxxxxxA}{\ensuremath{10.561\pm0.021}}        
\newcommand{\hatcurCCbbKmagxxxxxA}{\ensuremath{10.544\pm0.018}}        
\newcommand{\hatcurCCesoJmagxxxxxA}{\ensuremath{10.882\pm0.024}}       
\newcommand{\hatcurCCesoHmagxxxxxA}{\ensuremath{10.555\pm0.023}}       
\newcommand{\hatcurCCesoKmagxxxxxA}{\ensuremath{10.543\pm0.019}}       
\newcommand{\hatcurCCesoJHmagxxxxxA}{\ensuremath{0.327\pm0.031}}       
\newcommand{\hatcurCCesoJKmagxxxxxA}{\ensuremath{0.339\pm0.030}}       
\newcommand{\hatcurCCesoHKmagxxxxxA}{\ensuremath{0.012\pm0.030}}       
\newcommand{\hatcurLCdipxxxxxA}{\ensuremath{8.1}}                      
\newcommand{\hatcurLCrprstarxxxxxA}{\ensuremath{0.0782\pm0.0012}}      
\newcommand{\hatcurLCbsqxxxxxA}{\ensuremath{0.395_{-0.050}^{+0.041}}}  
\newcommand{\hatcurLCimpxxxxxA}{\ensuremath{0.629_{-0.041}^{+0.032}}}  
\newcommand{\hatcurLCzetaxxxxxA}{\ensuremath{14.710\pm0.077}}          
\newcommand{\hatcurLCdurxxxxxA}{\ensuremath{0.1531\pm0.0011}}          
\newcommand{\hatcurLCdurshortxxxxxA}{\ensuremath{0.1531}}              
\newcommand{\hatcurLCdurhrxxxxxA}{\ensuremath{3.674\pm0.026}}          
\newcommand{\hatcurLCdurhrshortxxxxxA}{\ensuremath{3.674}}             
\newcommand{\hatcurLCqxxxxxA}{\ensuremath{0.04900\pm0.00035}}          
\newcommand{\hatcurLCqshortxxxxxA}{\ensuremath{0.049}}                 
\newcommand{\hatcurLCingdurxxxxxA}{\ensuremath{0.0176\pm0.0013}}       
\newcommand{\hatcurLCPxxxxxA}{\ensuremath{3.1220109\pm0.0000065}}      
\newcommand{\hatcurLCPprecxxxxxA}{\ensuremath{3.1220109}}              
\newcommand{\hatcurLCPshortxxxxxA}{\ensuremath{3.1220}}                
\newcommand{\hatcurLCTxxxxxA}{\ensuremath{2456285.90993\pm0.00036}}    
\newcommand{\hatcurLCTAxxxxxA}{\ensuremath{2454793.5888\pm0.0030}}     
\newcommand{\hatcurLCTBxxxxxA}{\ensuremath{2456310.88602\pm0.00038}}   
\newcommand{\hatcurLChatnetmAxxxxxA}{\ensuremath{11.82901\pm0.00013}}  
\newcommand{\hatcurLCiblendAxxxxxA}{\ensuremath{0.72\pm0.14}}          
\newcommand{\hatcurLChatnetmBxxxxxA}{\ensuremath{11.82858\pm0.00013}}  
\newcommand{\hatcurLCiblendBxxxxxA}{\ensuremath{0.76\pm0.11}}          
\newcommand{\hatcurSMEiteffxxxxxA}{\ensuremath{6169\pm50}}             
\newcommand{\hatcurSMEizfehxxxxxA}{\ensuremath{-0.260\pm0.080}}        
\newcommand{\hatcurSMEizfehshortxxxxxA}{\ensuremath{-0.26}}            
\newcommand{\hatcurSMEiloggxxxxxA}{\ensuremath{3.84\pm0.10}}           
\newcommand{\hatcurSMEivsinxxxxxA}{\ensuremath{9.00\pm0.50}}           
\newcommand{\hatcurSMEivmacxxxxxA}{\ensuremath{0.0}}                   
\newcommand{\hatcurSMEivmicxxxxxA}{\ensuremath{0.0}}                   
\newcommand{\hatcurSMEiiteffxxxxxA}{\ensuremath{6280\pm49}}            
\newcommand{\hatcurSMEiizfehxxxxxA}{\ensuremath{-0.180\pm0.080}}       
\newcommand{\hatcurSMEiizfehshortxxxxxA}{\ensuremath{-0.18}}           
\newcommand{\hatcurSMEiiloggxxxxxA}{\ensuremath{4.060\pm0.030}}        
\newcommand{\hatcurSMEiivsinxxxxxA}{\ensuremath{8.90\pm0.50}}          
\newcommand{\hatcurTRESteffxxxxxA}{\ensuremath{6161\pm60}}             
\newcommand{\hatcurTRESzfehxxxxxA}{\ensuremath{-0.040\pm0.080}}        
\newcommand{\hatcurTRESloggxxxxxA}{\ensuremath{4.00\pm0.11}}           
\newcommand{\hatcurTRESvsinixxxxxA}{\ensuremath{9.55\pm0.51}}          
\newcommand{\hatcurTRESgammaxxxxxA}{\ensuremath{7.119\pm0.014}}        
\newcommand{\hatcurTRESnumspecxxxxxA}{\ensuremath{6}}                  
\newcommand{\hatcurTRESspanxxxxxA}{\ensuremath{NULL}}                  
\newcommand{\hatcurTRESrvrmsxxxxxA}{\ensuremath{0.14}}                 
\newcommand{\hatcurFIESteffxxxxxA}{\ensuremath{6089\pm54}}             
\newcommand{\hatcurFIESzfehxxxxxA}{\ensuremath{-0.282\pm0.080}}        
\newcommand{\hatcurFIESloggxxxxxA}{\ensuremath{3.80\pm0.10}}           
\newcommand{\hatcurFIESvsinixxxxxA}{\ensuremath{9.72\pm0.50}}          
\newcommand{\hatcurFIESgammaxxxxxA}{\ensuremath{7.014\pm0.075}}        
\newcommand{\hatcurFIESnumspecxxxxxA}{\ensuremath{5}}                  
\newcommand{\hatcurFIESspanxxxxxA}{\ensuremath{NULL}}                  
\newcommand{\hatcurFIESrvrmsxxxxxA}{\ensuremath{0.10}}                 
\newcommand{\hatcurLBizxxxxxA}{\ensuremath{0.1470}}                    
\newcommand{\hatcurLBiizxxxxxA}{\ensuremath{0.3467}}                   
\newcommand{\hatcurLBiixxxxxA}{\ensuremath{0.1965}}                    
\newcommand{\hatcurLBiiixxxxxA}{\ensuremath{0.3570}}                   
\newcommand{\hatcurLBiIxxxxxA}{\ensuremath{0.1794}}                    
\newcommand{\hatcurLBiiIxxxxxA}{\ensuremath{0.3541}}                   
\newcommand{\hatcurLBigxxxxxA}{\ensuremath{0.4220}}                    
\newcommand{\hatcurLBiigxxxxxA}{\ensuremath{0.3248}}                   
\newcommand{\hatcurLBirxxxxxA}{\ensuremath{0.2658}}                    
\newcommand{\hatcurLBiirxxxxxA}{\ensuremath{0.3694}}                   
\newcommand{\hatcurLBiRxxxxxA}{\ensuremath{0.2463}}                    
\newcommand{\hatcurLBiiRxxxxxA}{\ensuremath{0.3673}}                   
\newcommand{\hatcurISOmxxxxxA}{\ensuremath{1.273_{-0.115}^{+0.049}}}   
\newcommand{\hatcurISOmshortxxxxxA}{\ensuremath{1.27}}                 
\newcommand{\hatcurISOmlongxxxxxA}{\ensuremath{1.273_{-0.115}^{+0.049}}} 
\newcommand{\hatcurISOrxxxxxA}{\ensuremath{1.698\pm0.071}}             
\newcommand{\hatcurISOrshortxxxxxA}{\ensuremath{1.70}}                 
\newcommand{\hatcurISOrlongxxxxxA}{\ensuremath{1.698\pm0.071}}         
\newcommand{\hatcurISOrhoxxxxxA}{\ensuremath{0.357\pm0.037}}           
\newcommand{\hatcurISOrholongxxxxxA}{\ensuremath{0.357\pm0.037}}       
\newcommand{\hatcurISOloggxxxxxA}{\ensuremath{4.072\pm0.029}}          
\newcommand{\hatcurISOlumxxxxxA}{\ensuremath{4.01\pm0.38}}             
\newcommand{\hatcurISOlumshortxxxxxA}{\ensuremath{4.01}}               
\newcommand{\hatcurISOmvxxxxxA}{\ensuremath{3.27\pm0.11}}              
\newcommand{\hatcurISOvixxxxxA}{\ensuremath{0.552\pm0.013}}            
\newcommand{\hatcurISOagexxxxxA}{\ensuremath{3.37_{-0.27}^{+1.44}}}    
\newcommand{\hatcurISOsigmaxxxxxA}{\ensuremath{0.000600\pm0.000060}}   
\newcommand{\hatcurISOMJxxxxxA}{\ensuremath{2.360\pm0.097}}            
\newcommand{\hatcurISOMHxxxxxA}{\ensuremath{2.098\pm0.093}}            
\newcommand{\hatcurISOMKxxxxxA}{\ensuremath{2.052\pm0.093}}            
\newcommand{\hatcurISOJKxxxxxA}{\ensuremath{0.310\pm0.010}}            
\newcommand{\hatcurISOspecxxxxxA}{F}                                   
\newcommand{\hatcurRVKxxxxxA}{\ensuremath{161.3\pm5.6}}                
\newcommand{\hatcurRVrkxxxxxA}{\ensuremath{0\pm0}}                     
\newcommand{\hatcurRVrhxxxxxA}{\ensuremath{0\pm0}}                     
\newcommand{\hatcurRVkxxxxxA}{\ensuremath{0\pm0}}                      
\newcommand{\hatcurRVhxxxxxA}{\ensuremath{0\pm0}}                      
\newcommand{\hatcurRVtronexxxxxA}{\ensuremath{0\pm0}}                  
\newcommand{\hatcurRVtrtwoxxxxxA}{\ensuremath{0\pm0}}                  
\newcommand{\hatcurRVgammaAxxxxxA}{\ensuremath{-173\pm35}}             
\newcommand{\hatcurRVjitterAxxxxxA}{\ensuremath{68\pm38}}              
\newcommand{\hatcurRVfitrmsAxxxxxA}{\ensuremath{116.0}}                
\newcommand{\hatcurRVgammaBxxxxxA}{\ensuremath{-121\pm13}}             
\newcommand{\hatcurRVjitterBxxxxxA}{\ensuremath{0.00\pm0.91}}          
\newcommand{\hatcurRVfitrmsBxxxxxA}{\ensuremath{25.0}}                 
\newcommand{\hatcurRVgammaCxxxxxA}{\ensuremath{43.1\pm4.4}}            
\newcommand{\hatcurRVjitterCxxxxxA}{\ensuremath{0.7\pm7.1}}            
\newcommand{\hatcurRVfitrmsCxxxxxA}{\ensuremath{23.0}}                 
\newcommand{\hatcurRVeccenxxxxxA}{\ensuremath{0\pm0}}                  
\newcommand{\hatcurRVeccentwosiglimxxxxxA}{\ensuremath{<0.000}}        
\newcommand{\hatcurRVomegaxxxxxA}{\ensuremath{0\pm0}}                  
\newcommand{\hatcurPPixxxxxA}{\ensuremath{83.65\pm0.57}}               
\newcommand{\hatcurPPgxxxxxA}{\ensuremath{20.1\pm1.8}}                 
\newcommand{\hatcurPPloggxxxxxA}{\ensuremath{3.302\pm0.038}}           
\newcommand{\hatcurPParxxxxxA}{\ensuremath{5.68\pm0.19}}               
\newcommand{\hatcurPParelxxxxxA}{\ensuremath{0.04530_{-0.00140}^{+0.00058}}} 
\newcommand{\hatcurPPrhoxxxxxA}{\ensuremath{0.78\pm0.11}}              
\newcommand{\hatcurPPmxxxxxA}{\ensuremath{1.350\pm0.073}}              
\newcommand{\hatcurPPmshortxxxxxA}{\ensuremath{1.35}}                  
\newcommand{\hatcurPPmlongxxxxxA}{\ensuremath{1.350\pm0.073}}          
\newcommand{\hatcurPPmexxxxxA}{\ensuremath{429\pm23}}                  
\newcommand{\hatcurPPmeshortxxxxxA}{\ensuremath{429.1}}                
\newcommand{\hatcurPPmelongxxxxxA}{\ensuremath{429\pm23}}              
\newcommand{\hatcurPPrxxxxxA}{\ensuremath{1.288\pm0.064}}              
\newcommand{\hatcurPPrshortxxxxxA}{\ensuremath{1.29}}                  
\newcommand{\hatcurPPrlongxxxxxA}{\ensuremath{1.288\pm0.064}}          
\newcommand{\hatcurPPrexxxxxA}{\ensuremath{14.43\pm0.72}}              
\newcommand{\hatcurPPreshortxxxxxA}{\ensuremath{14.4}}                 
\newcommand{\hatcurPPrelongxxxxxA}{\ensuremath{14.43\pm0.72}}          
\newcommand{\hatcurPPmrcorrxxxxxA}{\ensuremath{0.48}}                  
\newcommand{\hatcurPPteffxxxxxA}{\ensuremath{1862\pm34}}               
\newcommand{\hatcurPPthetaxxxxxA}{\ensuremath{0.0751\pm0.0044}}        
\newcommand{\hatcurPPfluxperixxxxxA}{\ensuremath{2.71\pm0.20}}         
\newcommand{\hatcurPPfluxperidimxxxxxA}{\ensuremath{9}}                
\newcommand{\hatcurPPfluxapxxxxxA}{\ensuremath{2.71\pm0.20}}           
\newcommand{\hatcurPPfluxapdimxxxxxA}{\ensuremath{9}}                  
\newcommand{\hatcurPPfluxavgxxxxxA}{\ensuremath{2.71\pm0.20}}          
\newcommand{\hatcurPPfluxavgdimxxxxxA}{\ensuremath{9}}                 
\newcommand{\hatcurPPfluxavglogxxxxxA}{\ensuremath{9.433\pm0.032}}     
\newcommand{\hatcurXsecphasexxxxxA}{\ensuremath{0\pm0}}                
\newcommand{\hatcurXsecondaryxxxxxA}{\ensuremath{2456287.47093\pm0.00036}} 
\newcommand{\hatcurXsecdurxxxxxA}{\ensuremath{0.1531\pm0.0011}}        
\newcommand{\hatcurXsecingdurxxxxxA}{\ensuremath{0.0176\pm0.0013}}     
\newcommand{\hatcurPPphiconjxxxxxA}{\ensuremath{0\pm0}}                
\newcommand{\hatcurPPperixxxxxA}{\ensuremath{2456285.12942\pm0.00036}} 
\newcommand{\hatcurPPaequivxxxxxA}{\ensuremath{0.02240\pm0.00084}}     
\newcommand{\hatcurPPtcircxxxxxA}{\ensuremath{184_{-33}^{+42}}}        
\newcommand{\hatcurPPtinfallxxxxxA}{\ensuremath{164\pm29}}             
\newcommand{\hatcurXdistxxxxxA}{\ensuremath{499\pm22}}                 
\newcommand{\hatcurXAvxxxxxA}{\ensuremath{0.011_{-0.011}^{+0.056}}}    
\newcommand{\hatcurXdistredxxxxxA}{\ensuremath{497\pm21}}              
\newcommand{\hatcurXEBVxxxxxA}{\ensuremath{0.0040_{-0.0040}^{+0.0180}}} 
\newcommand{\hatcurXmvisoredxxxxxA}{\ensuremath{11.777\pm0.024}}       
\newcommand{\hatcurXmiisoredxxxxxA}{\ensuremath{11.211\pm0.016}}       
\newcommand{\hatcurXmjisoredxxxxxA}{\ensuremath{10.848\pm0.012}}       
\newcommand{\hatcurXmhisoredxxxxxA}{\ensuremath{10.584\pm0.013}}       
\newcommand{\hatcurXmkisoredxxxxxA}{\ensuremath{10.536\pm0.013}}       
\newcommand{\hatcurXviisoredxxxxxA}{\ensuremath{0.565_{-0.012}^{+0.016}}} 
\newcommand{\hatcurXvkisoredxxxxxA}{\ensuremath{1.240\pm0.028}}        
\newcommand{\hatcurXjhisoredxxxxxA}{\ensuremath{0.2640\pm0.0073}}      
\newcommand{\hatcurXjkisoredxxxxxA}{\ensuremath{0.3120\pm0.0079}}      
\newcommand{\hatcurCCpmraxxxxxA}{\ensuremath{10.20\pm0.80}}            
\newcommand{\hatcurCCpmdecxxxxxA}{\ensuremath{-4.6\pm1.6}}             
\newcommand{\hatcurCCpmxxxxxA}{\ensuremath{11.2\pm1.8}}                
\newcommand{\hatcurhtreccenxxxxxA}{HTR316-006}                         
\newcommand{\hatcurfieldeccenxxxxxA}{316}                              
\newcommand{\hatcurCCraeccenxxxxxA}{\ensuremath{07^{\mathrm h}52^{\mathrm m}15.20{\mathrm s}}}                       
\newcommand{\hatcurCCdececcenxxxxxA}{\ensuremath{+12{\arcdeg}08{\arcmin}21.9{\arcsec}}}                      
\newcommand{\hatcurCCmageccenxxxxxA}{11.762}                           
\newcommand{\hatcurCCtwomasseccenxxxxxA}{2MASS~07521521+1208218}       
\newcommand{\hatcurCCgsceccenxxxxxA}{GSC~0787-00340}                   
\newcommand{\hatcurCCtassmveccenxxxxxA}{\ensuremath{11.762\pm0.030}}   
\newcommand{\hatcurCCtassmvshorteccenxxxxxA}{\ensuremath{11.8}}        
\newcommand{\hatcurCCtassmBeccenxxxxxA}{\ensuremath{12.282\pm0.050}}   
\newcommand{\hatcurCCtassmBshorteccenxxxxxA}{\ensuremath{12.3}}        
\newcommand{\hatcurCCtassmIeccenxxxxxA}{\ensuremath{11.194\pm0.052}}   
\newcommand{\hatcurCCtassmIshorteccenxxxxxA}{\ensuremath{11.2}}        
\newcommand{\hatcurCCtassmgeccenxxxxxA}{\ensuremath{11.973\pm0.050}}   
\newcommand{\hatcurCCtassmgshorteccenxxxxxA}{\ensuremath{12.0}}        
\newcommand{\hatcurCCtassmreccenxxxxxA}{\ensuremath{11.650\pm0.030}}   
\newcommand{\hatcurCCtassmrshorteccenxxxxxA}{\ensuremath{11.7}}        
\newcommand{\hatcurCCtassmieccenxxxxxA}{\ensuremath{11.550\pm0.020}}   
\newcommand{\hatcurCCtassmishorteccenxxxxxA}{\ensuremath{11.6}}        
\newcommand{\hatcurCCtwomassJmageccenxxxxxA}{\ensuremath{10.816\pm0.021}} 
\newcommand{\hatcurCCtwomassHmageccenxxxxxA}{\ensuremath{10.545\pm0.020}} 
\newcommand{\hatcurCCtwomassKmageccenxxxxxA}{\ensuremath{10.500\pm0.018}} 
\newcommand{\hatcurCCcitJmageccenxxxxxA}{\ensuremath{10.836\pm0.021}}  
\newcommand{\hatcurCCcitHmageccenxxxxxA}{\ensuremath{10.541\pm0.021}}  
\newcommand{\hatcurCCcitKmageccenxxxxxA}{\ensuremath{10.524\pm0.018}}  
\newcommand{\hatcurCCbbJmageccenxxxxxA}{\ensuremath{10.880\pm0.023}}   
\newcommand{\hatcurCCbbHmageccenxxxxxA}{\ensuremath{10.561\pm0.021}}   
\newcommand{\hatcurCCbbKmageccenxxxxxA}{\ensuremath{10.544\pm0.018}}   
\newcommand{\hatcurCCesoJmageccenxxxxxA}{\ensuremath{10.882\pm0.024}}  
\newcommand{\hatcurCCesoHmageccenxxxxxA}{\ensuremath{10.555\pm0.023}}  
\newcommand{\hatcurCCesoKmageccenxxxxxA}{\ensuremath{10.543\pm0.019}}  
\newcommand{\hatcurCCesoJHmageccenxxxxxA}{\ensuremath{0.327\pm0.031}}  
\newcommand{\hatcurCCesoJKmageccenxxxxxA}{\ensuremath{0.339\pm0.030}}  
\newcommand{\hatcurCCesoHKmageccenxxxxxA}{\ensuremath{0.012\pm0.030}}  
\newcommand{\hatcurLCdipeccenxxxxxA}{\ensuremath{}}            
\newcommand{\hatcurLCrprstareccenxxxxxA}{\ensuremath{0.0777\pm0.0013}} 
\newcommand{\hatcurLCbsqeccenxxxxxA}{\ensuremath{0.379_{-0.054}^{+0.043}}} 
\newcommand{\hatcurLCimpeccenxxxxxA}{\ensuremath{0.616_{-0.046}^{+0.034}}} 
\newcommand{\hatcurLCzetaeccenxxxxxA}{\ensuremath{14.667\pm0.072}}     
\newcommand{\hatcurLCdureccenxxxxxA}{\ensuremath{0.1531\pm0.0012}}     
\newcommand{\hatcurLCdurshorteccenxxxxxA}{\ensuremath{0.1531}}         
\newcommand{\hatcurLCdurhreccenxxxxxA}{\ensuremath{3.674\pm0.029}}     
\newcommand{\hatcurLCdurhrshorteccenxxxxxA}{\ensuremath{3.674}}        
\newcommand{\hatcurLCqeccenxxxxxA}{\ensuremath{0.04900\pm0.00039}}     
\newcommand{\hatcurLCqshorteccenxxxxxA}{\ensuremath{0.049}}            
\newcommand{\hatcurLCingdureccenxxxxxA}{\ensuremath{0.0170\pm0.0015}}  
\newcommand{\hatcurLCPeccenxxxxxA}{\ensuremath{3.1220115\pm0.0000048}} 
\newcommand{\hatcurLCPprececcenxxxxxA}{\ensuremath{3.1220115}}         
\newcommand{\hatcurLCPshorteccenxxxxxA}{\ensuremath{3.1220}}           
\newcommand{\hatcurLCTeccenxxxxxA}{\ensuremath{2456267.17772\pm0.00035}} 
\newcommand{\hatcurLCTAeccenxxxxxA}{\ensuremath{2454793.5883\pm0.0022}} 
\newcommand{\hatcurLCTBeccenxxxxxA}{\ensuremath{2456310.88589\pm0.00038}} 
\newcommand{\hatcurLChatnetmAeccenxxxxxA}{\ensuremath{11.82896\pm0.00011}} 
\newcommand{\hatcurLCiblendAeccenxxxxxA}{\ensuremath{0.71\pm0.13}}     
\newcommand{\hatcurLChatnetmBeccenxxxxxA}{\ensuremath{11.82859\pm0.00013}} 
\newcommand{\hatcurLCiblendBeccenxxxxxA}{\ensuremath{0.767\pm0.096}}   
\newcommand{\hatcurSMEiteffeccenxxxxxA}{\ensuremath{6089\pm54}}        
\newcommand{\hatcurSMEizfeheccenxxxxxA}{\ensuremath{-0.282\pm0.080}}   
\newcommand{\hatcurSMEizfehshorteccenxxxxxA}{\ensuremath{-0.28}}       
\newcommand{\hatcurSMEiloggeccenxxxxxA}{\ensuremath{3.80\pm0.10}}      
\newcommand{\hatcurSMEivsineccenxxxxxA}{\ensuremath{9.72\pm0.50}}      
\newcommand{\hatcurSMEivmaceccenxxxxxA}{\ensuremath{0.0}}              
\newcommand{\hatcurSMEivmiceccenxxxxxA}{\ensuremath{0.0}}              
\newcommand{\hatcurTRESteffeccenxxxxxA}{\ensuremath{6161\pm60}}        
\newcommand{\hatcurTRESzfeheccenxxxxxA}{\ensuremath{-0.040\pm0.080}}   
\newcommand{\hatcurTRESloggeccenxxxxxA}{\ensuremath{4.00\pm0.11}}      
\newcommand{\hatcurTRESvsinieccenxxxxxA}{\ensuremath{9.55\pm0.51}}     
\newcommand{\hatcurTRESgammaeccenxxxxxA}{\ensuremath{7.119\pm0.014}}   
\newcommand{\hatcurTRESnumspececcenxxxxxA}{\ensuremath{6}}             
\newcommand{\hatcurTRESspaneccenxxxxxA}{\ensuremath{convertHATNetTime.sh}} 
\newcommand{\hatcurTRESrvrmseccenxxxxxA}{\ensuremath{0.14}}            
\newcommand{\hatcurFIESteffeccenxxxxxA}{\ensuremath{6089\pm54}}        
\newcommand{\hatcurFIESzfeheccenxxxxxA}{\ensuremath{-0.282\pm0.080}}   
\newcommand{\hatcurFIESloggeccenxxxxxA}{\ensuremath{3.80\pm0.10}}      
\newcommand{\hatcurFIESvsinieccenxxxxxA}{\ensuremath{9.72\pm0.50}}     
\newcommand{\hatcurFIESgammaeccenxxxxxA}{\ensuremath{7.014\pm0.075}}   
\newcommand{\hatcurFIESnumspececcenxxxxxA}{\ensuremath{5}}             
\newcommand{\hatcurFIESspaneccenxxxxxA}{\ensuremath{NULL}}             
\newcommand{\hatcurFIESrvrmseccenxxxxxA}{\ensuremath{0.10}}            
\newcommand{\hatcurLBizeccenxxxxxA}{\ensuremath{0.1630}}               
\newcommand{\hatcurLBiizeccenxxxxxA}{\ensuremath{0.3382}}              
\newcommand{\hatcurLBiieccenxxxxxA}{\ensuremath{0.2130}}               
\newcommand{\hatcurLBiiieccenxxxxxA}{\ensuremath{0.3482}}              
\newcommand{\hatcurLBiIeccenxxxxxA}{\ensuremath{0.1959}}               
\newcommand{\hatcurLBiiIeccenxxxxxA}{\ensuremath{0.3453}}              
\newcommand{\hatcurLBigeccenxxxxxA}{\ensuremath{0.4448}}               
\newcommand{\hatcurLBiigeccenxxxxxA}{\ensuremath{0.3098}}              
\newcommand{\hatcurLBireccenxxxxxA}{\ensuremath{0.2791}}               
\newcommand{\hatcurLBiireccenxxxxxA}{\ensuremath{0.3638}}              
\newcommand{\hatcurLBiReccenxxxxxA}{\ensuremath{0.2602}}               
\newcommand{\hatcurLBiiReccenxxxxxA}{\ensuremath{0.3611}}              
\newcommand{\hatcurISOmeccenxxxxxA}{\ensuremath{1.075\pm0.047}}        
\newcommand{\hatcurISOmshorteccenxxxxxA}{\ensuremath{1.08}}            
\newcommand{\hatcurISOmlongeccenxxxxxA}{\ensuremath{1.075\pm0.047}}    
\newcommand{\hatcurISOreccenxxxxxA}{\ensuremath{1.552\pm0.086}}        
\newcommand{\hatcurISOrshorteccenxxxxxA}{\ensuremath{1.55}}            
\newcommand{\hatcurISOrlongeccenxxxxxA}{\ensuremath{1.552\pm0.086}}    
\newcommand{\hatcurISOrhoeccenxxxxxA}{\ensuremath{0.408\pm0.061}}      
\newcommand{\hatcurISOrholongeccenxxxxxA}{\ensuremath{0.408\pm0.061}}  
\newcommand{\hatcurISOloggeccenxxxxxA}{\ensuremath{4.088\pm0.041}}     
\newcommand{\hatcurISOlumeccenxxxxxA}{\ensuremath{2.96\pm0.36}}        
\newcommand{\hatcurISOlumshorteccenxxxxxA}{\ensuremath{2.96}}          
\newcommand{\hatcurISOmveccenxxxxxA}{\ensuremath{3.63\pm0.14}}         
\newcommand{\hatcurISOvieccenxxxxxA}{\ensuremath{0.604\pm0.015}}       
\newcommand{\hatcurISOageeccenxxxxxA}{\ensuremath{6.27\pm0.84}}        
\newcommand{\hatcurISOsigmaeccenxxxxxA}{\ensuremath{0.000700\pm0.000086}} 
\newcommand{\hatcurISOMJeccenxxxxxA}{\ensuremath{2.62\pm0.12}}         
\newcommand{\hatcurISOMHeccenxxxxxA}{\ensuremath{2.32\pm0.12}}         
\newcommand{\hatcurISOMKeccenxxxxxA}{\ensuremath{2.27\pm0.12}}         
\newcommand{\hatcurISOJKeccenxxxxxA}{\ensuremath{0.350\pm0.010}}       
\newcommand{\hatcurISOspececcenxxxxxA}{F}                              
\newcommand{\hatcurRVKeccenxxxxxA}{\ensuremath{162.1\pm5.6}}           
\newcommand{\hatcurRVrkeccenxxxxxA}{\ensuremath{-0.209_{-0.054}^{+0.103}}} 
\newcommand{\hatcurRVrheccenxxxxxA}{\ensuremath{-0.132_{-0.082}^{+0.112}}} 
\newcommand{\hatcurRVkeccenxxxxxA}{\ensuremath{-0.051\pm0.029}}        
\newcommand{\hatcurRVheccenxxxxxA}{\ensuremath{-0.034\pm0.031}}        
\newcommand{\hatcurRVtroneeccenxxxxxA}{\ensuremath{0\pm0}}             
\newcommand{\hatcurRVtrtwoeccenxxxxxA}{\ensuremath{0\pm0}}             
\newcommand{\hatcurRVgammaAeccenxxxxxA}{\ensuremath{-177\pm34}}        
\newcommand{\hatcurRVjitterAeccenxxxxxA}{\ensuremath{69\pm18}}         
\newcommand{\hatcurRVfitrmsAeccenxxxxxA}{\ensuremath{96.1}}            
\newcommand{\hatcurRVgammaBeccenxxxxxA}{\ensuremath{-122\pm12}}        
\newcommand{\hatcurRVjitterBeccenxxxxxA}{\ensuremath{0.000\pm0.013}}   
\newcommand{\hatcurRVfitrmsBeccenxxxxxA}{\ensuremath{22.4}}            
\newcommand{\hatcurRVgammaCeccenxxxxxA}{\ensuremath{51.4\pm7.0}}       
\newcommand{\hatcurRVjitterCeccenxxxxxA}{\ensuremath{5.5\pm5.2}}       
\newcommand{\hatcurRVfitrmsCeccenxxxxxA}{\ensuremath{21.0}}            
\newcommand{\hatcurRVecceneccenxxxxxA}{\ensuremath{0.069\pm0.030}}     
\newcommand{\hatcurRVeccentwosiglimeccenxxxxxA}{\ensuremath{<0.115}}   
\newcommand{\hatcurRVomegaeccenxxxxxA}{\ensuremath{211\pm35}}          
\newcommand{\hatcurPPieccenxxxxxA}{\ensuremath{84.24\pm0.70}}          
\newcommand{\hatcurPPgeccenxxxxxA}{\ensuremath{22.1\pm3.1}}            
\newcommand{\hatcurPPloggeccenxxxxxA}{\ensuremath{3.344\pm0.059}}      
\newcommand{\hatcurPPareccenxxxxxA}{\ensuremath{5.94\pm0.29}}          
\newcommand{\hatcurPPareleccenxxxxxA}{\ensuremath{0.04283\pm0.00062}}  
\newcommand{\hatcurPPrhoeccenxxxxxA}{\ensuremath{0.94\pm0.20}}         
\newcommand{\hatcurPPmeccenxxxxxA}{\ensuremath{1.227\pm0.051}}         
\newcommand{\hatcurPPmshorteccenxxxxxA}{\ensuremath{1.23}}             
\newcommand{\hatcurPPmlongeccenxxxxxA}{\ensuremath{1.227\pm0.051}}     
\newcommand{\hatcurPPmeeccenxxxxxA}{\ensuremath{390\pm16}}             
\newcommand{\hatcurPPmeshorteccenxxxxxA}{\ensuremath{389.8}}           
\newcommand{\hatcurPPmelongeccenxxxxxA}{\ensuremath{390\pm16}}         
\newcommand{\hatcurPPreccenxxxxxA}{\ensuremath{1.172\pm0.079}}         
\newcommand{\hatcurPPrshorteccenxxxxxA}{\ensuremath{1.17}}             
\newcommand{\hatcurPPrlongeccenxxxxxA}{\ensuremath{1.172\pm0.079}}     
\newcommand{\hatcurPPreeccenxxxxxA}{\ensuremath{13.14\pm0.88}}         
\newcommand{\hatcurPPreshorteccenxxxxxA}{\ensuremath{13.1}}            
\newcommand{\hatcurPPrelongeccenxxxxxA}{\ensuremath{13.14\pm0.88}}     
\newcommand{\hatcurPPmrcorreccenxxxxxA}{\ensuremath{0.08}}             
\newcommand{\hatcurPPteffeccenxxxxxA}{\ensuremath{1769\pm46}}          
\newcommand{\hatcurPPthetaeccenxxxxxA}{\ensuremath{0.0828\pm0.0071}}   
\newcommand{\hatcurPPfluxperieccenxxxxxA}{\ensuremath{2.53\pm0.26}}    
\newcommand{\hatcurPPfluxperidimeccenxxxxxA}{\ensuremath{9}}           
\newcommand{\hatcurPPfluxapeccenxxxxxA}{\ensuremath{1.94\pm0.26}}      
\newcommand{\hatcurPPfluxapdimeccenxxxxxA}{\ensuremath{9}}             
\newcommand{\hatcurPPfluxavgeccenxxxxxA}{\ensuremath{2.21\pm0.23}}     
\newcommand{\hatcurPPfluxavgdimeccenxxxxxA}{\ensuremath{9}}            
\newcommand{\hatcurPPfluxavglogeccenxxxxxA}{\ensuremath{9.345\pm0.045}} 
\newcommand{\hatcurXsecphaseeccenxxxxxA}{\ensuremath{0.467\pm0.019}}   
\newcommand{\hatcurXsecondaryeccenxxxxxA}{\ensuremath{2456268.637\pm0.059}} 
\newcommand{\hatcurXsecdureccenxxxxxA}{\ensuremath{0.1478\pm0.0058}}   
\newcommand{\hatcurXsecingdureccenxxxxxA}{\ensuremath{0.0154\pm0.0021}} 
\newcommand{\hatcurPPphiconjeccenxxxxxA}{\ensuremath{-0.31\pm0.14}}    
\newcommand{\hatcurPPperieccenxxxxxA}{\ensuremath{2456268.15\pm0.44}}  
\newcommand{\hatcurPPaequiveccenxxxxxA}{\ensuremath{0.0249\pm0.0013}}  
\newcommand{\hatcurPPtcirceccenxxxxxA}{\ensuremath{238_{-61}^{+81}}}   
\newcommand{\hatcurPPtinfalleccenxxxxxA}{\ensuremath{192_{-38}^{+51}}} 
\newcommand{\hatcurXdisteccenxxxxxA}{\ensuremath{451\pm26}}            
\newcommand{\hatcurXAveccenxxxxxA}{\ensuremath{0.00000\pm0.00087}}     
\newcommand{\hatcurXdistredeccenxxxxxA}{\ensuremath{441\pm26}}         
\newcommand{\hatcurXEBVeccenxxxxxA}{\ensuremath{0.00000\pm0.00028}}    
\newcommand{\hatcurXmvisoredeccenxxxxxA}{\ensuremath{11.849\pm0.026}}  
\newcommand{\hatcurXmiisoredeccenxxxxxA}{\ensuremath{11.245\pm0.014}}  
\newcommand{\hatcurXmjisoredeccenxxxxxA}{\ensuremath{10.841\pm0.011}}  
\newcommand{\hatcurXmhisoredeccenxxxxxA}{\ensuremath{10.543\pm0.016}}  
\newcommand{\hatcurXmkisoredeccenxxxxxA}{\ensuremath{10.492\pm0.017}}  
\newcommand{\hatcurXviisoredeccenxxxxxA}{\ensuremath{0.604\pm0.015}}   
\newcommand{\hatcurXvkisoredeccenxxxxxA}{\ensuremath{1.358\pm0.037}}   
\newcommand{\hatcurXjhisoredeccenxxxxxA}{\ensuremath{0.2990\pm0.0093}} 
\newcommand{\hatcurXjkisoredeccenxxxxxA}{\ensuremath{0.350\pm0.011}}   
\newcommand{\hatcurCCpmraeccenxxxxxA}{\ensuremath{10.20\pm0.80}}       
\newcommand{\hatcurCCpmdececcenxxxxxA}{\ensuremath{-4.6\pm1.6}}        
\newcommand{\hatcurCCpmeccenxxxxxA}{\ensuremath{11.2\pm1.8}}           

\newcommand{\hatcurhtrxxxxxB}{HTR210-001}                              
\newcommand{\hatcurfieldxxxxxB}{210}                                   
\newcommand{\hatcurCCraxxxxxB}{\ensuremath{01^{\mathrm h}24^{\mathrm m}15.66{\mathrm s}}}                            
\newcommand{\hatcurCCdecxxxxxB}{\ensuremath{+32{\arcdeg}48{\arcmin}38.8{\arcsec}}}                           
\newcommand{\hatcurCCmagxxxxxB}{13.440}                                
\newcommand{\hatcurCCtwomassxxxxxB}{2MASS~01241564+3248387}            
\newcommand{\hatcurCCgscxxxxxB}{GSC~2296-00637}                        
\newcommand{\hatcurCCtassmvxxxxxB}{\ensuremath{13.440\pm0.040}}        
\newcommand{\hatcurCCtassmvshortxxxxxB}{\ensuremath{13.4}}             
\newcommand{\hatcurCCtassmBxxxxxB}{\ensuremath{14.261\pm0.070}}        
\newcommand{\hatcurCCtassmBshortxxxxxB}{\ensuremath{14.3}}             
\newcommand{\hatcurCCtassmIxxxxxB}{\ensuremath{12.67\pm0.12}}          
\newcommand{\hatcurCCtassmIshortxxxxxB}{\ensuremath{12.7}}             
\newcommand{\hatcurCCtassmgxxxxxB}{\ensuremath{13.839\pm0.050}}        
\newcommand{\hatcurCCtassmgshortxxxxxB}{\ensuremath{13.8}}             
\newcommand{\hatcurCCtassmrxxxxxB}{\ensuremath{13.194\pm0.030}}        
\newcommand{\hatcurCCtassmrshortxxxxxB}{\ensuremath{13.2}}             
\newcommand{\hatcurCCtassmixxxxxB}{\ensuremath{12.998\pm0.040}}        
\newcommand{\hatcurCCtassmishortxxxxxB}{\ensuremath{13.0}}             
\newcommand{\hatcurCCtwomassJmagxxxxxB}{\ensuremath{12.039\pm0.022}}   
\newcommand{\hatcurCCtwomassHmagxxxxxB}{\ensuremath{11.645\pm0.023}}   
\newcommand{\hatcurCCtwomassKmagxxxxxB}{\ensuremath{11.614\pm0.020}}   
\newcommand{\hatcurCCcitJmagxxxxxB}{\ensuremath{12.053\pm0.022}}       
\newcommand{\hatcurCCcitHmagxxxxxB}{\ensuremath{11.641\pm0.023}}       
\newcommand{\hatcurCCcitKmagxxxxxB}{\ensuremath{11.638\pm0.020}}       
\newcommand{\hatcurCCbbJmagxxxxxB}{\ensuremath{12.107\pm0.024}}        
\newcommand{\hatcurCCbbHmagxxxxxB}{\ensuremath{11.661\pm0.024}}        
\newcommand{\hatcurCCbbKmagxxxxxB}{\ensuremath{11.658\pm0.020}}        
\newcommand{\hatcurCCesoJmagxxxxxB}{\ensuremath{12.110\pm0.026}}       
\newcommand{\hatcurCCesoHmagxxxxxB}{\ensuremath{11.654\pm0.027}}       
\newcommand{\hatcurCCesoKmagxxxxxB}{\ensuremath{11.657\pm0.021}}       
\newcommand{\hatcurCCesoJHmagxxxxxB}{\ensuremath{0.455\pm0.035}}       
\newcommand{\hatcurCCesoJKmagxxxxxB}{\ensuremath{0.453\pm0.033}}       
\newcommand{\hatcurCCesoHKmagxxxxxB}{\ensuremath{-0.002\pm0.034}}      
\newcommand{\hatcurLCdipxxxxxB}{\ensuremath{18.5}}                     
\newcommand{\hatcurLCrprstarxxxxxB}{\ensuremath{0.1278\pm0.0020}}      
\newcommand{\hatcurLCbsqxxxxxB}{\ensuremath{0.077_{-0.052}^{+0.055}}}  
\newcommand{\hatcurLCimpxxxxxB}{\ensuremath{0.277_{-0.119}^{+0.085}}}  
\newcommand{\hatcurLCzetaxxxxxB}{\ensuremath{16.21\pm0.12}}            
\newcommand{\hatcurLCdurxxxxxB}{\ensuremath{0.1403\pm0.0016}}          
\newcommand{\hatcurLCdurshortxxxxxB}{\ensuremath{0.1403}}              
\newcommand{\hatcurLCdurhrxxxxxB}{\ensuremath{3.367\pm0.040}}          
\newcommand{\hatcurLCdurhrshortxxxxxB}{\ensuremath{3.367}}             
\newcommand{\hatcurLCqxxxxxB}{\ensuremath{0.03330\pm0.00039}}          
\newcommand{\hatcurLCqshortxxxxxB}{\ensuremath{0.033}}                 
\newcommand{\hatcurLCingdurxxxxxB}{\ensuremath{0.0170\pm0.0012}}       
\newcommand{\hatcurLCPxxxxxB}{\ensuremath{4.2180278\pm0.0000059}}      
\newcommand{\hatcurLCPprecxxxxxB}{\ensuremath{4.2180278}}              
\newcommand{\hatcurLCPshortxxxxxB}{\ensuremath{4.2180}}                
\newcommand{\hatcurLCTxxxxxB}{\ensuremath{2456194.12204\pm0.00040}}    
\newcommand{\hatcurLCTAxxxxxB}{\ensuremath{2454350.8440\pm0.0025}}     
\newcommand{\hatcurLCTBxxxxxB}{\ensuremath{2456244.73840\pm0.00042}}   
\newcommand{\hatcurLChatnetmAxxxxxB}{\ensuremath{12.92064\pm0.00031}}  
\newcommand{\hatcurLCiblendAxxxxxB}{\ensuremath{0.71\pm0.10}}          
\newcommand{\hatcurLChatnetmBxxxxxB}{\ensuremath{13.24968\pm0.00020}}  
\newcommand{\hatcurLCiblendBxxxxxB}{\ensuremath{0.808\pm0.066}}        
\newcommand{\hatcurLChatnetmCxxxxxB}{\ensuremath{13.24993\pm0.00021}}  
\newcommand{\hatcurLCiblendCxxxxxB}{\ensuremath{0.750\pm0.057}}        
\newcommand{\hatcurLChatnetmDxxxxxB}{\ensuremath{12.24114\pm0.00071}}  
\newcommand{\hatcurLCiblendDxxxxxB}{\ensuremath{0.0406\pm0.0042}}      
\newcommand{\hatcurSMEiteffxxxxxB}{\ensuremath{5449\pm50}}             
\newcommand{\hatcurSMEizfehxxxxxB}{\ensuremath{0.270\pm0.080}}         
\newcommand{\hatcurSMEizfehshortxxxxxB}{\ensuremath{0.27}}             
\newcommand{\hatcurSMEiloggxxxxxB}{\ensuremath{4.41\pm0.10}}           
\newcommand{\hatcurSMEivsinxxxxxB}{\ensuremath{1.70\pm0.50}}           
\newcommand{\hatcurSMEivmacxxxxxB}{\ensuremath{0.0}}                   
\newcommand{\hatcurSMEivmicxxxxxB}{\ensuremath{0.0}}                   
\newcommand{\hatcurSMEiiteffxxxxxB}{\ensuremath{6290\pm100}}           
\newcommand{\hatcurSMEiizfehxxxxxB}{\ensuremath{-0.08\pm0.10}}         
\newcommand{\hatcurSMEiizfehshortxxxxxB}{\ensuremath{-0.08}}           
\newcommand{\hatcurSMEiiloggxxxxxB}{\ensuremath{3.81\pm0.10}}          
\newcommand{\hatcurSMEiivsinxxxxxB}{\ensuremath{12.00\pm0.50}}         
\newcommand{\hatcurSMEiivmacxxxxxB}{\ensuremath{4.82}}                 
\newcommand{\hatcurSMEiivmicxxxxxB}{\ensuremath{0.85}}                 
\newcommand{\hatcurTRESteffxxxxxB}{\ensuremath{5250\pm120}}            
\newcommand{\hatcurTRESzfehxxxxxB}{\ensuremath{0.00\pm0.10}}           
\newcommand{\hatcurTRESloggxxxxxB}{\ensuremath{4.00\pm0.25}}           
\newcommand{\hatcurTRESvsinixxxxxB}{\ensuremath{4.0\pm2.0}}            
\newcommand{\hatcurTRESgammaxxxxxB}{\ensuremath{-26.99\pm0.10}}        
\newcommand{\hatcurTRESnumspecxxxxxB}{\ensuremath{1}}                  
\newcommand{\hatcurTRESspanxxxxxB}{\ensuremath{0}}                     
\newcommand{\hatcurTRESrvrmsxxxxxB}{\ensuremath{0.00}}                 
\newcommand{\hatcurFIESteffxxxxxB}{\ensuremath{5580\pm55}}             
\newcommand{\hatcurFIESzfehxxxxxB}{\ensuremath{0.320\pm0.080}}         
\newcommand{\hatcurFIESloggxxxxxB}{\ensuremath{4.44\pm0.10}}           
\newcommand{\hatcurFIESvsinixxxxxB}{\ensuremath{2.0\pm1.0}}            
\newcommand{\hatcurFIESgammaxxxxxB}{\ensuremath{-27.20\pm0.10}}        
\newcommand{\hatcurFIESnumspecxxxxxB}{\ensuremath{NULL}}               
\newcommand{\hatcurFIESspanxxxxxB}{\ensuremath{NULL}}                  
\newcommand{\hatcurFIESrvrmsxxxxxB}{\ensuremath{0.00}}                 
\newcommand{\hatcurAPOteffxxxxxB}{\ensuremath{5571\pm51}}            
\newcommand{\hatcurAPOzfehxxxxxB}{\ensuremath{0.420\pm0.080}}        
\newcommand{\hatcurAPOloggxxxxxB}{\ensuremath{4.70\pm0.10}}          
\newcommand{\hatcurAPOvsinixxxxxB}{\ensuremath{3.0\pm1.0}}           
\newcommand{\hatcurAPOgammaxxxxxB}{\ensuremath{-27.9\pm1.0}}         
\newcommand{\hatcurAPOnumspecxxxxxB}{\ensuremath{NULL}}              
\newcommand{\hatcurAPOspanxxxxxB}{\ensuremath{NULL}}                 
\newcommand{\hatcurAPOrvrmsxxxxxB}{\ensuremath{0.00}}                
\newcommand{\hatcurLBizxxxxxB}{\ensuremath{0.2579}}                    
\newcommand{\hatcurLBiizxxxxxB}{\ensuremath{0.3100}}                   
\newcommand{\hatcurLBiixxxxxB}{\ensuremath{0.3348}}                    
\newcommand{\hatcurLBiiixxxxxB}{\ensuremath{0.2989}}                   
\newcommand{\hatcurLBiIxxxxxB}{\ensuremath{0.3095}}                    
\newcommand{\hatcurLBiiIxxxxxB}{\ensuremath{0.3029}}                   
\newcommand{\hatcurLBigxxxxxB}{\ensuremath{0.6692}}                    
\newcommand{\hatcurLBiigxxxxxB}{\ensuremath{0.1480}}                   
\newcommand{\hatcurLBirxxxxxB}{\ensuremath{0.4443}}                    
\newcommand{\hatcurLBiirxxxxxB}{\ensuremath{0.2749}}                   
\newcommand{\hatcurLBiRxxxxxB}{\ensuremath{0.4140}}                    
\newcommand{\hatcurLBiiRxxxxxB}{\ensuremath{0.2824}}                   
\newcommand{\hatcurISOmxxxxxB}{\ensuremath{0.976\pm0.028}}             
\newcommand{\hatcurISOmshortxxxxxB}{\ensuremath{0.98}}                 
\newcommand{\hatcurISOmlongxxxxxB}{\ensuremath{0.976\pm0.028}}         
\newcommand{\hatcurISOrxxxxxB}{\ensuremath{1.041_{-0.029}^{+0.038}}}   
\newcommand{\hatcurISOrshortxxxxxB}{\ensuremath{1.04}}                 
\newcommand{\hatcurISOrlongxxxxxB}{\ensuremath{1.041_{-0.029}^{+0.038}}} 
\newcommand{\hatcurISOrhoxxxxxB}{\ensuremath{1.223_{-0.135}^{+0.100}}} 
\newcommand{\hatcurISOrholongxxxxxB}{\ensuremath{1.223_{-0.135}^{+0.100}}} 
\newcommand{\hatcurISOloggxxxxxB}{\ensuremath{4.392\pm0.027}}          
\newcommand{\hatcurISOlumxxxxxB}{\ensuremath{0.859\pm0.070}}           
\newcommand{\hatcurISOlumshortxxxxxB}{\ensuremath{0.86}}               
\newcommand{\hatcurISOmvxxxxxB}{\ensuremath{5.055\pm0.095}}            
\newcommand{\hatcurISOvixxxxxB}{\ensuremath{0.804\pm0.014}}            
\newcommand{\hatcurISOagexxxxxB}{\ensuremath{8.2\pm1.7}}               
\newcommand{\hatcurISOsigmaxxxxxB}{\ensuremath{0.000400\pm0.000042}}   
\newcommand{\hatcurISOMJxxxxxB}{\ensuremath{3.734\pm0.078}}            
\newcommand{\hatcurISOMHxxxxxB}{\ensuremath{3.335\pm0.072}}            
\newcommand{\hatcurISOMKxxxxxB}{\ensuremath{3.268\pm0.071}}            
\newcommand{\hatcurISOJKxxxxxB}{\ensuremath{0.470\pm0.010}}            
\newcommand{\hatcurISOspecxxxxxB}{G}                                   
\newcommand{\hatcurRVKxxxxxB}{\ensuremath{39.5\pm2.2}}                 
\newcommand{\hatcurRVrkxxxxxB}{\ensuremath{0\pm0}}                     
\newcommand{\hatcurRVrhxxxxxB}{\ensuremath{0\pm0}}                     
\newcommand{\hatcurRVkxxxxxB}{\ensuremath{0\pm0}}                      
\newcommand{\hatcurRVhxxxxxB}{\ensuremath{0\pm0}}                      
\newcommand{\hatcurRVtronexxxxxB}{\ensuremath{0\pm0}}                  
\newcommand{\hatcurRVtrtwoxxxxxB}{\ensuremath{0\pm0}}                  
\newcommand{\hatcurRVgammaAxxxxxB}{\ensuremath{-5.6\pm2.2}}            
\newcommand{\hatcurRVjitterAxxxxxB}{\ensuremath{4.3\pm1.2}}            
\newcommand{\hatcurRVfitrmsAxxxxxB}{\ensuremath{5.4}}                  
\newcommand{\hatcurRVgammaBxxxxxB}{\ensuremath{-1.4\pm2.9}}            
\newcommand{\hatcurRVjitterBxxxxxB}{\ensuremath{0.0\pm1.4}}            
\newcommand{\hatcurRVfitrmsBxxxxxB}{\ensuremath{9.2}}                  
\newcommand{\hatcurRVgammaCxxxxxB}{\ensuremath{13.25020\pm0.00023}}    
\newcommand{\hatcurRVjitterCxxxxxB}{\ensuremath{0.666\pm0.073}}        
\newcommand{\hatcurRVfitrmsCxxxxxB}{\ensuremath{.1fym}}                %
\newcommand{\hatcurRVeccenxxxxxB}{\ensuremath{0\pm0}}                  
\newcommand{\hatcurRVeccentwosiglimxxxxxB}{\ensuremath{<0.000}}        
\newcommand{\hatcurRVomegaxxxxxB}{\ensuremath{0\pm0}}                  
\newcommand{\hatcurPPixxxxxB}{\ensuremath{88.48\pm0.57}}               
\newcommand{\hatcurPPgxxxxxB}{\ensuremath{4.58\pm0.46}}                
\newcommand{\hatcurPPloggxxxxxB}{\ensuremath{2.661_{-0.051}^{+0.037}}} 
\newcommand{\hatcurPParxxxxxB}{\ensuremath{10.48_{-0.40}^{+0.28}}}     
\newcommand{\hatcurPParelxxxxxB}{\ensuremath{0.05069\pm0.00049}}       
\newcommand{\hatcurPPrhoxxxxxB}{\ensuremath{0.178\pm0.024}}            
\newcommand{\hatcurPPmxxxxxB}{\ensuremath{0.309\pm0.018}}              
\newcommand{\hatcurPPmshortxxxxxB}{\ensuremath{0.31}}                  
\newcommand{\hatcurPPmlongxxxxxB}{\ensuremath{0.309\pm0.018}}          
\newcommand{\hatcurPPmexxxxxB}{\ensuremath{98.3\pm5.7}}                
\newcommand{\hatcurPPmeshortxxxxxB}{\ensuremath{98.3}}                 
\newcommand{\hatcurPPmelongxxxxxB}{\ensuremath{98.3\pm5.7}}            
\newcommand{\hatcurPPrxxxxxB}{\ensuremath{1.293\pm0.054}}              
\newcommand{\hatcurPPrshortxxxxxB}{\ensuremath{1.29}}                  
\newcommand{\hatcurPPrlongxxxxxB}{\ensuremath{1.293\pm0.054}}          
\newcommand{\hatcurPPrexxxxxB}{\ensuremath{14.50\pm0.61}}              
\newcommand{\hatcurPPreshortxxxxxB}{\ensuremath{14.5}}                 
\newcommand{\hatcurPPrelongxxxxxB}{\ensuremath{14.50\pm0.61}}          
\newcommand{\hatcurPPmrcorrxxxxxB}{\ensuremath{-0.03}}                 
\newcommand{\hatcurPPteffxxxxxB}{\ensuremath{1192\pm21}}               
\newcommand{\hatcurPPthetaxxxxxB}{\ensuremath{0.0247\pm0.0018}}        
\newcommand{\hatcurPPfluxperixxxxxB}{\ensuremath{4.56\pm0.33}}         
\newcommand{\hatcurPPfluxperidimxxxxxB}{\ensuremath{8}}                
\newcommand{\hatcurPPfluxapxxxxxB}{\ensuremath{4.56\pm0.33}}           
\newcommand{\hatcurPPfluxapdimxxxxxB}{\ensuremath{8}}                  
\newcommand{\hatcurPPfluxavgxxxxxB}{\ensuremath{4.56\pm0.33}}          
\newcommand{\hatcurPPfluxavgdimxxxxxB}{\ensuremath{8}}                 
\newcommand{\hatcurPPfluxavglogxxxxxB}{\ensuremath{8.659\pm0.031}}     
\newcommand{\hatcurXsecphasexxxxxB}{\ensuremath{0\pm0}}                
\newcommand{\hatcurXsecondaryxxxxxB}{\ensuremath{2456196.23106\pm0.00040}} 
\newcommand{\hatcurXsecdurxxxxxB}{\ensuremath{0.1403\pm0.0016}}        
\newcommand{\hatcurXsecingdurxxxxxB}{\ensuremath{0.0170\pm0.0012}}     
\newcommand{\hatcurPPphiconjxxxxxB}{\ensuremath{0\pm0}}                
\newcommand{\hatcurPPperixxxxxB}{\ensuremath{2456193.06754\pm0.00040}} 
\newcommand{\hatcurPPaequivxxxxxB}{\ensuremath{0.0547\pm0.0019}}       
\newcommand{\hatcurPPtcircxxxxxB}{\ensuremath{131\pm27}}               
\newcommand{\hatcurPPtinfallxxxxxB}{\ensuremath{15900\pm2400}}         
\newcommand{\hatcurXdistxxxxxB}{\ensuremath{476\pm16}}                 
\newcommand{\hatcurXAvxxxxxB}{\ensuremath{0.012_{-0.012}^{+0.071}}}    
\newcommand{\hatcurXdistredxxxxxB}{\ensuremath{470\pm16}}              
\newcommand{\hatcurXEBVxxxxxB}{\ensuremath{0.0040_{-0.0040}^{+0.0230}}} 
\newcommand{\hatcurXmvisoredxxxxxB}{\ensuremath{13.452\pm0.032}}       
\newcommand{\hatcurXmiisoredxxxxxB}{\ensuremath{12.630\pm0.021}}       
\newcommand{\hatcurXmjisoredxxxxxB}{\ensuremath{12.104\pm0.014}}       
\newcommand{\hatcurXmhisoredxxxxxB}{\ensuremath{11.701\pm0.014}}       
\newcommand{\hatcurXmkisoredxxxxxB}{\ensuremath{11.631\pm0.015}}       
\newcommand{\hatcurXviisoredxxxxxB}{\ensuremath{0.819_{-0.014}^{+0.021}}} 
\newcommand{\hatcurXvkisoredxxxxxB}{\ensuremath{1.821\pm0.036}}        
\newcommand{\hatcurXjhisoredxxxxxB}{\ensuremath{0.403\pm0.010}}        
\newcommand{\hatcurXjkisoredxxxxxB}{\ensuremath{0.473\pm0.011}}        
\newcommand{\hatcurCCpmraxxxxxB}{\ensuremath{-9.8\pm1.4}}              
\newcommand{\hatcurCCpmdecxxxxxB}{\ensuremath{-16.8\pm2.1}}            
\newcommand{\hatcurCCpmxxxxxB}{\ensuremath{19.4\pm2.5}}                
\newcommand{\hatcurhtreccenxxxxxB}{HTR210-001}                         
\newcommand{\hatcurfieldeccenxxxxxB}{210}                              
\newcommand{\hatcurCCraeccenxxxxxB}{\ensuremath{01^{\mathrm h}24^{\mathrm m}15.66{\mathrm s}}}                       
\newcommand{\hatcurCCdececcenxxxxxB}{\ensuremath{+32{\arcdeg}48{\arcmin}38.8{\arcsec}}}                      
\newcommand{\hatcurCCmageccenxxxxxB}{13.440}                           
\newcommand{\hatcurCCtwomasseccenxxxxxB}{2MASS~01241564+3248387}       
\newcommand{\hatcurCCgsceccenxxxxxB}{GSC~2296-00637}                   
\newcommand{\hatcurCCtassmveccenxxxxxB}{\ensuremath{13.440\pm0.040}}   
\newcommand{\hatcurCCtassmvshorteccenxxxxxB}{\ensuremath{13.4}}        
\newcommand{\hatcurCCtassmBeccenxxxxxB}{\ensuremath{14.261\pm0.070}}   
\newcommand{\hatcurCCtassmBshorteccenxxxxxB}{\ensuremath{14.3}}        
\newcommand{\hatcurCCtassmIeccenxxxxxB}{\ensuremath{12.67\pm0.12}}     
\newcommand{\hatcurCCtassmIshorteccenxxxxxB}{\ensuremath{12.7}}        
\newcommand{\hatcurCCtassmgeccenxxxxxB}{\ensuremath{14.3980\pm0.0040}} 
\newcommand{\hatcurCCtassmgshorteccenxxxxxB}{\ensuremath{14.4}}        
\newcommand{\hatcurCCtassmreccenxxxxxB}{\ensuremath{14.792\pm0.010}}   
\newcommand{\hatcurCCtassmrshorteccenxxxxxB}{\ensuremath{14.8}}        
\newcommand{\hatcurCCtassmieccenxxxxxB}{\ensuremath{13.1930\pm0.0020}} 
\newcommand{\hatcurCCtassmishorteccenxxxxxB}{\ensuremath{13.2}}        
\newcommand{\hatcurCCtwomassJmageccenxxxxxB}{\ensuremath{12.039\pm0.022}} 
\newcommand{\hatcurCCtwomassHmageccenxxxxxB}{\ensuremath{11.645\pm0.023}} 
\newcommand{\hatcurCCtwomassKmageccenxxxxxB}{\ensuremath{11.614\pm0.020}} 
\newcommand{\hatcurCCcitJmageccenxxxxxB}{\ensuremath{12.053\pm0.022}}  
\newcommand{\hatcurCCcitHmageccenxxxxxB}{\ensuremath{11.641\pm0.023}}  
\newcommand{\hatcurCCcitKmageccenxxxxxB}{\ensuremath{11.638\pm0.020}}  
\newcommand{\hatcurCCbbJmageccenxxxxxB}{\ensuremath{12.107\pm0.024}}   
\newcommand{\hatcurCCbbHmageccenxxxxxB}{\ensuremath{11.661\pm0.024}}   
\newcommand{\hatcurCCbbKmageccenxxxxxB}{\ensuremath{11.658\pm0.020}}   
\newcommand{\hatcurCCesoJmageccenxxxxxB}{\ensuremath{12.110\pm0.026}}  
\newcommand{\hatcurCCesoHmageccenxxxxxB}{\ensuremath{11.654\pm0.027}}  
\newcommand{\hatcurCCesoKmageccenxxxxxB}{\ensuremath{11.657\pm0.021}}  
\newcommand{\hatcurCCesoJHmageccenxxxxxB}{\ensuremath{0.455\pm0.035}}  
\newcommand{\hatcurCCesoJKmageccenxxxxxB}{\ensuremath{0.453\pm0.033}}  
\newcommand{\hatcurCCesoHKmageccenxxxxxB}{\ensuremath{-0.002\pm0.034}} 
\newcommand{\hatcurLCdipeccenxxxxxB}{\ensuremath{18.5}}                
\newcommand{\hatcurLCrprstareccenxxxxxB}{\ensuremath{0.1278\pm0.0015}} 
\newcommand{\hatcurLCbsqeccenxxxxxB}{\ensuremath{0.070_{-0.045}^{+0.066}}} 
\newcommand{\hatcurLCimpeccenxxxxxB}{\ensuremath{0.26_{-0.11}^{+0.10}}} 
\newcommand{\hatcurLCzetaeccenxxxxxB}{\ensuremath{16.215\pm0.097}}     
\newcommand{\hatcurLCdureccenxxxxxB}{\ensuremath{0.1403\pm0.0013}}     
\newcommand{\hatcurLCdurshorteccenxxxxxB}{\ensuremath{0.1403}}         
\newcommand{\hatcurLCdurhreccenxxxxxB}{\ensuremath{3.367\pm0.032}}     
\newcommand{\hatcurLCdurhrshorteccenxxxxxB}{\ensuremath{3.367}}        
\newcommand{\hatcurLCqeccenxxxxxB}{\ensuremath{0.03330\pm0.00032}}     
\newcommand{\hatcurLCqshorteccenxxxxxB}{\ensuremath{0.033}}            
\newcommand{\hatcurLCingdureccenxxxxxB}{\ensuremath{0.0170\pm0.0012}}  
\newcommand{\hatcurLCPeccenxxxxxB}{\ensuremath{4.2180264\pm0.0000062}} 
\newcommand{\hatcurLCPprececcenxxxxxB}{\ensuremath{4.2180264}}         
\newcommand{\hatcurLCPshorteccenxxxxxB}{\ensuremath{4.2180}}           
\newcommand{\hatcurLCTeccenxxxxxB}{\ensuremath{2456206.77607\pm0.00036}} 
\newcommand{\hatcurLCTAeccenxxxxxB}{\ensuremath{2454350.8445\pm0.0027}} 
\newcommand{\hatcurLCTBeccenxxxxxB}{\ensuremath{2456244.73829\pm0.00039}} 
\newcommand{\hatcurLChatnetmAeccenxxxxxB}{\ensuremath{13.25026\pm0.00024}} 
\newcommand{\hatcurLCiblendAeccenxxxxxB}{\ensuremath{0.676\pm0.067}}   
\newcommand{\hatcurLChatnetmBeccenxxxxxB}{\ensuremath{12.92077\pm0.00030}} 
\newcommand{\hatcurLCiblendBeccenxxxxxB}{\ensuremath{0.71\pm0.12}}     
\newcommand{\hatcurLChatnetmCeccenxxxxxB}{\ensuremath{13.24964\pm0.00020}} 
\newcommand{\hatcurLCiblendCeccenxxxxxB}{\ensuremath{0.817\pm0.063}}   
\newcommand{\hatcurLChatnetmDeccenxxxxxB}{\ensuremath{13.24996\pm0.00019}} 
\newcommand{\hatcurLCiblendDeccenxxxxxB}{\ensuremath{0.747\pm0.059}}   
\newcommand{\hatcurSMEiteffeccenxxxxxB}{\ensuremath{5449\pm50}}        
\newcommand{\hatcurSMEizfeheccenxxxxxB}{\ensuremath{0.270\pm0.080}}    
\newcommand{\hatcurSMEizfehshorteccenxxxxxB}{\ensuremath{0.27}}        
\newcommand{\hatcurSMEiloggeccenxxxxxB}{\ensuremath{4.41\pm0.10}}      
\newcommand{\hatcurSMEivsineccenxxxxxB}{\ensuremath{1.70\pm0.50}}      
\newcommand{\hatcurSMEivmaceccenxxxxxB}{\ensuremath{0.0}}              
\newcommand{\hatcurSMEivmiceccenxxxxxB}{\ensuremath{0.0}}              
\newcommand{\hatcurSMEiiteffeccenxxxxxB}{\ensuremath{6290\pm100}}      
\newcommand{\hatcurSMEiizfeheccenxxxxxB}{\ensuremath{-0.08\pm0.10}}    
\newcommand{\hatcurSMEiizfehshorteccenxxxxxB}{\ensuremath{-0.08}}      
\newcommand{\hatcurSMEiiloggeccenxxxxxB}{\ensuremath{3.81\pm0.10}}     
\newcommand{\hatcurSMEiivsineccenxxxxxB}{\ensuremath{12.00\pm0.50}}    
\newcommand{\hatcurSMEiivmaceccenxxxxxB}{\ensuremath{4.82}}            
\newcommand{\hatcurSMEiivmiceccenxxxxxB}{\ensuremath{0.85}}            
\newcommand{\hatcurTRESteffeccenxxxxxB}{\ensuremath{5250\pm120}}       
\newcommand{\hatcurTRESzfeheccenxxxxxB}{\ensuremath{0.00\pm0.10}}      
\newcommand{\hatcurTRESloggeccenxxxxxB}{\ensuremath{4.00\pm0.25}}      
\newcommand{\hatcurTRESvsinieccenxxxxxB}{\ensuremath{4.0\pm2.0}}       
\newcommand{\hatcurTRESgammaeccenxxxxxB}{\ensuremath{-26.99\pm0.10}}   
\newcommand{\hatcurTRESnumspececcenxxxxxB}{\ensuremath{1}}             
\newcommand{\hatcurTRESspaneccenxxxxxB}{\ensuremath{0}}                
\newcommand{\hatcurTRESrvrmseccenxxxxxB}{\ensuremath{0.00}}            
\newcommand{\hatcurFIESteffeccenxxxxxB}{\ensuremath{5580\pm55}}        
\newcommand{\hatcurFIESzfeheccenxxxxxB}{\ensuremath{0.320\pm0.080}}    
\newcommand{\hatcurFIESloggeccenxxxxxB}{\ensuremath{4.44\pm0.10}}      
\newcommand{\hatcurFIESvsinieccenxxxxxB}{\ensuremath{2.0\pm1.0}}       
\newcommand{\hatcurFIESgammaeccenxxxxxB}{\ensuremath{-27.20\pm0.10}}   
\newcommand{\hatcurFIESnumspececcenxxxxxB}{\ensuremath{NULL}}          
\newcommand{\hatcurFIESspaneccenxxxxxB}{\ensuremath{NULL}}             
\newcommand{\hatcurFIESrvrmseccenxxxxxB}{\ensuremath{0.00}}            
\newcommand{\hatcurARCESteffeccenxxxxxB}{\ensuremath{5571\pm51}}       
\newcommand{\hatcurARCESzfeheccenxxxxxB}{\ensuremath{0.420\pm0.080}}   
\newcommand{\hatcurARCESloggeccenxxxxxB}{\ensuremath{4.70\pm0.10}}     
\newcommand{\hatcurARCESvsinieccenxxxxxB}{\ensuremath{3.0\pm1.0}}      
\newcommand{\hatcurARCESgammaeccenxxxxxB}{\ensuremath{-27.9\pm1.0}}    
\newcommand{\hatcurARCESnumspececcenxxxxxB}{\ensuremath{NULL}}         
\newcommand{\hatcurARCESspaneccenxxxxxB}{\ensuremath{NULL}}            
\newcommand{\hatcurARCESrvrmseccenxxxxxB}{\ensuremath{0.00}}           
\newcommand{\hatcurLBizeccenxxxxxB}{\ensuremath{0.2579}}               
\newcommand{\hatcurLBiizeccenxxxxxB}{\ensuremath{0.3100}}              
\newcommand{\hatcurLBiieccenxxxxxB}{\ensuremath{0.3348}}               
\newcommand{\hatcurLBiiieccenxxxxxB}{\ensuremath{0.2989}}              
\newcommand{\hatcurLBiIeccenxxxxxB}{\ensuremath{0.3095}}               
\newcommand{\hatcurLBiiIeccenxxxxxB}{\ensuremath{0.3029}}              
\newcommand{\hatcurLBigeccenxxxxxB}{\ensuremath{0.6692}}               
\newcommand{\hatcurLBiigeccenxxxxxB}{\ensuremath{0.1480}}              
\newcommand{\hatcurLBireccenxxxxxB}{\ensuremath{0.4443}}               
\newcommand{\hatcurLBiireccenxxxxxB}{\ensuremath{0.2749}}              
\newcommand{\hatcurLBiReccenxxxxxB}{\ensuremath{0.4140}}               
\newcommand{\hatcurLBiiReccenxxxxxB}{\ensuremath{0.2824}}              
\newcommand{\hatcurISOmeccenxxxxxB}{\ensuremath{0.982\pm0.028}}        
\newcommand{\hatcurISOmshorteccenxxxxxB}{\ensuremath{0.98}}            
\newcommand{\hatcurISOmlongeccenxxxxxB}{\ensuremath{0.982\pm0.028}}    
\newcommand{\hatcurISOreccenxxxxxB}{\ensuremath{1.004\pm0.049}}        
\newcommand{\hatcurISOrshorteccenxxxxxB}{\ensuremath{1.00}}            
\newcommand{\hatcurISOrlongeccenxxxxxB}{\ensuremath{1.004\pm0.049}}    
\newcommand{\hatcurISOrhoeccenxxxxxB}{\ensuremath{1.36_{-0.18}^{+0.25}}} 
\newcommand{\hatcurISOrholongeccenxxxxxB}{\ensuremath{1.36_{-0.18}^{+0.25}}} 
\newcommand{\hatcurISOloggeccenxxxxxB}{\ensuremath{4.425\pm0.045}}     
\newcommand{\hatcurISOlumeccenxxxxxB}{\ensuremath{0.795\pm0.088}}      
\newcommand{\hatcurISOlumshorteccenxxxxxB}{\ensuremath{0.79}}          
\newcommand{\hatcurISOmveccenxxxxxB}{\ensuremath{5.14\pm0.13}}         
\newcommand{\hatcurISOvieccenxxxxxB}{\ensuremath{0.804\pm0.014}}       
\newcommand{\hatcurISOageeccenxxxxxB}{\ensuremath{6.5\pm2.4}}          
\newcommand{\hatcurISOsigmaeccenxxxxxB}{\ensuremath{0.000400\pm0.000064}} 
\newcommand{\hatcurISOMJeccenxxxxxB}{\ensuremath{3.81\pm0.11}}         
\newcommand{\hatcurISOMHeccenxxxxxB}{\ensuremath{3.41\pm0.11}}         
\newcommand{\hatcurISOMKeccenxxxxxB}{\ensuremath{3.35\pm0.11}}         
\newcommand{\hatcurISOJKeccenxxxxxB}{\ensuremath{0.470\pm0.010}}       
\newcommand{\hatcurISOspececcenxxxxxB}{G}                              
\newcommand{\hatcurRVKeccenxxxxxB}{\ensuremath{39.9\pm1.9}}            
\newcommand{\hatcurRVrkeccenxxxxxB}{\ensuremath{-0.107_{-0.074}^{+0.098}}} 
\newcommand{\hatcurRVrheccenxxxxxB}{\ensuremath{-0.17\pm0.12}}         
\newcommand{\hatcurRVkeccenxxxxxB}{\ensuremath{-0.022\pm0.022}}        
\newcommand{\hatcurRVheccenxxxxxB}{\ensuremath{-0.035_{-0.050}^{+0.032}}} 
\newcommand{\hatcurRVtroneeccenxxxxxB}{\ensuremath{0\pm0}}             
\newcommand{\hatcurRVtrtwoeccenxxxxxB}{\ensuremath{0\pm0}}             
\newcommand{\hatcurRVgammaAeccenxxxxxB}{\ensuremath{-5.0\pm1.9}}       
\newcommand{\hatcurRVjitterAeccenxxxxxB}{\ensuremath{3.8\pm1.2}}       
\newcommand{\hatcurRVfitrmsAeccenxxxxxB}{\ensuremath{.1fym}}           %
\newcommand{\hatcurRVgammaBeccenxxxxxB}{\ensuremath{-2.1\pm2.7}}       
\newcommand{\hatcurRVjitterBeccenxxxxxB}{\ensuremath{0.00\pm0.68}}     
\newcommand{\hatcurRVfitrmsBeccenxxxxxB}{\ensuremath{.1fym}}           %
\newcommand{\hatcurRVecceneccenxxxxxB}{\ensuremath{0.050\pm0.036}}     
\newcommand{\hatcurRVeccentwosiglimeccenxxxxxB}{\ensuremath{<0.123}}   
\newcommand{\hatcurRVomegaeccenxxxxxB}{\ensuremath{240\pm45}}          
\newcommand{\hatcurPPieccenxxxxxB}{\ensuremath{88.69\pm0.56}}          
\newcommand{\hatcurPPgeccenxxxxxB}{\ensuremath{4.97_{-0.50}^{+0.66}}}  
\newcommand{\hatcurPPloggeccenxxxxxB}{\ensuremath{2.696\pm0.051}}      
\newcommand{\hatcurPPareccenxxxxxB}{\ensuremath{10.86\pm0.55}}         
\newcommand{\hatcurPPareleccenxxxxxB}{\ensuremath{0.05079\pm0.00048}}  
\newcommand{\hatcurPPrhoeccenxxxxxB}{\ensuremath{0.200\pm0.034}}       
\newcommand{\hatcurPPmeccenxxxxxB}{\ensuremath{0.313\pm0.016}}         
\newcommand{\hatcurPPmshorteccenxxxxxB}{\ensuremath{0.31}}             
\newcommand{\hatcurPPmlongeccenxxxxxB}{\ensuremath{0.313\pm0.016}}     
\newcommand{\hatcurPPmeeccenxxxxxB}{\ensuremath{99.4\pm5.0}}           
\newcommand{\hatcurPPmeshorteccenxxxxxB}{\ensuremath{99.4}}            
\newcommand{\hatcurPPmelongeccenxxxxxB}{\ensuremath{99.4\pm5.0}}       
\newcommand{\hatcurPPreccenxxxxxB}{\ensuremath{1.246\pm0.068}}         
\newcommand{\hatcurPPrshorteccenxxxxxB}{\ensuremath{1.25}}             
\newcommand{\hatcurPPrlongeccenxxxxxB}{\ensuremath{1.246\pm0.068}}     
\newcommand{\hatcurPPreeccenxxxxxB}{\ensuremath{13.97\pm0.76}}         
\newcommand{\hatcurPPreshorteccenxxxxxB}{\ensuremath{14.0}}            
\newcommand{\hatcurPPrelongeccenxxxxxB}{\ensuremath{13.97\pm0.76}}     
\newcommand{\hatcurPPmrcorreccenxxxxxB}{\ensuremath{0.01}}             
\newcommand{\hatcurPPteffeccenxxxxxB}{\ensuremath{1169\pm31}}          
\newcommand{\hatcurPPthetaeccenxxxxxB}{\ensuremath{0.0259\pm0.0018}}   
\newcommand{\hatcurPPfluxperieccenxxxxxB}{\ensuremath{4.68_{-0.32}^{+0.43}}} 
\newcommand{\hatcurPPfluxperidimeccenxxxxxB}{\ensuremath{8}}           
\newcommand{\hatcurPPfluxapeccenxxxxxB}{\ensuremath{3.88_{-0.70}^{+0.53}}} 
\newcommand{\hatcurPPfluxapdimeccenxxxxxB}{\ensuremath{8}}             
\newcommand{\hatcurPPfluxavgeccenxxxxxB}{\ensuremath{4.22\pm0.45}}     
\newcommand{\hatcurPPfluxavgdimeccenxxxxxB}{\ensuremath{8}}            
\newcommand{\hatcurPPfluxavglogeccenxxxxxB}{\ensuremath{8.626\pm0.046}} 
\newcommand{\hatcurXsecphaseeccenxxxxxB}{\ensuremath{0.486\pm0.014}}   
\newcommand{\hatcurXsecondaryeccenxxxxxB}{\ensuremath{2456208.828\pm0.060}} 
\newcommand{\hatcurXsecdureccenxxxxxB}{\ensuremath{0.1318\pm0.0100}}   
\newcommand{\hatcurXsecingdureccenxxxxxB}{\ensuremath{0.0157\pm0.0017}} 
\newcommand{\hatcurPPphiconjeccenxxxxxB}{\ensuremath{-0.375_{-0.077}^{+0.269}}} 
\newcommand{\hatcurPPperieccenxxxxxB}{\ensuremath{2456208.4\pm1.3}}    
\newcommand{\hatcurPPaequiveccenxxxxxB}{\ensuremath{0.0569\pm0.0031}}  
\newcommand{\hatcurPPtcirceccenxxxxxB}{\ensuremath{157\pm40}}          
\newcommand{\hatcurPPtinfalleccenxxxxxB}{\ensuremath{19000_{-4100}^{+6300}}} 
\newcommand{\hatcurXdisteccenxxxxxB}{\ensuremath{459\pm23}}            
\newcommand{\hatcurXAveccenxxxxxB}{\ensuremath{0.011_{-0.011}^{+0.071}}} 
\newcommand{\hatcurXdistredeccenxxxxxB}{\ensuremath{453\pm23}}         
\newcommand{\hatcurXEBVeccenxxxxxB}{\ensuremath{0.0040_{-0.0040}^{+0.0229}}} 
\newcommand{\hatcurXmvisoredeccenxxxxxB}{\ensuremath{13.452\pm0.032}}  
\newcommand{\hatcurXmiisoredeccenxxxxxB}{\ensuremath{12.630\pm0.021}}  
\newcommand{\hatcurXmjisoredeccenxxxxxB}{\ensuremath{12.104\pm0.014}}  
\newcommand{\hatcurXmhisoredeccenxxxxxB}{\ensuremath{11.700\pm0.014}}  
\newcommand{\hatcurXmkisoredeccenxxxxxB}{\ensuremath{11.631\pm0.015}}  
\newcommand{\hatcurXviisoredeccenxxxxxB}{\ensuremath{0.819_{-0.014}^{+0.020}}} 
\newcommand{\hatcurXvkisoredeccenxxxxxB}{\ensuremath{1.821\pm0.036}}   
\newcommand{\hatcurXjhisoredeccenxxxxxB}{\ensuremath{0.403\pm0.010}}   
\newcommand{\hatcurXjkisoredeccenxxxxxB}{\ensuremath{0.473\pm0.011}}   
\newcommand{\hatcurCCpmraeccenxxxxxB}{\ensuremath{-9.8\pm1.4}}         
\newcommand{\hatcurCCpmdececcenxxxxxB}{\ensuremath{-16.8\pm2.1}}       
\newcommand{\hatcurCCpmeccenxxxxxB}{\ensuremath{19.4\pm2.5}}           

\newcommand{\hatcurhtrxxxxxC}{HTR212-005}                              
\newcommand{\hatcurfieldxxxxxC}{212}                                   
\newcommand{\hatcurCCraxxxxxC}{\ensuremath{02^{\mathrm h}50^{\mathrm m}53.20{\mathrm s}}}                            
\newcommand{\hatcurCCdecxxxxxC}{\ensuremath{+29{\arcdeg}01{\arcmin}20.6{\arcsec}}}                           
\newcommand{\hatcurCCmagxxxxxC}{14.068}                                
\newcommand{\hatcurCCtwomassxxxxxC}{2MASS~02505320+2901206}            
\newcommand{\hatcurCCgscxxxxxC}{GSC~1793-01136}                        
\newcommand{\hatcurCCtassmvxxxxxC}{\ensuremath{14.068\pm0.020}}        
\newcommand{\hatcurCCtassmvshortxxxxxC}{\ensuremath{14.1}}             
\newcommand{\hatcurCCtassmBxxxxxC}{\ensuremath{15.183\pm0.050}}        
\newcommand{\hatcurCCtassmBshortxxxxxC}{\ensuremath{15.2}}             
\newcommand{\hatcurCCtassmIxxxxxC}{\ensuremath{13.02\pm0.17}}          
\newcommand{\hatcurCCtassmIshortxxxxxC}{\ensuremath{13.0}}             
\newcommand{\hatcurCCtassmgxxxxxC}{\ensuremath{14.631\pm0.060}}        
\newcommand{\hatcurCCtassmgshortxxxxxC}{\ensuremath{14.6}}             
\newcommand{\hatcurCCtassmrxxxxxC}{\ensuremath{13.677\pm0.080}}        
\newcommand{\hatcurCCtassmrshortxxxxxC}{\ensuremath{13.7}}             
\newcommand{\hatcurCCtassmixxxxxC}{\ensuremath{13.441\pm0.090}}        
\newcommand{\hatcurCCtassmishortxxxxxC}{\ensuremath{13.4}}             
\newcommand{\hatcurCCtwomassJmagxxxxxC}{\ensuremath{12.195\pm0.022}}   
\newcommand{\hatcurCCtwomassHmagxxxxxC}{\ensuremath{11.745\pm0.022}}   
\newcommand{\hatcurCCtwomassKmagxxxxxC}{\ensuremath{11.621\pm0.021}}   
\newcommand{\hatcurCCcitJmagxxxxxC}{\ensuremath{12.201\pm0.023}}       
\newcommand{\hatcurCCcitHmagxxxxxC}{\ensuremath{11.738\pm0.023}}       
\newcommand{\hatcurCCcitKmagxxxxxC}{\ensuremath{11.645\pm0.021}}       
\newcommand{\hatcurCCbbJmagxxxxxC}{\ensuremath{12.267\pm0.025}}        
\newcommand{\hatcurCCbbHmagxxxxxC}{\ensuremath{11.761\pm0.024}}        
\newcommand{\hatcurCCbbKmagxxxxxC}{\ensuremath{11.665\pm0.021}}        
\newcommand{\hatcurCCesoJmagxxxxxC}{\ensuremath{12.271\pm0.027}}       
\newcommand{\hatcurCCesoHmagxxxxxC}{\ensuremath{11.757\pm0.029}}       
\newcommand{\hatcurCCesoKmagxxxxxC}{\ensuremath{11.663\pm0.022}}       
\newcommand{\hatcurCCesoJHmagxxxxxC}{\ensuremath{0.514\pm0.018}}       
\newcommand{\hatcurCCesoJKmagxxxxxC}{\ensuremath{0.609\pm0.034}}       
\newcommand{\hatcurCCesoHKmagxxxxxC}{\ensuremath{0.094\pm0.036}}       
\newcommand{\hatcurLCdipxxxxxC}{\ensuremath{15.2}}                     
\newcommand{\hatcurLCrprstarxxxxxC}{\ensuremath{0.1161\pm0.0027}}      
\newcommand{\hatcurLCbsqxxxxxC}{\ensuremath{0.213_{-0.098}^{+0.096}}}  
\newcommand{\hatcurLCimpxxxxxC}{\ensuremath{0.461_{-0.122}^{+0.094}}}  
\newcommand{\hatcurLCzetaxxxxxC}{\ensuremath{22.88\pm0.27}}            
\newcommand{\hatcurLCdurxxxxxC}{\ensuremath{0.1003\pm0.0017}}          
\newcommand{\hatcurLCdurshortxxxxxC}{\ensuremath{0.1003}}              
\newcommand{\hatcurLCdurhrxxxxxC}{\ensuremath{2.406\pm0.040}}          
\newcommand{\hatcurLCdurhrshortxxxxxC}{\ensuremath{2.406}}             
\newcommand{\hatcurLCqxxxxxC}{\ensuremath{0.03640\pm0.00061}}          
\newcommand{\hatcurLCqshortxxxxxC}{\ensuremath{0.036}}                 
\newcommand{\hatcurLCingdurxxxxxC}{\ensuremath{0.0130\pm0.0018}}       
\newcommand{\hatcurLCPxxxxxC}{\ensuremath{2.7535953\pm0.0000094}}      
\newcommand{\hatcurLCPprecxxxxxC}{\ensuremath{2.7535953}}              
\newcommand{\hatcurLCPshortxxxxxC}{\ensuremath{2.7536}}                
\newcommand{\hatcurLCTxxxxxC}{\ensuremath{2455852.10326\pm0.00041}}    
\newcommand{\hatcurLCTAxxxxxC}{\ensuremath{2455417.0352\pm0.0014}}     
\newcommand{\hatcurLCTBxxxxxC}{\ensuremath{2455934.71112\pm0.00058}}   
\newcommand{\hatcurLChatnetmxxxxxC}{\ensuremath{13.61925\pm0.00023}}   
\newcommand{\hatcurLCiblendxxxxxC}{\ensuremath{0.664\pm0.085}}         
\newcommand{\hatcurSMEiteffxxxxxC}{\ensuremath{5014\pm50}}             
\newcommand{\hatcurSMEizfehxxxxxC}{\ensuremath{0.170\pm0.080}}         
\newcommand{\hatcurSMEizfehshortxxxxxC}{\ensuremath{0.17}}             
\newcommand{\hatcurSMEiloggxxxxxC}{\ensuremath{4.37\pm0.10}}           
\newcommand{\hatcurSMEivsinxxxxxC}{\ensuremath{1.50\pm0.50}}           
\newcommand{\hatcurSMEivmacxxxxxC}{\ensuremath{0.0}}                   
\newcommand{\hatcurSMEivmicxxxxxC}{\ensuremath{0.0}}                   
\newcommand{\hatcurSMEiiteffxxxxxC}{\ensuremath{5131\pm50}}            
\newcommand{\hatcurSMEiizfehxxxxxC}{\ensuremath{0.280\pm0.080}}        
\newcommand{\hatcurSMEiizfehshortxxxxxC}{\ensuremath{0.28}}            
\newcommand{\hatcurSMEiiloggxxxxxC}{\ensuremath{4.490\pm0.044}}        
\newcommand{\hatcurSMEiivsinxxxxxC}{\ensuremath{0.60\pm0.50}}          
\newcommand{\hatcurTRESteffxxxxxC}{\ensuremath{5000\pm100}}            
\newcommand{\hatcurTRESzfehxxxxxC}{\ensuremath{0.00\pm0.10}}           
\newcommand{\hatcurTRESloggxxxxxC}{\ensuremath{4.50\pm0.10}}           
\newcommand{\hatcurTRESvsinixxxxxC}{\ensuremath{2.00\pm0.50}}          
\newcommand{\hatcurTRESgammaxxxxxC}{\ensuremath{62.19\pm0.10}}         
\newcommand{\hatcurTRESnumspecxxxxxC}{\ensuremath{2}}                  
\newcommand{\hatcurTRESspanxxxxxC}{\ensuremath{NULL}}                  
\newcommand{\hatcurTRESrvrmsxxxxxC}{\ensuremath{0.00}}                 
\newcommand{\hatcurLBizxxxxxC}{\ensuremath{0.3028}}                    
\newcommand{\hatcurLBiizxxxxxC}{\ensuremath{0.2853}}                   
\newcommand{\hatcurLBiixxxxxC}{\ensuremath{0.3908}}                    
\newcommand{\hatcurLBiiixxxxxC}{\ensuremath{0.2628}}                   
\newcommand{\hatcurLBiIxxxxxC}{\ensuremath{0.3617}}                    
\newcommand{\hatcurLBiiIxxxxxC}{\ensuremath{0.2701}}                   
\newcommand{\hatcurLBigxxxxxC}{\ensuremath{0.7630}}                    
\newcommand{\hatcurLBiigxxxxxC}{\ensuremath{0.0708}}                   
\newcommand{\hatcurLBirxxxxxC}{\ensuremath{0.5191}}                    
\newcommand{\hatcurLBiirxxxxxC}{\ensuremath{0.2228}}                   
\newcommand{\hatcurLBiRxxxxxC}{\ensuremath{0.4838}}                    
\newcommand{\hatcurLBiiRxxxxxC}{\ensuremath{0.2347}}                   
\newcommand{\hatcurISOmxxxxxC}{\ensuremath{0.887\pm0.027}}             
\newcommand{\hatcurISOmshortxxxxxC}{\ensuremath{0.89}}                 
\newcommand{\hatcurISOmlongxxxxxC}{\ensuremath{0.887\pm0.027}}         
\newcommand{\hatcurISOrxxxxxC}{\ensuremath{0.893\pm0.047}}             
\newcommand{\hatcurISOrshortxxxxxC}{\ensuremath{0.89}}                 
\newcommand{\hatcurISOrlongxxxxxC}{\ensuremath{0.893\pm0.047}}         
\newcommand{\hatcurISOrhoxxxxxC}{\ensuremath{1.75\pm0.29}}             
\newcommand{\hatcurISOrholongxxxxxC}{\ensuremath{1.75\pm0.29}}         
\newcommand{\hatcurISOloggxxxxxC}{\ensuremath{4.483\pm0.051}}          
\newcommand{\hatcurISOlumxxxxxC}{\ensuremath{0.496\pm0.060}}           
\newcommand{\hatcurISOlumshortxxxxxC}{\ensuremath{0.50}}               
\newcommand{\hatcurISOmvxxxxxC}{\ensuremath{5.73\pm0.13}}              
\newcommand{\hatcurISOvixxxxxC}{\ensuremath{0.884\pm0.014}}            
\newcommand{\hatcurISOagexxxxxC}{\ensuremath{9.4\pm4.1}}               
\newcommand{\hatcurISOsigmaxxxxxC}{\ensuremath{0.00140\pm0.00017}}     
\newcommand{\hatcurISOMJxxxxxC}{\ensuremath{4.22\pm0.12}}              
\newcommand{\hatcurISOMHxxxxxC}{\ensuremath{3.77\pm0.12}}              
\newcommand{\hatcurISOMKxxxxxC}{\ensuremath{3.69\pm0.12}}              
\newcommand{\hatcurISOJKxxxxxC}{\ensuremath{0.540\pm0.010}}            
\newcommand{\hatcurISOspecxxxxxC}{G}                                   
\newcommand{\hatcurRVKxxxxxC}{\ensuremath{128.4\pm3.8}}                
\newcommand{\hatcurRVrkxxxxxC}{\ensuremath{0\pm0}}                     
\newcommand{\hatcurRVrhxxxxxC}{\ensuremath{0\pm0}}                     
\newcommand{\hatcurRVkxxxxxC}{\ensuremath{0\pm0}}                      
\newcommand{\hatcurRVhxxxxxC}{\ensuremath{0\pm0}}                      
\newcommand{\hatcurRVtronexxxxxC}{\ensuremath{0\pm0}}                  
\newcommand{\hatcurRVtrtwoxxxxxC}{\ensuremath{0\pm0}}                  
\newcommand{\hatcurRVgammaxxxxxC}{\ensuremath{-31.0\pm2.8}}            
\newcommand{\hatcurRVjitterxxxxxC}{\ensuremath{5.2\pm1.9}}             
\newcommand{\hatcurRVfitrmsxxxxxC}{\ensuremath{7.5}}                   
\newcommand{\hatcurRVeccenxxxxxC}{\ensuremath{0\pm0}}                  
\newcommand{\hatcurRVeccentwosiglimxxxxxC}{\ensuremath{<0.000}}        
\newcommand{\hatcurRVomegaxxxxxC}{\ensuremath{0\pm0}}                  
\newcommand{\hatcurPPixxxxxC}{\ensuremath{87.02\pm0.86}}               
\newcommand{\hatcurPPgxxxxxC}{\ensuremath{19.8\pm2.9}}                 
\newcommand{\hatcurPPloggxxxxxC}{\ensuremath{3.296\pm0.065}}           
\newcommand{\hatcurPParxxxxxC}{\ensuremath{8.89\pm0.49}}               
\newcommand{\hatcurPParelxxxxxC}{\ensuremath{0.03694\pm0.00038}}       
\newcommand{\hatcurPPrhoxxxxxC}{\ensuremath{0.98\pm0.21}}              
\newcommand{\hatcurPPmxxxxxC}{\ensuremath{0.818\pm0.029}}              
\newcommand{\hatcurPPmshortxxxxxC}{\ensuremath{0.82}}                  
\newcommand{\hatcurPPmlongxxxxxC}{\ensuremath{0.818\pm0.029}}          
\newcommand{\hatcurPPmexxxxxC}{\ensuremath{260.0\pm9.3}}               
\newcommand{\hatcurPPmeshortxxxxxC}{\ensuremath{260.0}}                
\newcommand{\hatcurPPmelongxxxxxC}{\ensuremath{260.0\pm9.3}}           
\newcommand{\hatcurPPrxxxxxC}{\ensuremath{1.009\pm0.072}}              
\newcommand{\hatcurPPrshortxxxxxC}{\ensuremath{1.01}}                  
\newcommand{\hatcurPPrlongxxxxxC}{\ensuremath{1.009\pm0.072}}          
\newcommand{\hatcurPPrexxxxxC}{\ensuremath{11.31\pm0.81}}              
\newcommand{\hatcurPPreshortxxxxxC}{\ensuremath{11.3}}                 
\newcommand{\hatcurPPrelongxxxxxC}{\ensuremath{11.31\pm0.81}}          
\newcommand{\hatcurPPmrcorrxxxxxC}{\ensuremath{-0.15}}                 
\newcommand{\hatcurPPteffxxxxxC}{\ensuremath{1218\pm37}}               
\newcommand{\hatcurPPthetaxxxxxC}{\ensuremath{0.0673\pm0.0050}}        
\newcommand{\hatcurPPfluxperixxxxxC}{\ensuremath{4.96\pm0.61}}         
\newcommand{\hatcurPPfluxperidimxxxxxC}{\ensuremath{8}}                
\newcommand{\hatcurPPfluxapxxxxxC}{\ensuremath{4.96\pm0.61}}           
\newcommand{\hatcurPPfluxapdimxxxxxC}{\ensuremath{8}}                  
\newcommand{\hatcurPPfluxavgxxxxxC}{\ensuremath{4.96\pm0.61}}          
\newcommand{\hatcurPPfluxavgdimxxxxxC}{\ensuremath{8}}                 
\newcommand{\hatcurPPfluxavglogxxxxxC}{\ensuremath{8.696\pm0.052}}     
\newcommand{\hatcurXsecphasexxxxxC}{\ensuremath{0\pm0}}                
\newcommand{\hatcurXsecondaryxxxxxC}{\ensuremath{2455853.48006\pm0.00041}} 
\newcommand{\hatcurXsecdurxxxxxC}{\ensuremath{0.1003\pm0.0017}}        
\newcommand{\hatcurXsecingdurxxxxxC}{\ensuremath{0.0130\pm0.0018}}     
\newcommand{\hatcurPPphiconjxxxxxC}{\ensuremath{0\pm0}}                
\newcommand{\hatcurPPperixxxxxC}{\ensuremath{2455851.41486\pm0.00041}} 
\newcommand{\hatcurPPaequivxxxxxC}{\ensuremath{0.0524\pm0.0031}}       
\newcommand{\hatcurPPtcircxxxxxC}{\ensuremath{175_{-58}^{+78}}}        
\newcommand{\hatcurPPtinfallxxxxxC}{\ensuremath{1570\pm450}}           
\newcommand{\hatcurXdistxxxxxC}{\ensuremath{393\pm22}}                 
\newcommand{\hatcurXAvxxxxxC}{\ensuremath{0.412\pm0.052}}              
\newcommand{\hatcurXdistredxxxxxC}{\ensuremath{385\pm21}}              
\newcommand{\hatcurXEBVxxxxxC}{\ensuremath{0.133\pm0.017}}             
\newcommand{\hatcurXmvisoredxxxxxC}{\ensuremath{14.069\pm0.020}}       
\newcommand{\hatcurXmiisoredxxxxxC}{\ensuremath{12.970\pm0.015}}       
\newcommand{\hatcurXmjisoredxxxxxC}{\ensuremath{12.266\pm0.014}}       
\newcommand{\hatcurXmhisoredxxxxxC}{\ensuremath{11.769\pm0.015}}       
\newcommand{\hatcurXmkisoredxxxxxC}{\ensuremath{11.661\pm0.016}}       
\newcommand{\hatcurXviisoredxxxxxC}{\ensuremath{1.099\pm0.019}}        
\newcommand{\hatcurXvkisoredxxxxxC}{\ensuremath{2.408\pm0.027}}        
\newcommand{\hatcurXjhisoredxxxxxC}{\ensuremath{0.4970\pm0.0057}}      
\newcommand{\hatcurXjkisoredxxxxxC}{\ensuremath{0.6050\pm0.0068}}      
\newcommand{\hatcurCCpmraxxxxxC}{\ensuremath{12.5\pm2.1}}              
\newcommand{\hatcurCCpmdecxxxxxC}{\ensuremath{-24.7\pm2.6}}            
\newcommand{\hatcurCCpmxxxxxC}{\ensuremath{27.7\pm3.3}}                
\newcommand{\hatcurhtreccenxxxxxC}{HTR212-005}                         
\newcommand{\hatcurfieldeccenxxxxxC}{212}                              
\newcommand{\hatcurCCraeccenxxxxxC}{\ensuremath{02^{\mathrm h}50^{\mathrm m}53.20{\mathrm s}}}                       
\newcommand{\hatcurCCdececcenxxxxxC}{\ensuremath{+29{\arcdeg}01{\arcmin}20.6{\arcsec}}}                      
\newcommand{\hatcurCCmageccenxxxxxC}{14.068}                           
\newcommand{\hatcurCCtwomasseccenxxxxxC}{2MASS~02505320+2901206}       
\newcommand{\hatcurCCgsceccenxxxxxC}{GSC~1793-01136}                   
\newcommand{\hatcurCCtassmveccenxxxxxC}{\ensuremath{14.068\pm0.020}}   
\newcommand{\hatcurCCtassmvshorteccenxxxxxC}{\ensuremath{14.1}}        
\newcommand{\hatcurCCtassmBeccenxxxxxC}{\ensuremath{15.183\pm0.050}}   
\newcommand{\hatcurCCtassmBshorteccenxxxxxC}{\ensuremath{15.2}}        
\newcommand{\hatcurCCtassmIeccenxxxxxC}{\ensuremath{13.02\pm0.17}}     
\newcommand{\hatcurCCtassmIshorteccenxxxxxC}{\ensuremath{13.0}}        
\newcommand{\hatcurCCtassmgeccenxxxxxC}{\ensuremath{15.5260\pm0.0070}} 
\newcommand{\hatcurCCtassmgshorteccenxxxxxC}{\ensuremath{15.5}}        
\newcommand{\hatcurCCtassmreccenxxxxxC}{\ensuremath{13.8690\pm0.0020}} 
\newcommand{\hatcurCCtassmrshorteccenxxxxxC}{\ensuremath{13.9}}        
\newcommand{\hatcurCCtassmieccenxxxxxC}{\ensuremath{13.5580\pm0.0010}} 
\newcommand{\hatcurCCtassmishorteccenxxxxxC}{\ensuremath{13.6}}        
\newcommand{\hatcurCCtwomassJmageccenxxxxxC}{\ensuremath{12.195\pm0.022}} 
\newcommand{\hatcurCCtwomassHmageccenxxxxxC}{\ensuremath{11.745\pm0.022}} 
\newcommand{\hatcurCCtwomassKmageccenxxxxxC}{\ensuremath{11.621\pm0.021}} 
\newcommand{\hatcurCCcitJmageccenxxxxxC}{\ensuremath{12.201\pm0.023}}  
\newcommand{\hatcurCCcitHmageccenxxxxxC}{\ensuremath{11.738\pm0.023}}  
\newcommand{\hatcurCCcitKmageccenxxxxxC}{\ensuremath{11.645\pm0.021}}  
\newcommand{\hatcurCCbbJmageccenxxxxxC}{\ensuremath{12.267\pm0.025}}   
\newcommand{\hatcurCCbbHmageccenxxxxxC}{\ensuremath{11.761\pm0.024}}   
\newcommand{\hatcurCCbbKmageccenxxxxxC}{\ensuremath{11.665\pm0.021}}   
\newcommand{\hatcurCCesoJmageccenxxxxxC}{\ensuremath{12.271\pm0.027}}  
\newcommand{\hatcurCCesoHmageccenxxxxxC}{\ensuremath{11.757\pm0.029}}  
\newcommand{\hatcurCCesoKmageccenxxxxxC}{\ensuremath{11.663\pm0.022}}  
\newcommand{\hatcurCCesoJHmageccenxxxxxC}{\ensuremath{0.514\pm0.018}}  
\newcommand{\hatcurCCesoJKmageccenxxxxxC}{\ensuremath{0.609\pm0.034}}  
\newcommand{\hatcurCCesoHKmageccenxxxxxC}{\ensuremath{0.094\pm0.036}}  
\newcommand{\hatcurLCdipeccenxxxxxC}{\ensuremath{15.2}}                
\newcommand{\hatcurLCrprstareccenxxxxxC}{\ensuremath{0.1156\pm0.0024}} 
\newcommand{\hatcurLCbsqeccenxxxxxC}{\ensuremath{0.188_{-0.098}^{+0.083}}} 
\newcommand{\hatcurLCimpeccenxxxxxC}{\ensuremath{0.433_{-0.134}^{+0.087}}} 
\newcommand{\hatcurLCzetaeccenxxxxxC}{\ensuremath{22.84\pm0.24}}       
\newcommand{\hatcurLCdureccenxxxxxC}{\ensuremath{0.0999\pm0.0015}}     
\newcommand{\hatcurLCdurshorteccenxxxxxC}{\ensuremath{0.0999}}         
\newcommand{\hatcurLCdurhreccenxxxxxC}{\ensuremath{2.398\pm0.036}}     
\newcommand{\hatcurLCdurhrshorteccenxxxxxC}{\ensuremath{2.398}}        
\newcommand{\hatcurLCqeccenxxxxxC}{\ensuremath{0.03630\pm0.00054}}     
\newcommand{\hatcurLCqshorteccenxxxxxC}{\ensuremath{0.036}}            
\newcommand{\hatcurLCingdureccenxxxxxC}{\ensuremath{0.0125\pm0.0015}}  
\newcommand{\hatcurLCPeccenxxxxxC}{\ensuremath{2.7535958\pm0.0000087}} 
\newcommand{\hatcurLCPprececcenxxxxxC}{\ensuremath{2.7535958}}         
\newcommand{\hatcurLCPshorteccenxxxxxC}{\ensuremath{2.7536}}           
\newcommand{\hatcurLCTeccenxxxxxC}{\ensuremath{2455854.85683\pm0.00040}} 
\newcommand{\hatcurLCTAeccenxxxxxC}{\ensuremath{2455417.0351\pm0.0013}} 
\newcommand{\hatcurLCTBeccenxxxxxC}{\ensuremath{2455934.71112\pm0.00055}} 
\newcommand{\hatcurLChatnetmeccenxxxxxC}{\ensuremath{13.61926\pm0.00023}} 
\newcommand{\hatcurLCiblendeccenxxxxxC}{\ensuremath{0.661\pm0.086}}    
\newcommand{\hatcurSMEiteffeccenxxxxxC}{\ensuremath{5014\pm50}}        
\newcommand{\hatcurSMEizfeheccenxxxxxC}{\ensuremath{0.170\pm0.080}}    
\newcommand{\hatcurSMEizfehshorteccenxxxxxC}{\ensuremath{0.17}}        
\newcommand{\hatcurSMEiloggeccenxxxxxC}{\ensuremath{4.37\pm0.10}}      
\newcommand{\hatcurSMEivsineccenxxxxxC}{\ensuremath{1.50\pm0.50}}      
\newcommand{\hatcurSMEivmaceccenxxxxxC}{\ensuremath{0.0}}              
\newcommand{\hatcurSMEivmiceccenxxxxxC}{\ensuremath{0.0}}              
\newcommand{\hatcurSMEiiteffeccenxxxxxC}{\ensuremath{5014\pm50}}       
\newcommand{\hatcurSMEiizfeheccenxxxxxC}{\ensuremath{0.170\pm0.080}}   
\newcommand{\hatcurSMEiizfehshorteccenxxxxxC}{\ensuremath{0.17}}       
\newcommand{\hatcurSMEiiloggeccenxxxxxC}{\ensuremath{4.37\pm0.10}}     
\newcommand{\hatcurSMEiivsineccenxxxxxC}{\ensuremath{1.50\pm0.50}}     
\newcommand{\hatcurTRESteffeccenxxxxxC}{\ensuremath{5000\pm100}}       
\newcommand{\hatcurTRESzfeheccenxxxxxC}{\ensuremath{0.00\pm0.10}}      
\newcommand{\hatcurTRESloggeccenxxxxxC}{\ensuremath{4.50\pm0.10}}      
\newcommand{\hatcurTRESvsinieccenxxxxxC}{\ensuremath{2.00\pm0.50}}     
\newcommand{\hatcurTRESgammaeccenxxxxxC}{\ensuremath{62.19\pm0.10}}    
\newcommand{\hatcurTRESnumspececcenxxxxxC}{\ensuremath{2}}             
\newcommand{\hatcurTRESspaneccenxxxxxC}{\ensuremath{NULL}}             
\newcommand{\hatcurTRESrvrmseccenxxxxxC}{\ensuremath{0.00}}            
\newcommand{\hatcurLBizeccenxxxxxC}{\ensuremath{0.3159}}               
\newcommand{\hatcurLBiizeccenxxxxxC}{\ensuremath{0.2747}}              
\newcommand{\hatcurLBiieccenxxxxxC}{\ensuremath{0.4050}}               
\newcommand{\hatcurLBiiieccenxxxxxC}{\ensuremath{0.2508}}              
\newcommand{\hatcurLBiIeccenxxxxxC}{\ensuremath{0.3753}}               
\newcommand{\hatcurLBiiIeccenxxxxxC}{\ensuremath{0.2586}}              
\newcommand{\hatcurLBigeccenxxxxxC}{\ensuremath{0.7874}}               
\newcommand{\hatcurLBiigeccenxxxxxC}{\ensuremath{0.0497}}              
\newcommand{\hatcurLBireccenxxxxxC}{\ensuremath{0.5385}}               
\newcommand{\hatcurLBiireccenxxxxxC}{\ensuremath{0.2071}}              
\newcommand{\hatcurLBiReccenxxxxxC}{\ensuremath{0.5017}}               
\newcommand{\hatcurLBiiReccenxxxxxC}{\ensuremath{0.2200}}              
\newcommand{\hatcurISOmeccenxxxxxC}{\ensuremath{0.826\pm0.027}}        
\newcommand{\hatcurISOmshorteccenxxxxxC}{\ensuremath{0.83}}            
\newcommand{\hatcurISOmlongeccenxxxxxC}{\ensuremath{0.826\pm0.027}}    
\newcommand{\hatcurISOreccenxxxxxC}{\ensuremath{0.858\pm0.044}}        
\newcommand{\hatcurISOrshorteccenxxxxxC}{\ensuremath{0.86}}            
\newcommand{\hatcurISOrlongeccenxxxxxC}{\ensuremath{0.858\pm0.044}}    
\newcommand{\hatcurISOrhoeccenxxxxxC}{\ensuremath{1.84\pm0.29}}        
\newcommand{\hatcurISOrholongeccenxxxxxC}{\ensuremath{1.84\pm0.29}}    
\newcommand{\hatcurISOloggeccenxxxxxC}{\ensuremath{4.486\pm0.046}}     
\newcommand{\hatcurISOlumeccenxxxxxC}{\ensuremath{0.418\pm0.051}}      
\newcommand{\hatcurISOlumshorteccenxxxxxC}{\ensuremath{0.42}}          
\newcommand{\hatcurISOmveccenxxxxxC}{\ensuremath{5.95\pm0.14}}         
\newcommand{\hatcurISOvieccenxxxxxC}{\ensuremath{0.907\pm0.015}}       
\newcommand{\hatcurISOageeccenxxxxxC}{\ensuremath{13.8\pm4.2}}         
\newcommand{\hatcurISOsigmaeccenxxxxxC}{\ensuremath{0.00140\pm0.00018}} 
\newcommand{\hatcurISOMJeccenxxxxxC}{\ensuremath{4.38\pm0.12}}         
\newcommand{\hatcurISOMHeccenxxxxxC}{\ensuremath{3.90\pm0.12}}         
\newcommand{\hatcurISOMKeccenxxxxxC}{\ensuremath{3.82\pm0.11}}         
\newcommand{\hatcurISOJKeccenxxxxxC}{\ensuremath{0.560\pm0.010}}       
\newcommand{\hatcurISOspececcenxxxxxC}{G}                              
\newcommand{\hatcurRVKeccenxxxxxC}{\ensuremath{128.6\pm4.2}}           
\newcommand{\hatcurRVrkeccenxxxxxC}{\ensuremath{-0.032\pm0.073}}       
\newcommand{\hatcurRVrheccenxxxxxC}{\ensuremath{-0.01\pm0.11}}         
\newcommand{\hatcurRVkeccenxxxxxC}{\ensuremath{-0.0030_{-0.0120}^{+0.0078}}} 
\newcommand{\hatcurRVheccenxxxxxC}{\ensuremath{-0.000\pm0.019}}        
\newcommand{\hatcurRVtroneeccenxxxxxC}{\ensuremath{0\pm0}}             
\newcommand{\hatcurRVtrtwoeccenxxxxxC}{\ensuremath{0\pm0}}             
\newcommand{\hatcurRVgammaeccenxxxxxC}{\ensuremath{-31.3\pm3.1}}       
\newcommand{\hatcurRVjittereccenxxxxxC}{\ensuremath{5.5\pm2.2}}        
\newcommand{\hatcurRVfitrmseccenxxxxxC}{\ensuremath{10.0}}             
\newcommand{\hatcurRVecceneccenxxxxxC}{\ensuremath{0.014\pm0.015}}     
\newcommand{\hatcurRVeccentwosiglimeccenxxxxxC}{\ensuremath{<0.047}}   
\newcommand{\hatcurRVomegaeccenxxxxxC}{\ensuremath{189\pm88}}          
\newcommand{\hatcurPPieccenxxxxxC}{\ensuremath{87.26_{-0.73}^{+0.95}}} 
\newcommand{\hatcurPPgeccenxxxxxC}{\ensuremath{20.7\pm2.9}}            
\newcommand{\hatcurPPloggeccenxxxxxC}{\ensuremath{3.316\pm0.060}}      
\newcommand{\hatcurPPareccenxxxxxC}{\ensuremath{9.03\pm0.47}}          
\newcommand{\hatcurPPareleccenxxxxxC}{\ensuremath{0.03607\pm0.00039}}  
\newcommand{\hatcurPPrhoeccenxxxxxC}{\ensuremath{1.07\pm0.22}}         
\newcommand{\hatcurPPmeccenxxxxxC}{\ensuremath{0.781\pm0.031}}         
\newcommand{\hatcurPPmshorteccenxxxxxC}{\ensuremath{0.78}}             
\newcommand{\hatcurPPmlongeccenxxxxxC}{\ensuremath{0.781\pm0.031}}     
\newcommand{\hatcurPPmeeccenxxxxxC}{\ensuremath{248.2\pm9.8}}          
\newcommand{\hatcurPPmeshorteccenxxxxxC}{\ensuremath{248.2}}           
\newcommand{\hatcurPPmelongeccenxxxxxC}{\ensuremath{248.2\pm9.8}}      
\newcommand{\hatcurPPreccenxxxxxC}{\ensuremath{0.965\pm0.064}}         
\newcommand{\hatcurPPrshorteccenxxxxxC}{\ensuremath{0.97}}             
\newcommand{\hatcurPPrlongeccenxxxxxC}{\ensuremath{0.965\pm0.064}}     
\newcommand{\hatcurPPreeccenxxxxxC}{\ensuremath{10.82\pm0.71}}         
\newcommand{\hatcurPPreshorteccenxxxxxC}{\ensuremath{10.8}}            
\newcommand{\hatcurPPrelongeccenxxxxxC}{\ensuremath{10.82\pm0.71}}     
\newcommand{\hatcurPPmrcorreccenxxxxxC}{\ensuremath{-0.04}}            
\newcommand{\hatcurPPteffeccenxxxxxC}{\ensuremath{1181\pm35}}          
\newcommand{\hatcurPPthetaeccenxxxxxC}{\ensuremath{0.0704\pm0.0052}}   
\newcommand{\hatcurPPfluxperieccenxxxxxC}{\ensuremath{4.54\pm0.56}}    
\newcommand{\hatcurPPfluxperidimeccenxxxxxC}{\ensuremath{8}}           
\newcommand{\hatcurPPfluxapeccenxxxxxC}{\ensuremath{4.25\pm0.52}}      
\newcommand{\hatcurPPfluxapdimeccenxxxxxC}{\ensuremath{8}}             
\newcommand{\hatcurPPfluxavgeccenxxxxxC}{\ensuremath{4.39\pm0.52}}     
\newcommand{\hatcurPPfluxavgdimeccenxxxxxC}{\ensuremath{8}}            
\newcommand{\hatcurPPfluxavglogeccenxxxxxC}{\ensuremath{8.643\pm0.051}} 
\newcommand{\hatcurXsecphaseeccenxxxxxC}{\ensuremath{0.4981\pm0.0073}} 
\newcommand{\hatcurXsecondaryeccenxxxxxC}{\ensuremath{2455856.228\pm0.020}} 
\newcommand{\hatcurXsecdureccenxxxxxC}{\ensuremath{0.0999\pm0.0034}}   
\newcommand{\hatcurXsecingdureccenxxxxxC}{\ensuremath{0.0124\pm0.0016}} 
\newcommand{\hatcurPPphiconjeccenxxxxxC}{\ensuremath{-0.10_{-0.30}^{+0.42}}} 
\newcommand{\hatcurPPperieccenxxxxxC}{\ensuremath{2455855.14\pm0.81}}  
\newcommand{\hatcurPPaequiveccenxxxxxC}{\ensuremath{0.0558\pm0.0032}}  
\newcommand{\hatcurPPtcirceccenxxxxxC}{\ensuremath{199_{-58}^{+83}}}   
\newcommand{\hatcurPPtinfalleccenxxxxxC}{\ensuremath{1660_{-390}^{+540}}} 
\newcommand{\hatcurXdisteccenxxxxxC}{\ensuremath{371\pm20}}            
\newcommand{\hatcurXAveccenxxxxxC}{\ensuremath{0.320\pm0.057}}         
\newcommand{\hatcurXdistredeccenxxxxxC}{\ensuremath{364\pm19}}         
\newcommand{\hatcurXEBVeccenxxxxxC}{\ensuremath{0.103\pm0.018}}        
\newcommand{\hatcurXmvisoredeccenxxxxxC}{\ensuremath{14.068\pm0.020}}  
\newcommand{\hatcurXmiisoredeccenxxxxxC}{\ensuremath{12.995\pm0.016}}  
\newcommand{\hatcurXmjisoredeccenxxxxxC}{\ensuremath{12.272\pm0.013}}  
\newcommand{\hatcurXmhisoredeccenxxxxxC}{\ensuremath{11.764\pm0.015}}  
\newcommand{\hatcurXmkisoredeccenxxxxxC}{\ensuremath{11.659\pm0.016}}  
\newcommand{\hatcurXviisoredeccenxxxxxC}{\ensuremath{1.073\pm0.019}}   
\newcommand{\hatcurXvkisoredeccenxxxxxC}{\ensuremath{2.409\pm0.027}}   
\newcommand{\hatcurXjhisoredeccenxxxxxC}{\ensuremath{0.5090\pm0.0056}} 
\newcommand{\hatcurXjkisoredeccenxxxxxC}{\ensuremath{0.6130\pm0.0064}} 
\newcommand{\hatcurCCpmraeccenxxxxxC}{\ensuremath{12.5\pm2.1}}         
\newcommand{\hatcurCCpmdececcenxxxxxC}{\ensuremath{-24.7\pm2.6}}       
\newcommand{\hatcurCCpmeccenxxxxxC}{\ensuremath{27.7\pm3.3}}           

\newcommand{\hatcurhtrxxxxxD}{HTR165-010}                              
\newcommand{\hatcurfieldxxxxxD}{165}                                   
\newcommand{\hatcurCCraxxxxxD}{\ensuremath{01^{\mathrm h}27^{\mathrm m}29.05{\mathrm s}}}                            
\newcommand{\hatcurCCdecxxxxxD}{\ensuremath{+38{\arcdeg}58{\arcmin}05.3{\arcsec}}}                           
\newcommand{\hatcurCCmagxxxxxD}{13.733}                                
\newcommand{\hatcurCCtwomassxxxxxD}{2MASS~01272906+3858053}            
\newcommand{\hatcurCCgscxxxxxD}{GSC~2813-01266}                        
\newcommand{\hatcurCCtassmvxxxxxD}{\ensuremath{13.73\pm0.18}}          
\newcommand{\hatcurCCtassmvshortxxxxxD}{\ensuremath{13.7}}             
\newcommand{\hatcurCCtassmBxxxxxD}{\ensuremath{0\pm0}}                 
\newcommand{\hatcurCCtassmBshortxxxxxD}{\ensuremath{.1fym}}            
\newcommand{\hatcurCCtassmIxxxxxD}{\ensuremath{13.13\pm0.12}}          
\newcommand{\hatcurCCtassmIshortxxxxxD}{\ensuremath{13.1}}             
\newcommand{\hatcurCCtassmgxxxxxD}{\ensuremath{100\pm100}}             
\newcommand{\hatcurCCtassmgshortxxxxxD}{\ensuremath{100.0}}            
\newcommand{\hatcurCCtassmrxxxxxD}{\ensuremath{100\pm100}}             
\newcommand{\hatcurCCtassmrshortxxxxxD}{\ensuremath{100.0}}            
\newcommand{\hatcurCCtassmixxxxxD}{\ensuremath{100\pm100}}             
\newcommand{\hatcurCCtassmishortxxxxxD}{\ensuremath{100.0}}            
\newcommand{\hatcurCCtwomassJmagxxxxxD}{\ensuremath{12.468\pm0.023}}   
\newcommand{\hatcurCCtwomassHmagxxxxxD}{\ensuremath{12.202\pm0.026}}   
\newcommand{\hatcurCCtwomassKmagxxxxxD}{\ensuremath{12.100\pm0.019}}   
\newcommand{\hatcurCCcitJmagxxxxxD}{\ensuremath{12.485\pm0.023}}       
\newcommand{\hatcurCCcitHmagxxxxxD}{\ensuremath{12.196\pm0.026}}       
\newcommand{\hatcurCCcitKmagxxxxxD}{\ensuremath{12.124\pm0.019}}       
\newcommand{\hatcurCCbbJmagxxxxxD}{\ensuremath{12.534\pm0.025}}        
\newcommand{\hatcurCCbbHmagxxxxxD}{\ensuremath{12.219\pm0.027}}        
\newcommand{\hatcurCCbbKmagxxxxxD}{\ensuremath{12.144\pm0.019}}        
\newcommand{\hatcurCCesoJmagxxxxxD}{\ensuremath{12.536\pm0.026}}       
\newcommand{\hatcurCCesoHmagxxxxxD}{\ensuremath{12.215\pm0.031}}       
\newcommand{\hatcurCCesoKmagxxxxxD}{\ensuremath{12.143\pm0.020}}       
\newcommand{\hatcurCCesoJHmagxxxxxD}{\ensuremath{0.321\pm0.039}}       
\newcommand{\hatcurCCesoJKmagxxxxxD}{\ensuremath{0.394\pm0.032}}       
\newcommand{\hatcurCCesoHKmagxxxxxD}{\ensuremath{0.072\pm0.037}}       
\newcommand{\hatcurLCdipxxxxxD}{\ensuremath{9.9}}                      
\newcommand{\hatcurLCrprstarxxxxxD}{\ensuremath{0.1120\pm0.0019}}      
\newcommand{\hatcurLCbsqxxxxxD}{\ensuremath{0.142_{-0.090}^{+0.099}}}  
\newcommand{\hatcurLCimpxxxxxD}{\ensuremath{0.38_{-0.15}^{+0.11}}}     
\newcommand{\hatcurLCzetaxxxxxD}{\ensuremath{19.42\pm0.17}}            
\newcommand{\hatcurLCdurxxxxxD}{\ensuremath{0.1164\pm0.0017}}          
\newcommand{\hatcurLCdurshortxxxxxD}{\ensuremath{0.1164}}              
\newcommand{\hatcurLCdurhrxxxxxD}{\ensuremath{2.793\pm0.040}}          
\newcommand{\hatcurLCdurhrshortxxxxxD}{\ensuremath{2.793}}             
\newcommand{\hatcurLCqxxxxxD}{\ensuremath{0.05930\pm0.00085}}          
\newcommand{\hatcurLCqshortxxxxxD}{\ensuremath{0.059}}                 
\newcommand{\hatcurLCingdurxxxxxD}{\ensuremath{0.0135\pm0.0016}}       
\newcommand{\hatcurLCPxxxxxD}{\ensuremath{1.9616241\pm0.0000039}}      
\newcommand{\hatcurLCPprecxxxxxD}{\ensuremath{1.9616241}}              
\newcommand{\hatcurLCPshortxxxxxD}{\ensuremath{1.9616}}                
\newcommand{\hatcurLCTxxxxxD}{\ensuremath{2455829.44781\pm0.00044}}    
\newcommand{\hatcurLCTAxxxxxD}{\ensuremath{2454348.4216\pm0.0030}}     
\newcommand{\hatcurLCTBxxxxxD}{\ensuremath{2455862.79543\pm0.00045}}   
\newcommand{\hatcurLChatnetmAxxxxxD}{\ensuremath{13.49185\pm0.00017}}  
\newcommand{\hatcurLCiblendAxxxxxD}{\ensuremath{0.730\pm0.055}}        
\newcommand{\hatcurLChatnetmBxxxxxD}{\ensuremath{13.31283\pm0.00029}}  
\newcommand{\hatcurLCiblendBxxxxxD}{\ensuremath{0.90\pm0.12}}          
\newcommand{\hatcurSMEiteffxxxxxD}{\ensuremath{5956\pm50}}             
\newcommand{\hatcurSMEizfehxxxxxD}{\ensuremath{0.000\pm0.080}}         
\newcommand{\hatcurSMEizfehshortxxxxxD}{\ensuremath{0.00}}             
\newcommand{\hatcurSMEiloggxxxxxD}{\ensuremath{4.32\pm0.10}}           
\newcommand{\hatcurSMEivsinxxxxxD}{\ensuremath{4.10\pm0.50}}           
\newcommand{\hatcurSMEivmacxxxxxD}{\ensuremath{0.0}}                   
\newcommand{\hatcurSMEivmicxxxxxD}{\ensuremath{0.0}}                   
\newcommand{\hatcurSMEiiteffxxxxxD}{\ensuremath{6290\pm100}}           
\newcommand{\hatcurSMEiizfehxxxxxD}{\ensuremath{-0.08\pm0.10}}         
\newcommand{\hatcurSMEiizfehshortxxxxxD}{\ensuremath{-0.08}}           
\newcommand{\hatcurSMEiiloggxxxxxD}{\ensuremath{3.81\pm0.10}}          
\newcommand{\hatcurSMEiivsinxxxxxD}{\ensuremath{12.00\pm0.50}}         
\newcommand{\hatcurSMEiivmacxxxxxD}{\ensuremath{4.82}}                 
\newcommand{\hatcurSMEiivmicxxxxxD}{\ensuremath{0.85}}                 
\newcommand{\hatcurTRESteffxxxxxD}{\ensuremath{5880\pm120}}            
\newcommand{\hatcurTRESzfehxxxxxD}{\ensuremath{0.00\pm0.10}}           
\newcommand{\hatcurTRESloggxxxxxD}{\ensuremath{4.25\pm0.25}}           
\newcommand{\hatcurTRESvsinixxxxxD}{\ensuremath{6.0\pm2.0}}            
\newcommand{\hatcurTRESgammaxxxxxD}{\ensuremath{-16.57\pm0.10}}        
\newcommand{\hatcurTRESnumspecxxxxxD}{\ensuremath{2}}                  
\newcommand{\hatcurTRESspanxxxxxD}{\ensuremath{NULL}}                  
\newcommand{\hatcurTRESrvrmsxxxxxD}{\ensuremath{0.00}}                 
\newcommand{\hatcurAPOteffxxxxxD}{\ensuremath{5916\pm50}}            
\newcommand{\hatcurAPOzfehxxxxxD}{\ensuremath{0.320\pm0.080}}        
\newcommand{\hatcurAPOloggxxxxxD}{\ensuremath{4.25\pm0.10}}          
\newcommand{\hatcurAPOvsinixxxxxD}{\ensuremath{4.7\pm1.0}}           
\newcommand{\hatcurAPOgammaxxxxxD}{\ensuremath{-16.9\pm1.0}}         
\newcommand{\hatcurAPOnumspecxxxxxD}{\ensuremath{3}}                 
\newcommand{\hatcurAPOspanxxxxxD}{\ensuremath{NULL}}                 
\newcommand{\hatcurAPOrvrmsxxxxxD}{\ensuremath{0.00}}                
\newcommand{\hatcurLBizxxxxxD}{\ensuremath{0.1829}}                    
\newcommand{\hatcurLBiizxxxxxD}{\ensuremath{0.3404}}                   
\newcommand{\hatcurLBiixxxxxD}{\ensuremath{0.2387}}                    
\newcommand{\hatcurLBiiixxxxxD}{\ensuremath{0.3447}}                   
\newcommand{\hatcurLBiIxxxxxD}{\ensuremath{0.2195}}                    
\newcommand{\hatcurLBiiIxxxxxD}{\ensuremath{0.3444}}                   
\newcommand{\hatcurLBigxxxxxD}{\ensuremath{0.5017}}                    
\newcommand{\hatcurLBiigxxxxxD}{\ensuremath{0.2734}}                   
\newcommand{\hatcurLBirxxxxxD}{\ensuremath{0.3197}}                    
\newcommand{\hatcurLBiirxxxxxD}{\ensuremath{0.3471}}                   
\newcommand{\hatcurLBiRxxxxxD}{\ensuremath{0.2971}}                    
\newcommand{\hatcurLBiiRxxxxxD}{\ensuremath{0.3476}}                   
\newcommand{\hatcurISOmxxxxxD}{\ensuremath{1.093\pm0.043}}             
\newcommand{\hatcurISOmshortxxxxxD}{\ensuremath{1.09}}                 
\newcommand{\hatcurISOmlongxxxxxD}{\ensuremath{1.093\pm0.043}}         
\newcommand{\hatcurISOrxxxxxD}{\ensuremath{1.209_{-0.062}^{+0.081}}}   
\newcommand{\hatcurISOrshortxxxxxD}{\ensuremath{1.21}}                 
\newcommand{\hatcurISOrlongxxxxxD}{\ensuremath{1.209_{-0.062}^{+0.081}}} 
\newcommand{\hatcurISOrhoxxxxxD}{\ensuremath{0.87\pm0.13}}             
\newcommand{\hatcurISOrholongxxxxxD}{\ensuremath{0.87\pm0.13}}         
\newcommand{\hatcurISOloggxxxxxD}{\ensuremath{4.310\pm0.043}}          
\newcommand{\hatcurISOlumxxxxxD}{\ensuremath{1.65_{-0.18}^{+0.24}}}    
\newcommand{\hatcurISOlumshortxxxxxD}{\ensuremath{1.65}}               
\newcommand{\hatcurISOmvxxxxxD}{\ensuremath{4.27\pm0.13}}              
\newcommand{\hatcurISOvixxxxxD}{\ensuremath{0.643\pm0.014}}            
\newcommand{\hatcurISOagexxxxxD}{\ensuremath{4.67_{-0.83}^{+1.45}}}    
\newcommand{\hatcurISOsigmaxxxxxD}{\ensuremath{0.00140\pm0.00015}}     
\newcommand{\hatcurISOMJxxxxxD}{\ensuremath{3.20\pm0.13}}              
\newcommand{\hatcurISOMHxxxxxD}{\ensuremath{2.89\pm0.12}}              
\newcommand{\hatcurISOMKxxxxxD}{\ensuremath{2.83\pm0.12}}              
\newcommand{\hatcurISOJKxxxxxD}{\ensuremath{0.370\pm0.010}}            
\newcommand{\hatcurISOspecxxxxxD}{G}                                   
\newcommand{\hatcurRVKxxxxxD}{\ensuremath{226.8\pm6.0}}                
\newcommand{\hatcurRVrkxxxxxD}{\ensuremath{0\pm0}}                     
\newcommand{\hatcurRVrhxxxxxD}{\ensuremath{0\pm0}}                     
\newcommand{\hatcurRVkxxxxxD}{\ensuremath{0\pm0}}                      
\newcommand{\hatcurRVhxxxxxD}{\ensuremath{0\pm0}}                      
\newcommand{\hatcurRVtronexxxxxD}{\ensuremath{0\pm0}}                  
\newcommand{\hatcurRVtrtwoxxxxxD}{\ensuremath{0\pm0}}                  
\newcommand{\hatcurRVgammaxxxxxD}{\ensuremath{-63.5\pm5.0}}            
\newcommand{\hatcurRVjitterxxxxxD}{\ensuremath{9.3\pm3.2}}             
\newcommand{\hatcurRVfitrmsxxxxxD}{\ensuremath{11.0}}                  
\newcommand{\hatcurRVeccenxxxxxD}{\ensuremath{0\pm0}}                  
\newcommand{\hatcurRVeccentwosiglimxxxxxD}{\ensuremath{<0.000}}        
\newcommand{\hatcurRVomegaxxxxxD}{\ensuremath{0\pm0}}                  
\newcommand{\hatcurPPixxxxxD}{\ensuremath{86.2\pm1.5}}                 
\newcommand{\hatcurPPgxxxxxD}{\ensuremath{21.1\pm2.6}}                 
\newcommand{\hatcurPPloggxxxxxD}{\ensuremath{3.325\pm0.055}}           
\newcommand{\hatcurPParxxxxxD}{\ensuremath{5.61\pm0.28}}               
\newcommand{\hatcurPParelxxxxxD}{\ensuremath{0.03159\pm0.00042}}       
\newcommand{\hatcurPPrhoxxxxxD}{\ensuremath{0.80\pm0.15}}              
\newcommand{\hatcurPPmxxxxxD}{\ensuremath{1.484\pm0.056}}              
\newcommand{\hatcurPPmshortxxxxxD}{\ensuremath{1.48}}                  
\newcommand{\hatcurPPmlongxxxxxD}{\ensuremath{1.484\pm0.056}}          
\newcommand{\hatcurPPmexxxxxD}{\ensuremath{472\pm18}}                  
\newcommand{\hatcurPPmeshortxxxxxD}{\ensuremath{471.6}}                
\newcommand{\hatcurPPmelongxxxxxD}{\ensuremath{472\pm18}}              
\newcommand{\hatcurPPrxxxxxD}{\ensuremath{1.318\pm0.091}}              
\newcommand{\hatcurPPrshortxxxxxD}{\ensuremath{1.32}}                  
\newcommand{\hatcurPPrlongxxxxxD}{\ensuremath{1.318\pm0.091}}          
\newcommand{\hatcurPPrexxxxxD}{\ensuremath{14.8\pm1.0}}                
\newcommand{\hatcurPPreshortxxxxxD}{\ensuremath{14.8}}                 
\newcommand{\hatcurPPrelongxxxxxD}{\ensuremath{14.8\pm1.0}}            
\newcommand{\hatcurPPmrcorrxxxxxD}{\ensuremath{0.34}}                  
\newcommand{\hatcurPPteffxxxxxD}{\ensuremath{1778\pm48}}               
\newcommand{\hatcurPPthetaxxxxxD}{\ensuremath{0.0649\pm0.0046}}        
\newcommand{\hatcurPPfluxperixxxxxD}{\ensuremath{2.26_{-0.22}^{+0.29}}} 
\newcommand{\hatcurPPfluxperidimxxxxxD}{\ensuremath{9}}                
\newcommand{\hatcurPPfluxapxxxxxD}{\ensuremath{2.26_{-0.22}^{+0.29}}}  
\newcommand{\hatcurPPfluxapdimxxxxxD}{\ensuremath{9}}                  
\newcommand{\hatcurPPfluxavgxxxxxD}{\ensuremath{2.26_{-0.22}^{+0.29}}} 
\newcommand{\hatcurPPfluxavgdimxxxxxD}{\ensuremath{9}}                 
\newcommand{\hatcurPPfluxavglogxxxxxD}{\ensuremath{9.354\pm0.046}}     
\newcommand{\hatcurXsecphasexxxxxD}{\ensuremath{0\pm0}}                
\newcommand{\hatcurXsecondaryxxxxxD}{\ensuremath{2455830.42863\pm0.00044}} 
\newcommand{\hatcurXsecdurxxxxxD}{\ensuremath{0.1164\pm0.0017}}        
\newcommand{\hatcurXsecingdurxxxxxD}{\ensuremath{0.0135\pm0.0016}}     
\newcommand{\hatcurPPphiconjxxxxxD}{\ensuremath{0\pm0}}                
\newcommand{\hatcurPPperixxxxxD}{\ensuremath{2455828.95741\pm0.00044}} 
\newcommand{\hatcurPPaequivxxxxxD}{\ensuremath{0.0246\pm0.0013}}       
\newcommand{\hatcurPPtcircxxxxxD}{\ensuremath{22.2\pm6.8}}             
\newcommand{\hatcurPPtinfallxxxxxD}{\ensuremath{76\pm18}}              
\newcommand{\hatcurXdistxxxxxD}{\ensuremath{727\pm43}}                 
\newcommand{\hatcurXAvxxxxxD}{\ensuremath{0.21\pm0.14}}                
\newcommand{\hatcurXdistredxxxxxD}{\ensuremath{719\pm43}}              
\newcommand{\hatcurXEBVxxxxxD}{\ensuremath{0.069\pm0.044}}             
\newcommand{\hatcurXmvisoredxxxxxD}{\ensuremath{13.77\pm0.11}}         
\newcommand{\hatcurXmiisoredxxxxxD}{\ensuremath{13.012\pm0.040}}       
\newcommand{\hatcurXmjisoredxxxxxD}{\ensuremath{12.545\pm0.019}}       
\newcommand{\hatcurXmhisoredxxxxxD}{\ensuremath{12.211\pm0.014}}       
\newcommand{\hatcurXmkisoredxxxxxD}{\ensuremath{12.142\pm0.017}}       
\newcommand{\hatcurXviisoredxxxxxD}{\ensuremath{0.755\pm0.069}}        
\newcommand{\hatcurXvkisoredxxxxxD}{\ensuremath{1.62\pm0.12}}          
\newcommand{\hatcurXjhisoredxxxxxD}{\ensuremath{0.335\pm0.013}}        
\newcommand{\hatcurXjkisoredxxxxxD}{\ensuremath{0.403\pm0.022}}        
\newcommand{\hatcurCCpmraxxxxxD}{\ensuremath{-1.1\pm1.8}}              
\newcommand{\hatcurCCpmdecxxxxxD}{\ensuremath{3.0\pm2.2}}              
\newcommand{\hatcurCCpmxxxxxD}{\ensuremath{3.2\pm2.8}}                 
\newcommand{\hatcurhtreccenxxxxxD}{HTR165-010}                         
\newcommand{\hatcurfieldeccenxxxxxD}{165}                              
\newcommand{\hatcurCCraeccenxxxxxD}{\ensuremath{01^{\mathrm h}27^{\mathrm m}29.05{\mathrm s}}}                       
\newcommand{\hatcurCCdececcenxxxxxD}{\ensuremath{+38{\arcdeg}58{\arcmin}05.3{\arcsec}}}                      
\newcommand{\hatcurCCmageccenxxxxxD}{13.733}                           
\newcommand{\hatcurCCtwomasseccenxxxxxD}{2MASS~01272906+3858053}       
\newcommand{\hatcurCCgsceccenxxxxxD}{GSC~2813-01266}                   
\newcommand{\hatcurCCtassmveccenxxxxxD}{\ensuremath{13.73\pm0.18}}     
\newcommand{\hatcurCCtassmvshorteccenxxxxxD}{\ensuremath{13.7}}        
\newcommand{\hatcurCCtassmBeccenxxxxxD}{\ensuremath{0\pm0}}            
\newcommand{\hatcurCCtassmBshorteccenxxxxxD}{\ensuremath{.1fym}}       
\newcommand{\hatcurCCtassmIeccenxxxxxD}{\ensuremath{13.13\pm0.12}}     
\newcommand{\hatcurCCtassmIshorteccenxxxxxD}{\ensuremath{13.1}}        
\newcommand{\hatcurCCtassmgeccenxxxxxD}{\ensuremath{16.959\pm0.017}}   
\newcommand{\hatcurCCtassmgshorteccenxxxxxD}{\ensuremath{17.0}}        
\newcommand{\hatcurCCtassmreccenxxxxxD}{\ensuremath{13.6150\pm0.0030}} 
\newcommand{\hatcurCCtassmrshorteccenxxxxxD}{\ensuremath{13.6}}        
\newcommand{\hatcurCCtassmieccenxxxxxD}{\ensuremath{13.5230\pm0.0020}} 
\newcommand{\hatcurCCtassmishorteccenxxxxxD}{\ensuremath{13.5}}        
\newcommand{\hatcurCCtwomassJmageccenxxxxxD}{\ensuremath{12.468\pm0.023}} 
\newcommand{\hatcurCCtwomassHmageccenxxxxxD}{\ensuremath{12.202\pm0.026}} 
\newcommand{\hatcurCCtwomassKmageccenxxxxxD}{\ensuremath{12.100\pm0.019}} 
\newcommand{\hatcurCCcitJmageccenxxxxxD}{\ensuremath{12.485\pm0.023}}  
\newcommand{\hatcurCCcitHmageccenxxxxxD}{\ensuremath{12.196\pm0.026}}  
\newcommand{\hatcurCCcitKmageccenxxxxxD}{\ensuremath{12.124\pm0.019}}  
\newcommand{\hatcurCCbbJmageccenxxxxxD}{\ensuremath{12.534\pm0.025}}   
\newcommand{\hatcurCCbbHmageccenxxxxxD}{\ensuremath{12.219\pm0.027}}   
\newcommand{\hatcurCCbbKmageccenxxxxxD}{\ensuremath{12.144\pm0.019}}   
\newcommand{\hatcurCCesoJmageccenxxxxxD}{\ensuremath{12.536\pm0.026}}  
\newcommand{\hatcurCCesoHmageccenxxxxxD}{\ensuremath{12.215\pm0.031}}  
\newcommand{\hatcurCCesoKmageccenxxxxxD}{\ensuremath{12.143\pm0.020}}  
\newcommand{\hatcurCCesoJHmageccenxxxxxD}{\ensuremath{0.321\pm0.039}}  
\newcommand{\hatcurCCesoJKmageccenxxxxxD}{\ensuremath{0.394\pm0.032}}  
\newcommand{\hatcurCCesoHKmageccenxxxxxD}{\ensuremath{0.072\pm0.037}}  
\newcommand{\hatcurLCdipeccenxxxxxD}{\ensuremath{9.9}}                 
\newcommand{\hatcurLCrprstareccenxxxxxD}{\ensuremath{0.1120\pm0.0019}} 
\newcommand{\hatcurLCbsqeccenxxxxxD}{\ensuremath{0.146_{-0.090}^{+0.094}}} 
\newcommand{\hatcurLCimpeccenxxxxxD}{\ensuremath{0.38_{-0.14}^{+0.11}}} 
\newcommand{\hatcurLCzetaeccenxxxxxD}{\ensuremath{19.40\pm0.17}}       
\newcommand{\hatcurLCdureccenxxxxxD}{\ensuremath{0.1165\pm0.0016}}     
\newcommand{\hatcurLCdurshorteccenxxxxxD}{\ensuremath{0.1165}}         
\newcommand{\hatcurLCdurhreccenxxxxxD}{\ensuremath{2.797\pm0.039}}     
\newcommand{\hatcurLCdurhrshorteccenxxxxxD}{\ensuremath{2.797}}        
\newcommand{\hatcurLCqeccenxxxxxD}{\ensuremath{0.05940\pm0.00082}}     
\newcommand{\hatcurLCqshorteccenxxxxxD}{\ensuremath{0.059}}            
\newcommand{\hatcurLCingdureccenxxxxxD}{\ensuremath{0.0135\pm0.0016}}  
\newcommand{\hatcurLCPeccenxxxxxD}{\ensuremath{1.9616242\pm0.0000037}} 
\newcommand{\hatcurLCPprececcenxxxxxD}{\ensuremath{1.9616242}}         
\newcommand{\hatcurLCPshorteccenxxxxxD}{\ensuremath{1.9616}}           
\newcommand{\hatcurLCTeccenxxxxxD}{\ensuremath{2455827.48617\pm0.00043}} 
\newcommand{\hatcurLCTAeccenxxxxxD}{\ensuremath{2454348.4216\pm0.0028}} 
\newcommand{\hatcurLCTBeccenxxxxxD}{\ensuremath{2455862.79540\pm0.00044}} 
\newcommand{\hatcurLChatnetmAeccenxxxxxD}{\ensuremath{13.49184\pm0.00017}} 
\newcommand{\hatcurLCiblendAeccenxxxxxD}{\ensuremath{0.736\pm0.053}}   
\newcommand{\hatcurLChatnetmBeccenxxxxxD}{\ensuremath{13.31283\pm0.00028}} 
\newcommand{\hatcurLCiblendBeccenxxxxxD}{\ensuremath{0.90\pm0.12}}     
\newcommand{\hatcurSMEiteffeccenxxxxxD}{\ensuremath{5956\pm50}}        
\newcommand{\hatcurSMEizfeheccenxxxxxD}{\ensuremath{0.000\pm0.080}}    
\newcommand{\hatcurSMEizfehshorteccenxxxxxD}{\ensuremath{0.00}}        
\newcommand{\hatcurSMEiloggeccenxxxxxD}{\ensuremath{4.32\pm0.10}}      
\newcommand{\hatcurSMEivsineccenxxxxxD}{\ensuremath{4.10\pm0.50}}      
\newcommand{\hatcurSMEivmaceccenxxxxxD}{\ensuremath{0.0}}              
\newcommand{\hatcurSMEivmiceccenxxxxxD}{\ensuremath{0.0}}              
\newcommand{\hatcurSMEiiteffeccenxxxxxD}{\ensuremath{6290\pm100}}      
\newcommand{\hatcurSMEiizfeheccenxxxxxD}{\ensuremath{-0.08\pm0.10}}    
\newcommand{\hatcurSMEiizfehshorteccenxxxxxD}{\ensuremath{-0.08}}      
\newcommand{\hatcurSMEiiloggeccenxxxxxD}{\ensuremath{3.81\pm0.10}}     
\newcommand{\hatcurSMEiivsineccenxxxxxD}{\ensuremath{12.00\pm0.50}}    
\newcommand{\hatcurSMEiivmaceccenxxxxxD}{\ensuremath{4.82}}            
\newcommand{\hatcurSMEiivmiceccenxxxxxD}{\ensuremath{0.85}}            
\newcommand{\hatcurTRESteffeccenxxxxxD}{\ensuremath{5880\pm120}}       
\newcommand{\hatcurTRESzfeheccenxxxxxD}{\ensuremath{0.00\pm0.10}}      
\newcommand{\hatcurTRESloggeccenxxxxxD}{\ensuremath{4.25\pm0.25}}      
\newcommand{\hatcurTRESvsinieccenxxxxxD}{\ensuremath{6.0\pm2.0}}       
\newcommand{\hatcurTRESgammaeccenxxxxxD}{\ensuremath{-16.57\pm0.10}}   
\newcommand{\hatcurTRESnumspececcenxxxxxD}{\ensuremath{2}}             
\newcommand{\hatcurTRESspaneccenxxxxxD}{\ensuremath{NULL}}             
\newcommand{\hatcurTRESrvrmseccenxxxxxD}{\ensuremath{0.00}}            
\newcommand{\hatcurARCESteffeccenxxxxxD}{\ensuremath{5916\pm50}}       
\newcommand{\hatcurARCESzfeheccenxxxxxD}{\ensuremath{0.320\pm0.080}}   
\newcommand{\hatcurARCESloggeccenxxxxxD}{\ensuremath{4.25\pm0.10}}     
\newcommand{\hatcurARCESvsinieccenxxxxxD}{\ensuremath{4.7\pm1.0}}      
\newcommand{\hatcurARCESgammaeccenxxxxxD}{\ensuremath{-16.9\pm1.0}}    
\newcommand{\hatcurARCESnumspececcenxxxxxD}{\ensuremath{3}}            
\newcommand{\hatcurARCESspaneccenxxxxxD}{\ensuremath{NULL}}            
\newcommand{\hatcurARCESrvrmseccenxxxxxD}{\ensuremath{0.00}}           
\newcommand{\hatcurLBizeccenxxxxxD}{\ensuremath{0.1829}}               
\newcommand{\hatcurLBiizeccenxxxxxD}{\ensuremath{0.3404}}              
\newcommand{\hatcurLBiieccenxxxxxD}{\ensuremath{0.2387}}               
\newcommand{\hatcurLBiiieccenxxxxxD}{\ensuremath{0.3447}}              
\newcommand{\hatcurLBiIeccenxxxxxD}{\ensuremath{0.2195}}               
\newcommand{\hatcurLBiiIeccenxxxxxD}{\ensuremath{0.3444}}              
\newcommand{\hatcurLBigeccenxxxxxD}{\ensuremath{0.5017}}               
\newcommand{\hatcurLBiigeccenxxxxxD}{\ensuremath{0.2734}}              
\newcommand{\hatcurLBireccenxxxxxD}{\ensuremath{0.3197}}               
\newcommand{\hatcurLBiireccenxxxxxD}{\ensuremath{0.3471}}              
\newcommand{\hatcurLBiReccenxxxxxD}{\ensuremath{0.2971}}               
\newcommand{\hatcurLBiiReccenxxxxxD}{\ensuremath{0.3476}}              
\newcommand{\hatcurISOmeccenxxxxxD}{\ensuremath{1.083\pm0.039}}        
\newcommand{\hatcurISOmshorteccenxxxxxD}{\ensuremath{1.08}}            
\newcommand{\hatcurISOmlongeccenxxxxxD}{\ensuremath{1.083\pm0.039}}    
\newcommand{\hatcurISOreccenxxxxxD}{\ensuremath{1.134\pm0.070}}        
\newcommand{\hatcurISOrshorteccenxxxxxD}{\ensuremath{1.13}}            
\newcommand{\hatcurISOrlongeccenxxxxxD}{\ensuremath{1.134\pm0.070}}    
\newcommand{\hatcurISOrhoeccenxxxxxD}{\ensuremath{1.04\pm0.17}}        
\newcommand{\hatcurISOrholongeccenxxxxxD}{\ensuremath{1.04\pm0.17}}    
\newcommand{\hatcurISOloggeccenxxxxxD}{\ensuremath{4.361\pm0.049}}     
\newcommand{\hatcurISOlumeccenxxxxxD}{\ensuremath{1.45\pm0.20}}        
\newcommand{\hatcurISOlumshorteccenxxxxxD}{\ensuremath{1.45}}          
\newcommand{\hatcurISOmveccenxxxxxD}{\ensuremath{4.40\pm0.14}}         
\newcommand{\hatcurISOvieccenxxxxxD}{\ensuremath{0.643\pm0.014}}       
\newcommand{\hatcurISOageeccenxxxxxD}{\ensuremath{4.0\pm1.3}}          
\newcommand{\hatcurISOsigmaeccenxxxxxD}{\ensuremath{0.00150\pm0.00019}} 
\newcommand{\hatcurISOMJeccenxxxxxD}{\ensuremath{3.34\pm0.14}}         
\newcommand{\hatcurISOMHeccenxxxxxD}{\ensuremath{3.02\pm0.13}}         
\newcommand{\hatcurISOMKeccenxxxxxD}{\ensuremath{2.97\pm0.13}}         
\newcommand{\hatcurISOJKeccenxxxxxD}{\ensuremath{0.370\pm0.010}}       
\newcommand{\hatcurISOspececcenxxxxxD}{G}                              
\newcommand{\hatcurRVKeccenxxxxxD}{\ensuremath{227.4\pm4.5}}           
\newcommand{\hatcurRVrkeccenxxxxxD}{\ensuremath{-0.09\pm0.13}}         
\newcommand{\hatcurRVrheccenxxxxxD}{\ensuremath{-0.236_{-0.063}^{+0.083}}} 
\newcommand{\hatcurRVkeccenxxxxxD}{\ensuremath{-0.021\pm0.041}}        
\newcommand{\hatcurRVheccenxxxxxD}{\ensuremath{-0.063\pm0.030}}        
\newcommand{\hatcurRVtroneeccenxxxxxD}{\ensuremath{0\pm0}}             
\newcommand{\hatcurRVtrtwoeccenxxxxxD}{\ensuremath{0\pm0}}             
\newcommand{\hatcurRVgammaeccenxxxxxD}{\ensuremath{-60.2\pm10.0}}      
\newcommand{\hatcurRVjittereccenxxxxxD}{\ensuremath{4.1\pm2.8}}        
\newcommand{\hatcurRVfitrmseccenxxxxxD}{\ensuremath{10.0}}             
\newcommand{\hatcurRVecceneccenxxxxxD}{\ensuremath{0.076\pm0.030}}     
\newcommand{\hatcurRVeccentwosiglimeccenxxxxxD}{\ensuremath{<0.134}}   
\newcommand{\hatcurRVomegaeccenxxxxxD}{\ensuremath{250\pm31}}          
\newcommand{\hatcurPPieccenxxxxxD}{\ensuremath{86.6\pm1.3}}            
\newcommand{\hatcurPPgeccenxxxxxD}{\ensuremath{23.9\pm3.2}}            
\newcommand{\hatcurPPloggeccenxxxxxD}{\ensuremath{3.378\pm0.060}}      
\newcommand{\hatcurPPareccenxxxxxD}{\ensuremath{5.96\pm0.34}}          
\newcommand{\hatcurPPareleccenxxxxxD}{\ensuremath{0.03149\pm0.00038}}  
\newcommand{\hatcurPPrhoeccenxxxxxD}{\ensuremath{0.97\pm0.20}}         
\newcommand{\hatcurPPmeccenxxxxxD}{\ensuremath{1.476\pm0.044}}         
\newcommand{\hatcurPPmshorteccenxxxxxD}{\ensuremath{1.48}}             
\newcommand{\hatcurPPmlongeccenxxxxxD}{\ensuremath{1.476\pm0.044}}     
\newcommand{\hatcurPPmeeccenxxxxxD}{\ensuremath{469\pm14}}             
\newcommand{\hatcurPPmeshorteccenxxxxxD}{\ensuremath{468.9}}           
\newcommand{\hatcurPPmelongeccenxxxxxD}{\ensuremath{469\pm14}}         
\newcommand{\hatcurPPreccenxxxxxD}{\ensuremath{1.235\pm0.089}}         
\newcommand{\hatcurPPrshorteccenxxxxxD}{\ensuremath{1.24}}             
\newcommand{\hatcurPPrlongeccenxxxxxD}{\ensuremath{1.235\pm0.089}}     
\newcommand{\hatcurPPreeccenxxxxxD}{\ensuremath{13.8\pm1.0}}           
\newcommand{\hatcurPPreshorteccenxxxxxD}{\ensuremath{13.8}}            
\newcommand{\hatcurPPrelongeccenxxxxxD}{\ensuremath{13.8\pm1.0}}       
\newcommand{\hatcurPPmrcorreccenxxxxxD}{\ensuremath{0.22}}             
\newcommand{\hatcurPPteffeccenxxxxxD}{\ensuremath{1727\pm51}}          
\newcommand{\hatcurPPthetaeccenxxxxxD}{\ensuremath{0.0693\pm0.0052}}   
\newcommand{\hatcurPPfluxperieccenxxxxxD}{\ensuremath{2.36_{-0.24}^{+0.32}}} 
\newcommand{\hatcurPPfluxperidimeccenxxxxxD}{\ensuremath{9}}           
\newcommand{\hatcurPPfluxapeccenxxxxxD}{\ensuremath{1.73\pm0.26}}      
\newcommand{\hatcurPPfluxapdimeccenxxxxxD}{\ensuremath{9}}             
\newcommand{\hatcurPPfluxavgeccenxxxxxD}{\ensuremath{2.01\pm0.25}}     
\newcommand{\hatcurPPfluxavgdimeccenxxxxxD}{\ensuremath{9}}            
\newcommand{\hatcurPPfluxavglogeccenxxxxxD}{\ensuremath{9.303\pm0.051}} 
\newcommand{\hatcurXsecphaseeccenxxxxxD}{\ensuremath{0.487\pm0.026}}   
\newcommand{\hatcurXsecondaryeccenxxxxxD}{\ensuremath{2455828.441\pm0.051}} 
\newcommand{\hatcurXsecdureccenxxxxxD}{\ensuremath{0.1045\pm0.0060}}   
\newcommand{\hatcurXsecingdureccenxxxxxD}{\ensuremath{0.0116\pm0.0014}} 
\newcommand{\hatcurPPphiconjeccenxxxxxD}{\ensuremath{-0.35_{-0.11}^{+0.80}}} 
\newcommand{\hatcurPPperieccenxxxxxD}{\ensuremath{2455828.17\pm0.85}}  
\newcommand{\hatcurPPaequiveccenxxxxxD}{\ensuremath{0.0261\pm0.0015}}  
\newcommand{\hatcurPPtcirceccenxxxxxD}{\ensuremath{28.9\pm9.4}}        
\newcommand{\hatcurPPtinfalleccenxxxxxD}{\ensuremath{103\pm29}}        
\newcommand{\hatcurXdisteccenxxxxxD}{\ensuremath{682\pm44}}            
\newcommand{\hatcurXAveccenxxxxxD}{\ensuremath{0.21\pm0.14}}           
\newcommand{\hatcurXdistredeccenxxxxxD}{\ensuremath{674\pm43}}         
\newcommand{\hatcurXEBVeccenxxxxxD}{\ensuremath{0.069\pm0.044}}        
\newcommand{\hatcurXmvisoredeccenxxxxxD}{\ensuremath{13.77\pm0.11}}    
\newcommand{\hatcurXmiisoredeccenxxxxxD}{\ensuremath{13.011\pm0.041}}  
\newcommand{\hatcurXmjisoredeccenxxxxxD}{\ensuremath{12.546\pm0.019}}  
\newcommand{\hatcurXmhisoredeccenxxxxxD}{\ensuremath{12.210\pm0.014}}  
\newcommand{\hatcurXmkisoredeccenxxxxxD}{\ensuremath{12.142\pm0.017}}  
\newcommand{\hatcurXviisoredeccenxxxxxD}{\ensuremath{0.755\pm0.069}}   
\newcommand{\hatcurXvkisoredeccenxxxxxD}{\ensuremath{1.62\pm0.12}}     
\newcommand{\hatcurXjhisoredeccenxxxxxD}{\ensuremath{0.335\pm0.013}}   
\newcommand{\hatcurXjkisoredeccenxxxxxD}{\ensuremath{0.403\pm0.022}}   
\newcommand{\hatcurCCpmraeccenxxxxxD}{\ensuremath{-1.1\pm1.8}}         
\newcommand{\hatcurCCpmdececcenxxxxxD}{\ensuremath{3.0\pm2.2}}         
\newcommand{\hatcurCCpmeccenxxxxxD}{\ensuremath{3.2\pm2.8}}            

\newcommand{\hatcurAPOgamma}[1]{\ifnum#1=51 %
\hatcurAPOgammaxxxxxB
\else
\ifnum#1=53 %
\hatcurAPOgammaxxxxxD
\else
??????\fi
\fi
}
\newcommand{\hatcurAPOlogg}[1]{\ifnum#1=51 %
\hatcurAPOloggxxxxxB
\else
\ifnum#1=53 %
\hatcurAPOloggxxxxxD
\else
??????\fi
\fi
}
\newcommand{\hatcurAPOnumspec}[1]{\ifnum#1=51 %
\hatcurAPOnumspecxxxxxB
\else
\ifnum#1=53 %
\hatcurAPOnumspecxxxxxD
\else
??????\fi
\fi
}
\newcommand{\hatcurAPOrvrms}[1]{\ifnum#1=51 %
\hatcurAPOrvrmsxxxxxB
\else
\ifnum#1=53 %
\hatcurAPOrvrmsxxxxxD
\else
??????\fi
\fi
}
\newcommand{\hatcurAPOspan}[1]{\ifnum#1=51 %
\hatcurAPOspanxxxxxB
\else
\ifnum#1=53 %
\hatcurAPOspanxxxxxD
\else
??????\fi
\fi
}
\newcommand{\hatcurAPOteff}[1]{\ifnum#1=51 %
\hatcurAPOteffxxxxxB
\else
\ifnum#1=53 %
\hatcurAPOteffxxxxxD
\else
??????\fi
\fi
}
\newcommand{\hatcurAPOvsini}[1]{\ifnum#1=51 %
\hatcurAPOvsinixxxxxB
\else
\ifnum#1=53 %
\hatcurAPOvsinixxxxxD
\else
??????\fi
\fi
}
\newcommand{\hatcurAPOzfeh}[1]{\ifnum#1=51 %
\hatcurAPOzfehxxxxxB
\else
\ifnum#1=53 %
\hatcurAPOzfehxxxxxD
\else
??????\fi
\fi
}
\newcommand{\hatcurARCESgammaeccen}[1]{\ifnum#1=51 %
\hatcurARCESgammaeccenxxxxxB
\else
\ifnum#1=53 %
\hatcurARCESgammaeccenxxxxxD
\else
??????\fi
\fi
}
\newcommand{\hatcurARCESloggeccen}[1]{\ifnum#1=51 %
\hatcurARCESloggeccenxxxxxB
\else
\ifnum#1=53 %
\hatcurARCESloggeccenxxxxxD
\else
??????\fi
\fi
}
\newcommand{\hatcurARCESnumspececcen}[1]{\ifnum#1=51 %
\hatcurARCESnumspececcenxxxxxB
\else
\ifnum#1=53 %
\hatcurARCESnumspececcenxxxxxD
\else
??????\fi
\fi
}
\newcommand{\hatcurARCESrvrmseccen}[1]{\ifnum#1=51 %
\hatcurARCESrvrmseccenxxxxxB
\else
\ifnum#1=53 %
\hatcurARCESrvrmseccenxxxxxD
\else
??????\fi
\fi
}
\newcommand{\hatcurARCESspaneccen}[1]{\ifnum#1=51 %
\hatcurARCESspaneccenxxxxxB
\else
\ifnum#1=53 %
\hatcurARCESspaneccenxxxxxD
\else
??????\fi
\fi
}
\newcommand{\hatcurARCESteffeccen}[1]{\ifnum#1=51 %
\hatcurARCESteffeccenxxxxxB
\else
\ifnum#1=53 %
\hatcurARCESteffeccenxxxxxD
\else
??????\fi
\fi
}
\newcommand{\hatcurARCESvsinieccen}[1]{\ifnum#1=51 %
\hatcurARCESvsinieccenxxxxxB
\else
\ifnum#1=53 %
\hatcurARCESvsinieccenxxxxxD
\else
??????\fi
\fi
}
\newcommand{\hatcurARCESzfeheccen}[1]{\ifnum#1=51 %
\hatcurARCESzfeheccenxxxxxB
\else
\ifnum#1=53 %
\hatcurARCESzfeheccenxxxxxD
\else
??????\fi
\fi
}
\newcommand{\hatcurCCbbHmag}[1]{\ifnum#1=50 %
\hatcurCCbbHmagxxxxxA
\else
\ifnum#1=51 %
\hatcurCCbbHmagxxxxxB
\else
\ifnum#1=52 %
\hatcurCCbbHmagxxxxxC
\else
\ifnum#1=53 %
\hatcurCCbbHmagxxxxxD
\else
??????\fi
\fi
\fi
\fi
}
\newcommand{\hatcurCCbbHmageccen}[1]{\ifnum#1=50 %
\hatcurCCbbHmageccenxxxxxA
\else
\ifnum#1=51 %
\hatcurCCbbHmageccenxxxxxB
\else
\ifnum#1=52 %
\hatcurCCbbHmageccenxxxxxC
\else
\ifnum#1=53 %
\hatcurCCbbHmageccenxxxxxD
\else
??????\fi
\fi
\fi
\fi
}
\newcommand{\hatcurCCbbJmag}[1]{\ifnum#1=50 %
\hatcurCCbbJmagxxxxxA
\else
\ifnum#1=51 %
\hatcurCCbbJmagxxxxxB
\else
\ifnum#1=52 %
\hatcurCCbbJmagxxxxxC
\else
\ifnum#1=53 %
\hatcurCCbbJmagxxxxxD
\else
??????\fi
\fi
\fi
\fi
}
\newcommand{\hatcurCCbbJmageccen}[1]{\ifnum#1=50 %
\hatcurCCbbJmageccenxxxxxA
\else
\ifnum#1=51 %
\hatcurCCbbJmageccenxxxxxB
\else
\ifnum#1=52 %
\hatcurCCbbJmageccenxxxxxC
\else
\ifnum#1=53 %
\hatcurCCbbJmageccenxxxxxD
\else
??????\fi
\fi
\fi
\fi
}
\newcommand{\hatcurCCbbKmag}[1]{\ifnum#1=50 %
\hatcurCCbbKmagxxxxxA
\else
\ifnum#1=51 %
\hatcurCCbbKmagxxxxxB
\else
\ifnum#1=52 %
\hatcurCCbbKmagxxxxxC
\else
\ifnum#1=53 %
\hatcurCCbbKmagxxxxxD
\else
??????\fi
\fi
\fi
\fi
}
\newcommand{\hatcurCCbbKmageccen}[1]{\ifnum#1=50 %
\hatcurCCbbKmageccenxxxxxA
\else
\ifnum#1=51 %
\hatcurCCbbKmageccenxxxxxB
\else
\ifnum#1=52 %
\hatcurCCbbKmageccenxxxxxC
\else
\ifnum#1=53 %
\hatcurCCbbKmageccenxxxxxD
\else
??????\fi
\fi
\fi
\fi
}
\newcommand{\hatcurCCcitHmag}[1]{\ifnum#1=50 %
\hatcurCCcitHmagxxxxxA
\else
\ifnum#1=51 %
\hatcurCCcitHmagxxxxxB
\else
\ifnum#1=52 %
\hatcurCCcitHmagxxxxxC
\else
\ifnum#1=53 %
\hatcurCCcitHmagxxxxxD
\else
??????\fi
\fi
\fi
\fi
}
\newcommand{\hatcurCCcitHmageccen}[1]{\ifnum#1=50 %
\hatcurCCcitHmageccenxxxxxA
\else
\ifnum#1=51 %
\hatcurCCcitHmageccenxxxxxB
\else
\ifnum#1=52 %
\hatcurCCcitHmageccenxxxxxC
\else
\ifnum#1=53 %
\hatcurCCcitHmageccenxxxxxD
\else
??????\fi
\fi
\fi
\fi
}
\newcommand{\hatcurCCcitJmag}[1]{\ifnum#1=50 %
\hatcurCCcitJmagxxxxxA
\else
\ifnum#1=51 %
\hatcurCCcitJmagxxxxxB
\else
\ifnum#1=52 %
\hatcurCCcitJmagxxxxxC
\else
\ifnum#1=53 %
\hatcurCCcitJmagxxxxxD
\else
??????\fi
\fi
\fi
\fi
}
\newcommand{\hatcurCCcitJmageccen}[1]{\ifnum#1=50 %
\hatcurCCcitJmageccenxxxxxA
\else
\ifnum#1=51 %
\hatcurCCcitJmageccenxxxxxB
\else
\ifnum#1=52 %
\hatcurCCcitJmageccenxxxxxC
\else
\ifnum#1=53 %
\hatcurCCcitJmageccenxxxxxD
\else
??????\fi
\fi
\fi
\fi
}
\newcommand{\hatcurCCcitKmag}[1]{\ifnum#1=50 %
\hatcurCCcitKmagxxxxxA
\else
\ifnum#1=51 %
\hatcurCCcitKmagxxxxxB
\else
\ifnum#1=52 %
\hatcurCCcitKmagxxxxxC
\else
\ifnum#1=53 %
\hatcurCCcitKmagxxxxxD
\else
??????\fi
\fi
\fi
\fi
}
\newcommand{\hatcurCCcitKmageccen}[1]{\ifnum#1=50 %
\hatcurCCcitKmageccenxxxxxA
\else
\ifnum#1=51 %
\hatcurCCcitKmageccenxxxxxB
\else
\ifnum#1=52 %
\hatcurCCcitKmageccenxxxxxC
\else
\ifnum#1=53 %
\hatcurCCcitKmageccenxxxxxD
\else
??????\fi
\fi
\fi
\fi
}
\newcommand{\hatcurCCdec}[1]{\ifnum#1=50 %
\hatcurCCdecxxxxxA
\else
\ifnum#1=51 %
\hatcurCCdecxxxxxB
\else
\ifnum#1=52 %
\hatcurCCdecxxxxxC
\else
\ifnum#1=53 %
\hatcurCCdecxxxxxD
\else
??????\fi
\fi
\fi
\fi
}
\newcommand{\hatcurCCdececcen}[1]{\ifnum#1=50 %
\hatcurCCdececcenxxxxxA
\else
\ifnum#1=51 %
\hatcurCCdececcenxxxxxB
\else
\ifnum#1=52 %
\hatcurCCdececcenxxxxxC
\else
\ifnum#1=53 %
\hatcurCCdececcenxxxxxD
\else
??????\fi
\fi
\fi
\fi
}
\newcommand{\hatcurCCesoHKmag}[1]{\ifnum#1=50 %
\hatcurCCesoHKmagxxxxxA
\else
\ifnum#1=51 %
\hatcurCCesoHKmagxxxxxB
\else
\ifnum#1=52 %
\hatcurCCesoHKmagxxxxxC
\else
\ifnum#1=53 %
\hatcurCCesoHKmagxxxxxD
\else
??????\fi
\fi
\fi
\fi
}
\newcommand{\hatcurCCesoHKmageccen}[1]{\ifnum#1=50 %
\hatcurCCesoHKmageccenxxxxxA
\else
\ifnum#1=51 %
\hatcurCCesoHKmageccenxxxxxB
\else
\ifnum#1=52 %
\hatcurCCesoHKmageccenxxxxxC
\else
\ifnum#1=53 %
\hatcurCCesoHKmageccenxxxxxD
\else
??????\fi
\fi
\fi
\fi
}
\newcommand{\hatcurCCesoHmag}[1]{\ifnum#1=50 %
\hatcurCCesoHmagxxxxxA
\else
\ifnum#1=51 %
\hatcurCCesoHmagxxxxxB
\else
\ifnum#1=52 %
\hatcurCCesoHmagxxxxxC
\else
\ifnum#1=53 %
\hatcurCCesoHmagxxxxxD
\else
??????\fi
\fi
\fi
\fi
}
\newcommand{\hatcurCCesoHmageccen}[1]{\ifnum#1=50 %
\hatcurCCesoHmageccenxxxxxA
\else
\ifnum#1=51 %
\hatcurCCesoHmageccenxxxxxB
\else
\ifnum#1=52 %
\hatcurCCesoHmageccenxxxxxC
\else
\ifnum#1=53 %
\hatcurCCesoHmageccenxxxxxD
\else
??????\fi
\fi
\fi
\fi
}
\newcommand{\hatcurCCesoJHmag}[1]{\ifnum#1=50 %
\hatcurCCesoJHmagxxxxxA
\else
\ifnum#1=51 %
\hatcurCCesoJHmagxxxxxB
\else
\ifnum#1=52 %
\hatcurCCesoJHmagxxxxxC
\else
\ifnum#1=53 %
\hatcurCCesoJHmagxxxxxD
\else
??????\fi
\fi
\fi
\fi
}
\newcommand{\hatcurCCesoJHmageccen}[1]{\ifnum#1=50 %
\hatcurCCesoJHmageccenxxxxxA
\else
\ifnum#1=51 %
\hatcurCCesoJHmageccenxxxxxB
\else
\ifnum#1=52 %
\hatcurCCesoJHmageccenxxxxxC
\else
\ifnum#1=53 %
\hatcurCCesoJHmageccenxxxxxD
\else
??????\fi
\fi
\fi
\fi
}
\newcommand{\hatcurCCesoJKmag}[1]{\ifnum#1=50 %
\hatcurCCesoJKmagxxxxxA
\else
\ifnum#1=51 %
\hatcurCCesoJKmagxxxxxB
\else
\ifnum#1=52 %
\hatcurCCesoJKmagxxxxxC
\else
\ifnum#1=53 %
\hatcurCCesoJKmagxxxxxD
\else
??????\fi
\fi
\fi
\fi
}
\newcommand{\hatcurCCesoJKmageccen}[1]{\ifnum#1=50 %
\hatcurCCesoJKmageccenxxxxxA
\else
\ifnum#1=51 %
\hatcurCCesoJKmageccenxxxxxB
\else
\ifnum#1=52 %
\hatcurCCesoJKmageccenxxxxxC
\else
\ifnum#1=53 %
\hatcurCCesoJKmageccenxxxxxD
\else
??????\fi
\fi
\fi
\fi
}
\newcommand{\hatcurCCesoJmag}[1]{\ifnum#1=50 %
\hatcurCCesoJmagxxxxxA
\else
\ifnum#1=51 %
\hatcurCCesoJmagxxxxxB
\else
\ifnum#1=52 %
\hatcurCCesoJmagxxxxxC
\else
\ifnum#1=53 %
\hatcurCCesoJmagxxxxxD
\else
??????\fi
\fi
\fi
\fi
}
\newcommand{\hatcurCCesoJmageccen}[1]{\ifnum#1=50 %
\hatcurCCesoJmageccenxxxxxA
\else
\ifnum#1=51 %
\hatcurCCesoJmageccenxxxxxB
\else
\ifnum#1=52 %
\hatcurCCesoJmageccenxxxxxC
\else
\ifnum#1=53 %
\hatcurCCesoJmageccenxxxxxD
\else
??????\fi
\fi
\fi
\fi
}
\newcommand{\hatcurCCesoKmag}[1]{\ifnum#1=50 %
\hatcurCCesoKmagxxxxxA
\else
\ifnum#1=51 %
\hatcurCCesoKmagxxxxxB
\else
\ifnum#1=52 %
\hatcurCCesoKmagxxxxxC
\else
\ifnum#1=53 %
\hatcurCCesoKmagxxxxxD
\else
??????\fi
\fi
\fi
\fi
}
\newcommand{\hatcurCCesoKmageccen}[1]{\ifnum#1=50 %
\hatcurCCesoKmageccenxxxxxA
\else
\ifnum#1=51 %
\hatcurCCesoKmageccenxxxxxB
\else
\ifnum#1=52 %
\hatcurCCesoKmageccenxxxxxC
\else
\ifnum#1=53 %
\hatcurCCesoKmageccenxxxxxD
\else
??????\fi
\fi
\fi
\fi
}
\newcommand{\hatcurCCgsc}[1]{\ifnum#1=50 %
\hatcurCCgscxxxxxA
\else
\ifnum#1=51 %
\hatcurCCgscxxxxxB
\else
\ifnum#1=52 %
\hatcurCCgscxxxxxC
\else
\ifnum#1=53 %
\hatcurCCgscxxxxxD
\else
??????\fi
\fi
\fi
\fi
}
\newcommand{\hatcurCCgsceccen}[1]{\ifnum#1=50 %
\hatcurCCgsceccenxxxxxA
\else
\ifnum#1=51 %
\hatcurCCgsceccenxxxxxB
\else
\ifnum#1=52 %
\hatcurCCgsceccenxxxxxC
\else
\ifnum#1=53 %
\hatcurCCgsceccenxxxxxD
\else
??????\fi
\fi
\fi
\fi
}
\newcommand{\hatcurCCmag}[1]{\ifnum#1=50 %
\hatcurCCmagxxxxxA
\else
\ifnum#1=51 %
\hatcurCCmagxxxxxB
\else
\ifnum#1=52 %
\hatcurCCmagxxxxxC
\else
\ifnum#1=53 %
\hatcurCCmagxxxxxD
\else
??????\fi
\fi
\fi
\fi
}
\newcommand{\hatcurCCmageccen}[1]{\ifnum#1=50 %
\hatcurCCmageccenxxxxxA
\else
\ifnum#1=51 %
\hatcurCCmageccenxxxxxB
\else
\ifnum#1=52 %
\hatcurCCmageccenxxxxxC
\else
\ifnum#1=53 %
\hatcurCCmageccenxxxxxD
\else
??????\fi
\fi
\fi
\fi
}
\newcommand{\hatcurCCpm}[1]{\ifnum#1=50 %
\hatcurCCpmxxxxxA
\else
\ifnum#1=51 %
\hatcurCCpmxxxxxB
\else
\ifnum#1=52 %
\hatcurCCpmxxxxxC
\else
\ifnum#1=53 %
\hatcurCCpmxxxxxD
\else
??????\fi
\fi
\fi
\fi
}
\newcommand{\hatcurCCpmdec}[1]{\ifnum#1=50 %
\hatcurCCpmdecxxxxxA
\else
\ifnum#1=51 %
\hatcurCCpmdecxxxxxB
\else
\ifnum#1=52 %
\hatcurCCpmdecxxxxxC
\else
\ifnum#1=53 %
\hatcurCCpmdecxxxxxD
\else
??????\fi
\fi
\fi
\fi
}
\newcommand{\hatcurCCpmdececcen}[1]{\ifnum#1=50 %
\hatcurCCpmdececcenxxxxxA
\else
\ifnum#1=51 %
\hatcurCCpmdececcenxxxxxB
\else
\ifnum#1=52 %
\hatcurCCpmdececcenxxxxxC
\else
\ifnum#1=53 %
\hatcurCCpmdececcenxxxxxD
\else
??????\fi
\fi
\fi
\fi
}
\newcommand{\hatcurCCpmeccen}[1]{\ifnum#1=50 %
\hatcurCCpmeccenxxxxxA
\else
\ifnum#1=51 %
\hatcurCCpmeccenxxxxxB
\else
\ifnum#1=52 %
\hatcurCCpmeccenxxxxxC
\else
\ifnum#1=53 %
\hatcurCCpmeccenxxxxxD
\else
??????\fi
\fi
\fi
\fi
}
\newcommand{\hatcurCCpmra}[1]{\ifnum#1=50 %
\hatcurCCpmraxxxxxA
\else
\ifnum#1=51 %
\hatcurCCpmraxxxxxB
\else
\ifnum#1=52 %
\hatcurCCpmraxxxxxC
\else
\ifnum#1=53 %
\hatcurCCpmraxxxxxD
\else
??????\fi
\fi
\fi
\fi
}
\newcommand{\hatcurCCpmraeccen}[1]{\ifnum#1=50 %
\hatcurCCpmraeccenxxxxxA
\else
\ifnum#1=51 %
\hatcurCCpmraeccenxxxxxB
\else
\ifnum#1=52 %
\hatcurCCpmraeccenxxxxxC
\else
\ifnum#1=53 %
\hatcurCCpmraeccenxxxxxD
\else
??????\fi
\fi
\fi
\fi
}
\newcommand{\hatcurCCra}[1]{\ifnum#1=50 %
\hatcurCCraxxxxxA
\else
\ifnum#1=51 %
\hatcurCCraxxxxxB
\else
\ifnum#1=52 %
\hatcurCCraxxxxxC
\else
\ifnum#1=53 %
\hatcurCCraxxxxxD
\else
??????\fi
\fi
\fi
\fi
}
\newcommand{\hatcurCCraeccen}[1]{\ifnum#1=50 %
\hatcurCCraeccenxxxxxA
\else
\ifnum#1=51 %
\hatcurCCraeccenxxxxxB
\else
\ifnum#1=52 %
\hatcurCCraeccenxxxxxC
\else
\ifnum#1=53 %
\hatcurCCraeccenxxxxxD
\else
??????\fi
\fi
\fi
\fi
}
\newcommand{\hatcurCCtassmB}[1]{\ifnum#1=50 %
\hatcurCCtassmBxxxxxA
\else
\ifnum#1=51 %
\hatcurCCtassmBxxxxxB
\else
\ifnum#1=52 %
\hatcurCCtassmBxxxxxC
\else
\ifnum#1=53 %
\hatcurCCtassmBxxxxxD
\else
??????\fi
\fi
\fi
\fi
}
\newcommand{\hatcurCCtassmBeccen}[1]{\ifnum#1=50 %
\hatcurCCtassmBeccenxxxxxA
\else
\ifnum#1=51 %
\hatcurCCtassmBeccenxxxxxB
\else
\ifnum#1=52 %
\hatcurCCtassmBeccenxxxxxC
\else
\ifnum#1=53 %
\hatcurCCtassmBeccenxxxxxD
\else
??????\fi
\fi
\fi
\fi
}
\newcommand{\hatcurCCtassmBshort}[1]{\ifnum#1=50 %
\hatcurCCtassmBshortxxxxxA
\else
\ifnum#1=51 %
\hatcurCCtassmBshortxxxxxB
\else
\ifnum#1=52 %
\hatcurCCtassmBshortxxxxxC
\else
\ifnum#1=53 %
\hatcurCCtassmBshortxxxxxD
\else
??????\fi
\fi
\fi
\fi
}
\newcommand{\hatcurCCtassmBshorteccen}[1]{\ifnum#1=50 %
\hatcurCCtassmBshorteccenxxxxxA
\else
\ifnum#1=51 %
\hatcurCCtassmBshorteccenxxxxxB
\else
\ifnum#1=52 %
\hatcurCCtassmBshorteccenxxxxxC
\else
\ifnum#1=53 %
\hatcurCCtassmBshorteccenxxxxxD
\else
??????\fi
\fi
\fi
\fi
}
\newcommand{\hatcurCCtassmg}[1]{\ifnum#1=50 %
\hatcurCCtassmgxxxxxA
\else
\ifnum#1=51 %
\hatcurCCtassmgxxxxxB
\else
\ifnum#1=52 %
\hatcurCCtassmgxxxxxC
\else
\ifnum#1=53 %
\hatcurCCtassmgxxxxxD
\else
??????\fi
\fi
\fi
\fi
}
\newcommand{\hatcurCCtassmgeccen}[1]{\ifnum#1=50 %
\hatcurCCtassmgeccenxxxxxA
\else
\ifnum#1=51 %
\hatcurCCtassmgeccenxxxxxB
\else
\ifnum#1=52 %
\hatcurCCtassmgeccenxxxxxC
\else
\ifnum#1=53 %
\hatcurCCtassmgeccenxxxxxD
\else
??????\fi
\fi
\fi
\fi
}
\newcommand{\hatcurCCtassmgshort}[1]{\ifnum#1=50 %
\hatcurCCtassmgshortxxxxxA
\else
\ifnum#1=51 %
\hatcurCCtassmgshortxxxxxB
\else
\ifnum#1=52 %
\hatcurCCtassmgshortxxxxxC
\else
\ifnum#1=53 %
\hatcurCCtassmgshortxxxxxD
\else
??????\fi
\fi
\fi
\fi
}
\newcommand{\hatcurCCtassmgshorteccen}[1]{\ifnum#1=50 %
\hatcurCCtassmgshorteccenxxxxxA
\else
\ifnum#1=51 %
\hatcurCCtassmgshorteccenxxxxxB
\else
\ifnum#1=52 %
\hatcurCCtassmgshorteccenxxxxxC
\else
\ifnum#1=53 %
\hatcurCCtassmgshorteccenxxxxxD
\else
??????\fi
\fi
\fi
\fi
}
\newcommand{\hatcurCCtassmi}[1]{\ifnum#1=50 %
\hatcurCCtassmixxxxxA
\else
\ifnum#1=51 %
\hatcurCCtassmixxxxxB
\else
\ifnum#1=52 %
\hatcurCCtassmixxxxxC
\else
\ifnum#1=53 %
\hatcurCCtassmixxxxxD
\else
??????\fi
\fi
\fi
\fi
}
\newcommand{\hatcurCCtassmI}[1]{\ifnum#1=50 %
\hatcurCCtassmIxxxxxA
\else
\ifnum#1=51 %
\hatcurCCtassmIxxxxxB
\else
\ifnum#1=52 %
\hatcurCCtassmIxxxxxC
\else
\ifnum#1=53 %
\hatcurCCtassmIxxxxxD
\else
??????\fi
\fi
\fi
\fi
}
\newcommand{\hatcurCCtassmieccen}[1]{\ifnum#1=50 %
\hatcurCCtassmieccenxxxxxA
\else
\ifnum#1=51 %
\hatcurCCtassmieccenxxxxxB
\else
\ifnum#1=52 %
\hatcurCCtassmieccenxxxxxC
\else
\ifnum#1=53 %
\hatcurCCtassmieccenxxxxxD
\else
??????\fi
\fi
\fi
\fi
}
\newcommand{\hatcurCCtassmIeccen}[1]{\ifnum#1=50 %
\hatcurCCtassmIeccenxxxxxA
\else
\ifnum#1=51 %
\hatcurCCtassmIeccenxxxxxB
\else
\ifnum#1=52 %
\hatcurCCtassmIeccenxxxxxC
\else
\ifnum#1=53 %
\hatcurCCtassmIeccenxxxxxD
\else
??????\fi
\fi
\fi
\fi
}
\newcommand{\hatcurCCtassmishort}[1]{\ifnum#1=50 %
\hatcurCCtassmishortxxxxxA
\else
\ifnum#1=51 %
\hatcurCCtassmishortxxxxxB
\else
\ifnum#1=52 %
\hatcurCCtassmishortxxxxxC
\else
\ifnum#1=53 %
\hatcurCCtassmishortxxxxxD
\else
??????\fi
\fi
\fi
\fi
}
\newcommand{\hatcurCCtassmIshort}[1]{\ifnum#1=50 %
\hatcurCCtassmIshortxxxxxA
\else
\ifnum#1=51 %
\hatcurCCtassmIshortxxxxxB
\else
\ifnum#1=52 %
\hatcurCCtassmIshortxxxxxC
\else
\ifnum#1=53 %
\hatcurCCtassmIshortxxxxxD
\else
??????\fi
\fi
\fi
\fi
}
\newcommand{\hatcurCCtassmishorteccen}[1]{\ifnum#1=50 %
\hatcurCCtassmishorteccenxxxxxA
\else
\ifnum#1=51 %
\hatcurCCtassmishorteccenxxxxxB
\else
\ifnum#1=52 %
\hatcurCCtassmishorteccenxxxxxC
\else
\ifnum#1=53 %
\hatcurCCtassmishorteccenxxxxxD
\else
??????\fi
\fi
\fi
\fi
}
\newcommand{\hatcurCCtassmIshorteccen}[1]{\ifnum#1=50 %
\hatcurCCtassmIshorteccenxxxxxA
\else
\ifnum#1=51 %
\hatcurCCtassmIshorteccenxxxxxB
\else
\ifnum#1=52 %
\hatcurCCtassmIshorteccenxxxxxC
\else
\ifnum#1=53 %
\hatcurCCtassmIshorteccenxxxxxD
\else
??????\fi
\fi
\fi
\fi
}
\newcommand{\hatcurCCtassmr}[1]{\ifnum#1=50 %
\hatcurCCtassmrxxxxxA
\else
\ifnum#1=51 %
\hatcurCCtassmrxxxxxB
\else
\ifnum#1=52 %
\hatcurCCtassmrxxxxxC
\else
\ifnum#1=53 %
\hatcurCCtassmrxxxxxD
\else
??????\fi
\fi
\fi
\fi
}
\newcommand{\hatcurCCtassmreccen}[1]{\ifnum#1=50 %
\hatcurCCtassmreccenxxxxxA
\else
\ifnum#1=51 %
\hatcurCCtassmreccenxxxxxB
\else
\ifnum#1=52 %
\hatcurCCtassmreccenxxxxxC
\else
\ifnum#1=53 %
\hatcurCCtassmreccenxxxxxD
\else
??????\fi
\fi
\fi
\fi
}
\newcommand{\hatcurCCtassmrshort}[1]{\ifnum#1=50 %
\hatcurCCtassmrshortxxxxxA
\else
\ifnum#1=51 %
\hatcurCCtassmrshortxxxxxB
\else
\ifnum#1=52 %
\hatcurCCtassmrshortxxxxxC
\else
\ifnum#1=53 %
\hatcurCCtassmrshortxxxxxD
\else
??????\fi
\fi
\fi
\fi
}
\newcommand{\hatcurCCtassmrshorteccen}[1]{\ifnum#1=50 %
\hatcurCCtassmrshorteccenxxxxxA
\else
\ifnum#1=51 %
\hatcurCCtassmrshorteccenxxxxxB
\else
\ifnum#1=52 %
\hatcurCCtassmrshorteccenxxxxxC
\else
\ifnum#1=53 %
\hatcurCCtassmrshorteccenxxxxxD
\else
??????\fi
\fi
\fi
\fi
}
\newcommand{\hatcurCCtassmv}[1]{\ifnum#1=50 %
\hatcurCCtassmvxxxxxA
\else
\ifnum#1=51 %
\hatcurCCtassmvxxxxxB
\else
\ifnum#1=52 %
\hatcurCCtassmvxxxxxC
\else
\ifnum#1=53 %
\hatcurCCtassmvxxxxxD
\else
??????\fi
\fi
\fi
\fi
}
\newcommand{\hatcurCCtassmveccen}[1]{\ifnum#1=50 %
\hatcurCCtassmveccenxxxxxA
\else
\ifnum#1=51 %
\hatcurCCtassmveccenxxxxxB
\else
\ifnum#1=52 %
\hatcurCCtassmveccenxxxxxC
\else
\ifnum#1=53 %
\hatcurCCtassmveccenxxxxxD
\else
??????\fi
\fi
\fi
\fi
}
\newcommand{\hatcurCCtassmvshort}[1]{\ifnum#1=50 %
\hatcurCCtassmvshortxxxxxA
\else
\ifnum#1=51 %
\hatcurCCtassmvshortxxxxxB
\else
\ifnum#1=52 %
\hatcurCCtassmvshortxxxxxC
\else
\ifnum#1=53 %
\hatcurCCtassmvshortxxxxxD
\else
??????\fi
\fi
\fi
\fi
}
\newcommand{\hatcurCCtassmvshorteccen}[1]{\ifnum#1=50 %
\hatcurCCtassmvshorteccenxxxxxA
\else
\ifnum#1=51 %
\hatcurCCtassmvshorteccenxxxxxB
\else
\ifnum#1=52 %
\hatcurCCtassmvshorteccenxxxxxC
\else
\ifnum#1=53 %
\hatcurCCtassmvshorteccenxxxxxD
\else
??????\fi
\fi
\fi
\fi
}
\newcommand{\hatcurCCtwomass}[1]{\ifnum#1=50 %
\hatcurCCtwomassxxxxxA
\else
\ifnum#1=51 %
\hatcurCCtwomassxxxxxB
\else
\ifnum#1=52 %
\hatcurCCtwomassxxxxxC
\else
\ifnum#1=53 %
\hatcurCCtwomassxxxxxD
\else
??????\fi
\fi
\fi
\fi
}
\newcommand{\hatcurCCtwomasseccen}[1]{\ifnum#1=50 %
\hatcurCCtwomasseccenxxxxxA
\else
\ifnum#1=51 %
\hatcurCCtwomasseccenxxxxxB
\else
\ifnum#1=52 %
\hatcurCCtwomasseccenxxxxxC
\else
\ifnum#1=53 %
\hatcurCCtwomasseccenxxxxxD
\else
??????\fi
\fi
\fi
\fi
}
\newcommand{\hatcurCCtwomassHmag}[1]{\ifnum#1=50 %
\hatcurCCtwomassHmagxxxxxA
\else
\ifnum#1=51 %
\hatcurCCtwomassHmagxxxxxB
\else
\ifnum#1=52 %
\hatcurCCtwomassHmagxxxxxC
\else
\ifnum#1=53 %
\hatcurCCtwomassHmagxxxxxD
\else
??????\fi
\fi
\fi
\fi
}
\newcommand{\hatcurCCtwomassHmageccen}[1]{\ifnum#1=50 %
\hatcurCCtwomassHmageccenxxxxxA
\else
\ifnum#1=51 %
\hatcurCCtwomassHmageccenxxxxxB
\else
\ifnum#1=52 %
\hatcurCCtwomassHmageccenxxxxxC
\else
\ifnum#1=53 %
\hatcurCCtwomassHmageccenxxxxxD
\else
??????\fi
\fi
\fi
\fi
}
\newcommand{\hatcurCCtwomassJmag}[1]{\ifnum#1=50 %
\hatcurCCtwomassJmagxxxxxA
\else
\ifnum#1=51 %
\hatcurCCtwomassJmagxxxxxB
\else
\ifnum#1=52 %
\hatcurCCtwomassJmagxxxxxC
\else
\ifnum#1=53 %
\hatcurCCtwomassJmagxxxxxD
\else
??????\fi
\fi
\fi
\fi
}
\newcommand{\hatcurCCtwomassJmageccen}[1]{\ifnum#1=50 %
\hatcurCCtwomassJmageccenxxxxxA
\else
\ifnum#1=51 %
\hatcurCCtwomassJmageccenxxxxxB
\else
\ifnum#1=52 %
\hatcurCCtwomassJmageccenxxxxxC
\else
\ifnum#1=53 %
\hatcurCCtwomassJmageccenxxxxxD
\else
??????\fi
\fi
\fi
\fi
}
\newcommand{\hatcurCCtwomassKmag}[1]{\ifnum#1=50 %
\hatcurCCtwomassKmagxxxxxA
\else
\ifnum#1=51 %
\hatcurCCtwomassKmagxxxxxB
\else
\ifnum#1=52 %
\hatcurCCtwomassKmagxxxxxC
\else
\ifnum#1=53 %
\hatcurCCtwomassKmagxxxxxD
\else
??????\fi
\fi
\fi
\fi
}
\newcommand{\hatcurCCtwomassKmageccen}[1]{\ifnum#1=50 %
\hatcurCCtwomassKmageccenxxxxxA
\else
\ifnum#1=51 %
\hatcurCCtwomassKmageccenxxxxxB
\else
\ifnum#1=52 %
\hatcurCCtwomassKmageccenxxxxxC
\else
\ifnum#1=53 %
\hatcurCCtwomassKmageccenxxxxxD
\else
??????\fi
\fi
\fi
\fi
}
\newcommand{\hatcurfield}[1]{\ifnum#1=50 %
\hatcurfieldxxxxxA
\else
\ifnum#1=51 %
\hatcurfieldxxxxxB
\else
\ifnum#1=52 %
\hatcurfieldxxxxxC
\else
\ifnum#1=53 %
\hatcurfieldxxxxxD
\else
??????\fi
\fi
\fi
\fi
}
\newcommand{\hatcurfieldeccen}[1]{\ifnum#1=50 %
\hatcurfieldeccenxxxxxA
\else
\ifnum#1=51 %
\hatcurfieldeccenxxxxxB
\else
\ifnum#1=52 %
\hatcurfieldeccenxxxxxC
\else
\ifnum#1=53 %
\hatcurfieldeccenxxxxxD
\else
??????\fi
\fi
\fi
\fi
}
\newcommand{\hatcurFIESgamma}[1]{\ifnum#1=50 %
\hatcurFIESgammaxxxxxA
\else
\ifnum#1=51 %
\hatcurFIESgammaxxxxxB
\else
??????\fi
\fi
}
\newcommand{\hatcurFIESgammaeccen}[1]{\ifnum#1=50 %
\hatcurFIESgammaeccenxxxxxA
\else
\ifnum#1=51 %
\hatcurFIESgammaeccenxxxxxB
\else
??????\fi
\fi
}
\newcommand{\hatcurFIESlogg}[1]{\ifnum#1=50 %
\hatcurFIESloggxxxxxA
\else
\ifnum#1=51 %
\hatcurFIESloggxxxxxB
\else
??????\fi
\fi
}
\newcommand{\hatcurFIESloggeccen}[1]{\ifnum#1=50 %
\hatcurFIESloggeccenxxxxxA
\else
\ifnum#1=51 %
\hatcurFIESloggeccenxxxxxB
\else
??????\fi
\fi
}
\newcommand{\hatcurFIESnumspec}[1]{\ifnum#1=50 %
\hatcurFIESnumspecxxxxxA
\else
\ifnum#1=51 %
\hatcurFIESnumspecxxxxxB
\else
??????\fi
\fi
}
\newcommand{\hatcurFIESnumspececcen}[1]{\ifnum#1=50 %
\hatcurFIESnumspececcenxxxxxA
\else
\ifnum#1=51 %
\hatcurFIESnumspececcenxxxxxB
\else
??????\fi
\fi
}
\newcommand{\hatcurFIESrvrms}[1]{\ifnum#1=50 %
\hatcurFIESrvrmsxxxxxA
\else
\ifnum#1=51 %
\hatcurFIESrvrmsxxxxxB
\else
??????\fi
\fi
}
\newcommand{\hatcurFIESrvrmseccen}[1]{\ifnum#1=50 %
\hatcurFIESrvrmseccenxxxxxA
\else
\ifnum#1=51 %
\hatcurFIESrvrmseccenxxxxxB
\else
??????\fi
\fi
}
\newcommand{\hatcurFIESspan}[1]{\ifnum#1=50 %
\hatcurFIESspanxxxxxA
\else
\ifnum#1=51 %
\hatcurFIESspanxxxxxB
\else
??????\fi
\fi
}
\newcommand{\hatcurFIESspaneccen}[1]{\ifnum#1=50 %
\hatcurFIESspaneccenxxxxxA
\else
\ifnum#1=51 %
\hatcurFIESspaneccenxxxxxB
\else
??????\fi
\fi
}
\newcommand{\hatcurFIESteff}[1]{\ifnum#1=50 %
\hatcurFIESteffxxxxxA
\else
\ifnum#1=51 %
\hatcurFIESteffxxxxxB
\else
??????\fi
\fi
}
\newcommand{\hatcurFIESteffeccen}[1]{\ifnum#1=50 %
\hatcurFIESteffeccenxxxxxA
\else
\ifnum#1=51 %
\hatcurFIESteffeccenxxxxxB
\else
??????\fi
\fi
}
\newcommand{\hatcurFIESvsini}[1]{\ifnum#1=50 %
\hatcurFIESvsinixxxxxA
\else
\ifnum#1=51 %
\hatcurFIESvsinixxxxxB
\else
??????\fi
\fi
}
\newcommand{\hatcurFIESvsinieccen}[1]{\ifnum#1=50 %
\hatcurFIESvsinieccenxxxxxA
\else
\ifnum#1=51 %
\hatcurFIESvsinieccenxxxxxB
\else
??????\fi
\fi
}
\newcommand{\hatcurFIESzfeh}[1]{\ifnum#1=50 %
\hatcurFIESzfehxxxxxA
\else
\ifnum#1=51 %
\hatcurFIESzfehxxxxxB
\else
??????\fi
\fi
}
\newcommand{\hatcurFIESzfeheccen}[1]{\ifnum#1=50 %
\hatcurFIESzfeheccenxxxxxA
\else
\ifnum#1=51 %
\hatcurFIESzfeheccenxxxxxB
\else
??????\fi
\fi
}
\newcommand{\hatcurhtr}[1]{\ifnum#1=50 %
\hatcurhtrxxxxxA
\else
\ifnum#1=51 %
\hatcurhtrxxxxxB
\else
\ifnum#1=52 %
\hatcurhtrxxxxxC
\else
\ifnum#1=53 %
\hatcurhtrxxxxxD
\else
??????\fi
\fi
\fi
\fi
}
\newcommand{\hatcurhtreccen}[1]{\ifnum#1=50 %
\hatcurhtreccenxxxxxA
\else
\ifnum#1=51 %
\hatcurhtreccenxxxxxB
\else
\ifnum#1=52 %
\hatcurhtreccenxxxxxC
\else
\ifnum#1=53 %
\hatcurhtreccenxxxxxD
\else
??????\fi
\fi
\fi
\fi
}
\newcommand{\hatcurISOage}[1]{\ifnum#1=50 %
\hatcurISOagexxxxxA
\else
\ifnum#1=51 %
\hatcurISOagexxxxxB
\else
\ifnum#1=52 %
\hatcurISOagexxxxxC
\else
\ifnum#1=53 %
\hatcurISOagexxxxxD
\else
??????\fi
\fi
\fi
\fi
}
\newcommand{\hatcurISOageeccen}[1]{\ifnum#1=50 %
\hatcurISOageeccenxxxxxA
\else
\ifnum#1=51 %
\hatcurISOageeccenxxxxxB
\else
\ifnum#1=52 %
\hatcurISOageeccenxxxxxC
\else
\ifnum#1=53 %
\hatcurISOageeccenxxxxxD
\else
??????\fi
\fi
\fi
\fi
}
\newcommand{\hatcurISOJK}[1]{\ifnum#1=50 %
\hatcurISOJKxxxxxA
\else
\ifnum#1=51 %
\hatcurISOJKxxxxxB
\else
\ifnum#1=52 %
\hatcurISOJKxxxxxC
\else
\ifnum#1=53 %
\hatcurISOJKxxxxxD
\else
??????\fi
\fi
\fi
\fi
}
\newcommand{\hatcurISOJKeccen}[1]{\ifnum#1=50 %
\hatcurISOJKeccenxxxxxA
\else
\ifnum#1=51 %
\hatcurISOJKeccenxxxxxB
\else
\ifnum#1=52 %
\hatcurISOJKeccenxxxxxC
\else
\ifnum#1=53 %
\hatcurISOJKeccenxxxxxD
\else
??????\fi
\fi
\fi
\fi
}
\newcommand{\hatcurISOlogg}[1]{\ifnum#1=50 %
\hatcurISOloggxxxxxA
\else
\ifnum#1=51 %
\hatcurISOloggxxxxxB
\else
\ifnum#1=52 %
\hatcurISOloggxxxxxC
\else
\ifnum#1=53 %
\hatcurISOloggxxxxxD
\else
??????\fi
\fi
\fi
\fi
}
\newcommand{\hatcurISOloggeccen}[1]{\ifnum#1=50 %
\hatcurISOloggeccenxxxxxA
\else
\ifnum#1=51 %
\hatcurISOloggeccenxxxxxB
\else
\ifnum#1=52 %
\hatcurISOloggeccenxxxxxC
\else
\ifnum#1=53 %
\hatcurISOloggeccenxxxxxD
\else
??????\fi
\fi
\fi
\fi
}
\newcommand{\hatcurISOlum}[1]{\ifnum#1=50 %
\hatcurISOlumxxxxxA
\else
\ifnum#1=51 %
\hatcurISOlumxxxxxB
\else
\ifnum#1=52 %
\hatcurISOlumxxxxxC
\else
\ifnum#1=53 %
\hatcurISOlumxxxxxD
\else
??????\fi
\fi
\fi
\fi
}
\newcommand{\hatcurISOlumeccen}[1]{\ifnum#1=50 %
\hatcurISOlumeccenxxxxxA
\else
\ifnum#1=51 %
\hatcurISOlumeccenxxxxxB
\else
\ifnum#1=52 %
\hatcurISOlumeccenxxxxxC
\else
\ifnum#1=53 %
\hatcurISOlumeccenxxxxxD
\else
??????\fi
\fi
\fi
\fi
}
\newcommand{\hatcurISOlumshort}[1]{\ifnum#1=50 %
\hatcurISOlumshortxxxxxA
\else
\ifnum#1=51 %
\hatcurISOlumshortxxxxxB
\else
\ifnum#1=52 %
\hatcurISOlumshortxxxxxC
\else
\ifnum#1=53 %
\hatcurISOlumshortxxxxxD
\else
??????\fi
\fi
\fi
\fi
}
\newcommand{\hatcurISOlumshorteccen}[1]{\ifnum#1=50 %
\hatcurISOlumshorteccenxxxxxA
\else
\ifnum#1=51 %
\hatcurISOlumshorteccenxxxxxB
\else
\ifnum#1=52 %
\hatcurISOlumshorteccenxxxxxC
\else
\ifnum#1=53 %
\hatcurISOlumshorteccenxxxxxD
\else
??????\fi
\fi
\fi
\fi
}
\newcommand{\hatcurISOm}[1]{\ifnum#1=50 %
\hatcurISOmxxxxxA
\else
\ifnum#1=51 %
\hatcurISOmxxxxxB
\else
\ifnum#1=52 %
\hatcurISOmxxxxxC
\else
\ifnum#1=53 %
\hatcurISOmxxxxxD
\else
??????\fi
\fi
\fi
\fi
}
\newcommand{\hatcurISOmeccen}[1]{\ifnum#1=50 %
\hatcurISOmeccenxxxxxA
\else
\ifnum#1=51 %
\hatcurISOmeccenxxxxxB
\else
\ifnum#1=52 %
\hatcurISOmeccenxxxxxC
\else
\ifnum#1=53 %
\hatcurISOmeccenxxxxxD
\else
??????\fi
\fi
\fi
\fi
}
\newcommand{\hatcurISOMH}[1]{\ifnum#1=50 %
\hatcurISOMHxxxxxA
\else
\ifnum#1=51 %
\hatcurISOMHxxxxxB
\else
\ifnum#1=52 %
\hatcurISOMHxxxxxC
\else
\ifnum#1=53 %
\hatcurISOMHxxxxxD
\else
??????\fi
\fi
\fi
\fi
}
\newcommand{\hatcurISOMHeccen}[1]{\ifnum#1=50 %
\hatcurISOMHeccenxxxxxA
\else
\ifnum#1=51 %
\hatcurISOMHeccenxxxxxB
\else
\ifnum#1=52 %
\hatcurISOMHeccenxxxxxC
\else
\ifnum#1=53 %
\hatcurISOMHeccenxxxxxD
\else
??????\fi
\fi
\fi
\fi
}
\newcommand{\hatcurISOMJ}[1]{\ifnum#1=50 %
\hatcurISOMJxxxxxA
\else
\ifnum#1=51 %
\hatcurISOMJxxxxxB
\else
\ifnum#1=52 %
\hatcurISOMJxxxxxC
\else
\ifnum#1=53 %
\hatcurISOMJxxxxxD
\else
??????\fi
\fi
\fi
\fi
}
\newcommand{\hatcurISOMJeccen}[1]{\ifnum#1=50 %
\hatcurISOMJeccenxxxxxA
\else
\ifnum#1=51 %
\hatcurISOMJeccenxxxxxB
\else
\ifnum#1=52 %
\hatcurISOMJeccenxxxxxC
\else
\ifnum#1=53 %
\hatcurISOMJeccenxxxxxD
\else
??????\fi
\fi
\fi
\fi
}
\newcommand{\hatcurISOMK}[1]{\ifnum#1=50 %
\hatcurISOMKxxxxxA
\else
\ifnum#1=51 %
\hatcurISOMKxxxxxB
\else
\ifnum#1=52 %
\hatcurISOMKxxxxxC
\else
\ifnum#1=53 %
\hatcurISOMKxxxxxD
\else
??????\fi
\fi
\fi
\fi
}
\newcommand{\hatcurISOMKeccen}[1]{\ifnum#1=50 %
\hatcurISOMKeccenxxxxxA
\else
\ifnum#1=51 %
\hatcurISOMKeccenxxxxxB
\else
\ifnum#1=52 %
\hatcurISOMKeccenxxxxxC
\else
\ifnum#1=53 %
\hatcurISOMKeccenxxxxxD
\else
??????\fi
\fi
\fi
\fi
}
\newcommand{\hatcurISOmlong}[1]{\ifnum#1=50 %
\hatcurISOmlongxxxxxA
\else
\ifnum#1=51 %
\hatcurISOmlongxxxxxB
\else
\ifnum#1=52 %
\hatcurISOmlongxxxxxC
\else
\ifnum#1=53 %
\hatcurISOmlongxxxxxD
\else
??????\fi
\fi
\fi
\fi
}
\newcommand{\hatcurISOmlongeccen}[1]{\ifnum#1=50 %
\hatcurISOmlongeccenxxxxxA
\else
\ifnum#1=51 %
\hatcurISOmlongeccenxxxxxB
\else
\ifnum#1=52 %
\hatcurISOmlongeccenxxxxxC
\else
\ifnum#1=53 %
\hatcurISOmlongeccenxxxxxD
\else
??????\fi
\fi
\fi
\fi
}
\newcommand{\hatcurISOmshort}[1]{\ifnum#1=50 %
\hatcurISOmshortxxxxxA
\else
\ifnum#1=51 %
\hatcurISOmshortxxxxxB
\else
\ifnum#1=52 %
\hatcurISOmshortxxxxxC
\else
\ifnum#1=53 %
\hatcurISOmshortxxxxxD
\else
??????\fi
\fi
\fi
\fi
}
\newcommand{\hatcurISOmshorteccen}[1]{\ifnum#1=50 %
\hatcurISOmshorteccenxxxxxA
\else
\ifnum#1=51 %
\hatcurISOmshorteccenxxxxxB
\else
\ifnum#1=52 %
\hatcurISOmshorteccenxxxxxC
\else
\ifnum#1=53 %
\hatcurISOmshorteccenxxxxxD
\else
??????\fi
\fi
\fi
\fi
}
\newcommand{\hatcurISOmv}[1]{\ifnum#1=50 %
\hatcurISOmvxxxxxA
\else
\ifnum#1=51 %
\hatcurISOmvxxxxxB
\else
\ifnum#1=52 %
\hatcurISOmvxxxxxC
\else
\ifnum#1=53 %
\hatcurISOmvxxxxxD
\else
??????\fi
\fi
\fi
\fi
}
\newcommand{\hatcurISOmveccen}[1]{\ifnum#1=50 %
\hatcurISOmveccenxxxxxA
\else
\ifnum#1=51 %
\hatcurISOmveccenxxxxxB
\else
\ifnum#1=52 %
\hatcurISOmveccenxxxxxC
\else
\ifnum#1=53 %
\hatcurISOmveccenxxxxxD
\else
??????\fi
\fi
\fi
\fi
}
\newcommand{\hatcurISOr}[1]{\ifnum#1=50 %
\hatcurISOrxxxxxA
\else
\ifnum#1=51 %
\hatcurISOrxxxxxB
\else
\ifnum#1=52 %
\hatcurISOrxxxxxC
\else
\ifnum#1=53 %
\hatcurISOrxxxxxD
\else
??????\fi
\fi
\fi
\fi
}
\newcommand{\hatcurISOreccen}[1]{\ifnum#1=50 %
\hatcurISOreccenxxxxxA
\else
\ifnum#1=51 %
\hatcurISOreccenxxxxxB
\else
\ifnum#1=52 %
\hatcurISOreccenxxxxxC
\else
\ifnum#1=53 %
\hatcurISOreccenxxxxxD
\else
??????\fi
\fi
\fi
\fi
}
\newcommand{\hatcurISOrho}[1]{\ifnum#1=50 %
\hatcurISOrhoxxxxxA
\else
\ifnum#1=51 %
\hatcurISOrhoxxxxxB
\else
\ifnum#1=52 %
\hatcurISOrhoxxxxxC
\else
\ifnum#1=53 %
\hatcurISOrhoxxxxxD
\else
??????\fi
\fi
\fi
\fi
}
\newcommand{\hatcurISOrhoeccen}[1]{\ifnum#1=50 %
\hatcurISOrhoeccenxxxxxA
\else
\ifnum#1=51 %
\hatcurISOrhoeccenxxxxxB
\else
\ifnum#1=52 %
\hatcurISOrhoeccenxxxxxC
\else
\ifnum#1=53 %
\hatcurISOrhoeccenxxxxxD
\else
??????\fi
\fi
\fi
\fi
}
\newcommand{\hatcurISOrholong}[1]{\ifnum#1=50 %
\hatcurISOrholongxxxxxA
\else
\ifnum#1=51 %
\hatcurISOrholongxxxxxB
\else
\ifnum#1=52 %
\hatcurISOrholongxxxxxC
\else
\ifnum#1=53 %
\hatcurISOrholongxxxxxD
\else
??????\fi
\fi
\fi
\fi
}
\newcommand{\hatcurISOrholongeccen}[1]{\ifnum#1=50 %
\hatcurISOrholongeccenxxxxxA
\else
\ifnum#1=51 %
\hatcurISOrholongeccenxxxxxB
\else
\ifnum#1=52 %
\hatcurISOrholongeccenxxxxxC
\else
\ifnum#1=53 %
\hatcurISOrholongeccenxxxxxD
\else
??????\fi
\fi
\fi
\fi
}
\newcommand{\hatcurISOrlong}[1]{\ifnum#1=50 %
\hatcurISOrlongxxxxxA
\else
\ifnum#1=51 %
\hatcurISOrlongxxxxxB
\else
\ifnum#1=52 %
\hatcurISOrlongxxxxxC
\else
\ifnum#1=53 %
\hatcurISOrlongxxxxxD
\else
??????\fi
\fi
\fi
\fi
}
\newcommand{\hatcurISOrlongeccen}[1]{\ifnum#1=50 %
\hatcurISOrlongeccenxxxxxA
\else
\ifnum#1=51 %
\hatcurISOrlongeccenxxxxxB
\else
\ifnum#1=52 %
\hatcurISOrlongeccenxxxxxC
\else
\ifnum#1=53 %
\hatcurISOrlongeccenxxxxxD
\else
??????\fi
\fi
\fi
\fi
}
\newcommand{\hatcurISOrshort}[1]{\ifnum#1=50 %
\hatcurISOrshortxxxxxA
\else
\ifnum#1=51 %
\hatcurISOrshortxxxxxB
\else
\ifnum#1=52 %
\hatcurISOrshortxxxxxC
\else
\ifnum#1=53 %
\hatcurISOrshortxxxxxD
\else
??????\fi
\fi
\fi
\fi
}
\newcommand{\hatcurISOrshorteccen}[1]{\ifnum#1=50 %
\hatcurISOrshorteccenxxxxxA
\else
\ifnum#1=51 %
\hatcurISOrshorteccenxxxxxB
\else
\ifnum#1=52 %
\hatcurISOrshorteccenxxxxxC
\else
\ifnum#1=53 %
\hatcurISOrshorteccenxxxxxD
\else
??????\fi
\fi
\fi
\fi
}
\newcommand{\hatcurISOsigma}[1]{\ifnum#1=50 %
\hatcurISOsigmaxxxxxA
\else
\ifnum#1=51 %
\hatcurISOsigmaxxxxxB
\else
\ifnum#1=52 %
\hatcurISOsigmaxxxxxC
\else
\ifnum#1=53 %
\hatcurISOsigmaxxxxxD
\else
??????\fi
\fi
\fi
\fi
}
\newcommand{\hatcurISOsigmaeccen}[1]{\ifnum#1=50 %
\hatcurISOsigmaeccenxxxxxA
\else
\ifnum#1=51 %
\hatcurISOsigmaeccenxxxxxB
\else
\ifnum#1=52 %
\hatcurISOsigmaeccenxxxxxC
\else
\ifnum#1=53 %
\hatcurISOsigmaeccenxxxxxD
\else
??????\fi
\fi
\fi
\fi
}
\newcommand{\hatcurISOspec}[1]{\ifnum#1=50 %
\hatcurISOspecxxxxxA
\else
\ifnum#1=51 %
\hatcurISOspecxxxxxB
\else
\ifnum#1=52 %
\hatcurISOspecxxxxxC
\else
\ifnum#1=53 %
\hatcurISOspecxxxxxD
\else
??????\fi
\fi
\fi
\fi
}
\newcommand{\hatcurISOspececcen}[1]{\ifnum#1=50 %
\hatcurISOspececcenxxxxxA
\else
\ifnum#1=51 %
\hatcurISOspececcenxxxxxB
\else
\ifnum#1=52 %
\hatcurISOspececcenxxxxxC
\else
\ifnum#1=53 %
\hatcurISOspececcenxxxxxD
\else
??????\fi
\fi
\fi
\fi
}
\newcommand{\hatcurISOvi}[1]{\ifnum#1=50 %
\hatcurISOvixxxxxA
\else
\ifnum#1=51 %
\hatcurISOvixxxxxB
\else
\ifnum#1=52 %
\hatcurISOvixxxxxC
\else
\ifnum#1=53 %
\hatcurISOvixxxxxD
\else
??????\fi
\fi
\fi
\fi
}
\newcommand{\hatcurISOvieccen}[1]{\ifnum#1=50 %
\hatcurISOvieccenxxxxxA
\else
\ifnum#1=51 %
\hatcurISOvieccenxxxxxB
\else
\ifnum#1=52 %
\hatcurISOvieccenxxxxxC
\else
\ifnum#1=53 %
\hatcurISOvieccenxxxxxD
\else
??????\fi
\fi
\fi
\fi
}
\newcommand{\hatcurLBig}[1]{\ifnum#1=50 %
\hatcurLBigxxxxxA
\else
\ifnum#1=51 %
\hatcurLBigxxxxxB
\else
\ifnum#1=52 %
\hatcurLBigxxxxxC
\else
\ifnum#1=53 %
\hatcurLBigxxxxxD
\else
??????\fi
\fi
\fi
\fi
}
\newcommand{\hatcurLBigeccen}[1]{\ifnum#1=50 %
\hatcurLBigeccenxxxxxA
\else
\ifnum#1=51 %
\hatcurLBigeccenxxxxxB
\else
\ifnum#1=52 %
\hatcurLBigeccenxxxxxC
\else
\ifnum#1=53 %
\hatcurLBigeccenxxxxxD
\else
??????\fi
\fi
\fi
\fi
}
\newcommand{\hatcurLBii}[1]{\ifnum#1=50 %
\hatcurLBiixxxxxA
\else
\ifnum#1=51 %
\hatcurLBiixxxxxB
\else
\ifnum#1=52 %
\hatcurLBiixxxxxC
\else
\ifnum#1=53 %
\hatcurLBiixxxxxD
\else
??????\fi
\fi
\fi
\fi
}
\newcommand{\hatcurLBiI}[1]{\ifnum#1=50 %
\hatcurLBiIxxxxxA
\else
\ifnum#1=51 %
\hatcurLBiIxxxxxB
\else
\ifnum#1=52 %
\hatcurLBiIxxxxxC
\else
\ifnum#1=53 %
\hatcurLBiIxxxxxD
\else
??????\fi
\fi
\fi
\fi
}
\newcommand{\hatcurLBiieccen}[1]{\ifnum#1=50 %
\hatcurLBiieccenxxxxxA
\else
\ifnum#1=51 %
\hatcurLBiieccenxxxxxB
\else
\ifnum#1=52 %
\hatcurLBiieccenxxxxxC
\else
\ifnum#1=53 %
\hatcurLBiieccenxxxxxD
\else
??????\fi
\fi
\fi
\fi
}
\newcommand{\hatcurLBiIeccen}[1]{\ifnum#1=50 %
\hatcurLBiIeccenxxxxxA
\else
\ifnum#1=51 %
\hatcurLBiIeccenxxxxxB
\else
\ifnum#1=52 %
\hatcurLBiIeccenxxxxxC
\else
\ifnum#1=53 %
\hatcurLBiIeccenxxxxxD
\else
??????\fi
\fi
\fi
\fi
}
\newcommand{\hatcurLBiig}[1]{\ifnum#1=50 %
\hatcurLBiigxxxxxA
\else
\ifnum#1=51 %
\hatcurLBiigxxxxxB
\else
\ifnum#1=52 %
\hatcurLBiigxxxxxC
\else
\ifnum#1=53 %
\hatcurLBiigxxxxxD
\else
??????\fi
\fi
\fi
\fi
}
\newcommand{\hatcurLBiigeccen}[1]{\ifnum#1=50 %
\hatcurLBiigeccenxxxxxA
\else
\ifnum#1=51 %
\hatcurLBiigeccenxxxxxB
\else
\ifnum#1=52 %
\hatcurLBiigeccenxxxxxC
\else
\ifnum#1=53 %
\hatcurLBiigeccenxxxxxD
\else
??????\fi
\fi
\fi
\fi
}
\newcommand{\hatcurLBiii}[1]{\ifnum#1=50 %
\hatcurLBiiixxxxxA
\else
\ifnum#1=51 %
\hatcurLBiiixxxxxB
\else
\ifnum#1=52 %
\hatcurLBiiixxxxxC
\else
\ifnum#1=53 %
\hatcurLBiiixxxxxD
\else
??????\fi
\fi
\fi
\fi
}
\newcommand{\hatcurLBiiI}[1]{\ifnum#1=50 %
\hatcurLBiiIxxxxxA
\else
\ifnum#1=51 %
\hatcurLBiiIxxxxxB
\else
\ifnum#1=52 %
\hatcurLBiiIxxxxxC
\else
\ifnum#1=53 %
\hatcurLBiiIxxxxxD
\else
??????\fi
\fi
\fi
\fi
}
\newcommand{\hatcurLBiiieccen}[1]{\ifnum#1=50 %
\hatcurLBiiieccenxxxxxA
\else
\ifnum#1=51 %
\hatcurLBiiieccenxxxxxB
\else
\ifnum#1=52 %
\hatcurLBiiieccenxxxxxC
\else
\ifnum#1=53 %
\hatcurLBiiieccenxxxxxD
\else
??????\fi
\fi
\fi
\fi
}
\newcommand{\hatcurLBiiIeccen}[1]{\ifnum#1=50 %
\hatcurLBiiIeccenxxxxxA
\else
\ifnum#1=51 %
\hatcurLBiiIeccenxxxxxB
\else
\ifnum#1=52 %
\hatcurLBiiIeccenxxxxxC
\else
\ifnum#1=53 %
\hatcurLBiiIeccenxxxxxD
\else
??????\fi
\fi
\fi
\fi
}
\newcommand{\hatcurLBiir}[1]{\ifnum#1=50 %
\hatcurLBiirxxxxxA
\else
\ifnum#1=51 %
\hatcurLBiirxxxxxB
\else
\ifnum#1=52 %
\hatcurLBiirxxxxxC
\else
\ifnum#1=53 %
\hatcurLBiirxxxxxD
\else
??????\fi
\fi
\fi
\fi
}
\newcommand{\hatcurLBiiR}[1]{\ifnum#1=50 %
\hatcurLBiiRxxxxxA
\else
\ifnum#1=51 %
\hatcurLBiiRxxxxxB
\else
\ifnum#1=52 %
\hatcurLBiiRxxxxxC
\else
\ifnum#1=53 %
\hatcurLBiiRxxxxxD
\else
??????\fi
\fi
\fi
\fi
}
\newcommand{\hatcurLBiireccen}[1]{\ifnum#1=50 %
\hatcurLBiireccenxxxxxA
\else
\ifnum#1=51 %
\hatcurLBiireccenxxxxxB
\else
\ifnum#1=52 %
\hatcurLBiireccenxxxxxC
\else
\ifnum#1=53 %
\hatcurLBiireccenxxxxxD
\else
??????\fi
\fi
\fi
\fi
}
\newcommand{\hatcurLBiiReccen}[1]{\ifnum#1=50 %
\hatcurLBiiReccenxxxxxA
\else
\ifnum#1=51 %
\hatcurLBiiReccenxxxxxB
\else
\ifnum#1=52 %
\hatcurLBiiReccenxxxxxC
\else
\ifnum#1=53 %
\hatcurLBiiReccenxxxxxD
\else
??????\fi
\fi
\fi
\fi
}
\newcommand{\hatcurLBiiz}[1]{\ifnum#1=50 %
\hatcurLBiizxxxxxA
\else
\ifnum#1=51 %
\hatcurLBiizxxxxxB
\else
\ifnum#1=52 %
\hatcurLBiizxxxxxC
\else
\ifnum#1=53 %
\hatcurLBiizxxxxxD
\else
??????\fi
\fi
\fi
\fi
}
\newcommand{\hatcurLBiizeccen}[1]{\ifnum#1=50 %
\hatcurLBiizeccenxxxxxA
\else
\ifnum#1=51 %
\hatcurLBiizeccenxxxxxB
\else
\ifnum#1=52 %
\hatcurLBiizeccenxxxxxC
\else
\ifnum#1=53 %
\hatcurLBiizeccenxxxxxD
\else
??????\fi
\fi
\fi
\fi
}
\newcommand{\hatcurLBir}[1]{\ifnum#1=50 %
\hatcurLBirxxxxxA
\else
\ifnum#1=51 %
\hatcurLBirxxxxxB
\else
\ifnum#1=52 %
\hatcurLBirxxxxxC
\else
\ifnum#1=53 %
\hatcurLBirxxxxxD
\else
??????\fi
\fi
\fi
\fi
}
\newcommand{\hatcurLBiR}[1]{\ifnum#1=50 %
\hatcurLBiRxxxxxA
\else
\ifnum#1=51 %
\hatcurLBiRxxxxxB
\else
\ifnum#1=52 %
\hatcurLBiRxxxxxC
\else
\ifnum#1=53 %
\hatcurLBiRxxxxxD
\else
??????\fi
\fi
\fi
\fi
}
\newcommand{\hatcurLBireccen}[1]{\ifnum#1=50 %
\hatcurLBireccenxxxxxA
\else
\ifnum#1=51 %
\hatcurLBireccenxxxxxB
\else
\ifnum#1=52 %
\hatcurLBireccenxxxxxC
\else
\ifnum#1=53 %
\hatcurLBireccenxxxxxD
\else
??????\fi
\fi
\fi
\fi
}
\newcommand{\hatcurLBiReccen}[1]{\ifnum#1=50 %
\hatcurLBiReccenxxxxxA
\else
\ifnum#1=51 %
\hatcurLBiReccenxxxxxB
\else
\ifnum#1=52 %
\hatcurLBiReccenxxxxxC
\else
\ifnum#1=53 %
\hatcurLBiReccenxxxxxD
\else
??????\fi
\fi
\fi
\fi
}
\newcommand{\hatcurLBiz}[1]{\ifnum#1=50 %
\hatcurLBizxxxxxA
\else
\ifnum#1=51 %
\hatcurLBizxxxxxB
\else
\ifnum#1=52 %
\hatcurLBizxxxxxC
\else
\ifnum#1=53 %
\hatcurLBizxxxxxD
\else
??????\fi
\fi
\fi
\fi
}
\newcommand{\hatcurLBizeccen}[1]{\ifnum#1=50 %
\hatcurLBizeccenxxxxxA
\else
\ifnum#1=51 %
\hatcurLBizeccenxxxxxB
\else
\ifnum#1=52 %
\hatcurLBizeccenxxxxxC
\else
\ifnum#1=53 %
\hatcurLBizeccenxxxxxD
\else
??????\fi
\fi
\fi
\fi
}
\newcommand{\hatcurLCbsq}[1]{\ifnum#1=50 %
\hatcurLCbsqxxxxxA
\else
\ifnum#1=51 %
\hatcurLCbsqxxxxxB
\else
\ifnum#1=52 %
\hatcurLCbsqxxxxxC
\else
\ifnum#1=53 %
\hatcurLCbsqxxxxxD
\else
??????\fi
\fi
\fi
\fi
}
\newcommand{\hatcurLCbsqeccen}[1]{\ifnum#1=50 %
\hatcurLCbsqeccenxxxxxA
\else
\ifnum#1=51 %
\hatcurLCbsqeccenxxxxxB
\else
\ifnum#1=52 %
\hatcurLCbsqeccenxxxxxC
\else
\ifnum#1=53 %
\hatcurLCbsqeccenxxxxxD
\else
??????\fi
\fi
\fi
\fi
}
\newcommand{\hatcurLCdip}[1]{\ifnum#1=50 %
\hatcurLCdipxxxxxA
\else
\ifnum#1=51 %
\hatcurLCdipxxxxxB
\else
\ifnum#1=52 %
\hatcurLCdipxxxxxC
\else
\ifnum#1=53 %
\hatcurLCdipxxxxxD
\else
??????\fi
\fi
\fi
\fi
}
\newcommand{\hatcurLCdipeccen}[1]{\ifnum#1=50 %
\hatcurLCdipeccenxxxxxA
\else
\ifnum#1=51 %
\hatcurLCdipeccenxxxxxB
\else
\ifnum#1=52 %
\hatcurLCdipeccenxxxxxC
\else
\ifnum#1=53 %
\hatcurLCdipeccenxxxxxD
\else
??????\fi
\fi
\fi
\fi
}
\newcommand{\hatcurLCdur}[1]{\ifnum#1=50 %
\hatcurLCdurxxxxxA
\else
\ifnum#1=51 %
\hatcurLCdurxxxxxB
\else
\ifnum#1=52 %
\hatcurLCdurxxxxxC
\else
\ifnum#1=53 %
\hatcurLCdurxxxxxD
\else
??????\fi
\fi
\fi
\fi
}
\newcommand{\hatcurLCdureccen}[1]{\ifnum#1=50 %
\hatcurLCdureccenxxxxxA
\else
\ifnum#1=51 %
\hatcurLCdureccenxxxxxB
\else
\ifnum#1=52 %
\hatcurLCdureccenxxxxxC
\else
\ifnum#1=53 %
\hatcurLCdureccenxxxxxD
\else
??????\fi
\fi
\fi
\fi
}
\newcommand{\hatcurLCdurhr}[1]{\ifnum#1=50 %
\hatcurLCdurhrxxxxxA
\else
\ifnum#1=51 %
\hatcurLCdurhrxxxxxB
\else
\ifnum#1=52 %
\hatcurLCdurhrxxxxxC
\else
\ifnum#1=53 %
\hatcurLCdurhrxxxxxD
\else
??????\fi
\fi
\fi
\fi
}
\newcommand{\hatcurLCdurhreccen}[1]{\ifnum#1=50 %
\hatcurLCdurhreccenxxxxxA
\else
\ifnum#1=51 %
\hatcurLCdurhreccenxxxxxB
\else
\ifnum#1=52 %
\hatcurLCdurhreccenxxxxxC
\else
\ifnum#1=53 %
\hatcurLCdurhreccenxxxxxD
\else
??????\fi
\fi
\fi
\fi
}
\newcommand{\hatcurLCdurhrshort}[1]{\ifnum#1=50 %
\hatcurLCdurhrshortxxxxxA
\else
\ifnum#1=51 %
\hatcurLCdurhrshortxxxxxB
\else
\ifnum#1=52 %
\hatcurLCdurhrshortxxxxxC
\else
\ifnum#1=53 %
\hatcurLCdurhrshortxxxxxD
\else
??????\fi
\fi
\fi
\fi
}
\newcommand{\hatcurLCdurhrshorteccen}[1]{\ifnum#1=50 %
\hatcurLCdurhrshorteccenxxxxxA
\else
\ifnum#1=51 %
\hatcurLCdurhrshorteccenxxxxxB
\else
\ifnum#1=52 %
\hatcurLCdurhrshorteccenxxxxxC
\else
\ifnum#1=53 %
\hatcurLCdurhrshorteccenxxxxxD
\else
??????\fi
\fi
\fi
\fi
}
\newcommand{\hatcurLCdurshort}[1]{\ifnum#1=50 %
\hatcurLCdurshortxxxxxA
\else
\ifnum#1=51 %
\hatcurLCdurshortxxxxxB
\else
\ifnum#1=52 %
\hatcurLCdurshortxxxxxC
\else
\ifnum#1=53 %
\hatcurLCdurshortxxxxxD
\else
??????\fi
\fi
\fi
\fi
}
\newcommand{\hatcurLCdurshorteccen}[1]{\ifnum#1=50 %
\hatcurLCdurshorteccenxxxxxA
\else
\ifnum#1=51 %
\hatcurLCdurshorteccenxxxxxB
\else
\ifnum#1=52 %
\hatcurLCdurshorteccenxxxxxC
\else
\ifnum#1=53 %
\hatcurLCdurshorteccenxxxxxD
\else
??????\fi
\fi
\fi
\fi
}
\newcommand{\hatcurLChatnetm}[1]{\ifnum#1=52 %
\hatcurLChatnetmxxxxxC
\else
??????\fi
}
\newcommand{\hatcurLChatnetmA}[1]{\ifnum#1=50 %
\hatcurLChatnetmAxxxxxA
\else
\ifnum#1=51 %
\hatcurLChatnetmAxxxxxB
\else
\ifnum#1=53 %
\hatcurLChatnetmAxxxxxD
\else
??????\fi
\fi
\fi
}
\newcommand{\hatcurLChatnetmAeccen}[1]{\ifnum#1=50 %
\hatcurLChatnetmAeccenxxxxxA
\else
\ifnum#1=51 %
\hatcurLChatnetmAeccenxxxxxB
\else
\ifnum#1=53 %
\hatcurLChatnetmAeccenxxxxxD
\else
??????\fi
\fi
\fi
}
\newcommand{\hatcurLChatnetmB}[1]{\ifnum#1=50 %
\hatcurLChatnetmBxxxxxA
\else
\ifnum#1=51 %
\hatcurLChatnetmBxxxxxB
\else
\ifnum#1=53 %
\hatcurLChatnetmBxxxxxD
\else
??????\fi
\fi
\fi
}
\newcommand{\hatcurLChatnetmBeccen}[1]{\ifnum#1=50 %
\hatcurLChatnetmBeccenxxxxxA
\else
\ifnum#1=51 %
\hatcurLChatnetmBeccenxxxxxB
\else
\ifnum#1=53 %
\hatcurLChatnetmBeccenxxxxxD
\else
??????\fi
\fi
\fi
}
\newcommand{\hatcurLChatnetmC}[1]{\ifnum#1=51 %
\hatcurLChatnetmCxxxxxB
\else
??????\fi
}
\newcommand{\hatcurLChatnetmCeccen}[1]{\ifnum#1=51 %
\hatcurLChatnetmCeccenxxxxxB
\else
??????\fi
}
\newcommand{\hatcurLChatnetmD}[1]{\ifnum#1=51 %
\hatcurLChatnetmDxxxxxB
\else
??????\fi
}
\newcommand{\hatcurLChatnetmDeccen}[1]{\ifnum#1=51 %
\hatcurLChatnetmDeccenxxxxxB
\else
??????\fi
}
\newcommand{\hatcurLChatnetmeccen}[1]{\ifnum#1=52 %
\hatcurLChatnetmeccenxxxxxC
\else
??????\fi
}
\newcommand{\hatcurLCiblend}[1]{\ifnum#1=52 %
\hatcurLCiblendxxxxxC
\else
??????\fi
}
\newcommand{\hatcurLCiblendA}[1]{\ifnum#1=50 %
\hatcurLCiblendAxxxxxA
\else
\ifnum#1=51 %
\hatcurLCiblendAxxxxxB
\else
\ifnum#1=53 %
\hatcurLCiblendAxxxxxD
\else
??????\fi
\fi
\fi
}
\newcommand{\hatcurLCiblendAeccen}[1]{\ifnum#1=50 %
\hatcurLCiblendAeccenxxxxxA
\else
\ifnum#1=51 %
\hatcurLCiblendAeccenxxxxxB
\else
\ifnum#1=53 %
\hatcurLCiblendAeccenxxxxxD
\else
??????\fi
\fi
\fi
}
\newcommand{\hatcurLCiblendB}[1]{\ifnum#1=50 %
\hatcurLCiblendBxxxxxA
\else
\ifnum#1=51 %
\hatcurLCiblendBxxxxxB
\else
\ifnum#1=53 %
\hatcurLCiblendBxxxxxD
\else
??????\fi
\fi
\fi
}
\newcommand{\hatcurLCiblendBeccen}[1]{\ifnum#1=50 %
\hatcurLCiblendBeccenxxxxxA
\else
\ifnum#1=51 %
\hatcurLCiblendBeccenxxxxxB
\else
\ifnum#1=53 %
\hatcurLCiblendBeccenxxxxxD
\else
??????\fi
\fi
\fi
}
\newcommand{\hatcurLCiblendC}[1]{\ifnum#1=51 %
\hatcurLCiblendCxxxxxB
\else
??????\fi
}
\newcommand{\hatcurLCiblendCeccen}[1]{\ifnum#1=51 %
\hatcurLCiblendCeccenxxxxxB
\else
??????\fi
}
\newcommand{\hatcurLCiblendD}[1]{\ifnum#1=51 %
\hatcurLCiblendDxxxxxB
\else
??????\fi
}
\newcommand{\hatcurLCiblendDeccen}[1]{\ifnum#1=51 %
\hatcurLCiblendDeccenxxxxxB
\else
??????\fi
}
\newcommand{\hatcurLCiblendeccen}[1]{\ifnum#1=52 %
\hatcurLCiblendeccenxxxxxC
\else
??????\fi
}
\newcommand{\hatcurLCimp}[1]{\ifnum#1=50 %
\hatcurLCimpxxxxxA
\else
\ifnum#1=51 %
\hatcurLCimpxxxxxB
\else
\ifnum#1=52 %
\hatcurLCimpxxxxxC
\else
\ifnum#1=53 %
\hatcurLCimpxxxxxD
\else
??????\fi
\fi
\fi
\fi
}
\newcommand{\hatcurLCimpeccen}[1]{\ifnum#1=50 %
\hatcurLCimpeccenxxxxxA
\else
\ifnum#1=51 %
\hatcurLCimpeccenxxxxxB
\else
\ifnum#1=52 %
\hatcurLCimpeccenxxxxxC
\else
\ifnum#1=53 %
\hatcurLCimpeccenxxxxxD
\else
??????\fi
\fi
\fi
\fi
}
\newcommand{\hatcurLCingdur}[1]{\ifnum#1=50 %
\hatcurLCingdurxxxxxA
\else
\ifnum#1=51 %
\hatcurLCingdurxxxxxB
\else
\ifnum#1=52 %
\hatcurLCingdurxxxxxC
\else
\ifnum#1=53 %
\hatcurLCingdurxxxxxD
\else
??????\fi
\fi
\fi
\fi
}
\newcommand{\hatcurLCingdureccen}[1]{\ifnum#1=50 %
\hatcurLCingdureccenxxxxxA
\else
\ifnum#1=51 %
\hatcurLCingdureccenxxxxxB
\else
\ifnum#1=52 %
\hatcurLCingdureccenxxxxxC
\else
\ifnum#1=53 %
\hatcurLCingdureccenxxxxxD
\else
??????\fi
\fi
\fi
\fi
}
\newcommand{\hatcurLCP}[1]{\ifnum#1=50 %
\hatcurLCPxxxxxA
\else
\ifnum#1=51 %
\hatcurLCPxxxxxB
\else
\ifnum#1=52 %
\hatcurLCPxxxxxC
\else
\ifnum#1=53 %
\hatcurLCPxxxxxD
\else
??????\fi
\fi
\fi
\fi
}
\newcommand{\hatcurLCPeccen}[1]{\ifnum#1=50 %
\hatcurLCPeccenxxxxxA
\else
\ifnum#1=51 %
\hatcurLCPeccenxxxxxB
\else
\ifnum#1=52 %
\hatcurLCPeccenxxxxxC
\else
\ifnum#1=53 %
\hatcurLCPeccenxxxxxD
\else
??????\fi
\fi
\fi
\fi
}
\newcommand{\hatcurLCPprec}[1]{\ifnum#1=50 %
\hatcurLCPprecxxxxxA
\else
\ifnum#1=51 %
\hatcurLCPprecxxxxxB
\else
\ifnum#1=52 %
\hatcurLCPprecxxxxxC
\else
\ifnum#1=53 %
\hatcurLCPprecxxxxxD
\else
??????\fi
\fi
\fi
\fi
}
\newcommand{\hatcurLCPprececcen}[1]{\ifnum#1=50 %
\hatcurLCPprececcenxxxxxA
\else
\ifnum#1=51 %
\hatcurLCPprececcenxxxxxB
\else
\ifnum#1=52 %
\hatcurLCPprececcenxxxxxC
\else
\ifnum#1=53 %
\hatcurLCPprececcenxxxxxD
\else
??????\fi
\fi
\fi
\fi
}
\newcommand{\hatcurLCPshort}[1]{\ifnum#1=50 %
\hatcurLCPshortxxxxxA
\else
\ifnum#1=51 %
\hatcurLCPshortxxxxxB
\else
\ifnum#1=52 %
\hatcurLCPshortxxxxxC
\else
\ifnum#1=53 %
\hatcurLCPshortxxxxxD
\else
??????\fi
\fi
\fi
\fi
}
\newcommand{\hatcurLCPshorteccen}[1]{\ifnum#1=50 %
\hatcurLCPshorteccenxxxxxA
\else
\ifnum#1=51 %
\hatcurLCPshorteccenxxxxxB
\else
\ifnum#1=52 %
\hatcurLCPshorteccenxxxxxC
\else
\ifnum#1=53 %
\hatcurLCPshorteccenxxxxxD
\else
??????\fi
\fi
\fi
\fi
}
\newcommand{\hatcurLCq}[1]{\ifnum#1=50 %
\hatcurLCqxxxxxA
\else
\ifnum#1=51 %
\hatcurLCqxxxxxB
\else
\ifnum#1=52 %
\hatcurLCqxxxxxC
\else
\ifnum#1=53 %
\hatcurLCqxxxxxD
\else
??????\fi
\fi
\fi
\fi
}
\newcommand{\hatcurLCqeccen}[1]{\ifnum#1=50 %
\hatcurLCqeccenxxxxxA
\else
\ifnum#1=51 %
\hatcurLCqeccenxxxxxB
\else
\ifnum#1=52 %
\hatcurLCqeccenxxxxxC
\else
\ifnum#1=53 %
\hatcurLCqeccenxxxxxD
\else
??????\fi
\fi
\fi
\fi
}
\newcommand{\hatcurLCqshort}[1]{\ifnum#1=50 %
\hatcurLCqshortxxxxxA
\else
\ifnum#1=51 %
\hatcurLCqshortxxxxxB
\else
\ifnum#1=52 %
\hatcurLCqshortxxxxxC
\else
\ifnum#1=53 %
\hatcurLCqshortxxxxxD
\else
??????\fi
\fi
\fi
\fi
}
\newcommand{\hatcurLCqshorteccen}[1]{\ifnum#1=50 %
\hatcurLCqshorteccenxxxxxA
\else
\ifnum#1=51 %
\hatcurLCqshorteccenxxxxxB
\else
\ifnum#1=52 %
\hatcurLCqshorteccenxxxxxC
\else
\ifnum#1=53 %
\hatcurLCqshorteccenxxxxxD
\else
??????\fi
\fi
\fi
\fi
}
\newcommand{\hatcurLCrprstar}[1]{\ifnum#1=50 %
\hatcurLCrprstarxxxxxA
\else
\ifnum#1=51 %
\hatcurLCrprstarxxxxxB
\else
\ifnum#1=52 %
\hatcurLCrprstarxxxxxC
\else
\ifnum#1=53 %
\hatcurLCrprstarxxxxxD
\else
??????\fi
\fi
\fi
\fi
}
\newcommand{\hatcurLCrprstareccen}[1]{\ifnum#1=50 %
\hatcurLCrprstareccenxxxxxA
\else
\ifnum#1=51 %
\hatcurLCrprstareccenxxxxxB
\else
\ifnum#1=52 %
\hatcurLCrprstareccenxxxxxC
\else
\ifnum#1=53 %
\hatcurLCrprstareccenxxxxxD
\else
??????\fi
\fi
\fi
\fi
}
\newcommand{\hatcurLCT}[1]{\ifnum#1=50 %
\hatcurLCTxxxxxA
\else
\ifnum#1=51 %
\hatcurLCTxxxxxB
\else
\ifnum#1=52 %
\hatcurLCTxxxxxC
\else
\ifnum#1=53 %
\hatcurLCTxxxxxD
\else
??????\fi
\fi
\fi
\fi
}
\newcommand{\hatcurLCTA}[1]{\ifnum#1=50 %
\hatcurLCTAxxxxxA
\else
\ifnum#1=51 %
\hatcurLCTAxxxxxB
\else
\ifnum#1=52 %
\hatcurLCTAxxxxxC
\else
\ifnum#1=53 %
\hatcurLCTAxxxxxD
\else
??????\fi
\fi
\fi
\fi
}
\newcommand{\hatcurLCTAeccen}[1]{\ifnum#1=50 %
\hatcurLCTAeccenxxxxxA
\else
\ifnum#1=51 %
\hatcurLCTAeccenxxxxxB
\else
\ifnum#1=52 %
\hatcurLCTAeccenxxxxxC
\else
\ifnum#1=53 %
\hatcurLCTAeccenxxxxxD
\else
??????\fi
\fi
\fi
\fi
}
\newcommand{\hatcurLCTB}[1]{\ifnum#1=50 %
\hatcurLCTBxxxxxA
\else
\ifnum#1=51 %
\hatcurLCTBxxxxxB
\else
\ifnum#1=52 %
\hatcurLCTBxxxxxC
\else
\ifnum#1=53 %
\hatcurLCTBxxxxxD
\else
??????\fi
\fi
\fi
\fi
}
\newcommand{\hatcurLCTBeccen}[1]{\ifnum#1=50 %
\hatcurLCTBeccenxxxxxA
\else
\ifnum#1=51 %
\hatcurLCTBeccenxxxxxB
\else
\ifnum#1=52 %
\hatcurLCTBeccenxxxxxC
\else
\ifnum#1=53 %
\hatcurLCTBeccenxxxxxD
\else
??????\fi
\fi
\fi
\fi
}
\newcommand{\hatcurLCTeccen}[1]{\ifnum#1=50 %
\hatcurLCTeccenxxxxxA
\else
\ifnum#1=51 %
\hatcurLCTeccenxxxxxB
\else
\ifnum#1=52 %
\hatcurLCTeccenxxxxxC
\else
\ifnum#1=53 %
\hatcurLCTeccenxxxxxD
\else
??????\fi
\fi
\fi
\fi
}
\newcommand{\hatcurLCzeta}[1]{\ifnum#1=50 %
\hatcurLCzetaxxxxxA
\else
\ifnum#1=51 %
\hatcurLCzetaxxxxxB
\else
\ifnum#1=52 %
\hatcurLCzetaxxxxxC
\else
\ifnum#1=53 %
\hatcurLCzetaxxxxxD
\else
??????\fi
\fi
\fi
\fi
}
\newcommand{\hatcurLCzetaeccen}[1]{\ifnum#1=50 %
\hatcurLCzetaeccenxxxxxA
\else
\ifnum#1=51 %
\hatcurLCzetaeccenxxxxxB
\else
\ifnum#1=52 %
\hatcurLCzetaeccenxxxxxC
\else
\ifnum#1=53 %
\hatcurLCzetaeccenxxxxxD
\else
??????\fi
\fi
\fi
\fi
}
\newcommand{\hatcurPPaequiv}[1]{\ifnum#1=50 %
\hatcurPPaequivxxxxxA
\else
\ifnum#1=51 %
\hatcurPPaequivxxxxxB
\else
\ifnum#1=52 %
\hatcurPPaequivxxxxxC
\else
\ifnum#1=53 %
\hatcurPPaequivxxxxxD
\else
??????\fi
\fi
\fi
\fi
}
\newcommand{\hatcurPPaequiveccen}[1]{\ifnum#1=50 %
\hatcurPPaequiveccenxxxxxA
\else
\ifnum#1=51 %
\hatcurPPaequiveccenxxxxxB
\else
\ifnum#1=52 %
\hatcurPPaequiveccenxxxxxC
\else
\ifnum#1=53 %
\hatcurPPaequiveccenxxxxxD
\else
??????\fi
\fi
\fi
\fi
}
\newcommand{\hatcurPPar}[1]{\ifnum#1=50 %
\hatcurPParxxxxxA
\else
\ifnum#1=51 %
\hatcurPParxxxxxB
\else
\ifnum#1=52 %
\hatcurPParxxxxxC
\else
\ifnum#1=53 %
\hatcurPParxxxxxD
\else
??????\fi
\fi
\fi
\fi
}
\newcommand{\hatcurPPareccen}[1]{\ifnum#1=50 %
\hatcurPPareccenxxxxxA
\else
\ifnum#1=51 %
\hatcurPPareccenxxxxxB
\else
\ifnum#1=52 %
\hatcurPPareccenxxxxxC
\else
\ifnum#1=53 %
\hatcurPPareccenxxxxxD
\else
??????\fi
\fi
\fi
\fi
}
\newcommand{\hatcurPParel}[1]{\ifnum#1=50 %
\hatcurPParelxxxxxA
\else
\ifnum#1=51 %
\hatcurPParelxxxxxB
\else
\ifnum#1=52 %
\hatcurPParelxxxxxC
\else
\ifnum#1=53 %
\hatcurPParelxxxxxD
\else
??????\fi
\fi
\fi
\fi
}
\newcommand{\hatcurPPareleccen}[1]{\ifnum#1=50 %
\hatcurPPareleccenxxxxxA
\else
\ifnum#1=51 %
\hatcurPPareleccenxxxxxB
\else
\ifnum#1=52 %
\hatcurPPareleccenxxxxxC
\else
\ifnum#1=53 %
\hatcurPPareleccenxxxxxD
\else
??????\fi
\fi
\fi
\fi
}
\newcommand{\hatcurPPfluxap}[1]{\ifnum#1=50 %
\hatcurPPfluxapxxxxxA
\else
\ifnum#1=51 %
\hatcurPPfluxapxxxxxB
\else
\ifnum#1=52 %
\hatcurPPfluxapxxxxxC
\else
\ifnum#1=53 %
\hatcurPPfluxapxxxxxD
\else
??????\fi
\fi
\fi
\fi
}
\newcommand{\hatcurPPfluxapdim}[1]{\ifnum#1=50 %
\hatcurPPfluxapdimxxxxxA
\else
\ifnum#1=51 %
\hatcurPPfluxapdimxxxxxB
\else
\ifnum#1=52 %
\hatcurPPfluxapdimxxxxxC
\else
\ifnum#1=53 %
\hatcurPPfluxapdimxxxxxD
\else
??????\fi
\fi
\fi
\fi
}
\newcommand{\hatcurPPfluxapdimeccen}[1]{\ifnum#1=50 %
\hatcurPPfluxapdimeccenxxxxxA
\else
\ifnum#1=51 %
\hatcurPPfluxapdimeccenxxxxxB
\else
\ifnum#1=52 %
\hatcurPPfluxapdimeccenxxxxxC
\else
\ifnum#1=53 %
\hatcurPPfluxapdimeccenxxxxxD
\else
??????\fi
\fi
\fi
\fi
}
\newcommand{\hatcurPPfluxapeccen}[1]{\ifnum#1=50 %
\hatcurPPfluxapeccenxxxxxA
\else
\ifnum#1=51 %
\hatcurPPfluxapeccenxxxxxB
\else
\ifnum#1=52 %
\hatcurPPfluxapeccenxxxxxC
\else
\ifnum#1=53 %
\hatcurPPfluxapeccenxxxxxD
\else
??????\fi
\fi
\fi
\fi
}
\newcommand{\hatcurPPfluxavg}[1]{\ifnum#1=50 %
\hatcurPPfluxavgxxxxxA
\else
\ifnum#1=51 %
\hatcurPPfluxavgxxxxxB
\else
\ifnum#1=52 %
\hatcurPPfluxavgxxxxxC
\else
\ifnum#1=53 %
\hatcurPPfluxavgxxxxxD
\else
??????\fi
\fi
\fi
\fi
}
\newcommand{\hatcurPPfluxavgdim}[1]{\ifnum#1=50 %
\hatcurPPfluxavgdimxxxxxA
\else
\ifnum#1=51 %
\hatcurPPfluxavgdimxxxxxB
\else
\ifnum#1=52 %
\hatcurPPfluxavgdimxxxxxC
\else
\ifnum#1=53 %
\hatcurPPfluxavgdimxxxxxD
\else
??????\fi
\fi
\fi
\fi
}
\newcommand{\hatcurPPfluxavgdimeccen}[1]{\ifnum#1=50 %
\hatcurPPfluxavgdimeccenxxxxxA
\else
\ifnum#1=51 %
\hatcurPPfluxavgdimeccenxxxxxB
\else
\ifnum#1=52 %
\hatcurPPfluxavgdimeccenxxxxxC
\else
\ifnum#1=53 %
\hatcurPPfluxavgdimeccenxxxxxD
\else
??????\fi
\fi
\fi
\fi
}
\newcommand{\hatcurPPfluxavgeccen}[1]{\ifnum#1=50 %
\hatcurPPfluxavgeccenxxxxxA
\else
\ifnum#1=51 %
\hatcurPPfluxavgeccenxxxxxB
\else
\ifnum#1=52 %
\hatcurPPfluxavgeccenxxxxxC
\else
\ifnum#1=53 %
\hatcurPPfluxavgeccenxxxxxD
\else
??????\fi
\fi
\fi
\fi
}
\newcommand{\hatcurPPfluxavglog}[1]{\ifnum#1=50 %
\hatcurPPfluxavglogxxxxxA
\else
\ifnum#1=51 %
\hatcurPPfluxavglogxxxxxB
\else
\ifnum#1=52 %
\hatcurPPfluxavglogxxxxxC
\else
\ifnum#1=53 %
\hatcurPPfluxavglogxxxxxD
\else
??????\fi
\fi
\fi
\fi
}
\newcommand{\hatcurPPfluxavglogeccen}[1]{\ifnum#1=50 %
\hatcurPPfluxavglogeccenxxxxxA
\else
\ifnum#1=51 %
\hatcurPPfluxavglogeccenxxxxxB
\else
\ifnum#1=52 %
\hatcurPPfluxavglogeccenxxxxxC
\else
\ifnum#1=53 %
\hatcurPPfluxavglogeccenxxxxxD
\else
??????\fi
\fi
\fi
\fi
}
\newcommand{\hatcurPPfluxperi}[1]{\ifnum#1=50 %
\hatcurPPfluxperixxxxxA
\else
\ifnum#1=51 %
\hatcurPPfluxperixxxxxB
\else
\ifnum#1=52 %
\hatcurPPfluxperixxxxxC
\else
\ifnum#1=53 %
\hatcurPPfluxperixxxxxD
\else
??????\fi
\fi
\fi
\fi
}
\newcommand{\hatcurPPfluxperidim}[1]{\ifnum#1=50 %
\hatcurPPfluxperidimxxxxxA
\else
\ifnum#1=51 %
\hatcurPPfluxperidimxxxxxB
\else
\ifnum#1=52 %
\hatcurPPfluxperidimxxxxxC
\else
\ifnum#1=53 %
\hatcurPPfluxperidimxxxxxD
\else
??????\fi
\fi
\fi
\fi
}
\newcommand{\hatcurPPfluxperidimeccen}[1]{\ifnum#1=50 %
\hatcurPPfluxperidimeccenxxxxxA
\else
\ifnum#1=51 %
\hatcurPPfluxperidimeccenxxxxxB
\else
\ifnum#1=52 %
\hatcurPPfluxperidimeccenxxxxxC
\else
\ifnum#1=53 %
\hatcurPPfluxperidimeccenxxxxxD
\else
??????\fi
\fi
\fi
\fi
}
\newcommand{\hatcurPPfluxperieccen}[1]{\ifnum#1=50 %
\hatcurPPfluxperieccenxxxxxA
\else
\ifnum#1=51 %
\hatcurPPfluxperieccenxxxxxB
\else
\ifnum#1=52 %
\hatcurPPfluxperieccenxxxxxC
\else
\ifnum#1=53 %
\hatcurPPfluxperieccenxxxxxD
\else
??????\fi
\fi
\fi
\fi
}
\newcommand{\hatcurPPg}[1]{\ifnum#1=50 %
\hatcurPPgxxxxxA
\else
\ifnum#1=51 %
\hatcurPPgxxxxxB
\else
\ifnum#1=52 %
\hatcurPPgxxxxxC
\else
\ifnum#1=53 %
\hatcurPPgxxxxxD
\else
??????\fi
\fi
\fi
\fi
}
\newcommand{\hatcurPPgeccen}[1]{\ifnum#1=50 %
\hatcurPPgeccenxxxxxA
\else
\ifnum#1=51 %
\hatcurPPgeccenxxxxxB
\else
\ifnum#1=52 %
\hatcurPPgeccenxxxxxC
\else
\ifnum#1=53 %
\hatcurPPgeccenxxxxxD
\else
??????\fi
\fi
\fi
\fi
}
\newcommand{\hatcurPPi}[1]{\ifnum#1=50 %
\hatcurPPixxxxxA
\else
\ifnum#1=51 %
\hatcurPPixxxxxB
\else
\ifnum#1=52 %
\hatcurPPixxxxxC
\else
\ifnum#1=53 %
\hatcurPPixxxxxD
\else
??????\fi
\fi
\fi
\fi
}
\newcommand{\hatcurPPieccen}[1]{\ifnum#1=50 %
\hatcurPPieccenxxxxxA
\else
\ifnum#1=51 %
\hatcurPPieccenxxxxxB
\else
\ifnum#1=52 %
\hatcurPPieccenxxxxxC
\else
\ifnum#1=53 %
\hatcurPPieccenxxxxxD
\else
??????\fi
\fi
\fi
\fi
}
\newcommand{\hatcurPPlogg}[1]{\ifnum#1=50 %
\hatcurPPloggxxxxxA
\else
\ifnum#1=51 %
\hatcurPPloggxxxxxB
\else
\ifnum#1=52 %
\hatcurPPloggxxxxxC
\else
\ifnum#1=53 %
\hatcurPPloggxxxxxD
\else
??????\fi
\fi
\fi
\fi
}
\newcommand{\hatcurPPloggeccen}[1]{\ifnum#1=50 %
\hatcurPPloggeccenxxxxxA
\else
\ifnum#1=51 %
\hatcurPPloggeccenxxxxxB
\else
\ifnum#1=52 %
\hatcurPPloggeccenxxxxxC
\else
\ifnum#1=53 %
\hatcurPPloggeccenxxxxxD
\else
??????\fi
\fi
\fi
\fi
}
\newcommand{\hatcurPPm}[1]{\ifnum#1=50 %
\hatcurPPmxxxxxA
\else
\ifnum#1=51 %
\hatcurPPmxxxxxB
\else
\ifnum#1=52 %
\hatcurPPmxxxxxC
\else
\ifnum#1=53 %
\hatcurPPmxxxxxD
\else
??????\fi
\fi
\fi
\fi
}
\newcommand{\hatcurPPme}[1]{\ifnum#1=50 %
\hatcurPPmexxxxxA
\else
\ifnum#1=51 %
\hatcurPPmexxxxxB
\else
\ifnum#1=52 %
\hatcurPPmexxxxxC
\else
\ifnum#1=53 %
\hatcurPPmexxxxxD
\else
??????\fi
\fi
\fi
\fi
}
\newcommand{\hatcurPPmeccen}[1]{\ifnum#1=50 %
\hatcurPPmeccenxxxxxA
\else
\ifnum#1=51 %
\hatcurPPmeccenxxxxxB
\else
\ifnum#1=52 %
\hatcurPPmeccenxxxxxC
\else
\ifnum#1=53 %
\hatcurPPmeccenxxxxxD
\else
??????\fi
\fi
\fi
\fi
}
\newcommand{\hatcurPPmeeccen}[1]{\ifnum#1=50 %
\hatcurPPmeeccenxxxxxA
\else
\ifnum#1=51 %
\hatcurPPmeeccenxxxxxB
\else
\ifnum#1=52 %
\hatcurPPmeeccenxxxxxC
\else
\ifnum#1=53 %
\hatcurPPmeeccenxxxxxD
\else
??????\fi
\fi
\fi
\fi
}
\newcommand{\hatcurPPmelong}[1]{\ifnum#1=50 %
\hatcurPPmelongxxxxxA
\else
\ifnum#1=51 %
\hatcurPPmelongxxxxxB
\else
\ifnum#1=52 %
\hatcurPPmelongxxxxxC
\else
\ifnum#1=53 %
\hatcurPPmelongxxxxxD
\else
??????\fi
\fi
\fi
\fi
}
\newcommand{\hatcurPPmelongeccen}[1]{\ifnum#1=50 %
\hatcurPPmelongeccenxxxxxA
\else
\ifnum#1=51 %
\hatcurPPmelongeccenxxxxxB
\else
\ifnum#1=52 %
\hatcurPPmelongeccenxxxxxC
\else
\ifnum#1=53 %
\hatcurPPmelongeccenxxxxxD
\else
??????\fi
\fi
\fi
\fi
}
\newcommand{\hatcurPPmeshort}[1]{\ifnum#1=50 %
\hatcurPPmeshortxxxxxA
\else
\ifnum#1=51 %
\hatcurPPmeshortxxxxxB
\else
\ifnum#1=52 %
\hatcurPPmeshortxxxxxC
\else
\ifnum#1=53 %
\hatcurPPmeshortxxxxxD
\else
??????\fi
\fi
\fi
\fi
}
\newcommand{\hatcurPPmeshorteccen}[1]{\ifnum#1=50 %
\hatcurPPmeshorteccenxxxxxA
\else
\ifnum#1=51 %
\hatcurPPmeshorteccenxxxxxB
\else
\ifnum#1=52 %
\hatcurPPmeshorteccenxxxxxC
\else
\ifnum#1=53 %
\hatcurPPmeshorteccenxxxxxD
\else
??????\fi
\fi
\fi
\fi
}
\newcommand{\hatcurPPmlong}[1]{\ifnum#1=50 %
\hatcurPPmlongxxxxxA
\else
\ifnum#1=51 %
\hatcurPPmlongxxxxxB
\else
\ifnum#1=52 %
\hatcurPPmlongxxxxxC
\else
\ifnum#1=53 %
\hatcurPPmlongxxxxxD
\else
??????\fi
\fi
\fi
\fi
}
\newcommand{\hatcurPPmlongeccen}[1]{\ifnum#1=50 %
\hatcurPPmlongeccenxxxxxA
\else
\ifnum#1=51 %
\hatcurPPmlongeccenxxxxxB
\else
\ifnum#1=52 %
\hatcurPPmlongeccenxxxxxC
\else
\ifnum#1=53 %
\hatcurPPmlongeccenxxxxxD
\else
??????\fi
\fi
\fi
\fi
}
\newcommand{\hatcurPPmrcorr}[1]{\ifnum#1=50 %
\hatcurPPmrcorrxxxxxA
\else
\ifnum#1=51 %
\hatcurPPmrcorrxxxxxB
\else
\ifnum#1=52 %
\hatcurPPmrcorrxxxxxC
\else
\ifnum#1=53 %
\hatcurPPmrcorrxxxxxD
\else
??????\fi
\fi
\fi
\fi
}
\newcommand{\hatcurPPmrcorreccen}[1]{\ifnum#1=50 %
\hatcurPPmrcorreccenxxxxxA
\else
\ifnum#1=51 %
\hatcurPPmrcorreccenxxxxxB
\else
\ifnum#1=52 %
\hatcurPPmrcorreccenxxxxxC
\else
\ifnum#1=53 %
\hatcurPPmrcorreccenxxxxxD
\else
??????\fi
\fi
\fi
\fi
}
\newcommand{\hatcurPPmshort}[1]{\ifnum#1=50 %
\hatcurPPmshortxxxxxA
\else
\ifnum#1=51 %
\hatcurPPmshortxxxxxB
\else
\ifnum#1=52 %
\hatcurPPmshortxxxxxC
\else
\ifnum#1=53 %
\hatcurPPmshortxxxxxD
\else
??????\fi
\fi
\fi
\fi
}
\newcommand{\hatcurPPmshorteccen}[1]{\ifnum#1=50 %
\hatcurPPmshorteccenxxxxxA
\else
\ifnum#1=51 %
\hatcurPPmshorteccenxxxxxB
\else
\ifnum#1=52 %
\hatcurPPmshorteccenxxxxxC
\else
\ifnum#1=53 %
\hatcurPPmshorteccenxxxxxD
\else
??????\fi
\fi
\fi
\fi
}
\newcommand{\hatcurPPperi}[1]{\ifnum#1=50 %
\hatcurPPperixxxxxA
\else
\ifnum#1=51 %
\hatcurPPperixxxxxB
\else
\ifnum#1=52 %
\hatcurPPperixxxxxC
\else
\ifnum#1=53 %
\hatcurPPperixxxxxD
\else
??????\fi
\fi
\fi
\fi
}
\newcommand{\hatcurPPperieccen}[1]{\ifnum#1=50 %
\hatcurPPperieccenxxxxxA
\else
\ifnum#1=51 %
\hatcurPPperieccenxxxxxB
\else
\ifnum#1=52 %
\hatcurPPperieccenxxxxxC
\else
\ifnum#1=53 %
\hatcurPPperieccenxxxxxD
\else
??????\fi
\fi
\fi
\fi
}
\newcommand{\hatcurPPphiconj}[1]{\ifnum#1=50 %
\hatcurPPphiconjxxxxxA
\else
\ifnum#1=51 %
\hatcurPPphiconjxxxxxB
\else
\ifnum#1=52 %
\hatcurPPphiconjxxxxxC
\else
\ifnum#1=53 %
\hatcurPPphiconjxxxxxD
\else
??????\fi
\fi
\fi
\fi
}
\newcommand{\hatcurPPphiconjeccen}[1]{\ifnum#1=50 %
\hatcurPPphiconjeccenxxxxxA
\else
\ifnum#1=51 %
\hatcurPPphiconjeccenxxxxxB
\else
\ifnum#1=52 %
\hatcurPPphiconjeccenxxxxxC
\else
\ifnum#1=53 %
\hatcurPPphiconjeccenxxxxxD
\else
??????\fi
\fi
\fi
\fi
}
\newcommand{\hatcurPPr}[1]{\ifnum#1=50 %
\hatcurPPrxxxxxA
\else
\ifnum#1=51 %
\hatcurPPrxxxxxB
\else
\ifnum#1=52 %
\hatcurPPrxxxxxC
\else
\ifnum#1=53 %
\hatcurPPrxxxxxD
\else
??????\fi
\fi
\fi
\fi
}
\newcommand{\hatcurPPre}[1]{\ifnum#1=50 %
\hatcurPPrexxxxxA
\else
\ifnum#1=51 %
\hatcurPPrexxxxxB
\else
\ifnum#1=52 %
\hatcurPPrexxxxxC
\else
\ifnum#1=53 %
\hatcurPPrexxxxxD
\else
??????\fi
\fi
\fi
\fi
}
\newcommand{\hatcurPPreccen}[1]{\ifnum#1=50 %
\hatcurPPreccenxxxxxA
\else
\ifnum#1=51 %
\hatcurPPreccenxxxxxB
\else
\ifnum#1=52 %
\hatcurPPreccenxxxxxC
\else
\ifnum#1=53 %
\hatcurPPreccenxxxxxD
\else
??????\fi
\fi
\fi
\fi
}
\newcommand{\hatcurPPreeccen}[1]{\ifnum#1=50 %
\hatcurPPreeccenxxxxxA
\else
\ifnum#1=51 %
\hatcurPPreeccenxxxxxB
\else
\ifnum#1=52 %
\hatcurPPreeccenxxxxxC
\else
\ifnum#1=53 %
\hatcurPPreeccenxxxxxD
\else
??????\fi
\fi
\fi
\fi
}
\newcommand{\hatcurPPrelong}[1]{\ifnum#1=50 %
\hatcurPPrelongxxxxxA
\else
\ifnum#1=51 %
\hatcurPPrelongxxxxxB
\else
\ifnum#1=52 %
\hatcurPPrelongxxxxxC
\else
\ifnum#1=53 %
\hatcurPPrelongxxxxxD
\else
??????\fi
\fi
\fi
\fi
}
\newcommand{\hatcurPPrelongeccen}[1]{\ifnum#1=50 %
\hatcurPPrelongeccenxxxxxA
\else
\ifnum#1=51 %
\hatcurPPrelongeccenxxxxxB
\else
\ifnum#1=52 %
\hatcurPPrelongeccenxxxxxC
\else
\ifnum#1=53 %
\hatcurPPrelongeccenxxxxxD
\else
??????\fi
\fi
\fi
\fi
}
\newcommand{\hatcurPPreshort}[1]{\ifnum#1=50 %
\hatcurPPreshortxxxxxA
\else
\ifnum#1=51 %
\hatcurPPreshortxxxxxB
\else
\ifnum#1=52 %
\hatcurPPreshortxxxxxC
\else
\ifnum#1=53 %
\hatcurPPreshortxxxxxD
\else
??????\fi
\fi
\fi
\fi
}
\newcommand{\hatcurPPreshorteccen}[1]{\ifnum#1=50 %
\hatcurPPreshorteccenxxxxxA
\else
\ifnum#1=51 %
\hatcurPPreshorteccenxxxxxB
\else
\ifnum#1=52 %
\hatcurPPreshorteccenxxxxxC
\else
\ifnum#1=53 %
\hatcurPPreshorteccenxxxxxD
\else
??????\fi
\fi
\fi
\fi
}
\newcommand{\hatcurPPrho}[1]{\ifnum#1=50 %
\hatcurPPrhoxxxxxA
\else
\ifnum#1=51 %
\hatcurPPrhoxxxxxB
\else
\ifnum#1=52 %
\hatcurPPrhoxxxxxC
\else
\ifnum#1=53 %
\hatcurPPrhoxxxxxD
\else
??????\fi
\fi
\fi
\fi
}
\newcommand{\hatcurPPrhoeccen}[1]{\ifnum#1=50 %
\hatcurPPrhoeccenxxxxxA
\else
\ifnum#1=51 %
\hatcurPPrhoeccenxxxxxB
\else
\ifnum#1=52 %
\hatcurPPrhoeccenxxxxxC
\else
\ifnum#1=53 %
\hatcurPPrhoeccenxxxxxD
\else
??????\fi
\fi
\fi
\fi
}
\newcommand{\hatcurPPrlong}[1]{\ifnum#1=50 %
\hatcurPPrlongxxxxxA
\else
\ifnum#1=51 %
\hatcurPPrlongxxxxxB
\else
\ifnum#1=52 %
\hatcurPPrlongxxxxxC
\else
\ifnum#1=53 %
\hatcurPPrlongxxxxxD
\else
??????\fi
\fi
\fi
\fi
}
\newcommand{\hatcurPPrlongeccen}[1]{\ifnum#1=50 %
\hatcurPPrlongeccenxxxxxA
\else
\ifnum#1=51 %
\hatcurPPrlongeccenxxxxxB
\else
\ifnum#1=52 %
\hatcurPPrlongeccenxxxxxC
\else
\ifnum#1=53 %
\hatcurPPrlongeccenxxxxxD
\else
??????\fi
\fi
\fi
\fi
}
\newcommand{\hatcurPPrshort}[1]{\ifnum#1=50 %
\hatcurPPrshortxxxxxA
\else
\ifnum#1=51 %
\hatcurPPrshortxxxxxB
\else
\ifnum#1=52 %
\hatcurPPrshortxxxxxC
\else
\ifnum#1=53 %
\hatcurPPrshortxxxxxD
\else
??????\fi
\fi
\fi
\fi
}
\newcommand{\hatcurPPrshorteccen}[1]{\ifnum#1=50 %
\hatcurPPrshorteccenxxxxxA
\else
\ifnum#1=51 %
\hatcurPPrshorteccenxxxxxB
\else
\ifnum#1=52 %
\hatcurPPrshorteccenxxxxxC
\else
\ifnum#1=53 %
\hatcurPPrshorteccenxxxxxD
\else
??????\fi
\fi
\fi
\fi
}
\newcommand{\hatcurPPtcirc}[1]{\ifnum#1=50 %
\hatcurPPtcircxxxxxA
\else
\ifnum#1=51 %
\hatcurPPtcircxxxxxB
\else
\ifnum#1=52 %
\hatcurPPtcircxxxxxC
\else
\ifnum#1=53 %
\hatcurPPtcircxxxxxD
\else
??????\fi
\fi
\fi
\fi
}
\newcommand{\hatcurPPtcirceccen}[1]{\ifnum#1=50 %
\hatcurPPtcirceccenxxxxxA
\else
\ifnum#1=51 %
\hatcurPPtcirceccenxxxxxB
\else
\ifnum#1=52 %
\hatcurPPtcirceccenxxxxxC
\else
\ifnum#1=53 %
\hatcurPPtcirceccenxxxxxD
\else
??????\fi
\fi
\fi
\fi
}
\newcommand{\hatcurPPteff}[1]{\ifnum#1=50 %
\hatcurPPteffxxxxxA
\else
\ifnum#1=51 %
\hatcurPPteffxxxxxB
\else
\ifnum#1=52 %
\hatcurPPteffxxxxxC
\else
\ifnum#1=53 %
\hatcurPPteffxxxxxD
\else
??????\fi
\fi
\fi
\fi
}
\newcommand{\hatcurPPteffeccen}[1]{\ifnum#1=50 %
\hatcurPPteffeccenxxxxxA
\else
\ifnum#1=51 %
\hatcurPPteffeccenxxxxxB
\else
\ifnum#1=52 %
\hatcurPPteffeccenxxxxxC
\else
\ifnum#1=53 %
\hatcurPPteffeccenxxxxxD
\else
??????\fi
\fi
\fi
\fi
}
\newcommand{\hatcurPPtheta}[1]{\ifnum#1=50 %
\hatcurPPthetaxxxxxA
\else
\ifnum#1=51 %
\hatcurPPthetaxxxxxB
\else
\ifnum#1=52 %
\hatcurPPthetaxxxxxC
\else
\ifnum#1=53 %
\hatcurPPthetaxxxxxD
\else
??????\fi
\fi
\fi
\fi
}
\newcommand{\hatcurPPthetaeccen}[1]{\ifnum#1=50 %
\hatcurPPthetaeccenxxxxxA
\else
\ifnum#1=51 %
\hatcurPPthetaeccenxxxxxB
\else
\ifnum#1=52 %
\hatcurPPthetaeccenxxxxxC
\else
\ifnum#1=53 %
\hatcurPPthetaeccenxxxxxD
\else
??????\fi
\fi
\fi
\fi
}
\newcommand{\hatcurPPtinfall}[1]{\ifnum#1=50 %
\hatcurPPtinfallxxxxxA
\else
\ifnum#1=51 %
\hatcurPPtinfallxxxxxB
\else
\ifnum#1=52 %
\hatcurPPtinfallxxxxxC
\else
\ifnum#1=53 %
\hatcurPPtinfallxxxxxD
\else
??????\fi
\fi
\fi
\fi
}
\newcommand{\hatcurPPtinfalleccen}[1]{\ifnum#1=50 %
\hatcurPPtinfalleccenxxxxxA
\else
\ifnum#1=51 %
\hatcurPPtinfalleccenxxxxxB
\else
\ifnum#1=52 %
\hatcurPPtinfalleccenxxxxxC
\else
\ifnum#1=53 %
\hatcurPPtinfalleccenxxxxxD
\else
??????\fi
\fi
\fi
\fi
}
\newcommand{\hatcurRVeccen}[1]{\ifnum#1=50 %
\hatcurRVeccenxxxxxA
\else
\ifnum#1=51 %
\hatcurRVeccenxxxxxB
\else
\ifnum#1=52 %
\hatcurRVeccenxxxxxC
\else
\ifnum#1=53 %
\hatcurRVeccenxxxxxD
\else
??????\fi
\fi
\fi
\fi
}
\newcommand{\hatcurRVecceneccen}[1]{\ifnum#1=50 %
\hatcurRVecceneccenxxxxxA
\else
\ifnum#1=51 %
\hatcurRVecceneccenxxxxxB
\else
\ifnum#1=52 %
\hatcurRVecceneccenxxxxxC
\else
\ifnum#1=53 %
\hatcurRVecceneccenxxxxxD
\else
??????\fi
\fi
\fi
\fi
}
\newcommand{\hatcurRVeccentwosiglim}[1]{\ifnum#1=50 %
\hatcurRVeccentwosiglimxxxxxA
\else
\ifnum#1=51 %
\hatcurRVeccentwosiglimxxxxxB
\else
\ifnum#1=52 %
\hatcurRVeccentwosiglimxxxxxC
\else
\ifnum#1=53 %
\hatcurRVeccentwosiglimxxxxxD
\else
??????\fi
\fi
\fi
\fi
}
\newcommand{\hatcurRVeccentwosiglimeccen}[1]{\ifnum#1=50 %
\hatcurRVeccentwosiglimeccenxxxxxA
\else
\ifnum#1=51 %
\hatcurRVeccentwosiglimeccenxxxxxB
\else
\ifnum#1=52 %
\hatcurRVeccentwosiglimeccenxxxxxC
\else
\ifnum#1=53 %
\hatcurRVeccentwosiglimeccenxxxxxD
\else
??????\fi
\fi
\fi
\fi
}
\newcommand{\hatcurRVfitrms}[1]{\ifnum#1=52 %
\hatcurRVfitrmsxxxxxC
\else
\ifnum#1=53 %
\hatcurRVfitrmsxxxxxD
\else
??????\fi
\fi
}
\newcommand{\hatcurRVfitrmsA}[1]{\ifnum#1=50 %
\hatcurRVfitrmsAxxxxxA
\else
\ifnum#1=51 %
\hatcurRVfitrmsAxxxxxB
\else
??????\fi
\fi
}
\newcommand{\hatcurRVfitrmsAeccen}[1]{\ifnum#1=50 %
\hatcurRVfitrmsAeccenxxxxxA
\else
\ifnum#1=51 %
\hatcurRVfitrmsAeccenxxxxxB
\else
??????\fi
\fi
}
\newcommand{\hatcurRVfitrmsB}[1]{\ifnum#1=50 %
\hatcurRVfitrmsBxxxxxA
\else
\ifnum#1=51 %
\hatcurRVfitrmsBxxxxxB
\else
??????\fi
\fi
}
\newcommand{\hatcurRVfitrmsBeccen}[1]{\ifnum#1=50 %
\hatcurRVfitrmsBeccenxxxxxA
\else
\ifnum#1=51 %
\hatcurRVfitrmsBeccenxxxxxB
\else
??????\fi
\fi
}
\newcommand{\hatcurRVfitrmsC}[1]{\ifnum#1=50 %
\hatcurRVfitrmsCxxxxxA
\else
\ifnum#1=51 %
\hatcurRVfitrmsCxxxxxB
\else
??????\fi
\fi
}
\newcommand{\hatcurRVfitrmsCeccen}[1]{\ifnum#1=50 %
\hatcurRVfitrmsCeccenxxxxxA
\else
??????\fi
}
\newcommand{\hatcurRVfitrmseccen}[1]{\ifnum#1=52 %
\hatcurRVfitrmseccenxxxxxC
\else
\ifnum#1=53 %
\hatcurRVfitrmseccenxxxxxD
\else
??????\fi
\fi
}
\newcommand{\hatcurRVgamma}[1]{\ifnum#1=52 %
\hatcurRVgammaxxxxxC
\else
\ifnum#1=53 %
\hatcurRVgammaxxxxxD
\else
??????\fi
\fi
}
\newcommand{\hatcurRVgammaA}[1]{\ifnum#1=50 %
\hatcurRVgammaAxxxxxA
\else
\ifnum#1=51 %
\hatcurRVgammaAxxxxxB
\else
??????\fi
\fi
}
\newcommand{\hatcurRVgammaAeccen}[1]{\ifnum#1=50 %
\hatcurRVgammaAeccenxxxxxA
\else
\ifnum#1=51 %
\hatcurRVgammaAeccenxxxxxB
\else
??????\fi
\fi
}
\newcommand{\hatcurRVgammaB}[1]{\ifnum#1=50 %
\hatcurRVgammaBxxxxxA
\else
\ifnum#1=51 %
\hatcurRVgammaBxxxxxB
\else
??????\fi
\fi
}
\newcommand{\hatcurRVgammaBeccen}[1]{\ifnum#1=50 %
\hatcurRVgammaBeccenxxxxxA
\else
\ifnum#1=51 %
\hatcurRVgammaBeccenxxxxxB
\else
??????\fi
\fi
}
\newcommand{\hatcurRVgammaC}[1]{\ifnum#1=50 %
\hatcurRVgammaCxxxxxA
\else
\ifnum#1=51 %
\hatcurRVgammaCxxxxxB
\else
??????\fi
\fi
}
\newcommand{\hatcurRVgammaCeccen}[1]{\ifnum#1=50 %
\hatcurRVgammaCeccenxxxxxA
\else
??????\fi
}
\newcommand{\hatcurRVgammaeccen}[1]{\ifnum#1=52 %
\hatcurRVgammaeccenxxxxxC
\else
\ifnum#1=53 %
\hatcurRVgammaeccenxxxxxD
\else
??????\fi
\fi
}
\newcommand{\hatcurRVh}[1]{\ifnum#1=50 %
\hatcurRVhxxxxxA
\else
\ifnum#1=51 %
\hatcurRVhxxxxxB
\else
\ifnum#1=52 %
\hatcurRVhxxxxxC
\else
\ifnum#1=53 %
\hatcurRVhxxxxxD
\else
??????\fi
\fi
\fi
\fi
}
\newcommand{\hatcurRVheccen}[1]{\ifnum#1=50 %
\hatcurRVheccenxxxxxA
\else
\ifnum#1=51 %
\hatcurRVheccenxxxxxB
\else
\ifnum#1=52 %
\hatcurRVheccenxxxxxC
\else
\ifnum#1=53 %
\hatcurRVheccenxxxxxD
\else
??????\fi
\fi
\fi
\fi
}
\newcommand{\hatcurRVjitter}[1]{\ifnum#1=52 %
\hatcurRVjitterxxxxxC
\else
\ifnum#1=53 %
\hatcurRVjitterxxxxxD
\else
??????\fi
\fi
}
\newcommand{\hatcurRVjitterA}[1]{\ifnum#1=50 %
\hatcurRVjitterAxxxxxA
\else
\ifnum#1=51 %
\hatcurRVjitterAxxxxxB
\else
??????\fi
\fi
}
\newcommand{\hatcurRVjitterAeccen}[1]{\ifnum#1=50 %
\hatcurRVjitterAeccenxxxxxA
\else
\ifnum#1=51 %
\hatcurRVjitterAeccenxxxxxB
\else
??????\fi
\fi
}
\newcommand{\hatcurRVjitterB}[1]{\ifnum#1=50 %
\hatcurRVjitterBxxxxxA
\else
\ifnum#1=51 %
\hatcurRVjitterBxxxxxB
\else
??????\fi
\fi
}
\newcommand{\hatcurRVjitterBeccen}[1]{\ifnum#1=50 %
\hatcurRVjitterBeccenxxxxxA
\else
\ifnum#1=51 %
\hatcurRVjitterBeccenxxxxxB
\else
??????\fi
\fi
}
\newcommand{\hatcurRVjitterC}[1]{\ifnum#1=50 %
\hatcurRVjitterCxxxxxA
\else
\ifnum#1=51 %
\hatcurRVjitterCxxxxxB
\else
??????\fi
\fi
}
\newcommand{\hatcurRVjitterCeccen}[1]{\ifnum#1=50 %
\hatcurRVjitterCeccenxxxxxA
\else
??????\fi
}
\newcommand{\hatcurRVjittereccen}[1]{\ifnum#1=52 %
\hatcurRVjittereccenxxxxxC
\else
\ifnum#1=53 %
\hatcurRVjittereccenxxxxxD
\else
??????\fi
\fi
}
\newcommand{\hatcurRVk}[1]{\ifnum#1=50 %
\hatcurRVkxxxxxA
\else
\ifnum#1=51 %
\hatcurRVkxxxxxB
\else
\ifnum#1=52 %
\hatcurRVkxxxxxC
\else
\ifnum#1=53 %
\hatcurRVkxxxxxD
\else
??????\fi
\fi
\fi
\fi
}
\newcommand{\hatcurRVK}[1]{\ifnum#1=50 %
\hatcurRVKxxxxxA
\else
\ifnum#1=51 %
\hatcurRVKxxxxxB
\else
\ifnum#1=52 %
\hatcurRVKxxxxxC
\else
\ifnum#1=53 %
\hatcurRVKxxxxxD
\else
??????\fi
\fi
\fi
\fi
}
\newcommand{\hatcurRVkeccen}[1]{\ifnum#1=50 %
\hatcurRVkeccenxxxxxA
\else
\ifnum#1=51 %
\hatcurRVkeccenxxxxxB
\else
\ifnum#1=52 %
\hatcurRVkeccenxxxxxC
\else
\ifnum#1=53 %
\hatcurRVkeccenxxxxxD
\else
??????\fi
\fi
\fi
\fi
}
\newcommand{\hatcurRVKeccen}[1]{\ifnum#1=50 %
\hatcurRVKeccenxxxxxA
\else
\ifnum#1=51 %
\hatcurRVKeccenxxxxxB
\else
\ifnum#1=52 %
\hatcurRVKeccenxxxxxC
\else
\ifnum#1=53 %
\hatcurRVKeccenxxxxxD
\else
??????\fi
\fi
\fi
\fi
}
\newcommand{\hatcurRVomega}[1]{\ifnum#1=50 %
\hatcurRVomegaxxxxxA
\else
\ifnum#1=51 %
\hatcurRVomegaxxxxxB
\else
\ifnum#1=52 %
\hatcurRVomegaxxxxxC
\else
\ifnum#1=53 %
\hatcurRVomegaxxxxxD
\else
??????\fi
\fi
\fi
\fi
}
\newcommand{\hatcurRVomegaeccen}[1]{\ifnum#1=50 %
\hatcurRVomegaeccenxxxxxA
\else
\ifnum#1=51 %
\hatcurRVomegaeccenxxxxxB
\else
\ifnum#1=52 %
\hatcurRVomegaeccenxxxxxC
\else
\ifnum#1=53 %
\hatcurRVomegaeccenxxxxxD
\else
??????\fi
\fi
\fi
\fi
}
\newcommand{\hatcurRVrh}[1]{\ifnum#1=50 %
\hatcurRVrhxxxxxA
\else
\ifnum#1=51 %
\hatcurRVrhxxxxxB
\else
\ifnum#1=52 %
\hatcurRVrhxxxxxC
\else
\ifnum#1=53 %
\hatcurRVrhxxxxxD
\else
??????\fi
\fi
\fi
\fi
}
\newcommand{\hatcurRVrheccen}[1]{\ifnum#1=50 %
\hatcurRVrheccenxxxxxA
\else
\ifnum#1=51 %
\hatcurRVrheccenxxxxxB
\else
\ifnum#1=52 %
\hatcurRVrheccenxxxxxC
\else
\ifnum#1=53 %
\hatcurRVrheccenxxxxxD
\else
??????\fi
\fi
\fi
\fi
}
\newcommand{\hatcurRVrk}[1]{\ifnum#1=50 %
\hatcurRVrkxxxxxA
\else
\ifnum#1=51 %
\hatcurRVrkxxxxxB
\else
\ifnum#1=52 %
\hatcurRVrkxxxxxC
\else
\ifnum#1=53 %
\hatcurRVrkxxxxxD
\else
??????\fi
\fi
\fi
\fi
}
\newcommand{\hatcurRVrkeccen}[1]{\ifnum#1=50 %
\hatcurRVrkeccenxxxxxA
\else
\ifnum#1=51 %
\hatcurRVrkeccenxxxxxB
\else
\ifnum#1=52 %
\hatcurRVrkeccenxxxxxC
\else
\ifnum#1=53 %
\hatcurRVrkeccenxxxxxD
\else
??????\fi
\fi
\fi
\fi
}
\newcommand{\hatcurRVtrone}[1]{\ifnum#1=50 %
\hatcurRVtronexxxxxA
\else
\ifnum#1=51 %
\hatcurRVtronexxxxxB
\else
\ifnum#1=52 %
\hatcurRVtronexxxxxC
\else
\ifnum#1=53 %
\hatcurRVtronexxxxxD
\else
??????\fi
\fi
\fi
\fi
}
\newcommand{\hatcurRVtroneeccen}[1]{\ifnum#1=50 %
\hatcurRVtroneeccenxxxxxA
\else
\ifnum#1=51 %
\hatcurRVtroneeccenxxxxxB
\else
\ifnum#1=52 %
\hatcurRVtroneeccenxxxxxC
\else
\ifnum#1=53 %
\hatcurRVtroneeccenxxxxxD
\else
??????\fi
\fi
\fi
\fi
}
\newcommand{\hatcurRVtrtwo}[1]{\ifnum#1=50 %
\hatcurRVtrtwoxxxxxA
\else
\ifnum#1=51 %
\hatcurRVtrtwoxxxxxB
\else
\ifnum#1=52 %
\hatcurRVtrtwoxxxxxC
\else
\ifnum#1=53 %
\hatcurRVtrtwoxxxxxD
\else
??????\fi
\fi
\fi
\fi
}
\newcommand{\hatcurRVtrtwoeccen}[1]{\ifnum#1=50 %
\hatcurRVtrtwoeccenxxxxxA
\else
\ifnum#1=51 %
\hatcurRVtrtwoeccenxxxxxB
\else
\ifnum#1=52 %
\hatcurRVtrtwoeccenxxxxxC
\else
\ifnum#1=53 %
\hatcurRVtrtwoeccenxxxxxD
\else
??????\fi
\fi
\fi
\fi
}
\newcommand{\hatcurSMEiilogg}[1]{\ifnum#1=50 %
\hatcurSMEiiloggxxxxxA
\else
\ifnum#1=51 %
\hatcurSMEiiloggxxxxxB
\else
\ifnum#1=52 %
\hatcurSMEiiloggxxxxxC
\else
\ifnum#1=53 %
\hatcurSMEiiloggxxxxxD
\else
??????\fi
\fi
\fi
\fi
}
\newcommand{\hatcurSMEiiloggeccen}[1]{\ifnum#1=51 %
\hatcurSMEiiloggeccenxxxxxB
\else
\ifnum#1=52 %
\hatcurSMEiiloggeccenxxxxxC
\else
\ifnum#1=53 %
\hatcurSMEiiloggeccenxxxxxD
\else
??????\fi
\fi
\fi
}
\newcommand{\hatcurSMEiiteff}[1]{\ifnum#1=50 %
\hatcurSMEiiteffxxxxxA
\else
\ifnum#1=51 %
\hatcurSMEiiteffxxxxxB
\else
\ifnum#1=52 %
\hatcurSMEiiteffxxxxxC
\else
\ifnum#1=53 %
\hatcurSMEiiteffxxxxxD
\else
??????\fi
\fi
\fi
\fi
}
\newcommand{\hatcurSMEiiteffeccen}[1]{\ifnum#1=51 %
\hatcurSMEiiteffeccenxxxxxB
\else
\ifnum#1=52 %
\hatcurSMEiiteffeccenxxxxxC
\else
\ifnum#1=53 %
\hatcurSMEiiteffeccenxxxxxD
\else
??????\fi
\fi
\fi
}
\newcommand{\hatcurSMEiivmac}[1]{\ifnum#1=51 %
\hatcurSMEiivmacxxxxxB
\else
\ifnum#1=53 %
\hatcurSMEiivmacxxxxxD
\else
??????\fi
\fi
}
\newcommand{\hatcurSMEiivmaceccen}[1]{\ifnum#1=51 %
\hatcurSMEiivmaceccenxxxxxB
\else
\ifnum#1=53 %
\hatcurSMEiivmaceccenxxxxxD
\else
??????\fi
\fi
}
\newcommand{\hatcurSMEiivmic}[1]{\ifnum#1=51 %
\hatcurSMEiivmicxxxxxB
\else
\ifnum#1=53 %
\hatcurSMEiivmicxxxxxD
\else
??????\fi
\fi
}
\newcommand{\hatcurSMEiivmiceccen}[1]{\ifnum#1=51 %
\hatcurSMEiivmiceccenxxxxxB
\else
\ifnum#1=53 %
\hatcurSMEiivmiceccenxxxxxD
\else
??????\fi
\fi
}
\newcommand{\hatcurSMEiivsin}[1]{\ifnum#1=50 %
\hatcurSMEiivsinxxxxxA
\else
\ifnum#1=51 %
\hatcurSMEiivsinxxxxxB
\else
\ifnum#1=52 %
\hatcurSMEiivsinxxxxxC
\else
\ifnum#1=53 %
\hatcurSMEiivsinxxxxxD
\else
??????\fi
\fi
\fi
\fi
}
\newcommand{\hatcurSMEiivsineccen}[1]{\ifnum#1=51 %
\hatcurSMEiivsineccenxxxxxB
\else
\ifnum#1=52 %
\hatcurSMEiivsineccenxxxxxC
\else
\ifnum#1=53 %
\hatcurSMEiivsineccenxxxxxD
\else
??????\fi
\fi
\fi
}
\newcommand{\hatcurSMEiizfeh}[1]{\ifnum#1=50 %
\hatcurSMEiizfehxxxxxA
\else
\ifnum#1=51 %
\hatcurSMEiizfehxxxxxB
\else
\ifnum#1=52 %
\hatcurSMEiizfehxxxxxC
\else
\ifnum#1=53 %
\hatcurSMEiizfehxxxxxD
\else
??????\fi
\fi
\fi
\fi
}
\newcommand{\hatcurSMEiizfeheccen}[1]{\ifnum#1=51 %
\hatcurSMEiizfeheccenxxxxxB
\else
\ifnum#1=52 %
\hatcurSMEiizfeheccenxxxxxC
\else
\ifnum#1=53 %
\hatcurSMEiizfeheccenxxxxxD
\else
??????\fi
\fi
\fi
}
\newcommand{\hatcurSMEiizfehshort}[1]{\ifnum#1=50 %
\hatcurSMEiizfehshortxxxxxA
\else
\ifnum#1=51 %
\hatcurSMEiizfehshortxxxxxB
\else
\ifnum#1=52 %
\hatcurSMEiizfehshortxxxxxC
\else
\ifnum#1=53 %
\hatcurSMEiizfehshortxxxxxD
\else
??????\fi
\fi
\fi
\fi
}
\newcommand{\hatcurSMEiizfehshorteccen}[1]{\ifnum#1=51 %
\hatcurSMEiizfehshorteccenxxxxxB
\else
\ifnum#1=52 %
\hatcurSMEiizfehshorteccenxxxxxC
\else
\ifnum#1=53 %
\hatcurSMEiizfehshorteccenxxxxxD
\else
??????\fi
\fi
\fi
}
\newcommand{\hatcurSMEilogg}[1]{\ifnum#1=50 %
\hatcurSMEiloggxxxxxA
\else
\ifnum#1=51 %
\hatcurSMEiloggxxxxxB
\else
\ifnum#1=52 %
\hatcurSMEiloggxxxxxC
\else
\ifnum#1=53 %
\hatcurSMEiloggxxxxxD
\else
??????\fi
\fi
\fi
\fi
}
\newcommand{\hatcurSMEiloggeccen}[1]{\ifnum#1=50 %
\hatcurSMEiloggeccenxxxxxA
\else
\ifnum#1=51 %
\hatcurSMEiloggeccenxxxxxB
\else
\ifnum#1=52 %
\hatcurSMEiloggeccenxxxxxC
\else
\ifnum#1=53 %
\hatcurSMEiloggeccenxxxxxD
\else
??????\fi
\fi
\fi
\fi
}
\newcommand{\hatcurSMEiteff}[1]{\ifnum#1=50 %
\hatcurSMEiteffxxxxxA
\else
\ifnum#1=51 %
\hatcurSMEiteffxxxxxB
\else
\ifnum#1=52 %
\hatcurSMEiteffxxxxxC
\else
\ifnum#1=53 %
\hatcurSMEiteffxxxxxD
\else
??????\fi
\fi
\fi
\fi
}
\newcommand{\hatcurSMEiteffeccen}[1]{\ifnum#1=50 %
\hatcurSMEiteffeccenxxxxxA
\else
\ifnum#1=51 %
\hatcurSMEiteffeccenxxxxxB
\else
\ifnum#1=52 %
\hatcurSMEiteffeccenxxxxxC
\else
\ifnum#1=53 %
\hatcurSMEiteffeccenxxxxxD
\else
??????\fi
\fi
\fi
\fi
}
\newcommand{\hatcurSMEivmac}[1]{\ifnum#1=50 %
\hatcurSMEivmacxxxxxA
\else
\ifnum#1=51 %
\hatcurSMEivmacxxxxxB
\else
\ifnum#1=52 %
\hatcurSMEivmacxxxxxC
\else
\ifnum#1=53 %
\hatcurSMEivmacxxxxxD
\else
??????\fi
\fi
\fi
\fi
}
\newcommand{\hatcurSMEivmaceccen}[1]{\ifnum#1=50 %
\hatcurSMEivmaceccenxxxxxA
\else
\ifnum#1=51 %
\hatcurSMEivmaceccenxxxxxB
\else
\ifnum#1=52 %
\hatcurSMEivmaceccenxxxxxC
\else
\ifnum#1=53 %
\hatcurSMEivmaceccenxxxxxD
\else
??????\fi
\fi
\fi
\fi
}
\newcommand{\hatcurSMEivmic}[1]{\ifnum#1=50 %
\hatcurSMEivmicxxxxxA
\else
\ifnum#1=51 %
\hatcurSMEivmicxxxxxB
\else
\ifnum#1=52 %
\hatcurSMEivmicxxxxxC
\else
\ifnum#1=53 %
\hatcurSMEivmicxxxxxD
\else
??????\fi
\fi
\fi
\fi
}
\newcommand{\hatcurSMEivmiceccen}[1]{\ifnum#1=50 %
\hatcurSMEivmiceccenxxxxxA
\else
\ifnum#1=51 %
\hatcurSMEivmiceccenxxxxxB
\else
\ifnum#1=52 %
\hatcurSMEivmiceccenxxxxxC
\else
\ifnum#1=53 %
\hatcurSMEivmiceccenxxxxxD
\else
??????\fi
\fi
\fi
\fi
}
\newcommand{\hatcurSMEivsin}[1]{\ifnum#1=50 %
\hatcurSMEivsinxxxxxA
\else
\ifnum#1=51 %
\hatcurSMEivsinxxxxxB
\else
\ifnum#1=52 %
\hatcurSMEivsinxxxxxC
\else
\ifnum#1=53 %
\hatcurSMEivsinxxxxxD
\else
??????\fi
\fi
\fi
\fi
}
\newcommand{\hatcurSMEivsineccen}[1]{\ifnum#1=50 %
\hatcurSMEivsineccenxxxxxA
\else
\ifnum#1=51 %
\hatcurSMEivsineccenxxxxxB
\else
\ifnum#1=52 %
\hatcurSMEivsineccenxxxxxC
\else
\ifnum#1=53 %
\hatcurSMEivsineccenxxxxxD
\else
??????\fi
\fi
\fi
\fi
}
\newcommand{\hatcurSMEizfeh}[1]{\ifnum#1=50 %
\hatcurSMEizfehxxxxxA
\else
\ifnum#1=51 %
\hatcurSMEizfehxxxxxB
\else
\ifnum#1=52 %
\hatcurSMEizfehxxxxxC
\else
\ifnum#1=53 %
\hatcurSMEizfehxxxxxD
\else
??????\fi
\fi
\fi
\fi
}
\newcommand{\hatcurSMEizfeheccen}[1]{\ifnum#1=50 %
\hatcurSMEizfeheccenxxxxxA
\else
\ifnum#1=51 %
\hatcurSMEizfeheccenxxxxxB
\else
\ifnum#1=52 %
\hatcurSMEizfeheccenxxxxxC
\else
\ifnum#1=53 %
\hatcurSMEizfeheccenxxxxxD
\else
??????\fi
\fi
\fi
\fi
}
\newcommand{\hatcurSMEizfehshort}[1]{\ifnum#1=50 %
\hatcurSMEizfehshortxxxxxA
\else
\ifnum#1=51 %
\hatcurSMEizfehshortxxxxxB
\else
\ifnum#1=52 %
\hatcurSMEizfehshortxxxxxC
\else
\ifnum#1=53 %
\hatcurSMEizfehshortxxxxxD
\else
??????\fi
\fi
\fi
\fi
}
\newcommand{\hatcurSMEizfehshorteccen}[1]{\ifnum#1=50 %
\hatcurSMEizfehshorteccenxxxxxA
\else
\ifnum#1=51 %
\hatcurSMEizfehshorteccenxxxxxB
\else
\ifnum#1=52 %
\hatcurSMEizfehshorteccenxxxxxC
\else
\ifnum#1=53 %
\hatcurSMEizfehshorteccenxxxxxD
\else
??????\fi
\fi
\fi
\fi
}
\newcommand{\hatcurTRESgamma}[1]{\ifnum#1=50 %
\hatcurTRESgammaxxxxxA
\else
\ifnum#1=51 %
\hatcurTRESgammaxxxxxB
\else
\ifnum#1=52 %
\hatcurTRESgammaxxxxxC
\else
\ifnum#1=53 %
\hatcurTRESgammaxxxxxD
\else
??????\fi
\fi
\fi
\fi
}
\newcommand{\hatcurTRESgammaeccen}[1]{\ifnum#1=50 %
\hatcurTRESgammaeccenxxxxxA
\else
\ifnum#1=51 %
\hatcurTRESgammaeccenxxxxxB
\else
\ifnum#1=52 %
\hatcurTRESgammaeccenxxxxxC
\else
\ifnum#1=53 %
\hatcurTRESgammaeccenxxxxxD
\else
??????\fi
\fi
\fi
\fi
}
\newcommand{\hatcurTRESlogg}[1]{\ifnum#1=50 %
\hatcurTRESloggxxxxxA
\else
\ifnum#1=51 %
\hatcurTRESloggxxxxxB
\else
\ifnum#1=52 %
\hatcurTRESloggxxxxxC
\else
\ifnum#1=53 %
\hatcurTRESloggxxxxxD
\else
??????\fi
\fi
\fi
\fi
}
\newcommand{\hatcurTRESloggeccen}[1]{\ifnum#1=50 %
\hatcurTRESloggeccenxxxxxA
\else
\ifnum#1=51 %
\hatcurTRESloggeccenxxxxxB
\else
\ifnum#1=52 %
\hatcurTRESloggeccenxxxxxC
\else
\ifnum#1=53 %
\hatcurTRESloggeccenxxxxxD
\else
??????\fi
\fi
\fi
\fi
}
\newcommand{\hatcurTRESnumspec}[1]{\ifnum#1=50 %
\hatcurTRESnumspecxxxxxA
\else
\ifnum#1=51 %
\hatcurTRESnumspecxxxxxB
\else
\ifnum#1=52 %
\hatcurTRESnumspecxxxxxC
\else
\ifnum#1=53 %
\hatcurTRESnumspecxxxxxD
\else
??????\fi
\fi
\fi
\fi
}
\newcommand{\hatcurTRESnumspececcen}[1]{\ifnum#1=50 %
\hatcurTRESnumspececcenxxxxxA
\else
\ifnum#1=51 %
\hatcurTRESnumspececcenxxxxxB
\else
\ifnum#1=52 %
\hatcurTRESnumspececcenxxxxxC
\else
\ifnum#1=53 %
\hatcurTRESnumspececcenxxxxxD
\else
??????\fi
\fi
\fi
\fi
}
\newcommand{\hatcurTRESrvrms}[1]{\ifnum#1=50 %
\hatcurTRESrvrmsxxxxxA
\else
\ifnum#1=51 %
\hatcurTRESrvrmsxxxxxB
\else
\ifnum#1=52 %
\hatcurTRESrvrmsxxxxxC
\else
\ifnum#1=53 %
\hatcurTRESrvrmsxxxxxD
\else
??????\fi
\fi
\fi
\fi
}
\newcommand{\hatcurTRESrvrmseccen}[1]{\ifnum#1=50 %
\hatcurTRESrvrmseccenxxxxxA
\else
\ifnum#1=51 %
\hatcurTRESrvrmseccenxxxxxB
\else
\ifnum#1=52 %
\hatcurTRESrvrmseccenxxxxxC
\else
\ifnum#1=53 %
\hatcurTRESrvrmseccenxxxxxD
\else
??????\fi
\fi
\fi
\fi
}
\newcommand{\hatcurTRESspan}[1]{\ifnum#1=50 %
\hatcurTRESspanxxxxxA
\else
\ifnum#1=51 %
\hatcurTRESspanxxxxxB
\else
\ifnum#1=52 %
\hatcurTRESspanxxxxxC
\else
\ifnum#1=53 %
\hatcurTRESspanxxxxxD
\else
??????\fi
\fi
\fi
\fi
}
\newcommand{\hatcurTRESspaneccen}[1]{\ifnum#1=50 %
\hatcurTRESspaneccenxxxxxA
\else
\ifnum#1=51 %
\hatcurTRESspaneccenxxxxxB
\else
\ifnum#1=52 %
\hatcurTRESspaneccenxxxxxC
\else
\ifnum#1=53 %
\hatcurTRESspaneccenxxxxxD
\else
??????\fi
\fi
\fi
\fi
}
\newcommand{\hatcurTRESteff}[1]{\ifnum#1=50 %
\hatcurTRESteffxxxxxA
\else
\ifnum#1=51 %
\hatcurTRESteffxxxxxB
\else
\ifnum#1=52 %
\hatcurTRESteffxxxxxC
\else
\ifnum#1=53 %
\hatcurTRESteffxxxxxD
\else
??????\fi
\fi
\fi
\fi
}
\newcommand{\hatcurTRESteffeccen}[1]{\ifnum#1=50 %
\hatcurTRESteffeccenxxxxxA
\else
\ifnum#1=51 %
\hatcurTRESteffeccenxxxxxB
\else
\ifnum#1=52 %
\hatcurTRESteffeccenxxxxxC
\else
\ifnum#1=53 %
\hatcurTRESteffeccenxxxxxD
\else
??????\fi
\fi
\fi
\fi
}
\newcommand{\hatcurTRESvsini}[1]{\ifnum#1=50 %
\hatcurTRESvsinixxxxxA
\else
\ifnum#1=51 %
\hatcurTRESvsinixxxxxB
\else
\ifnum#1=52 %
\hatcurTRESvsinixxxxxC
\else
\ifnum#1=53 %
\hatcurTRESvsinixxxxxD
\else
??????\fi
\fi
\fi
\fi
}
\newcommand{\hatcurTRESvsinieccen}[1]{\ifnum#1=50 %
\hatcurTRESvsinieccenxxxxxA
\else
\ifnum#1=51 %
\hatcurTRESvsinieccenxxxxxB
\else
\ifnum#1=52 %
\hatcurTRESvsinieccenxxxxxC
\else
\ifnum#1=53 %
\hatcurTRESvsinieccenxxxxxD
\else
??????\fi
\fi
\fi
\fi
}
\newcommand{\hatcurTRESzfeh}[1]{\ifnum#1=50 %
\hatcurTRESzfehxxxxxA
\else
\ifnum#1=51 %
\hatcurTRESzfehxxxxxB
\else
\ifnum#1=52 %
\hatcurTRESzfehxxxxxC
\else
\ifnum#1=53 %
\hatcurTRESzfehxxxxxD
\else
??????\fi
\fi
\fi
\fi
}
\newcommand{\hatcurTRESzfeheccen}[1]{\ifnum#1=50 %
\hatcurTRESzfeheccenxxxxxA
\else
\ifnum#1=51 %
\hatcurTRESzfeheccenxxxxxB
\else
\ifnum#1=52 %
\hatcurTRESzfeheccenxxxxxC
\else
\ifnum#1=53 %
\hatcurTRESzfeheccenxxxxxD
\else
??????\fi
\fi
\fi
\fi
}
\newcommand{\hatcurXAv}[1]{\ifnum#1=50 %
\hatcurXAvxxxxxA
\else
\ifnum#1=51 %
\hatcurXAvxxxxxB
\else
\ifnum#1=52 %
\hatcurXAvxxxxxC
\else
\ifnum#1=53 %
\hatcurXAvxxxxxD
\else
??????\fi
\fi
\fi
\fi
}
\newcommand{\hatcurXAveccen}[1]{\ifnum#1=50 %
\hatcurXAveccenxxxxxA
\else
\ifnum#1=51 %
\hatcurXAveccenxxxxxB
\else
\ifnum#1=52 %
\hatcurXAveccenxxxxxC
\else
\ifnum#1=53 %
\hatcurXAveccenxxxxxD
\else
??????\fi
\fi
\fi
\fi
}
\newcommand{\hatcurXdist}[1]{\ifnum#1=50 %
\hatcurXdistxxxxxA
\else
\ifnum#1=51 %
\hatcurXdistxxxxxB
\else
\ifnum#1=52 %
\hatcurXdistxxxxxC
\else
\ifnum#1=53 %
\hatcurXdistxxxxxD
\else
??????\fi
\fi
\fi
\fi
}
\newcommand{\hatcurXdisteccen}[1]{\ifnum#1=50 %
\hatcurXdisteccenxxxxxA
\else
\ifnum#1=51 %
\hatcurXdisteccenxxxxxB
\else
\ifnum#1=52 %
\hatcurXdisteccenxxxxxC
\else
\ifnum#1=53 %
\hatcurXdisteccenxxxxxD
\else
??????\fi
\fi
\fi
\fi
}
\newcommand{\hatcurXdistred}[1]{\ifnum#1=50 %
\hatcurXdistredxxxxxA
\else
\ifnum#1=51 %
\hatcurXdistredxxxxxB
\else
\ifnum#1=52 %
\hatcurXdistredxxxxxC
\else
\ifnum#1=53 %
\hatcurXdistredxxxxxD
\else
??????\fi
\fi
\fi
\fi
}
\newcommand{\hatcurXdistredeccen}[1]{\ifnum#1=50 %
\hatcurXdistredeccenxxxxxA
\else
\ifnum#1=51 %
\hatcurXdistredeccenxxxxxB
\else
\ifnum#1=52 %
\hatcurXdistredeccenxxxxxC
\else
\ifnum#1=53 %
\hatcurXdistredeccenxxxxxD
\else
??????\fi
\fi
\fi
\fi
}
\newcommand{\hatcurXEBV}[1]{\ifnum#1=50 %
\hatcurXEBVxxxxxA
\else
\ifnum#1=51 %
\hatcurXEBVxxxxxB
\else
\ifnum#1=52 %
\hatcurXEBVxxxxxC
\else
\ifnum#1=53 %
\hatcurXEBVxxxxxD
\else
??????\fi
\fi
\fi
\fi
}
\newcommand{\hatcurXEBVeccen}[1]{\ifnum#1=50 %
\hatcurXEBVeccenxxxxxA
\else
\ifnum#1=51 %
\hatcurXEBVeccenxxxxxB
\else
\ifnum#1=52 %
\hatcurXEBVeccenxxxxxC
\else
\ifnum#1=53 %
\hatcurXEBVeccenxxxxxD
\else
??????\fi
\fi
\fi
\fi
}
\newcommand{\hatcurXjhisored}[1]{\ifnum#1=50 %
\hatcurXjhisoredxxxxxA
\else
\ifnum#1=51 %
\hatcurXjhisoredxxxxxB
\else
\ifnum#1=52 %
\hatcurXjhisoredxxxxxC
\else
\ifnum#1=53 %
\hatcurXjhisoredxxxxxD
\else
??????\fi
\fi
\fi
\fi
}
\newcommand{\hatcurXjhisoredeccen}[1]{\ifnum#1=50 %
\hatcurXjhisoredeccenxxxxxA
\else
\ifnum#1=51 %
\hatcurXjhisoredeccenxxxxxB
\else
\ifnum#1=52 %
\hatcurXjhisoredeccenxxxxxC
\else
\ifnum#1=53 %
\hatcurXjhisoredeccenxxxxxD
\else
??????\fi
\fi
\fi
\fi
}
\newcommand{\hatcurXjkisored}[1]{\ifnum#1=50 %
\hatcurXjkisoredxxxxxA
\else
\ifnum#1=51 %
\hatcurXjkisoredxxxxxB
\else
\ifnum#1=52 %
\hatcurXjkisoredxxxxxC
\else
\ifnum#1=53 %
\hatcurXjkisoredxxxxxD
\else
??????\fi
\fi
\fi
\fi
}
\newcommand{\hatcurXjkisoredeccen}[1]{\ifnum#1=50 %
\hatcurXjkisoredeccenxxxxxA
\else
\ifnum#1=51 %
\hatcurXjkisoredeccenxxxxxB
\else
\ifnum#1=52 %
\hatcurXjkisoredeccenxxxxxC
\else
\ifnum#1=53 %
\hatcurXjkisoredeccenxxxxxD
\else
??????\fi
\fi
\fi
\fi
}
\newcommand{\hatcurXmhisored}[1]{\ifnum#1=50 %
\hatcurXmhisoredxxxxxA
\else
\ifnum#1=51 %
\hatcurXmhisoredxxxxxB
\else
\ifnum#1=52 %
\hatcurXmhisoredxxxxxC
\else
\ifnum#1=53 %
\hatcurXmhisoredxxxxxD
\else
??????\fi
\fi
\fi
\fi
}
\newcommand{\hatcurXmhisoredeccen}[1]{\ifnum#1=50 %
\hatcurXmhisoredeccenxxxxxA
\else
\ifnum#1=51 %
\hatcurXmhisoredeccenxxxxxB
\else
\ifnum#1=52 %
\hatcurXmhisoredeccenxxxxxC
\else
\ifnum#1=53 %
\hatcurXmhisoredeccenxxxxxD
\else
??????\fi
\fi
\fi
\fi
}
\newcommand{\hatcurXmiisored}[1]{\ifnum#1=50 %
\hatcurXmiisoredxxxxxA
\else
\ifnum#1=51 %
\hatcurXmiisoredxxxxxB
\else
\ifnum#1=52 %
\hatcurXmiisoredxxxxxC
\else
\ifnum#1=53 %
\hatcurXmiisoredxxxxxD
\else
??????\fi
\fi
\fi
\fi
}
\newcommand{\hatcurXmiisoredeccen}[1]{\ifnum#1=50 %
\hatcurXmiisoredeccenxxxxxA
\else
\ifnum#1=51 %
\hatcurXmiisoredeccenxxxxxB
\else
\ifnum#1=52 %
\hatcurXmiisoredeccenxxxxxC
\else
\ifnum#1=53 %
\hatcurXmiisoredeccenxxxxxD
\else
??????\fi
\fi
\fi
\fi
}
\newcommand{\hatcurXmjisored}[1]{\ifnum#1=50 %
\hatcurXmjisoredxxxxxA
\else
\ifnum#1=51 %
\hatcurXmjisoredxxxxxB
\else
\ifnum#1=52 %
\hatcurXmjisoredxxxxxC
\else
\ifnum#1=53 %
\hatcurXmjisoredxxxxxD
\else
??????\fi
\fi
\fi
\fi
}
\newcommand{\hatcurXmjisoredeccen}[1]{\ifnum#1=50 %
\hatcurXmjisoredeccenxxxxxA
\else
\ifnum#1=51 %
\hatcurXmjisoredeccenxxxxxB
\else
\ifnum#1=52 %
\hatcurXmjisoredeccenxxxxxC
\else
\ifnum#1=53 %
\hatcurXmjisoredeccenxxxxxD
\else
??????\fi
\fi
\fi
\fi
}
\newcommand{\hatcurXmkisored}[1]{\ifnum#1=50 %
\hatcurXmkisoredxxxxxA
\else
\ifnum#1=51 %
\hatcurXmkisoredxxxxxB
\else
\ifnum#1=52 %
\hatcurXmkisoredxxxxxC
\else
\ifnum#1=53 %
\hatcurXmkisoredxxxxxD
\else
??????\fi
\fi
\fi
\fi
}
\newcommand{\hatcurXmkisoredeccen}[1]{\ifnum#1=50 %
\hatcurXmkisoredeccenxxxxxA
\else
\ifnum#1=51 %
\hatcurXmkisoredeccenxxxxxB
\else
\ifnum#1=52 %
\hatcurXmkisoredeccenxxxxxC
\else
\ifnum#1=53 %
\hatcurXmkisoredeccenxxxxxD
\else
??????\fi
\fi
\fi
\fi
}
\newcommand{\hatcurXmvisored}[1]{\ifnum#1=50 %
\hatcurXmvisoredxxxxxA
\else
\ifnum#1=51 %
\hatcurXmvisoredxxxxxB
\else
\ifnum#1=52 %
\hatcurXmvisoredxxxxxC
\else
\ifnum#1=53 %
\hatcurXmvisoredxxxxxD
\else
??????\fi
\fi
\fi
\fi
}
\newcommand{\hatcurXmvisoredeccen}[1]{\ifnum#1=50 %
\hatcurXmvisoredeccenxxxxxA
\else
\ifnum#1=51 %
\hatcurXmvisoredeccenxxxxxB
\else
\ifnum#1=52 %
\hatcurXmvisoredeccenxxxxxC
\else
\ifnum#1=53 %
\hatcurXmvisoredeccenxxxxxD
\else
??????\fi
\fi
\fi
\fi
}
\newcommand{\hatcurXsecdur}[1]{\ifnum#1=50 %
\hatcurXsecdurxxxxxA
\else
\ifnum#1=51 %
\hatcurXsecdurxxxxxB
\else
\ifnum#1=52 %
\hatcurXsecdurxxxxxC
\else
\ifnum#1=53 %
\hatcurXsecdurxxxxxD
\else
??????\fi
\fi
\fi
\fi
}
\newcommand{\hatcurXsecdureccen}[1]{\ifnum#1=50 %
\hatcurXsecdureccenxxxxxA
\else
\ifnum#1=51 %
\hatcurXsecdureccenxxxxxB
\else
\ifnum#1=52 %
\hatcurXsecdureccenxxxxxC
\else
\ifnum#1=53 %
\hatcurXsecdureccenxxxxxD
\else
??????\fi
\fi
\fi
\fi
}
\newcommand{\hatcurXsecingdur}[1]{\ifnum#1=50 %
\hatcurXsecingdurxxxxxA
\else
\ifnum#1=51 %
\hatcurXsecingdurxxxxxB
\else
\ifnum#1=52 %
\hatcurXsecingdurxxxxxC
\else
\ifnum#1=53 %
\hatcurXsecingdurxxxxxD
\else
??????\fi
\fi
\fi
\fi
}
\newcommand{\hatcurXsecingdureccen}[1]{\ifnum#1=50 %
\hatcurXsecingdureccenxxxxxA
\else
\ifnum#1=51 %
\hatcurXsecingdureccenxxxxxB
\else
\ifnum#1=52 %
\hatcurXsecingdureccenxxxxxC
\else
\ifnum#1=53 %
\hatcurXsecingdureccenxxxxxD
\else
??????\fi
\fi
\fi
\fi
}
\newcommand{\hatcurXsecondary}[1]{\ifnum#1=50 %
\hatcurXsecondaryxxxxxA
\else
\ifnum#1=51 %
\hatcurXsecondaryxxxxxB
\else
\ifnum#1=52 %
\hatcurXsecondaryxxxxxC
\else
\ifnum#1=53 %
\hatcurXsecondaryxxxxxD
\else
??????\fi
\fi
\fi
\fi
}
\newcommand{\hatcurXsecondaryeccen}[1]{\ifnum#1=50 %
\hatcurXsecondaryeccenxxxxxA
\else
\ifnum#1=51 %
\hatcurXsecondaryeccenxxxxxB
\else
\ifnum#1=52 %
\hatcurXsecondaryeccenxxxxxC
\else
\ifnum#1=53 %
\hatcurXsecondaryeccenxxxxxD
\else
??????\fi
\fi
\fi
\fi
}
\newcommand{\hatcurXsecphase}[1]{\ifnum#1=50 %
\hatcurXsecphasexxxxxA
\else
\ifnum#1=51 %
\hatcurXsecphasexxxxxB
\else
\ifnum#1=52 %
\hatcurXsecphasexxxxxC
\else
\ifnum#1=53 %
\hatcurXsecphasexxxxxD
\else
??????\fi
\fi
\fi
\fi
}
\newcommand{\hatcurXsecphaseeccen}[1]{\ifnum#1=50 %
\hatcurXsecphaseeccenxxxxxA
\else
\ifnum#1=51 %
\hatcurXsecphaseeccenxxxxxB
\else
\ifnum#1=52 %
\hatcurXsecphaseeccenxxxxxC
\else
\ifnum#1=53 %
\hatcurXsecphaseeccenxxxxxD
\else
??????\fi
\fi
\fi
\fi
}
\newcommand{\hatcurXviisored}[1]{\ifnum#1=50 %
\hatcurXviisoredxxxxxA
\else
\ifnum#1=51 %
\hatcurXviisoredxxxxxB
\else
\ifnum#1=52 %
\hatcurXviisoredxxxxxC
\else
\ifnum#1=53 %
\hatcurXviisoredxxxxxD
\else
??????\fi
\fi
\fi
\fi
}
\newcommand{\hatcurXviisoredeccen}[1]{\ifnum#1=50 %
\hatcurXviisoredeccenxxxxxA
\else
\ifnum#1=51 %
\hatcurXviisoredeccenxxxxxB
\else
\ifnum#1=52 %
\hatcurXviisoredeccenxxxxxC
\else
\ifnum#1=53 %
\hatcurXviisoredeccenxxxxxD
\else
??????\fi
\fi
\fi
\fi
}
\newcommand{\hatcurXvkisored}[1]{\ifnum#1=50 %
\hatcurXvkisoredxxxxxA
\else
\ifnum#1=51 %
\hatcurXvkisoredxxxxxB
\else
\ifnum#1=52 %
\hatcurXvkisoredxxxxxC
\else
\ifnum#1=53 %
\hatcurXvkisoredxxxxxD
\else
??????\fi
\fi
\fi
\fi
}
\newcommand{\hatcurXvkisoredeccen}[1]{\ifnum#1=50 %
\hatcurXvkisoredeccenxxxxxA
\else
\ifnum#1=51 %
\hatcurXvkisoredeccenxxxxxB
\else
\ifnum#1=52 %
\hatcurXvkisoredeccenxxxxxC
\else
\ifnum#1=53 %
\hatcurXvkisoredeccenxxxxxD
\else
??????\fi
\fi
\fi
\fi
}

\newcommand{\hatcurxxxxxA}{HAT-P-50}
\newcommand{\hatcurbxxxxxA}{HAT-P-50b}
\newcommand{\hatcurcxxxxxA}{HAT-P-50c}

\newcommand{\hatcurCCtwomassshortxxxxxA}{07521521+1208218}

\newcommand{\hatcurplanetnumxxxxxA}{50}

\newcommand{\hatcurRVgammaabsxxxxxA}{\ensuremath{6.25\pm0.10}}                           

\newcommand{\hatcurRVgammarelxxxxxA}{\hatcurRVgamma{\hatcurplanetnumxxxxxA}}                           

\newcommand{\hatcurCCtassvixxxxxA}{\ensuremath{0.527\pm0.12}}                  

\newcommand{\hatcurSMEversionxxxxxA}{ii}                                       

\newcommand{\hatcurisoshortxxxxxA}{YY}
\newcommand{\hatcurisofullxxxxxA}{Yonsei-Yale (YY)}
\newcommand{\hatcurisocitexxxxxA}{yi:2001}

\newcommand{\hatcurlumindxxxxxA}{\arstar}

\newcommand{\hatcurjhkfilsetxxxxxA}{ESO}

\newcommand{\hatcurSMEteffxxxxxA}{\ifthenelse{\equal{\hatcurSMEversionxxxxxA}{i}}{\hatcurSMEiteff{\hatcurplanetnumxxxxxA}}{\hatcurSMEiiteff{\hatcurplanetnumxxxxxA}}}
\newcommand{\hatcurSMEzfehxxxxxA}{\ifthenelse{\equal{\hatcurSMEversionxxxxxA}{i}}{\hatcurSMEizfeh{\hatcurplanetnumxxxxxA}}{\hatcurSMEiizfeh{\hatcurplanetnumxxxxxA}}}
\newcommand{\hatcurSMEzfehshortxxxxxA}{\ifthenelse{\equal{\hatcurSMEversionxxxxxA}{i}}{\hatcurSMEizfehshort{\hatcurplanetnumxxxxxA}}{\hatcurSMEiizfehshort{\hatcurplanetnumxxxxxA}}}
\newcommand{\hatcurSMEloggxxxxxA}{\ifthenelse{\equal{\hatcurSMEversionxxxxxA}{i}}{\hatcurSMEilogg{\hatcurplanetnumxxxxxA}}{\hatcurSMEiilogg{\hatcurplanetnumxxxxxA}}}
\newcommand{\hatcurSMEvsinxxxxxA}{\ifthenelse{\equal{\hatcurSMEversionxxxxxA}{i}}{\hatcurSMEivsin{\hatcurplanetnumxxxxxA}}{\hatcurSMEiivsin{\hatcurplanetnumxxxxxA}}}
\newcommand{\hatcurSMEvmacxxxxxA}{\ifthenelse{\equal{\hatcurSMEversionxxxxxA}{i}}{\hatcurSMEivmac{\hatcurplanetnumxxxxxA}}{\hatcurSMEiivmac{\hatcurplanetnumxxxxxA}}}
\newcommand{\hatcurSMEvmicxxxxxA}{\ifthenelse{\equal{\hatcurSMEversionxxxxxA}{i}}{\hatcurSMEivmic{\hatcurplanetnumxxxxxA}}{\hatcurSMEiivmic{\hatcurplanetnumxxxxxA}}}

\newcommand{\hatcurxxxxxB}{HAT-P-51}
\newcommand{\hatcurbxxxxxB}{HAT-P-51b}
\newcommand{\hatcurcxxxxxB}{HAT-P-51c}

\newcommand{\hatcurplanetnumxxxxxB}{51}

\newcommand{\hatcurCCtwomassshortxxxxxB}{01241564+3248387}

\newcommand{\hatcurlogRprimeHKxxxxxB}{\ensuremath{-5.057\pm0.050}}               

\newcommand{\hatcurRVgammaabsxxxxxB}{\ensuremath{-27.56\pm0.10}}                           

\newcommand{\hatcurRVgammarelxxxxxB}{\hatcurRVgamma{\hatcurplanetnumxxxxxB}}                           

\newcommand{\hatcurCCtassvixxxxxB}{\ensuremath{0.527\pm0.12}}                  

\newcommand{\hatcurSMEversionxxxxxB}{i}                                       

\newcommand{\hatcurisoshortxxxxxB}{YY}
\newcommand{\hatcurisofullxxxxxB}{Yonsei-Yale (YY)}
\newcommand{\hatcurisocitexxxxxB}{yi:2001}

\newcommand{\hatcurlumindxxxxxB}{\arstar}

\newcommand{\hatcurjhkfilsetxxxxxB}{ESO}

\newcommand{\hatcurSMEteffxxxxxB}{\ifthenelse{\equal{\hatcurSMEversionxxxxxB}{i}}{\hatcurSMEiteff{\hatcurplanetnumxxxxxB}}{\hatcurSMEiiteff{\hatcurplanetnumxxxxxB}}}
\newcommand{\hatcurSMEzfehxxxxxB}{\ifthenelse{\equal{\hatcurSMEversionxxxxxB}{i}}{\hatcurSMEizfeh{\hatcurplanetnumxxxxxB}}{\hatcurSMEiizfeh{\hatcurplanetnumxxxxxB}}}
\newcommand{\hatcurSMEzfehshortxxxxxB}{\ifthenelse{\equal{\hatcurSMEversionxxxxxB}{i}}{\hatcurSMEizfehshort{\hatcurplanetnumxxxxxB}}{\hatcurSMEiizfehshort{\hatcurplanetnumxxxxxB}}}
\newcommand{\hatcurSMEloggxxxxxB}{\ifthenelse{\equal{\hatcurSMEversionxxxxxB}{i}}{\hatcurSMEilogg{\hatcurplanetnumxxxxxB}}{\hatcurSMEiilogg{\hatcurplanetnumxxxxxB}}}
\newcommand{\hatcurSMEvsinxxxxxB}{\ifthenelse{\equal{\hatcurSMEversionxxxxxB}{i}}{\hatcurSMEivsin{\hatcurplanetnumxxxxxB}}{\hatcurSMEiivsin{\hatcurplanetnumxxxxxB}}}
\newcommand{\hatcurSMEvmacxxxxxB}{\ifthenelse{\equal{\hatcurSMEversionxxxxxB}{i}}{\hatcurSMEivmac{\hatcurplanetnumxxxxxB}}{\hatcurSMEiivmac{\hatcurplanetnumxxxxxB}}}
\newcommand{\hatcurSMEvmicxxxxxB}{\ifthenelse{\equal{\hatcurSMEversionxxxxxB}{i}}{\hatcurSMEivmic{\hatcurplanetnumxxxxxB}}{\hatcurSMEiivmic{\hatcurplanetnumxxxxxB}}}

\newcommand{\hatcurxxxxxC}{HAT-P-52}
\newcommand{\hatcurbxxxxxC}{HAT-P-52b}
\newcommand{\hatcurcxxxxxC}{HAT-P-52c}

\newcommand{\hatcurplanetnumxxxxxC}{52}

\newcommand{\hatcurCCtwomassshortxxxxxC}{02505320+2901206}

\newcommand{\hatcurlogRprimeHKxxxxxC}{\ensuremath{-5.154\pm0.089}}

\newcommand{\hatcurRVgammaabsxxxxxC}{\ensuremath{61.50\pm0.10}}                           

\newcommand{\hatcurRVgammarelxxxxxC}{\hatcurRVgamma{\hatcurplanetnumxxxxxC}}                           

\newcommand{\hatcurCCtassvixxxxxC}{\ensuremath{0.527\pm0.12}}                  

\newcommand{\hatcurSMEversionxxxxxC}{ii}                                       

\newcommand{\hatcurisoshortxxxxxC}{YY}
\newcommand{\hatcurisofullxxxxxC}{Yonsei-Yale (YY)}
\newcommand{\hatcurisocitexxxxxC}{yi:2001}

\newcommand{\hatcurlumindxxxxxC}{\arstar}

\newcommand{\hatcurjhkfilsetxxxxxC}{ESO}

\newcommand{\hatcurSMEteffxxxxxC}{\ifthenelse{\equal{\hatcurSMEversionxxxxxC}{i}}{\hatcurSMEiteff{\hatcurplanetnumxxxxxC}}{\hatcurSMEiiteff{\hatcurplanetnumxxxxxC}}}
\newcommand{\hatcurSMEzfehxxxxxC}{\ifthenelse{\equal{\hatcurSMEversionxxxxxC}{i}}{\hatcurSMEizfeh{\hatcurplanetnumxxxxxC}}{\hatcurSMEiizfeh{\hatcurplanetnumxxxxxC}}}
\newcommand{\hatcurSMEzfehshortxxxxxC}{\ifthenelse{\equal{\hatcurSMEversionxxxxxC}{i}}{\hatcurSMEizfehshort{\hatcurplanetnumxxxxxC}}{\hatcurSMEiizfehshort{\hatcurplanetnumxxxxxC}}}
\newcommand{\hatcurSMEloggxxxxxC}{\ifthenelse{\equal{\hatcurSMEversionxxxxxC}{i}}{\hatcurSMEilogg{\hatcurplanetnumxxxxxC}}{\hatcurSMEiilogg{\hatcurplanetnumxxxxxC}}}
\newcommand{\hatcurSMEvsinxxxxxC}{\ifthenelse{\equal{\hatcurSMEversionxxxxxC}{i}}{\hatcurSMEivsin{\hatcurplanetnumxxxxxC}}{\hatcurSMEiivsin{\hatcurplanetnumxxxxxC}}}
\newcommand{\hatcurSMEvmacxxxxxC}{\ifthenelse{\equal{\hatcurSMEversionxxxxxC}{i}}{\hatcurSMEivmac{\hatcurplanetnumxxxxxC}}{\hatcurSMEiivmac{\hatcurplanetnumxxxxxC}}}
\newcommand{\hatcurSMEvmicxxxxxC}{\ifthenelse{\equal{\hatcurSMEversionxxxxxC}{i}}{\hatcurSMEivmic{\hatcurplanetnumxxxxxC}}{\hatcurSMEiivmic{\hatcurplanetnumxxxxxC}}}

\newcommand{\hatcurxxxxxD}{HAT-P-53}
\newcommand{\hatcurbxxxxxD}{HAT-P-53b}
\newcommand{\hatcurcxxxxxD}{HAT-P-53c}

\newcommand{\hatcurplanetnumxxxxxD}{53}

\newcommand{\hatcurCCtwomassshortxxxxxD}{01272906+3858053}

\newcommand{\hatcurlogRprimeHKxxxxxD}{\ensuremath{-4.919\pm0.042}}

\newcommand{\hatcurRVgammaabsxxxxxD}{\ensuremath{-16.99\pm0.10}}                           

\newcommand{\hatcurRVgammarelxxxxxD}{\hatcurRVgamma{\hatcurplanetnumxxxxxD}}                           

\newcommand{\hatcurCCtassvixxxxxD}{\ensuremath{0.527\pm0.12}}                  

\newcommand{\hatcurSMEversionxxxxxD}{i}                                       

\newcommand{\hatcurisoshortxxxxxD}{YY}
\newcommand{\hatcurisofullxxxxxD}{Yonsei-Yale (YY)}
\newcommand{\hatcurisocitexxxxxD}{yi:2001}

\newcommand{\hatcurlumindxxxxxD}{\arstar}

\newcommand{\hatcurjhkfilsetxxxxxD}{ESO}

\newcommand{\hatcurSMEteffxxxxxD}{\ifthenelse{\equal{\hatcurSMEversionxxxxxD}{i}}{\hatcurSMEiteff{\hatcurplanetnumxxxxxD}}{\hatcurSMEiiteff{\hatcurplanetnumxxxxxD}}}
\newcommand{\hatcurSMEzfehxxxxxD}{\ifthenelse{\equal{\hatcurSMEversionxxxxxD}{i}}{\hatcurSMEizfeh{\hatcurplanetnumxxxxxD}}{\hatcurSMEiizfeh{\hatcurplanetnumxxxxxD}}}
\newcommand{\hatcurSMEzfehshortxxxxxD}{\ifthenelse{\equal{\hatcurSMEversionxxxxxD}{i}}{\hatcurSMEizfehshort{\hatcurplanetnumxxxxxD}}{\hatcurSMEiizfehshort{\hatcurplanetnumxxxxxD}}}
\newcommand{\hatcurSMEloggxxxxxD}{\ifthenelse{\equal{\hatcurSMEversionxxxxxD}{i}}{\hatcurSMEilogg{\hatcurplanetnumxxxxxD}}{\hatcurSMEiilogg{\hatcurplanetnumxxxxxD}}}
\newcommand{\hatcurSMEvsinxxxxxD}{\ifthenelse{\equal{\hatcurSMEversionxxxxxD}{i}}{\hatcurSMEivsin{\hatcurplanetnumxxxxxD}}{\hatcurSMEiivsin{\hatcurplanetnumxxxxxD}}}
\newcommand{\hatcurSMEvmacxxxxxD}{\ifthenelse{\equal{\hatcurSMEversionxxxxxD}{i}}{\hatcurSMEivmac{\hatcurplanetnumxxxxxD}}{\hatcurSMEiivmac{\hatcurplanetnumxxxxxD}}}
\newcommand{\hatcurSMEvmicxxxxxD}{\ifthenelse{\equal{\hatcurSMEversionxxxxxD}{i}}{\hatcurSMEivmic{\hatcurplanetnumxxxxxD}}{\hatcurSMEiivmic{\hatcurplanetnumxxxxxD}}}

\newcommand{\hatcur}[1]{\ifnum#1=50 %
\hatcurxxxxxA
\else
\ifnum#1=51 %
\hatcurxxxxxB
\else
\ifnum#1=52 %
\hatcurxxxxxC
\else
\ifnum#1=53 %
\hatcurxxxxxD
\else
??????\fi
\fi
\fi
\fi
}
\newcommand{\hatcurb}[1]{\ifnum#1=50 %
\hatcurbxxxxxA
\else
\ifnum#1=51 %
\hatcurbxxxxxB
\else
\ifnum#1=52 %
\hatcurbxxxxxC
\else
\ifnum#1=53 %
\hatcurbxxxxxD
\else
??????\fi
\fi
\fi
\fi
}
\newcommand{\hatcurc}[1]{\ifnum#1=50 %
\hatcurcxxxxxA
\else
\ifnum#1=51 %
\hatcurcxxxxxB
\else
\ifnum#1=52 %
\hatcurcxxxxxC
\else
\ifnum#1=53 %
\hatcurcxxxxxD
\else
??????\fi
\fi
\fi
\fi
}
\newcommand{\hatcurCCtwomassshort}[1]{\ifnum#1=50 %
\hatcurCCtwomassshortxxxxxA
\else
\ifnum#1=51 %
\hatcurCCtwomassshortxxxxxB
\else
\ifnum#1=52 %
\hatcurCCtwomassshortxxxxxC
\else
\ifnum#1=53 %
\hatcurCCtwomassshortxxxxxD
\else
??????\fi
\fi
\fi
\fi
}
\newcommand{\hatcurCCtassvi}[1]{\ifnum#1=50 %
\hatcurCCtassvixxxxxA
\else
\ifnum#1=51 %
\hatcurCCtassvixxxxxB
\else
\ifnum#1=52 %
\hatcurCCtassvixxxxxC
\else
\ifnum#1=53 %
\hatcurCCtassvixxxxxD
\else
??????\fi
\fi
\fi
\fi
}
\newcommand{\hatcurisocite}[1]{\ifnum#1=50 %
\hatcurisocitexxxxxA
\else
\ifnum#1=51 %
\hatcurisocitexxxxxB
\else
\ifnum#1=52 %
\hatcurisocitexxxxxC
\else
\ifnum#1=53 %
\hatcurisocitexxxxxD
\else
??????\fi
\fi
\fi
\fi
}
\newcommand{\hatcurisofull}[1]{\ifnum#1=50 %
\hatcurisofullxxxxxA
\else
\ifnum#1=51 %
\hatcurisofullxxxxxB
\else
\ifnum#1=52 %
\hatcurisofullxxxxxC
\else
\ifnum#1=53 %
\hatcurisofullxxxxxD
\else
??????\fi
\fi
\fi
\fi
}
\newcommand{\hatcurisoshort}[1]{\ifnum#1=50 %
\hatcurisoshortxxxxxA
\else
\ifnum#1=51 %
\hatcurisoshortxxxxxB
\else
\ifnum#1=52 %
\hatcurisoshortxxxxxC
\else
\ifnum#1=53 %
\hatcurisoshortxxxxxD
\else
??????\fi
\fi
\fi
\fi
}
\newcommand{\hatcurjhkfilset}[1]{\ifnum#1=50 %
\hatcurjhkfilsetxxxxxA
\else
\ifnum#1=51 %
\hatcurjhkfilsetxxxxxB
\else
\ifnum#1=52 %
\hatcurjhkfilsetxxxxxC
\else
\ifnum#1=53 %
\hatcurjhkfilsetxxxxxD
\else
??????\fi
\fi
\fi
\fi
}
\newcommand{\hatcurlumind}[1]{\ifnum#1=50 %
\hatcurlumindxxxxxA
\else
\ifnum#1=51 %
\hatcurlumindxxxxxB
\else
\ifnum#1=52 %
\hatcurlumindxxxxxC
\else
\ifnum#1=53 %
\hatcurlumindxxxxxD
\else
??????\fi
\fi
\fi
\fi
}
\newcommand{\hatcurplanetnum}[1]{\ifnum#1=50 %
\hatcurplanetnumxxxxxA
\else
\ifnum#1=51 %
\hatcurplanetnumxxxxxB
\else
\ifnum#1=52 %
\hatcurplanetnumxxxxxC
\else
\ifnum#1=53 %
\hatcurplanetnumxxxxxD
\else
??????\fi
\fi
\fi
\fi
}
\newcommand{\hatcurlogRprimeHK}[1]{\ifnum#1=51 %
\hatcurlogRprimeHKxxxxxB
\else
\ifnum#1=52 %
\hatcurlogRprimeHKxxxxxC
\else
\ifnum#1=53 %
\hatcurlogRprimeHKxxxxxD
\else
??????\fi
\fi
\fi
}
\newcommand{\hatcurRVgammaabs}[1]{\ifnum#1=50 %
\hatcurRVgammaabsxxxxxA
\else
\ifnum#1=51 %
\hatcurRVgammaabsxxxxxB
\else
\ifnum#1=52 %
\hatcurRVgammaabsxxxxxC
\else
\ifnum#1=53 %
\hatcurRVgammaabsxxxxxD
\else
??????\fi
\fi
\fi
\fi
}
\newcommand{\hatcurRVgammarel}[1]{\ifnum#1=50 %
\hatcurRVgammarelxxxxxA
\else
\ifnum#1=51 %
\hatcurRVgammarelxxxxxB
\else
\ifnum#1=52 %
\hatcurRVgammarelxxxxxC
\else
\ifnum#1=53 %
\hatcurRVgammarelxxxxxD
\else
??????\fi
\fi
\fi
\fi
}
\newcommand{\hatcurSMElogg}[1]{\ifnum#1=50 %
\hatcurSMEloggxxxxxA
\else
\ifnum#1=51 %
\hatcurSMEloggxxxxxB
\else
\ifnum#1=52 %
\hatcurSMEloggxxxxxC
\else
\ifnum#1=53 %
\hatcurSMEloggxxxxxD
\else
??????\fi
\fi
\fi
\fi
}
\newcommand{\hatcurSMEteff}[1]{\ifnum#1=50 %
\hatcurSMEteffxxxxxA
\else
\ifnum#1=51 %
\hatcurSMEteffxxxxxB
\else
\ifnum#1=52 %
\hatcurSMEteffxxxxxC
\else
\ifnum#1=53 %
\hatcurSMEteffxxxxxD
\else
??????\fi
\fi
\fi
\fi
}
\newcommand{\hatcurSMEversion}[1]{\ifnum#1=50 %
\hatcurSMEversionxxxxxA
\else
\ifnum#1=51 %
\hatcurSMEversionxxxxxB
\else
\ifnum#1=52 %
\hatcurSMEversionxxxxxC
\else
\ifnum#1=53 %
\hatcurSMEversionxxxxxD
\else
??????\fi
\fi
\fi
\fi
}
\newcommand{\hatcurSMEvmac}[1]{\ifnum#1=50 %
\hatcurSMEvmacxxxxxA
\else
\ifnum#1=51 %
\hatcurSMEvmacxxxxxB
\else
\ifnum#1=52 %
\hatcurSMEvmacxxxxxC
\else
\ifnum#1=53 %
\hatcurSMEvmacxxxxxD
\else
??????\fi
\fi
\fi
\fi
}
\newcommand{\hatcurSMEvmic}[1]{\ifnum#1=50 %
\hatcurSMEvmicxxxxxA
\else
\ifnum#1=51 %
\hatcurSMEvmicxxxxxB
\else
\ifnum#1=52 %
\hatcurSMEvmicxxxxxC
\else
\ifnum#1=53 %
\hatcurSMEvmicxxxxxD
\else
??????\fi
\fi
\fi
\fi
}
\newcommand{\hatcurSMEvsin}[1]{\ifnum#1=50 %
\hatcurSMEvsinxxxxxA
\else
\ifnum#1=51 %
\hatcurSMEvsinxxxxxB
\else
\ifnum#1=52 %
\hatcurSMEvsinxxxxxC
\else
\ifnum#1=53 %
\hatcurSMEvsinxxxxxD
\else
??????\fi
\fi
\fi
\fi
}
\newcommand{\hatcurSMEzfeh}[1]{\ifnum#1=50 %
\hatcurSMEzfehxxxxxA
\else
\ifnum#1=51 %
\hatcurSMEzfehxxxxxB
\else
\ifnum#1=52 %
\hatcurSMEzfehxxxxxC
\else
\ifnum#1=53 %
\hatcurSMEzfehxxxxxD
\else
??????\fi
\fi
\fi
\fi
}
\newcommand{\hatcurSMEzfehshort}[1]{\ifnum#1=50 %
\hatcurSMEzfehshortxxxxxA
\else
\ifnum#1=51 %
\hatcurSMEzfehshortxxxxxB
\else
\ifnum#1=52 %
\hatcurSMEzfehshortxxxxxC
\else
\ifnum#1=53 %
\hatcurSMEzfehshortxxxxxD
\else
??????\fi
\fi
\fi
\fi
}

\newcounter{planetcounter}

\newboolean{emulateapj}
\setboolean{emulateapj}{true}

\newboolean{rvtablelong}
\setboolean{rvtablelong}{true}

\newboolean{astroph}
\setboolean{astroph}{true}

\shortauthors{Hartman et al.}
\shorttitle{\hatcur{50}\lowercase{b}--\hatcur{53}\lowercase{b}}
\ifthenelse{\boolean{emulateapj}}{
    \newcommand{\titledag}{$\dagger$}
}{
    \newcommand{\titledag}{\dagger}
}

\begin{document}

\title{
\hatcur{50}\lowercase{b}, \hatcur{51}\lowercase{b}, \hatcur{52}\lowercase{b}, and \hatcur{53}\lowercase{b}: Three Transiting Hot Jupiters and a Transiting Hot Saturn from the HATNet Survey.\altaffilmark{\titledag}
}

\author{
    J.~D.~Hartman\altaffilmark{1},
    W.~Bhatti\altaffilmark{1},
    G.~\'A.~Bakos\altaffilmark{1,2},
    A.~Bieryla\altaffilmark{3},
    G.~Kov\'acs\altaffilmark{4},
    D.~W.~Latham\altaffilmark{3},
    Z.~Csubry\altaffilmark{1},
    M.~de~Val-Borro\altaffilmark{1},
    K.~Penev\altaffilmark{1},
    L.~A.~Buchhave\altaffilmark{3,5},
    G.~Torres\altaffilmark{3},
    A.~W.~Howard\altaffilmark{6},
    G.~W.~Marcy\altaffilmark{7},
    J.~A.~Johnson\altaffilmark{3},
    H.~Isaacson\altaffilmark{7},
    B.~Sato\altaffilmark{8},
    I.~Boisse\altaffilmark{9},
    E.~Falco\altaffilmark{3},
    M.~E.~Everett\altaffilmark{10},
    T.~Szklenar\altaffilmark{11},
    B.~J.~Fulton\altaffilmark{6},
    A.~Shporer\altaffilmark{12},
    T.~Kov\'acs\altaffilmark{4,1,14},
    T.~Hansen\altaffilmark{13},
    B.~B\'eky\altaffilmark{3},
    R.~W.~Noyes\altaffilmark{3},
    J.~L\'az\'ar\altaffilmark{11},
    I.~Papp\altaffilmark{11},
    P.~S\'ari\altaffilmark{11}
}
\altaffiltext{1}{Department of Astrophysical Sciences, Princeton
  University, Princeton, NJ 08544; email: jhartman@astro.princeton.edu}

\altaffiltext{2}{Sloan and Packard Fellow}

\altaffiltext{3}{Harvard-Smithsonian Center for Astrophysics,
    Cambridge, MA 02138}

\altaffiltext{4}{Konkoly Observatory, Budapest, Hungary}

\altaffiltext{5}{Niels Bohr Institute, University of Copenhagen, DK-2100, Denmark, and Centre for Star and Planet Formation, National History Museum of Denmark, DK-1350 Copenhagen}

\altaffiltext{6}{Institute for Astronomy, University of Hawaii,
	Honolulu, HI 96822}

\altaffiltext{7}{Department of Astronomy, University of California,
    Berkeley, CA}

\altaffiltext{8}{Tokyo Institute of Technology, 2-12-1 Ookayama, Meguro-ku, Tokyo 152-8550, Japan}

\altaffiltext{9}{Aix Marseille Universit\'e, CNRS, LAM (Laboratoire d'Astrophysique de Marseille) UMR 7326, 13388, Marseille, France}

\altaffiltext{10}{National Optical Astronomy Observatory, 950 N.~Cherry Ave, Tucson, AZ 85719}

\altaffiltext{11}{Hungarian Astronomical Association, Budapest, 
    Hungary}

\altaffiltext{12}{Jet Propulsion Laboratory, California Institute of Technology, 4800 Oak Grove Drive, Pasadena, CA 91109}

\altaffiltext{13}{Landessternwarte, ZAH, K\"{o}nigstuhl 12, D-69117 Heidelberg, Germany}

\altaffiltext{14}{Fulbright Fellow}

\altaffiltext{$\dagger$}{
    Based on observations obtained with the Hungarian-made
    Automated Telescope Network. Based on observations
    obtained at the W.~M.~Keck Observatory, which is operated by the
    University of California and the California Institute of
    Technology. Keck time has been granted by NOAO (A245Hr) and NASA
    (N154Hr, N130Hr).  Based on data collected at Subaru
    Telescope, which is operated by the National Astronomical
    Observatory of Japan. Based on observations made with the
    Nordic Optical Telescope, operated on the island of La Palma
    jointly by Denmark, Finland, Iceland, Norway, Sweden, in the
    Spanish Observatorio del Roque de los Muchachos of the Instituto
    de Astrof\'isica de Canarias. Based on observations obtained
    with the Tillinghast Reflector 1.5\,m telescope and the 1.2\,m
    telescope, both operated by the Smithsonian Astrophysical
    Observatory at the Fred Lawrence Whipple Observatory in AZ. Based
    on radial velocities obtained with the Sophie spectrograph
    mounted on the 1.93\,m telescope at Observatoire de Haute-Provence. Based on observations obtained with
    facilities of the Las Cumbres Observatory Global Telescope.
}


\begin{abstract}

\setcounter{footnote}{10}
We report the discovery and characterization of four transiting
exoplanets by the HATNet survey. The planet \hatcurb{50} has a mass of
$\hatcurPPmshort{50}$\,\mjup\ and radius of
$\hatcurPPrshort{50}$\,\rjup, and orbits a bright ($V =
\hatcurCCtassmvshort{50}$\,mag) $M = \hatcurISOmshort{50}$\,\msun, $R
= \hatcurISOrshort{50}$\,\rsun\ star every
$P=\hatcurLCPshort{50}$\,days. The planet \hatcurb{51} has a mass of
$\hatcurPPmshort{51}$\,\mjup\ and radius of
$\hatcurPPrshort{51}$\,\rjup, and orbits a $V =
\hatcurCCtassmvshort{51}$\,mag, $M = \hatcurISOmshort{51}$\,\msun, $R
= \hatcurISOrshort{51}$\,\rsun\ star with a period of
$P=\hatcurLCPshort{51}$\,days. The planet \hatcurb{52} has a mass of
$\hatcurPPmshort{52}$\,\mjup\ and radius of
$\hatcurPPrshort{52}$\,\rjup, and orbits a $V =
\hatcurCCtassmvshort{52}$\,mag, $M = \hatcurISOmshort{52}$\,\msun, $R
= \hatcurISOrshort{52}$\,\rsun\ star with a period of
$P=\hatcurLCPshort{52}$\,days. The planet \hatcurb{53} has a mass of
$\hatcurPPmshort{53}$\,\mjup\ and radius of
$\hatcurPPrshort{53}$\,\rjup, and orbits a $V =
\hatcurCCtassmvshort{53}$\,mag, $M = \hatcurISOmshort{53}$\,\msun, $R
= \hatcurISOrshort{53}$\,\rsun\ star with a period of
$P=\hatcurLCPshort{53}$\,days. All four planets are consistent with
having circular orbits and have masses and radii measured to better
than 10\% precision. The low stellar jitter and favorable
$\rpl/\rstar$ ratio for \hatcur{51} make it a promising target for
measuring the Rossiter-McLaughlin effect for a Saturn-mass planet.
\setcounter{footnote}{0}
\end{abstract}

\keywords{
    planetary systems ---
    stars: individual (
\setcounter{planetcounter}{1}
\hatcur{50},
\hatcurCCgsc{50}\loopcommanoperiod
\setcounter{planetcounter}{2}
\hatcur{51},
\hatcurCCgsc{51}\loopcommanoperiod
\setcounter{planetcounter}{3}
\hatcur{52},
\hatcurCCgsc{52}\loopcommanoperiod
\setcounter{planetcounter}{4}
\hatcur{53},
\hatcurCCgsc{53}\loopcommanoperiod
) 
    techniques: spectroscopic, photometric
}


\section{Introduction}
\label{sec:introduction}

Transiting exoplanets (TEPs) are important objects for studying the
physical properties of planets outside the solar system. By combining
time-series photometry of a transit with time-series radial velocity
(RV) observations of the star spanning the planetary orbit, it is
possible to accurately measure the mass and radius of a transiting
planet relative to those of the host star. Leveraging stellar
evolution models to estimate the stellar mass and radius given
observable parameters such as the effective temperature, metallicity
and bulk density of the star, then allows the physical mass and radius
of the planet, as well as its orbital separation, to be
determined. Other properties of the system such as the orbital
eccentricity and obliquity \citep[e.g.][]{queloz:2000}, and properties
of the planetary atmosphere (e.g.~emission or transmission spectra)
may also be accessible for transiting planets
\citep[e.g.][]{charbonneau:2002}. Motivated by the wealth of physical
information that may be measured for these objects, there has been a
significant effort over the past 15 years to discover and characterize
many TEPs. The aim of this effort is to explore the diversity of
exoplanets, and to identify statistically robust relations between
their physical parameters, which in turn inform theories of planet
formation and evolution \citep[e.g.][]{guillot:2006,burrows:2007,beky:2011:hat27,laughlin:2011,enoch:2012}.

Largely thanks to the ultra-high-precision photometric time-series
observations from the NASA {\em Kepler} mission, we now know of over
4000 high-quality {\em candidate} transiting exoplanets
\citep[e.g.][]{mullally:2015}. Some 51 of the {\em Kepler} candidates
have been confirmed through measuring the RV orbital wobble of their
host stars, while a further 845 have masses estimated through transit
time variations, or have been statistically validated as being very
unlikely to be anything other than transiting
planets\footnote{\url{http://exoplanets.org} accessed 2015 Feb
  18}. The majority of the candidates from {\em Kepler} are, however,
too small and/or orbiting stars that are too faint to allow their
masses and orbital eccentricities to be determined using existing
spectroscopic facilities. For most of these planets, all we can
determine at present are their radii, orbital periods, and a
constraint on their eccentricities using the so-called photo-eccentric
effect \citep[e.g.][]{dawson:2012}.

Most of the TEPs with spectroscopically determined masses have been
discovered by wide-field ground-based transit surveys such as HATNet
\citep{bakos:2004:hatnet}, HATSouth \citep{bakos:2013:hatsouth}, WASP
\citep{pollacco:2006:wasp}, XO \citep{mccullough:2005}, TrES
\citep{alonso:2004}, and KELT \citep{pepper:2007}, among others. These
surveys cover a greater area of the sky than has been surveyed so far
by {\em Kepler} or its successor mission {\em K2}, and have thereby
monitored more bright stars which may host TEPs amenable to
confirmation spectroscopy. In this paper we present the discovery and
characterization of four new transiting short-period gas-giant planets
by the HATNet survey.

The HATNet survey, which began operations in 2004, has to date
searched 17\% of the $4\pi$ steradian celestial sphere for planets. A
total of 5.5 million stars have been observed. The stars have from 2400
to 21000 high-cadence photometric observations (5th and 95th
percentiles; the median is 7200) spanning a few months to several
years. The point-to-point RMS precision of the observations ranges
from $\sim 3$\,mmag for stars with $r \sim 9$ to $\sim 2$\% for stars
with $r \sim 13.3$ (depending on sky conditions and the density of
stars in the field being observed). Based on these observations we
have identified $\sim 2000$ candidate TEPs, the majority of which are
false positives. The stars are generally bright (the median magnitude
of the candidates is $V = 12.7$\,mag) so that it has been possible to
carry out spectroscopic and/or photometric follow-up observations for
the majority of these objects. Based on this follow-up, 1468
candidates have been rejected as false positives (the transit signal
is probably real, but not due to a planet), 189 have been rejected as
false alarms (the identified transit signal was not real), while more
than 50 confirmed and well-characterized planets (including those
presented here) have been announced. Some $\sim 350$ candidates are
currently active.

The four new planets announced in this paper have properties that are
typical of short-period gas-giant planets. While they do not, in
themselves, reveal new properties of exoplanets, they will contribute to
our statistical understanding of planetary systems in the
Galaxy.

In the next section we describe the observations used to confirm the
new TEPs. In \refsecl{analysis} we discuss the analysis carried out to
rule out false positive blend scenarios and determine the physical
parameters of the planetary systems. We place these planets into
context with the other known transiting planets in
\refsecl{discussion}.

\section{Observations}
\label{sec:obs}

The discovery of all four transiting planet systems followed the
general observational procedures described by \citet{latham:2009:hat8}
and \citet{bakos:2010:hat11}. Here we summarize the observations of
each system, and our methods for reducing the raw data to
scientifically interesting measurements.

\subsection{Photometric detection}
\label{sec:detection}

The four TEPs presented here were initially identified as candidate
TEPs based on observations made with the HATNet wide-field photometric
instruments \citep{bakos:2004:hatnet}. This network consists of six
identical fully-automated instruments, with four at Fred Lawrence
Whipple Observatory (FLWO) in AZ, and two on the roof of the
Submillimeter Array Hangar Building at Mauna Kea Observatory (MKO) in
HI. The light-gathering elements of each instrument include an 11\,cm
diameter telephoto lens, a Sloan $r$ filter, and a
4K$\times$4K front-side-illuminated CCD camera. Observations made in
2007 and early 2008 were done using a Cousins $R$ filter. The
instruments have a field of view of $10\fdg 6 \times 10\fdg 6$ and a
pixel scale of 9\arcsec\,pixel$^{-1}$ at the center of an
image. Observations are fully automated with the typical procedure
being to continuously monitor a given field while it is above
30$^{\circ}$ elevation taking exposures of 180\,s (prior to 2010
December~an exposure time of 300\,s was used). The fields have been
defined by tiling the sky into 838 $7^{\circ} \times 7^{\circ}$
pointings. Because each tile is smaller than the field of view, there
is overlap between neighboring fields and a given source may be
observed in multiple (up to four) fields.

\reftabl{photobs} lists the HATNet observations which contributed to
the discovery of each system. All four objects were observed using
multiple HAT instruments, and three of the four objects are in
overlapping fields. In some cases the observations date back to 2007,
and may span as many as 3.5 years. \hatcur{51}, in particular, has
been observed extensively with HATNet, having more than 27,000
individual photometric measurements.

The raw HATNet images were reduced to systematic-noise-filtered light
curves following \citet{bakos:2010:hat11} and making use of aperture
and image subtraction photometry tools from
\citet{pal:2009:thesis}. The filtering includes decorrelating the
individual light curves against various instrumental parameters (we
refer to this procedure as External Parameter Decorrelation, or EPD)
including the image position of the source, the sub-pixel position,
the background flux, the local scatter in the background flux, and the
shape of the PSF. Following EPD we make use of the Trend Filtering
Algorithm \citep[TFA;][]{kovacs:2005:TFA} in non-reconstructive
mode. The data for each HATNet field were reduced independently, with
EPD applied separately to each instrument, and TFA applied globally to
all observations from a given field (with an option to perform a
complete TFA filtering, using data from all telescopes and all
fields).

Light curves were searched for periodic box-shaped transits using the
Box-fitting Least Squares algorithm
\citep[BLS;][]{kovacs:2002:BLS}. Candidates were selected using a
variety of automated cuts (e.g.~on the S/N, differences in depth
between even and odd transits, among others) and a final by-eye
inspection. Figure~\ref{fig:hatnet} shows the phase-folded,
trend-filtered light curves from HATNet for the four newly discovered
planetary systems. 

We used BLS to search the residual light curves for additional
planetary transits, but did not detect any additional signals. We also
calculated the Discrete Fourier Transform (DFT, see
\citealp{deeming:1975}, and using the method of \citealp{kurtz:1985}
for a fast recursive evaluation of the trigonometric functions) for
each of the light curves, after subtracting the best-fit transit
models, to search for any continuous periodic variations. Such
variations may be due to the rotation of spotted stars, for
example. For \hatcur{50}, -52 and -53 we can rule out signals in the
frequency range 0 to 50\,d$^{-1}$ with an amplitude above 0.6\,mmag,
1.3\,mmag and 1.3\,mmag, respectively. For \hatcur{51} we also do not
find a significant Fourier component. Curiously, the highest peak in
the frequency spectrum is within 1.3\% of the first harmonic of the
orbital frequency. We do not have a physical explanation of this near
coincidence, if it is a real signal, but we can exclude the
possibility of tidal distortion due to the well-demonstrated
sub-stellar nature of the companion (see Section~\ref{sec:blend}). It
may perhaps be a signature of stellar activity. After subtracting this
low amplitude (1.3\,mmag) component, the next highest peak in the
frequency spectrum has an amplitude of 1.0\,mmag.

%
%
\ifthenelse{\boolean{emulateapj}}{
    \begin{figure*}[!ht]
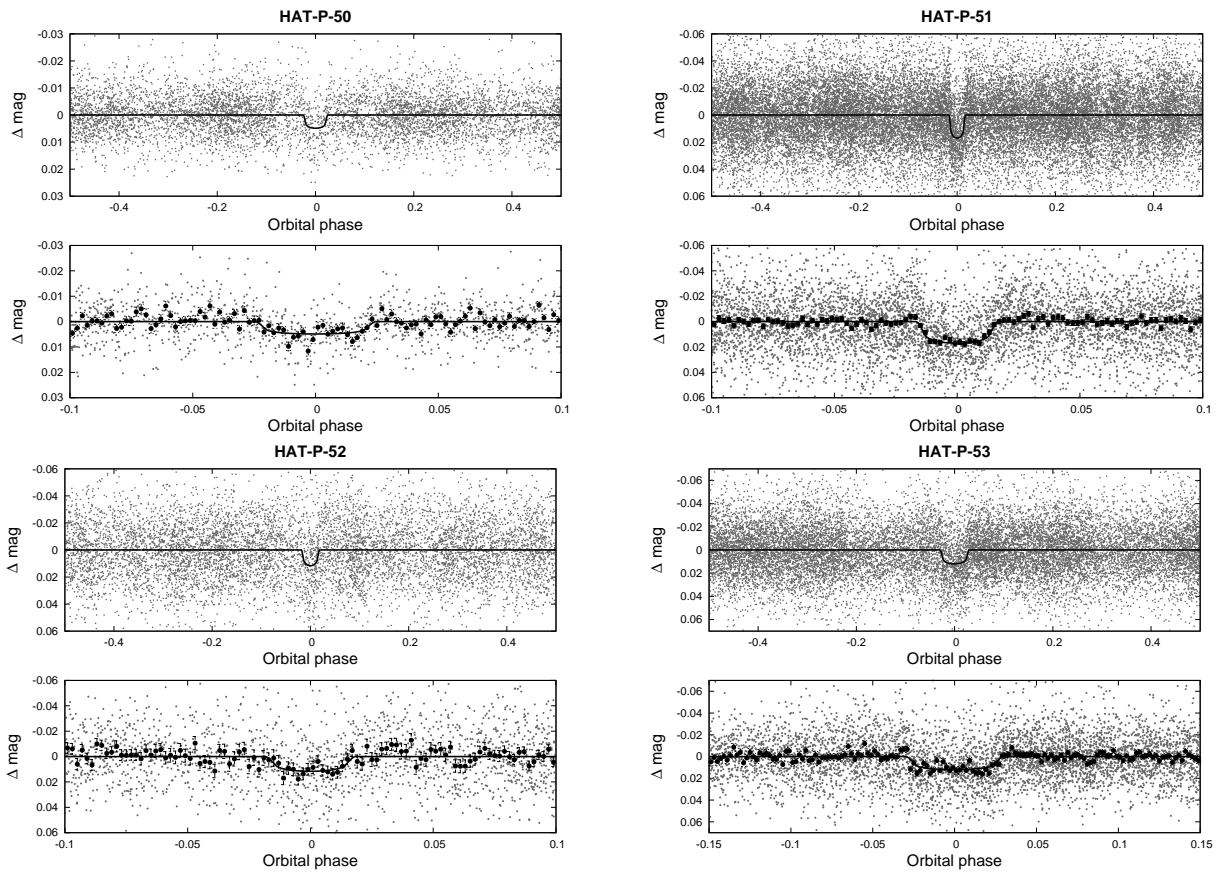

}{
    \begin{figure}[!ht]
}
\plottwo{\hatcurhtr{50}-hatnet.eps}{\hatcurhtr{51}-hatnet.eps}
\plottwo{\hatcurhtr{52}-hatnet.eps}{\hatcurhtr{53}-hatnet.eps}
\caption[]{
    Phase-folded unbinned HATNet light curves for \hatcur{50} (upper
    left), \hatcur{51} (upper right), \hatcur{52} (lower left), and
    \hatcur{53} (lower right). In each case we show two panels. The
    top panel shows the full light curve, while the bottom panel shows
    the light curve zoomed-in on the transit. The solid lines show the
    model fits to the light curves. The dark filled circles in the
    bottom panels show the light curves binned in phase with a bin
    size of 0.002.
\label{fig:hatnet}}
\ifthenelse{\boolean{emulateapj}}{
    \end{figure*}
}{
    \end{figure}
}

\ifthenelse{\boolean{emulateapj}}{
    \begin{deluxetable*}{llrrrr}
}{
    \begin{deluxetable}{llrrrr}
}
\tablewidth{0pc}
\tabletypesize{\scriptsize}
\tablecaption{
    Summary of photometric observations
    \label{tab:photobs}
}
\tablehead{
    \multicolumn{1}{c}{Instrument/Field\tablenotemark{a}} &
    \multicolumn{1}{c}{Date(s)} &
    \multicolumn{1}{c}{\# Images} &
    \multicolumn{1}{c}{Cadence\tablenotemark{b}} &
    \multicolumn{1}{c}{Filter} &
    \multicolumn{1}{c}{Precision\tablenotemark{c}} \\
    \multicolumn{1}{c}{} &
    \multicolumn{1}{c}{} &
    \multicolumn{1}{c}{} &
    \multicolumn{1}{c}{(sec)} &
    \multicolumn{1}{c}{} &
    \multicolumn{1}{c}{(mmag)}
}
\startdata
\sidehead{\textbf{\hatcur{50}}}
~~~~HAT-10/G316 & 2008 Nov--2009 May & 3214 & 352 & Sloan~$r$ & 7.5 \\
~~~~HAT-5/G364 & 2009 May & 21 & 411 & Sloan~$r$ & 10.6 \\
~~~~HAT-9/G364 & 2008 Dec--2009 May & 3159 & 352 & Sloan~$r$ & 7.5 \\
~~~~BOS & 2012 Feb 15 & 105 & 149 & Sloan~$i$ & 2.1 \\
~~~~Keplercam & 2012 Feb 18 & 443 & 54 & Sloan~$i$ & 1.2 \\
~~~~BOS & 2012 Feb 21 & 81 & 140 & Sloan~$i$ & 2.5 \\
~~~~BOS & 2012 Apr 08 & 61 & 143 & Sloan~$i$ & 1.6 \\
~~~~Keplercam & 2012 Nov 28 & 462 & 44 & Sloan~$i$ & 1.8 \\
~~~~Keplercam & 2012 Dec 23 & 277 & 45 & Sloan~$i$ & 2.3 \\
~~~~Keplercam & 2013 Jan 14 & 427 & 45 & Sloan~$i$ & 1.4 \\
~~~~Keplercam & 2013 Jan 17 & 380 & 45 & Sloan~$i$ & 1.6 \\
\sidehead{\textbf{\hatcur{51}}}
~~~~HAT-6/G164 & 2007 Sep--2008 Feb & 3652 & 349 & Cousins $R$ & 30.3 \\
~~~~HAT-9/G164 & 2007 Sep--2008 Feb & 2767 & 349 & Cousins $R$ & 25.9 \\
~~~~HAT-10/G165 & 2010 Sep--2011 Jan & 4215 & 230 & Sloan~$r$ & 24.3 \\
~~~~HAT-5/G165 & 2010 Nov--2011 Feb & 4142 & 354 & Sloan~$r$ & 24.1 \\
~~~~HAT-8/G165 & 2010 Nov--2011 Feb & 2240 & 238 & Sloan~$r$ & 23.6 \\
~~~~HAT-6/G209 & 2010 Nov--2011 Feb & 3794 & 351 & Sloan~$r$ & 18.4 \\
~~~~HAT-9/G209 & 2010 Nov--2011 Feb & 2151 & 352 & Sloan~$r$ & 18.0 \\
~~~~HAT-7/G210 & 2010 Nov--2011 Jan & 4047 & 229 & Sloan~$r$ & 19.1 \\
~~~~Keplercam & 2011 Oct 21 & 88 & 134 & Sloan~$i$ & 1.9 \\
~~~~Keplercam & 2012 Jan 05 & 92 & 133 & Sloan~$i$ & 2.7 \\
~~~~Keplercam & 2012 Oct 05 & 171 & 134 & Sloan~$i$ & 2.2 \\
~~~~Keplercam & 2012 Oct 26 & 137 & 134 & Sloan~$i$ & 2.6 \\
~~~~Keplercam & 2012 Nov 12 & 111 & 134 & Sloan~$i$ & 3.2 \\
\sidehead{\textbf{\hatcur{52}}}
~~~~HAT-5/G212 & 2010 Sep--2010 Nov & 2270 & 347 & Sloan~$r$ & 19.5 \\
~~~~HAT-8/G212 & 2010 Aug--2010 Nov & 5999 & 232 & Sloan~$r$ & 22.4 \\
~~~~Keplercam & 2010 Dec 23 & 101 & 134 & Sloan~$i$ & 2.0 \\
~~~~Keplercam & 2011 Sep 05 & 90 & 133 & Sloan~$i$ & 2.7 \\
~~~~Keplercam & 2011 Sep 27 & 188 & 134 & Sloan~$i$ & 2.3 \\
~~~~Keplercam & 2011 Nov 21 & 82 & 133 & Sloan~$i$ & 2.5 \\
~~~~Keplercam & 2012 Jan 07 & 64 & 194 & Sloan~$i$ & 3.0 \\
\sidehead{\textbf{\hatcur{53}}}
~~~~HAT-6/G164 & 2007 Sep--2008 Feb & 3653 & 349 & Cousins $R$ & 26.4 \\
~~~~HAT-9/G164 & 2007 Sep--2008 Feb & 2764 & 349 & Cousins $R$ & 24.5 \\
~~~~HAT-10/G165 & 2010 Sep--2011 Jan & 4234 & 230 & Sloan~$r$ & 19.3 \\
~~~~HAT-5/G165 & 2010 Nov--2011 Feb & 4134 & 354 & Sloan~$r$ & 19.4 \\
~~~~HAT-8/G165 & 2010 Nov--2011 Feb & 2240 & 238 & Sloan~$r$ & 20.4 \\
~~~~Keplercam & 2011 Oct 19 & 158 & 134 & Sloan~$i$ & 1.9 \\
~~~~Keplercam & 2011 Oct 27 & 381 & 73 & Sloan~$i$ & 2.5 \\
\enddata
\tablenotetext{a}{
    For HATNet data we list the HAT station and field name from which
    the observations are taken. HAT-5, -6, -7, and -10 are located at
    FLWO in Arizona, while HAT-8 and -9 are located at MKO in
    Hawaii. Each field corresponds to one of 838 fixed pointings used
    to cover the full 4$\pi$ celestial sphere. All data from a given
    HATNet field are reduced together, while detrending through
    External Parameter Decorrelation (EPD) is done independently for
    each unique field+station combination.
}
\tablenotetext{b}{
    The mode time between consecutive images rounded to the nearest
    second. Due to weather, the day--night cycle, guiding and focus
    corrections, and other factors, the cadence is only approximately
    uniform over short timescales.
}
\tablenotetext{c}{
    The RMS of the residuals from the best-fit model.
}
\ifthenelse{\boolean{emulateapj}}{
    \end{deluxetable*}
}{
    \end{deluxetable}
}

\subsection{Spectroscopic Observations}
\label{sec:obsspec}

Follow-up spectroscopic observations were carried out using six
different facilities. The aim of these observations was to aid in
ruling out false positives, determine the atmospheric parameters of
the host stars, and to confirm the planets by measuring the RV orbital
variations induced by the transiting planets. The facilities used for
each system are summarized in \reftabl{specobs}, and include the
Tillinghast Reflector Echelle Spectrograph
\citep[TRES;][]{furesz:2008} on the 1.5\,m Tillinghast Reflector at
FLWO; the Astrophysical Research Consortium Echelle Spectrometer
\citep[ARCES;][]{wang:2003} on the ARC~3.5\,m telescope at Apache
Point Observatory (APO) in New Mexico; the FIbre-fed \'Echelle
Spectrograph (FIES) at the 2.5\,m Nordic Optical Telescope (NOT) at La
Palma, Spain \citep{djupvik:2010}; the SOPHIE Spectrograph on the
1.93\,m telescope at OHP \citep{bouchy:2009} in France; HIRES
\citep{vogt:1994} on the Keck-I telescope in Hawaii together with the
I$_{\rm 2}$ absorption cell; and the High-Dispersion Spectrograph
\citep[HDS;][]{noguchi:2002} with the I$_{\rm 2}$ absorption cell
\citep{kambe:2002} on the Subaru telescope in Hawaii.

The TRES observations were used for reconnaissance (i.e.~ruling out
false positives with lower S/N spectra) for \hatcur{51}, \hatcur{52}
and \hatcur{53}. For \hatcur{50} they were used both for
reconnaissance and for measuring the orbital variation due to the
planet. The raw echelle images were reduced to extracted spectra and
analyzed to measure RVs and stellar atmospheric parameters following
\cite{buchhave:2010:hat16}. Observations of standard stars were made
during each observing run and are used to correct the velocities from
each run to the IAU system. Because these corrections are known for
TRES, we adopt the TRES measurements for the systemic $\gamma$
velocity of each object listed in \reftabl{stellar}. The uncertainty
on the absolute calibration is $\sim 0.1$\,\kms\ and is dominated by
the uncertainty in the absolute velocities of the standard stars.

The ARCES observations of \hatcur{51} and \hatcur{53} were used
exclusively for reconnaissance (based on observations of standard stars the RV precision of this instrument is
limited to $\sim 500$\,\ms). Observations were reduced to wavelength
calibrated spectra using the {\sc echelle} package in
IRAF\footnote{
IRAF is distributed by the National Optical Astronomy Observatories,
which are operated by the Association of Universities for Research in
Astronomy, Inc., under cooperative agreement with the National Science
Foundation.
}. For the wavelength calibration we made use of ThAr
lamp spectra obtained before or after each science exposure, and with
the same pointing as the science exposure. Each spectrum was analyzed
to measure the RV of the star, its surface gravity, effective
temperature, projected equatorial rotation velocity, and metallicity
using the Stellar Parameter Classification
\citep[SPC;][]{buchhave:2012:spc} procedure, which cross-correlates the
observed spectrum against a set of synthetic template spectra.

The single FIES spectrum obtained for \hatcur{51} was used for
reconnaissance, and was reduced and analyzed following
\cite{buchhave:2010:hat16}.

SOPHIE observations of \hatcur{51} were collected in high-efficiency
mode with the aim of confirming the planet by measuring the RV orbital
wobble of its host star. The SOPHIE observations were reduced and
analyzed following \cite{boisse:2013:hat42hat43}. Based on these
observations we determined that \hatcurb{51} is a Saturn-mass planet,
and that the $\sim 40$\,\ms\ precision of the SOPHIE observations for
this object was insufficient to accurately determine the planetary
mass. The precision in this case was limited due to significant
contamination from scattered moon light, for uncontaminated spectra
significantly higher precision may be obtained from the same S/N. We
do not include these data in the analysis of \hatcur{51}.

HDS observations were collected for \hatcur{50} and \hatcur{51} in
order to confirm these TEP systems and characterize the planetary
orbits. The observations were extracted and reduced to relative RVs in
the solar system barycentric frame following
\cite{sato:2002,sato:2012:hat38}, while spectral line bisector spans (BSs)
were computed following \citet{torres:2007:hat3}.

HIRES observations were collected for \hatcur{51}, \hatcur{52} and
\hatcur{53}. The observations have an RV precision of 5--10\,\ms\ and
are used here to characterize the orbital variations and to determine
the stellar atmospheric parameters. The data were reduced to relative
RVs in the barycentric frame following
\cite{butler:1996}. Spectral-line bisector spans (BSs) were computed
following \citet{torres:2007:hat3}, and $S$ activity indices were
calculated following \citet{isaacson:2010}. These were transformed to
$\log_{10}R^{\prime}_{\rm HK}$ values following \citet{noyes:1984}.

Based on the reconnaissance TRES, FIES and ARCES observations we find
that none of the four targets shows evidence of being a composite
system. All have radial velocity variations below 1\,\kms, and all are
dwarf stars. The effective temperatures, projected rotation
velocities, and surface gravities estimated from these spectra are
consistent with the higher precision values presented in
\reftabl{stellar}.

The high-precision RV measurements for all objects are seen to vary in
phase with the transit ephemerides. These are shown in
\reffigl{rvbis}. In this same figure we also show the phased BS
measurements, which in all cases are consistent with no variation in
phase with the ephemerides. The data are listed in \reftabl{rvs} at
the end of the paper.

\ifthenelse{\boolean{emulateapj}}{
    \begin{deluxetable*}{llrrrr}
}{
    \begin{deluxetable}{llrrrr}
}
\tablewidth{0pc}
\tabletypesize{\scriptsize}
\tablecaption{
    Summary of spectroscopy observations
    \label{tab:specobs}
}
\tablehead{
    \multicolumn{1}{c}{Instrument}          &
    \multicolumn{1}{c}{Date(s)}             &
    \multicolumn{1}{c}{\# Spec.}   &
    \multicolumn{1}{c}{Res.}          &
    \multicolumn{1}{c}{S/N Range \tablenotemark{a}}           &
    \multicolumn{1}{c}{RV Precision\tablenotemark{b}} \\
    &
    &
    &
    \multicolumn{1}{c}{$\Delta \lambda$/$\lambda$/1000} &
    &
    \multicolumn{1}{c}{(\ms)}
}
\startdata
\sidehead{\textbf{\hatcur{50}}}
~~~~TRES & 2010 Dec--2012 Feb & 5 & 44 & 24.8--36.1 & \hatcurRVfitrmsA{50} \\
~~~~FIES & 2012 Mar 13--17 & 5 & 67 & 31.0--69.9 & \hatcurRVfitrmsB{50} \\
~~~~HDS  & 2012 Feb 7 & 3 & 60 & 271--283 & $\cdots$ \\
~~~~HDS+I2  & 2012 Feb--2012 Sep & 20 & 60 & 84--166 & \hatcurRVfitrmsC{50} \\
\sidehead{\textbf{\hatcur{51}}}
~~~~FIES & 2011 Aug 4 & 1 & 46 & 27.4 & $\cdots$ \\
~~~~ARCES & 2011 Sep 19 & 1 & 31.5 & 20.6 & $\cdots$ \\
~~~~TRES & 2011 Sep 21 & 1 & 44 & 20.9 & $\cdots$ \\
~~~~SOPHIE & 2011 Dec 4--12 & 4 & 39 & 23--28 & $37$ \\
~~~~HIRES & 2011 Oct--Nov & 2 & 55 & 83--94 & $\cdots$ \\
~~~~HIRES+I2 & 2011 Oct--2012 Feb & 6 & 55 & 59--80 & \hatcurRVfitrmsA{51} \\
~~~~HDS   & 2012 Feb 9 & 4 & 60 & 52--56 & $\cdots$ \\
~~~~HDS+I2   & 2012 Feb 7--10 & 20 & 60 & 26--53 & \hatcurRVfitrmsB{51} \\
\sidehead{\textbf{\hatcur{52}}}
~~~~TRES & 2010 Dec--2011 Jan & 2 & 44 & 19.1--20.4 & 300 \\
~~~~HIRES & 2011 Oct 19 & 1 & 55 & 66 & $\cdots$ \\
~~~~HIRES+I2 & 2011 Feb--2012 Jul & 7 & 55 & 26--59 & \hatcurRVfitrms{52} \\
\sidehead{\textbf{\hatcur{53}}}
~~~~TRES & 2011 Sep 18--19 & 2 & 44 & 30.6--30.7 & $80$ \\
~~~~ARCES & 2011 Sep 19--20 & 2 & 31.5 & 19.8--20.1 & $380$ \\
~~~~HIRES & 2011 Nov 14 & 1 & 55 & 90 & $\cdots$ \\
~~~~HIRES+I2 & 2011 Nov--2012 Feb & 6 & 55 & 62--79 & \hatcurRVfitrms{53} \\
\enddata 
\tablenotetext{a}{The signal-to-noise ratio per resolution element near $5180$\,\AA.}
\tablenotetext{b}{
    The RMS of the RV residuals from the best-fit orbit, or the RMS of
    the RVs for reconnaissance observations. We do not give an
    estimate for template spectra (listed as HIRES or HDS without I2
    included), or for cases where only a single spectrum was obtained
    with a given instrument.
}
\ifthenelse{\boolean{emulateapj}}{
    \end{deluxetable*}
}{
    \end{deluxetable}
}

%
\setcounter{planetcounter}{1}
%
\ifthenelse{\boolean{emulateapj}}{
    \begin{figure*} [ht]
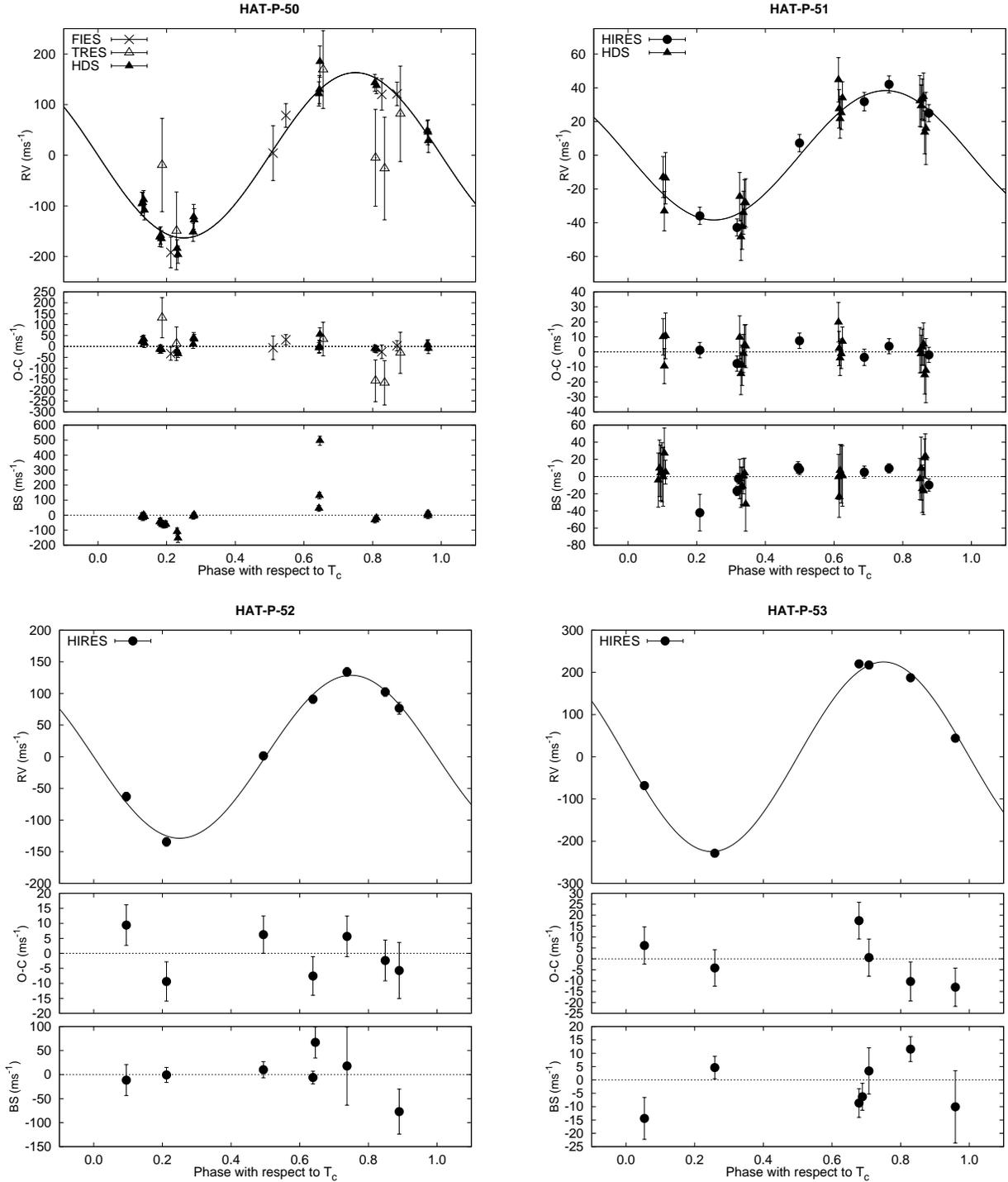

}{
    \begin{figure}[ht]
}
\plottwo{\hatcurhtr{50}-rv.eps}{\hatcurhtr{51}-rv.eps}
\plottwo{\hatcurhtr{52}-rv.eps}{\hatcurhtr{53}-rv.eps}
\caption{
    Phased high-precision RV measurements for \hbox{\hatcur{50}{}} (upper left), \hbox{\hatcur{51}{}} (upper right), \hbox{\hatcur{52}{}} (lower left), and \hbox{\hatcur{53}{}} (lower right). In each case we show three panels. The top panel shows the phased measurements together with our best-fit model (see \reftabl{planetparam}) for each system. Zero-phase corresponds to the time of mid-transit. The center-of-mass velocity has been subtracted. The second panel shows the velocity $O\!-\!C$ residuals from the best fit. The error bars include the jitter terms listed in \reftabl{planetparam} added in quadrature to the formal errors for each instrument. The third panel shows the bisector spans (BS), with the mean value subtracted. The symbols used for each instrument are indicated in the top panel for each planet. Note the different vertical scales of the panels.
}
\label{fig:rvbis}
\ifthenelse{\boolean{emulateapj}}{
    \end{figure*}
}{
    \end{figure}
}

\subsection{Photometric follow-up observations}
\label{sec:phot}

%
\setcounter{planetcounter}{1}
%
\begin{figure*}[!ht]
\plotone{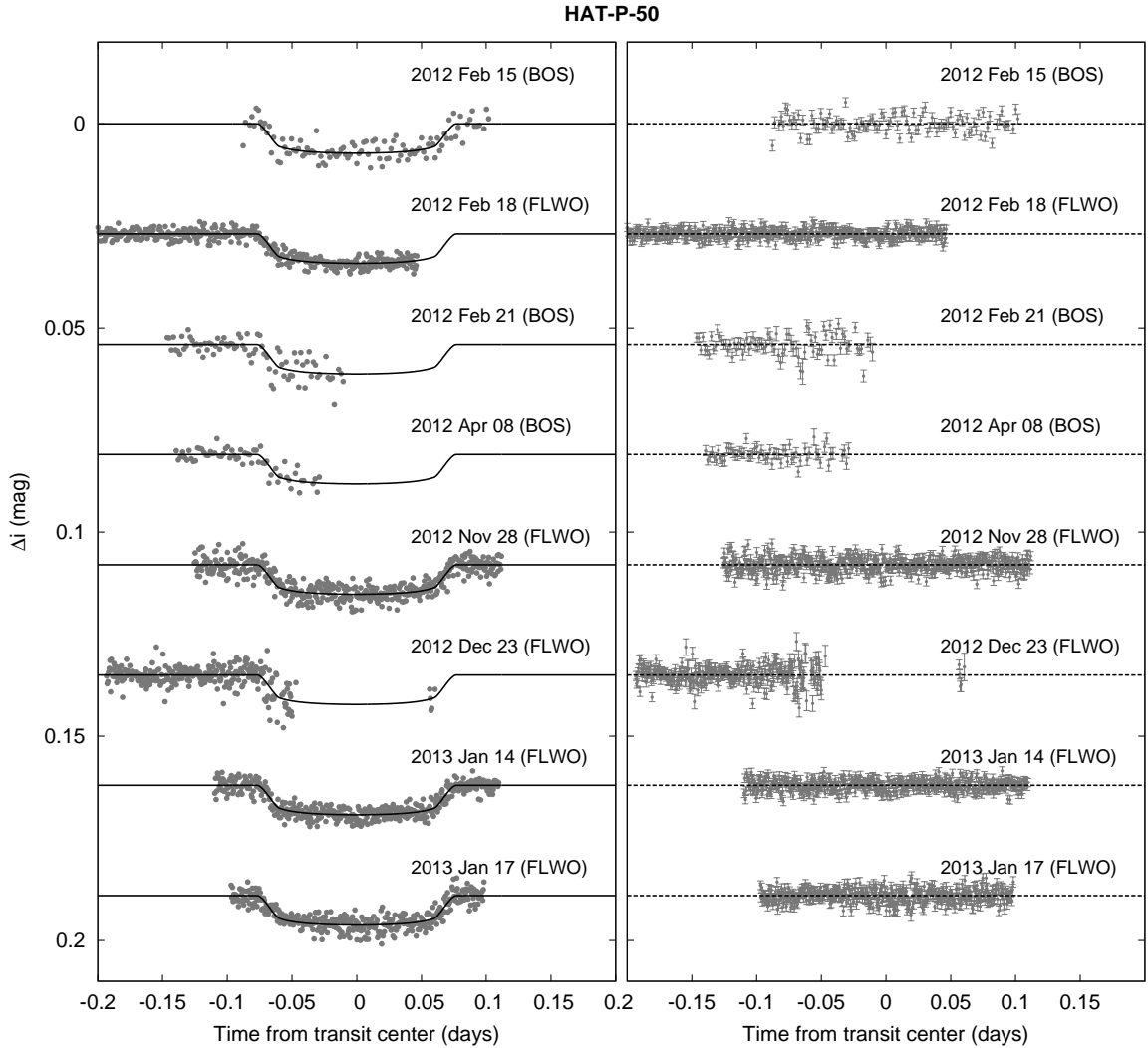}
\caption{
    Left: Unbinned transit \lcs{} for \hatcur{50}, acquired with the
    CCD imager on the BOS 0.8\,m telescope, and Keplercam on the
    \flwof{} telescope.  The light curves have been EPD and TFA
    processed, as described in \refsec{globmod}.
    The dates of the events and instruments used are indicated.
    Curves after the first are displaced vertically for clarity.  Our
    best fit from the global modeling described in \refsecl{globmod}
    is shown by the solid lines.  Right: Residuals from the fits in
    the same order as the left panel.  The error bars represent the
    photon and background shot noise, plus the readout noise.
}
\label{fig:lc50}
\end{figure*}
\setcounter{planetcounter}{2}
%
\begin{figure*}[!ht]
\plotone{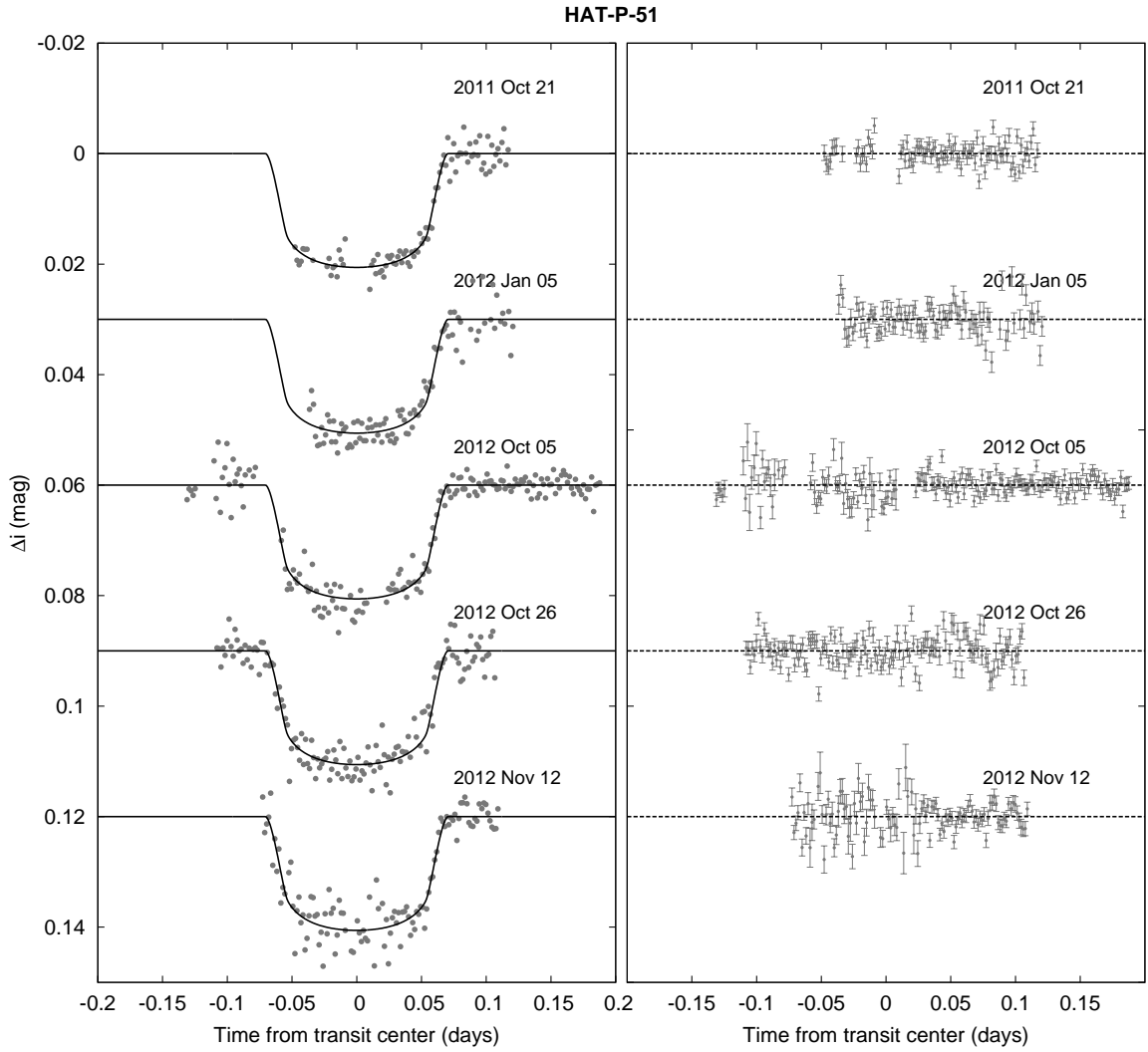}
\caption{
    Similar to \reffigl{lc50}; here we show the follow-up
    \lcs{} for \hatcur{51}. All light curves were obtained with Keplercam on the
    \flwof{} telescope.
}
\label{fig:lc51}
\end{figure*}
\setcounter{planetcounter}{3}
%
\begin{figure*}[!ht]
\plotone{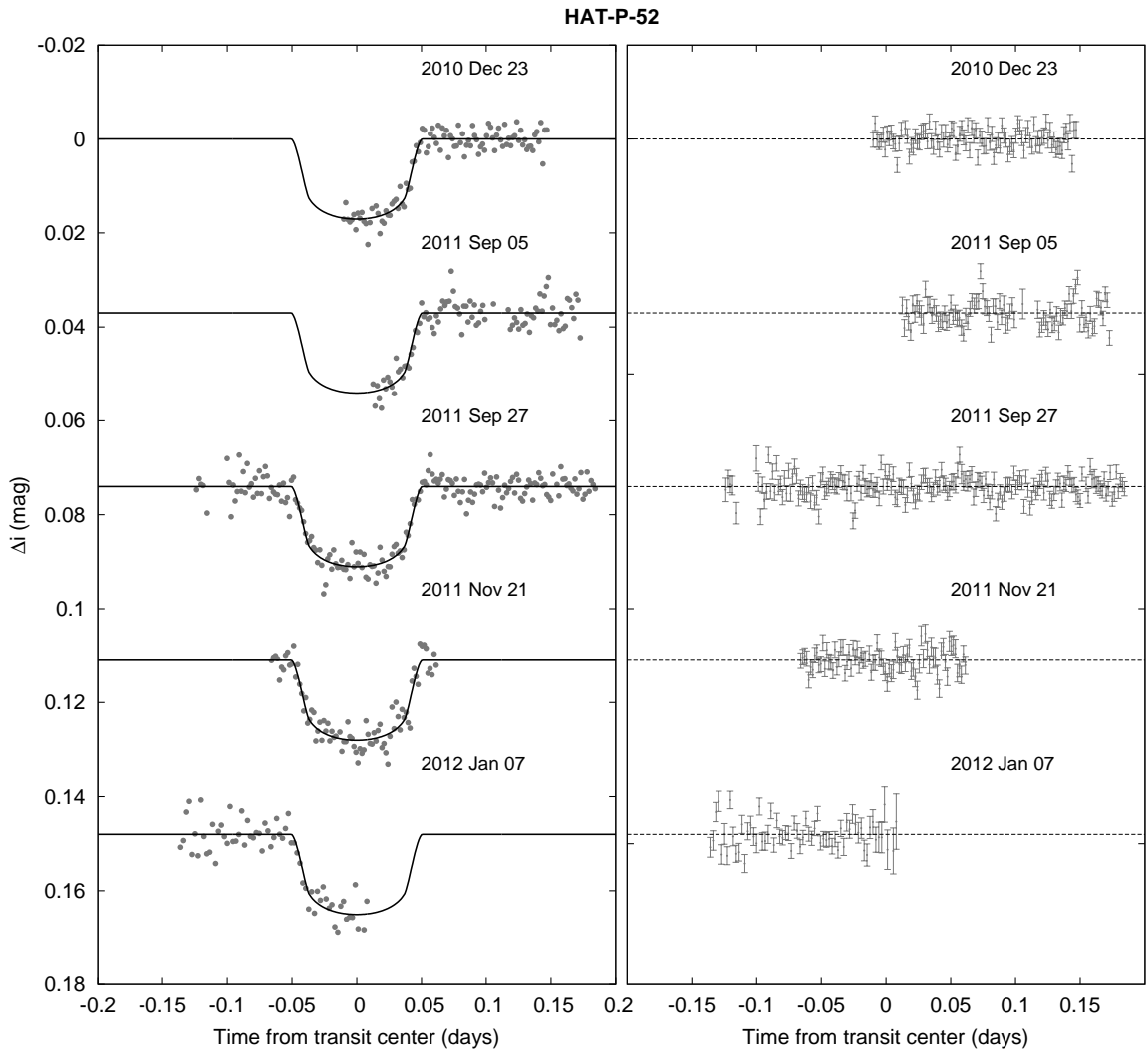}
\caption{
    Similar to \reffigl{lc50}; here we show the follow-up
    \lcs{} for \hatcur{52}. All light curves were obtained with Keplercam on the
    \flwof{} telescope.
}
\label{fig:lc52}
\end{figure*}
\setcounter{planetcounter}{4}
%
\begin{figure*}[!ht]
\plotone{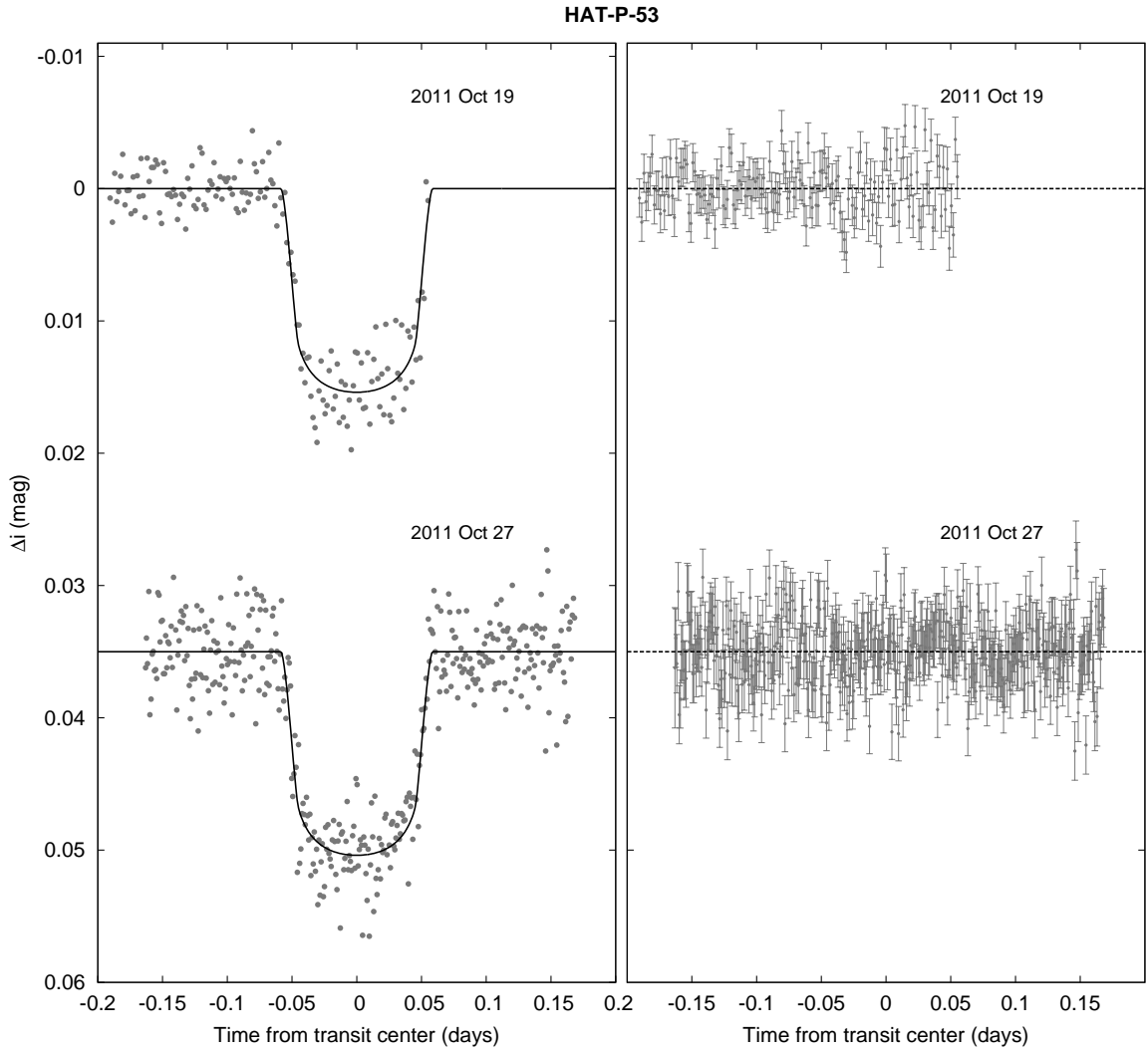}
\caption{
    Similar to \reffigl{lc50}; here we show the follow-up
    \lcs{} for \hatcur{53}. All light curves were obtained with Keplercam on the
    \flwof{} telescope.
}
\label{fig:lc53}
\end{figure*}

Additional time-series photometric measurements were obtained for all
four of the systems using Keplercam on the FLWO~1.2\,m
telescope. These observations were carried out during the planetary
transits to aid in ruling out blended eclipsing
binary false positive scenarios, and to refine the light curve
parameters (i.e.~the orbital period, the planet to star radius ratio,
the impact parameter and the transit duration). For \hatcur{50} we
also obtained follow-up photometry with the CCD imager on the Byrne
Observatory at Sedgwick (BOS) 0.8\,m telescope, located at Sedgwick
Reserve in Santa Ynez Valley, CA, and operated by the Las Cumbres
Observatory Global Telescope institute \citep[LCOGT;][]{brown:2013}. The events monitored
with each instrument, together with the number of images obtained, the
cadence, filter used and photometric precision are listed in
\reftabl{photobs}.

We applied standard CCD calibration procedures to the Keplercam and
BOS images and then reduced these to light curves using the aperture
photometry methods described by \citet{bakos:2010:hat11}. In doing
this we made use of the stellar centroid positions measured directly
from a set of registered and stacked frames rather than relying on
catalog positions for astrometry as done in
\citet{bakos:2010:hat11}. All sources in the images, save the target
TEP system, were used in performing the ensemble magnitude
calibration. We corrected for additional systematic trends in the data
by including the EPD and TFA noise filtering models in the fitting
mentioned in \refsecl{globmod}. The resulting trend filtered light curves for
\hatcur{50} through \hatcur{53} are shown in
Figures~\ref{fig:lc50}--\ref{fig:lc53}, respectively. All photometric
measurements made for the four objects are available in
machine-readable form in \reftabl{phfu}.


\clearpage

%
%
\ifthenelse{\boolean{emulateapj}}{
    \begin{deluxetable*}{llrrrrl}
}{
    \begin{deluxetable}{llrrrrl}
}
\tablewidth{0pc}
\tablecaption{
    Light curve data for \hatcur{50}--\hatcur{53}\label{tab:phfu}.
}
\tablehead{
    \colhead{Object\tablenotemark{a}} &
    \colhead{BJD\tablenotemark{b}} & 
    \colhead{Mag\tablenotemark{c}} & 
    \colhead{\ensuremath{\sigma_{\rm Mag}}} &
    \colhead{Mag(orig)\tablenotemark{d}} & 
    \colhead{Filter} &
    \colhead{Instrument} \\
    \colhead{} &
    \colhead{\hbox{~~~~(2,400,000$+$)~~~~}} & 
    \colhead{} & 
    \colhead{} &
    \colhead{} & 
    \colhead{} &
    \colhead{}
}
\startdata
HAT-P-50 & $ 54863.83699 $ & $   0.00928 $ & $   0.00393 $ & $ \cdots $ & $ r$ &     HATNet\\
HAT-P-50 & $ 54935.64419 $ & $  -0.00034 $ & $   0.00517 $ & $ \cdots $ & $ r$ &     HATNet\\
HAT-P-50 & $ 54838.86248 $ & $   0.00069 $ & $   0.00443 $ & $ \cdots $ & $ r$ &     HATNet\\
HAT-P-50 & $ 54888.81494 $ & $  -0.00915 $ & $   0.00442 $ & $ \cdots $ & $ r$ &     HATNet\\
HAT-P-50 & $ 54792.03291 $ & $   0.00668 $ & $   0.00450 $ & $ \cdots $ & $ r$ &     HATNet\\
HAT-P-50 & $ 54910.67022 $ & $   0.00400 $ & $   0.00418 $ & $ \cdots $ & $ r$ &     HATNet\\
HAT-P-50 & $ 54863.84110 $ & $   0.00001 $ & $   0.00388 $ & $ \cdots $ & $ r$ &     HATNet\\
HAT-P-50 & $ 54935.64826 $ & $   0.00737 $ & $   0.00652 $ & $ \cdots $ & $ r$ &     HATNet\\
HAT-P-50 & $ 54838.86690 $ & $   0.00070 $ & $   0.00433 $ & $ \cdots $ & $ r$ &     HATNet\\
HAT-P-50 & $ 54888.81903 $ & $   0.00426 $ & $   0.00441 $ & $ \cdots $ & $ r$ &     HATNet\\

\enddata
\tablenotetext{a}{
    Either HAT-P-50, HAT-P-51, HAT-P-52, or HAT-P-53.
}
\tablenotetext{b}{
    Barycentric Julian Date is computed directly from the UTC time
    without correction for leap seconds.
}
\tablenotetext{c}{
    The out-of-transit level has been subtracted. These magnitudes have
    been subjected to the EPD and TFA procedures, carried out
    simultaneously with the transit fit.
}
\tablenotetext{d}{
    Raw magnitude values without application of the EPD and TFA
    procedures. These are provided only for the follow-up
    observations. For HATNet, the transits are only detectable after
    applying the noise filtering methods.
}
\tablecomments{
    This table is available in a machine-readable form in the online
    journal.  A portion is shown here for guidance regarding its form
    and content.
}
\ifthenelse{\boolean{emulateapj}}{
    \end{deluxetable*}
}{
    \end{deluxetable}
}

\section{Analysis}
\label{sec:analysis}

\subsection{Properties of the parent star}
\label{sec:stelparam}

The stellar atmospheric parameters that we adopted for the analysis,
including the effective temperature $\teffstar$, the surface gravity
$\loggstar$, the metallicity \feh\ and the projected equatorial
rotation velocity \vsini, were determined for each system using
SPC. For \hatcur{50} we applied this to the TRES and FIES spectra
(applying to the individual spectra and adopting the average parameter
values) while for the other three systems we used the Keck/HIRES
I$_{2}$-free template spectra.

We used the Yonsei-Yale \citep[Y2;][]{yi:2001} theoretical stellar
models to determine physical parameters of the stars, such as their
masses, radii, luminosities and ages, based on the measured
atmospheric parameters together with the bulk stellar densities
$\rhostar$ determined from our modelling of the light curves and RV
measurements (\refsecl{globmod}). We generated a chain of
$\teffstar$, $\feh$ and $\rhostar$ values for each object, where the
$\rhostar$ values are taken from the output of the MCMC procedure used
to fit the light curves and RVs, while we assume uncorrelated Gaussian
distributions for $\teffstar$ and $\feh$. For each value in the chain
we interpolate the Y2 models to find a combination of $\mstar$, age
and $\feh$ which matches the three input parameters (we assume
solar-scaled abundances without $\alpha$-element
enhancement). Combinations of $\teffstar$, $\feh$ and $\rhostar$ that
do not match to a stellar model are rejected. In doing this we also
reject the corresponding link in the LC+RV MCMC chain so that the
final planetary parameters are restricted to regions of parameter
space allowed by the stellar evolution models. The stellar models also
provide other parameters such as $\rstar$ and $\lstar$ for a given
$\mstar$, age and $\feh$ combination. The result is a posterior chain
of stellar parameters for each star. We use the chains to calculate
the median and 68.3\% confidence interval for each of the stellar
parameters. These are listed in \reftabl{stellar}. We compare the
measured $\teffstar$ and $\rhostar$ values for each system to the
model isochrones in \reffigl{iso}.

For \hatcur{50} and \hatcur{52} we found that the median $\loggstar$
values determined from this procedure differed significantly from the
values estimated from the spectra. For these stars we carried out a
second iteration of SPC fixing $\loggstar$ to the values determined
from the stellar evolution modelling. We then performed a second
iteration of the LC+RV modelling, with revised limb darkening
parameters, followed by a second iteration of the stellar evolution
modelling. The $\loggstar$ values had converged after this
iteration. For \hatcur{51} and \hatcur{53} a second iteration of SPC
was not needed. 

Distances are determined for each system by comparing the measured
broad-band photometry listed in \reftabl{stellar} to the magnitudes
predicted in each filter by the models. We allow for extinction
assuming a $R_{V} = 3.1$ extinction law from \citet{cardelli:1989}. 

Based on this modelling we find that \hatcur{50} has a mass of
\hatcurISOmlong{50}\,\msun, a radius of \hatcurISOrlong{50}\,\rsun, an
age of \hatcurISOage{50}\,Gyr, and is at a distance of
\hatcurXdistred{50}\,pc. \hatcur{51} has a mass of
\hatcurISOmlong{51}\,\msun, a radius of \hatcurISOrlong{51}\,\rsun, an
age of \hatcurISOage{51}\,Gyr, and is at a distance of
\hatcurXdistred{51}\,pc. \hatcur{52} has a mass of
\hatcurISOmlong{52}\,\msun, a radius of \hatcurISOrlong{52}\,\rsun, an
age of \hatcurISOage{52}\,Gyr, and is at a distance of
\hatcurXdistred{52}\,pc. Finally, \hatcur{53} has a mass of
\hatcurISOmlong{53}\,\msun, a radius of \hatcurISOrlong{53}\,\rsun, an
age of \hatcurISOage{53}\,Gyr, and is at a distance of
\hatcurXdistred{53}\,pc.

For \hatcur{51}, -52 and -53 we used the Keck/HIRES spectra to
determine median $\log_{10}R^{\prime}_{\rm HK}$ activity indices. We
find that all three stars are inactive in the Ca II HK region,
consistent with their slow rotation and lack of photometric
variability.

\ifthenelse{\boolean{emulateapj}}{
    \begin{figure*}[!ht]
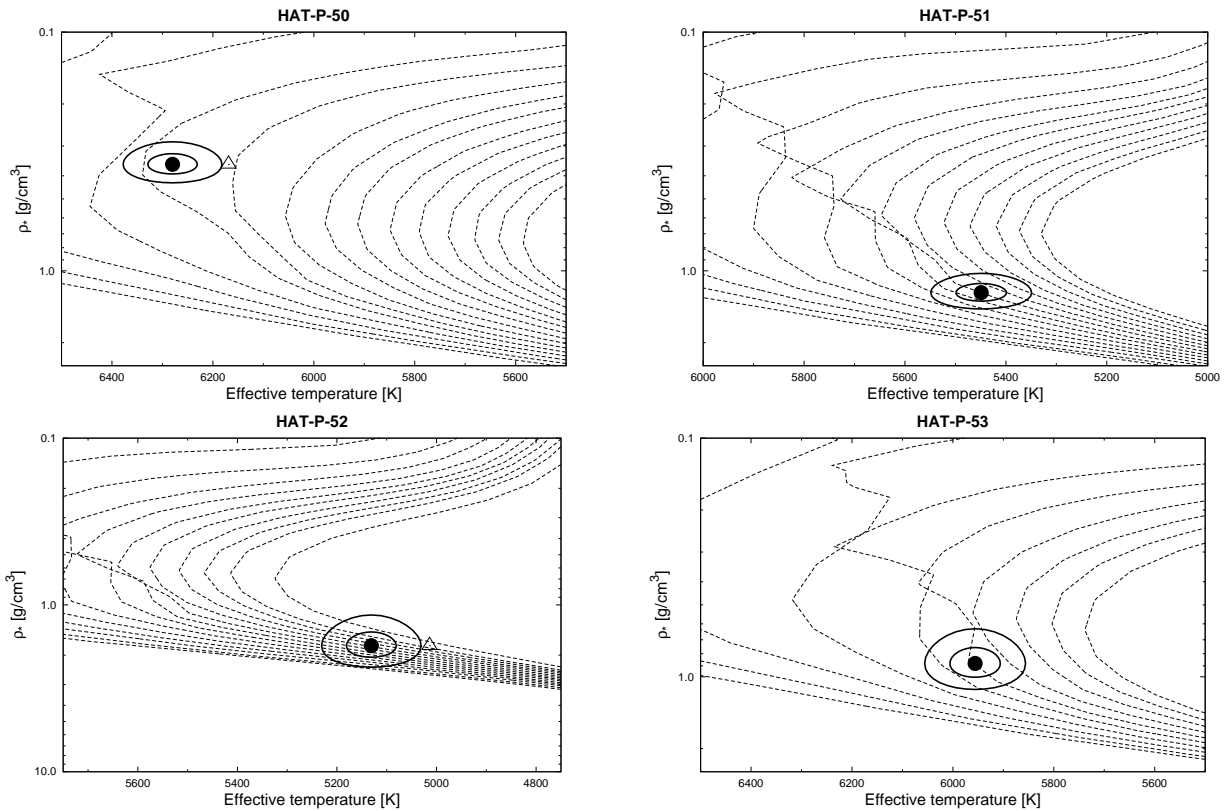

}{
    \begin{figure}[!ht]
}
\plottwo{\hatcurhtr{50}-iso-rho.eps}{\hatcurhtr{51}-iso-rho.eps}
\plottwo{\hatcurhtr{52}-iso-rho.eps}{\hatcurhtr{53}-iso-rho.eps}
\caption{
    Model isochrones from \cite{\hatcurisocite{50}} for the measured
    metallicities of \hatcur{50} (upper left), \hatcur{51} (upper right), \hatcur{52} (lower left) and \hatcur{53} (lower right). In each case we show models for several different ages, with younger models being on the left and older ones on the right. For \hatcur{50}, \hatcur{51} and \hatcur{52} we show ages of 0.2\,Gyr and 1.0 to 14.0\,Gyr in 1.0\,Gyr increments. For \hatcur{53} we show ages of 0.2\,Gyr and 1.0 to 9.0\,Gyr in 1.0\,Gyr increments.  The
    adopted values of $\teffstar$ and \rhostar\ are shown together with
    their 1$\sigma$ and 2$\sigma$ confidence ellipsoids.  The initial
    values of \teffstar\ and \rhostar\ from the first SPC and \lc\
    analyses are represented with a triangle for \hatcur{50} and \hatcur{52}. Note the logarithmic vertical axes, and the different scales used in each panel.
}
\label{fig:iso}
\ifthenelse{\boolean{emulateapj}}{
    \end{figure*}
}{
    \end{figure}
}

%
%
\ifthenelse{\boolean{emulateapj}}{
    \begin{deluxetable*}{lccccl}
}{
    \begin{deluxetable}{lccccl}
}
\tablewidth{0pc}
\tabletypesize{\scriptsize}
\tablecaption{
    Stellar parameters for \hatcur{50}--\hatcur{53}
    \label{tab:stellar}
}
\tablehead{
    \multicolumn{1}{c}{} &
    \multicolumn{1}{c}{\bf HAT-P-50} &
    \multicolumn{1}{c}{\bf HAT-P-51} &
    \multicolumn{1}{c}{\bf HAT-P-52} &
    \multicolumn{1}{c}{\bf HAT-P-53} &
    \multicolumn{1}{c}{} \\
    \multicolumn{1}{c}{~~~~~~~~Parameter~~~~~~~~} &
    \multicolumn{1}{c}{Value}                     &
    \multicolumn{1}{c}{Value}                     &
    \multicolumn{1}{c}{Value}                     &
    \multicolumn{1}{c}{Value}                     &
    \multicolumn{1}{c}{Source}
}
\startdata
\noalign{\vskip -3pt}
\sidehead{Astrometric properties and cross-identifications}
~~~~2MASS-ID\dotfill               & \hatcurCCtwomassshort{50}  & \hatcurCCtwomassshort{51} & \hatcurCCtwomassshort{52} & \hatcurCCtwomassshort{53} & \\
~~~~GSC-ID\dotfill                 & \hatcurCCgsc{50}      & \hatcurCCgsc{51}     & \hatcurCCgsc{52}     & \hatcurCCgsc{53}     & \\
~~~~R.A. (J2000)\dotfill            & \hatcurCCra{50}       & \hatcurCCra{51}    & \hatcurCCra{52}    & \hatcurCCra{53}    & 2MASS\\
~~~~Dec. (J2000)\dotfill            & \hatcurCCdec{50}      & \hatcurCCdec{51}   & \hatcurCCdec{52}   & \hatcurCCdec{53}   & 2MASS\\
~~~~$\mu_{\rm R.A.}$ (\masy)              & \hatcurCCpmra{50}     & \hatcurCCpmra{51}  & \hatcurCCpmra{52}  & \hatcurCCpmra{53}  & UCAC4\\
~~~~$\mu_{\rm Dec.}$ (\masy)              & \hatcurCCpmdec{50}    & \hatcurCCpmdec{51} & \hatcurCCpmdec{52} & \hatcurCCpmdec{53} & UCAC4\\
\sidehead{Spectroscopic properties}
~~~~$\teffstar$ (K)\dotfill         &  \hatcurSMEteff{50}   & \hatcurSMEteff{51} & \hatcurSMEteff{52} & \hatcurSMEteff{53} & SPC\tablenotemark{a}\\
~~~~$\feh$\dotfill                  &  \hatcurSMEzfeh{50}   & \hatcurSMEzfeh{51} & \hatcurSMEzfeh{52} & \hatcurSMEzfeh{53} & SPC                 \\
~~~~$\vsini$ (\kms)\dotfill         &  \hatcurSMEvsin{50}   & \hatcurSMEvsin{51} & \hatcurSMEvsin{52} & \hatcurSMEvsin{53} & SPC                 \\
~~~~$\vmac$ (\kms)\dotfill          &  1.0   & 1.0 & 1.0 & 1.0 & Assumed              \\
~~~~$\vmic$ (\kms)\dotfill          &  2.0   & 2.0 & 2.0 & 2.0 & Assumed              \\
~~~~$\gamma_{\rm RV}$ (\kms)\dotfill&  \hatcurRVgammaabs{50}  & \hatcurRVgammaabs{51} & \hatcurRVgammaabs{52} & \hatcurRVgammaabs{53} & TRES  \\
~~~~$\log_{10} R^{\prime}_{\rm HK}$ \tablenotemark{b}  & $\cdots$ & \hatcurlogRprimeHK{51} & \hatcurlogRprimeHK{52} & \hatcurlogRprimeHK{53} & HIRES \\
\sidehead{Photometric properties}
~~~~$B$ (mag)\dotfill               &  \hatcurCCtassmB{50}  & \hatcurCCtassmB{51} & \hatcurCCtassmB{52} & $\cdots$ & APASS,TASS\tablenotemark{c} \\
~~~~$V$ (mag)\dotfill               &  \hatcurCCtassmv{50}  & \hatcurCCtassmv{51} & \hatcurCCtassmv{52} & \hatcurCCtassmv{53} & APASS,TASS\tablenotemark{c} \\
~~~~$I_{C}$ (mag)\dotfill               &  \hatcurCCtassmI{50}  & \hatcurCCtassmI{51} & \hatcurCCtassmI{52} & \hatcurCCtassmI{53} & TASS\\
~~~~$g$ (mag)\dotfill               &  \hatcurCCtassmg{50}  & \hatcurCCtassmg{51} & \hatcurCCtassmg{52} & $\cdots$ & APASS \\
~~~~$r$ (mag)\dotfill               &  \hatcurCCtassmr{50}  & \hatcurCCtassmr{51} & \hatcurCCtassmr{52} & $\cdots$ & APASS \\
~~~~$i$ (mag)\dotfill               &  \hatcurCCtassmi{50}  & \hatcurCCtassmi{51} & \hatcurCCtassmi{52} & $\cdots$ & APASS \\
~~~~$J$ (mag)\dotfill               &  \hatcurCCtwomassJmag{50} & \hatcurCCtwomassJmag{51} & \hatcurCCtwomassJmag{52} & \hatcurCCtwomassJmag{53} & 2MASS           \\
~~~~$H$ (mag)\dotfill               &  \hatcurCCtwomassHmag{50} & \hatcurCCtwomassHmag{51} & \hatcurCCtwomassHmag{52} & \hatcurCCtwomassHmag{53} & 2MASS           \\
~~~~$K_s$ (mag)\dotfill             &  \hatcurCCtwomassKmag{50} & \hatcurCCtwomassKmag{51} & \hatcurCCtwomassKmag{52} & \hatcurCCtwomassKmag{53} & 2MASS           \\
\sidehead{Derived properties}
~~~~$\mstar$ ($\msun$)\dotfill      &  \hatcurISOmlong{50}   & \hatcurISOmlong{51} & \hatcurISOmlong{52} & \hatcurISOmlong{53} & YY+$a/R_{\star}$+SPC \tablenotemark{d}\\
~~~~$\rstar$ ($\rsun$)\dotfill      &  \hatcurISOrlong{50}   & \hatcurISOrlong{51} & \hatcurISOrlong{52} & \hatcurISOrlong{53} & YY+$a/R_{\star}$+SPC         \\
~~~~$\loggstar$ (cgs)\dotfill       &  \hatcurISOlogg{50}    & \hatcurISOlogg{51} & \hatcurISOlogg{52} & \hatcurISOlogg{53} & YY+$a/R_{\star}$+SPC         \\
~~~~$\rhostar$ (\gmcc)\dotfill       &  \hatcurISOrho{50}    & \hatcurISOrho{51} & \hatcurISOrho{52} & \hatcurISOrho{53} & YY+$a/R_{\star}$+SPC         \\
~~~~$\lstar$ ($\lsun$)\dotfill      &  \hatcurISOlum{50}     & \hatcurISOlum{51} & \hatcurISOlum{52} & \hatcurISOlum{53} & YY+$a/R_{\star}$+SPC         \\
~~~~$M_V$ (mag)\dotfill             &  \hatcurISOmv{50}      & \hatcurISOmv{51} & \hatcurISOmv{52} & \hatcurISOmv{53} & YY+$a/R_{\star}$+SPC         \\
~~~~$M_K$ (mag,\hatcurjhkfilset{50})\dotfill &  \hatcurISOMK{50} & \hatcurISOMK{51} & \hatcurISOMK{52} & \hatcurISOMK{53} & YY+$a/R_{\star}$+SPC         \\
~~~~Age (Gyr)\dotfill               &  \hatcurISOage{50}     & \hatcurISOage{51} & \hatcurISOage{52} & \hatcurISOage{53} & YY+$a/R_{\star}$+SPC         \\
~~~~$A_{V}$ (mag)\dotfill               &  \hatcurXAv{50}     & \hatcurXAv{51} & \hatcurXAv{52} & \hatcurXAv{53} & YY+$a/R_{\star}$+SPC         \\
~~~~Distance (pc)\dotfill           &  \hatcurXdistred{50}\phn  & \hatcurXdistred{51} & \hatcurXdistred{52} & \hatcurXdistred{53} & YY+$a/R_{\star}$+SPC\\ [-1.5ex]
\enddata
\tablenotetext{a}{
    SPC = ``Stellar Parameter Classification'' routine for the
    analysis of high-resolution spectra \citep{buchhave:2012:spc}, applied
    to the TRES and FIES spectra of \hatcur{50}, and to the Keck/HIRES
    I$_{2}$-free template spectra of \hatcur{51}, \hatcur{52} and
    \hatcur{53}. These parameters rely primarily on SPC, but have a
    small dependence also on the iterative analysis incorporating the
    isochrone search and global modeling of the data, as described in
    the text.
}
\tablenotetext{b}{
    The median of the $\log_{10}R^{\prime}_{\rm HK}$ values measured
    from the individual Keck/HIRES spectra for each target. The
    uncertainty is the standard error on the median.
}
\tablenotetext{c}{
    From APASS DR6 for \hatcur{50}, \hatcur{51} and \hatcur{52} as listed in the UCAC 4 catalog \citep{zacharias:2012:ucac4}. From TASS Mark IV \citep{droege:2006:tass} for \hatcur{53}.
}
\tablenotetext{d}{
    \hatcurisoshort{50}+\hatcurlumind{50}+SPC = Based on the \hatcurisoshort{50}
    isochrones \citep{\hatcurisocite{50}}, \hatcurlumind{50} as a luminosity
    indicator, and the SPC results.
}
\ifthenelse{\boolean{emulateapj}}{
    \end{deluxetable*}
}{
    \end{deluxetable}
}

\subsection{Excluding blend scenarios}
\label{sec:blend}

In order to exclude blend scenarios we carried out an analysis
following \citep{hartman:2012:hat39hat41}. We attempt to fit the
available photometry (light curves, and catalog broad-band magnitudes
calibrated to standard systems) for each object using a combination of
three stars (two eclipsing, with a third diluting the eclipse signal)
with properties taken from stellar evolution models. 

For \hatcur{50}, \hatcur{51} and \hatcur{53} we find that a model
consisting of a planet transiting an isolated star provides a better
(lower $\chi^2$) fit to the data than any of the blend models
tested. For \hatcur{50} the best-fit blend model is excluded with
$1.5\sigma$ confidence, while for both \hatcur{51} and \hatcur{53} it
is excluded with $2\sigma$ confidence. We also simulated
cross-correlation functions, RVs and BS measurements for the blend
models tested, and found that any model that could plausibly fit the
photometry for these systems (i.e.~provides a fit that is no more
than $5\sigma$ worse than the single star+planet model) would be
easily identified as a composite stellar system based on the
spectroscopy.  We therefore conclude that all three of these objects
are transiting planet systems.

\begin{figure}[!ht]
\plotone{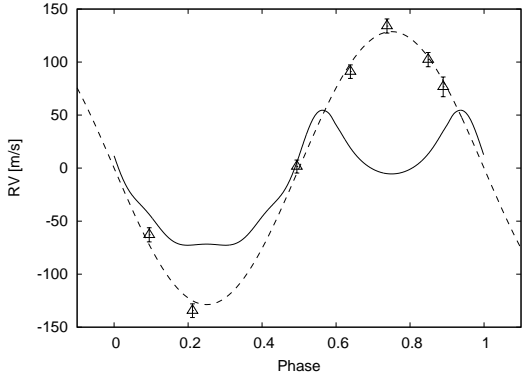}
\caption[]{
    Phase-folded Keck/HIRES RVs for \hatcur{52} compared to the
    best-fit planetary orbit (dashed curve) and the best-fit RVs from
    a blended stellar eclipsing binary model (solid curve). This same
    blend model has BS variations that are below the observed scatter,
    and fits the photometric data, however it does not reproduce the
    form of the RV variation, which is well described by a transiting
    planet. The blend-model fit shown here consists of a
    $0.88$\,\msun\ foreground star blended with a
    $0.84$\,\msun+$0.13$\,\msun\ eclipsing binary that has a distance
    modulus $0.35$\,mag greater than the foreground star, and which
    has a $\gamma$ velocity that is $70$\,\kms\ different from that of
    the foreground star. Reducing the difference in $\gamma$
    velocities creates a more symmetric RV variation, but also results
    in the RV variation going to $0$\,\ms\ at both phases 0.25 and
    0.75. All other blend models simulated have RV variations that
    provide even poorer fits to the observations.
\label{fig:simblendRV}}
\end{figure}

For \hatcur{52} we similarly find that the planet+star model provides
a better fit to the data than any blend model tested, however the
best-fit blend model differs by only $0.5\sigma$ from the planet+star
model. We also find that there is a range of models consisting of a
blend between a bright foreground star, and a background stellar
eclipsing binary that is between $0.35$\,mag and $4$\,mag further in
distance modulus than the foreground star, which cannot be ruled out
based on the photometry or BS spans. For these models the simulated BS
variations have a scatter that is below the 43\,\ms\ scatter in the
Keck/HIRES data, if we allow for a difference in the $\gamma$
velocities of the foreground star and background binary. We find,
however, that similar to the case of HAT-P-49
\citep{bieryla:2014:hat49}, the expected form of the RV variations in
these blends is significantly different from the observed sinusoidal
variation, even though the overall amplitude of the variations is
comparable (Fig.~\ref{fig:simblendRV}). We conclude that although the
photometry and BS measurements for \hatcur{52} can be fit by a blended
stellar eclipsing binary model, the RV observations cannot be.

While we can rule out the possibility that any of these objects is a
blended stellar eclipsing binary system, we cannot rule out the
possibility that one or more of these transiting planet systems also
has a stellar companion.  For \hatcur{50}, we find that models
including a faint companion with $M > 0.6$\,\msun\ provide a slightly
worse fit to the data than models without a companion. The difference
in $\chi^2$ is small, however, and we can only rule out a stellar
companion with $M > 1.2$\,\msun\ at greater than $3\sigma$
confidence. For \hatcur{51}, models with a companion having $M >
0.5$\,\msun have a slightly worse fit to the data, but we can only
rule out companions with $M > 0.95$\,\msun\ at greater than
$3\sigma$ confidence. For \hatcur{52}, companions with $M >
0.35$\,\msun\ provide a slightly worse fit to the data, but all
companions up to the mass of \hatcur{52} are permitted to within
3$\sigma$.  For \hatcur{53} companions with $M > 0.7$\,\msun\ provide
a slightly worse fit, but all companions up to the mass of \hatcur{52}
are permitted to within 3$\sigma$.

\subsection{Global modeling of the data}
\label{sec:globmod}

We modeled the HATNet photometry, the follow-up photometry, and the
high-precision RV measurements using the procedure described in detail
by \citet{pal:2008:hat7} and \citet{bakos:2010:hat11} with
modifications described by \citet{hartman:2012:hat39hat41}. This
procedure makes use of the differential evolution Markov Chain Monte
Carlo (DEMCMC) method \citep{terbraak:2006,eastman:2013} to explore
the fitness landscape and produce posterior parameter
distributions. We allowed for RV jitter which we varied as a free
parameter in the fit for each planet. We adopted independent jitters
for each instrument as the methods for estimating the ``formal''
errors differ between reduction methods and instruments. For HIRES we
made use of an empirical prior on the jitter as discussed in
\citet{hartman:2014:hat44hat46}, while for the other instruments we
used a Jeffreys prior (i.e.~the prior probability for parameter
$\sigma$ is $\propto 1/\sigma$). We fixed the limb darkening
coefficients using the tabulation by \citet{claret:2004} and the
stellar atmospheric parameters given in \reftabl{stellar}.

The resulting parameters for each system are listed in
\reftabl{planetparam}. We find that \hatcurb{50} is a hot Jupiter with
a mass of \hatcurPPmlong{50}\,\mjup\ and radius of
\hatcurPPrlong{50}\,\rjup, \hatcurb{51} is a hot Saturn with a mass of
\hatcurPPmlong{51}\,\mjup\ and radius of \hatcurPPrlong{51}\,\rjup,
while \hatcurb{52} and \hatcurb{53} are hot Jupiters with masses of
\hatcurPPmlong{52}\,\mjup\ and \hatcurPPmlong{53}\,\mjup, and radii of
\hatcurPPrlong{52}\,\rjup\ and \hatcurPPrlong{53}\,\rjup,
respectively. We fit all systems both allowing the eccentricity to
vary and fixing it to zero. We find that all four systems are
consistent with no eccentricity (the 95\% confidence upper limits on
the eccentricity when it is allowed to vary are $e
\hatcurRVeccentwosiglimeccen{50}$, $\hatcurRVeccentwosiglimeccen{51}$,
$\hatcurRVeccentwosiglimeccen{52}$, and
$\hatcurRVeccentwosiglimeccen{53}$ for \hatcurb{50} through
\hatcurb{53}, respectively). We therefore adopted the parameters for a
fixed circular orbit in all cases.

%
\ifthenelse{\boolean{emulateapj}}{
    \begin{deluxetable*}{lcccc}
}{
    \begin{deluxetable}{lcccc}
}
\tabletypesize{\scriptsize}
\tablecaption{Orbital and planetary parameters for \hatcurb{50}--\hatcurb{53}\label{tab:planetparam}}
\tablehead{
    \multicolumn{1}{c}{} &
    \multicolumn{1}{c}{\bf HAT-P-50b} &
    \multicolumn{1}{c}{\bf HAT-P-51b} &
    \multicolumn{1}{c}{\bf HAT-P-52b} &
    \multicolumn{1}{c}{\bf HAT-P-53b} \\
    \multicolumn{1}{c}{~~~~~~~~~~~~~~~Parameter~~~~~~~~~~~~~~~} &
    \multicolumn{1}{c}{Value} &
    \multicolumn{1}{c}{Value} &
    \multicolumn{1}{c}{Value} &
    \multicolumn{1}{c}{Value}
}
\startdata
\noalign{\vskip -3pt}
\sidehead{\Lc{} parameters}
~~~$P$ (days)             \dotfill    & $\hatcurLCP{50}$ & $\hatcurLCP{51}$ & $\hatcurLCP{52}$ & $\hatcurLCP{53}$              \\
~~~$T_c$ (${\rm BJD}$)    
      \tablenotemark{a}   \dotfill    & $\hatcurLCT{50}$ & $\hatcurLCT{51}$ & $\hatcurLCT{52}$ & $\hatcurLCT{53}$              \\
~~~$T_{14}$ (days)
      \tablenotemark{a}   \dotfill    & $\hatcurLCdur{50}$ & $\hatcurLCdur{51}$ & $\hatcurLCdur{52}$ & $\hatcurLCdur{53}$            \\
~~~$T_{12} = T_{34}$ (days)
      \tablenotemark{a}   \dotfill    & $\hatcurLCingdur{50}$ & $\hatcurLCingdur{51}$ & $\hatcurLCingdur{52}$ & $\hatcurLCingdur{53}$         \\
~~~$\arstar$              \dotfill    & $\hatcurPPar{50}$ & $\hatcurPPar{51}$ & $\hatcurPPar{52}$ & $\hatcurPPar{53}$             \\
~~~$\zrstar$ \tablenotemark{b}             \dotfill    & $\hatcurLCzeta{50}$\phn & $\hatcurLCzeta{51}$\phn & $\hatcurLCzeta{52}$\phn & $\hatcurLCzeta{53}$\phn       \\
~~~$\rpl/\rstar$          \dotfill    & $\hatcurLCrprstar{50}$ & $\hatcurLCrprstar{51}$ & $\hatcurLCrprstar{52}$ & $\hatcurLCrprstar{53}$        \\
~~~$b^2$                  \dotfill    & $\hatcurLCbsq{50}$ & $\hatcurLCbsq{51}$ & $\hatcurLCbsq{52}$ & $\hatcurLCbsq{53}$            \\
~~~$b \equiv a \cos i/\rstar$
                          \dotfill    & $\hatcurLCimp{50}$ & $\hatcurLCimp{51}$ & $\hatcurLCimp{52}$ & $\hatcurLCimp{53}$            \\
~~~$i$ (deg)              \dotfill    & $\hatcurPPi{50}$\phn & $\hatcurPPi{51}$\phn & $\hatcurPPi{52}$\phn & $\hatcurPPi{53}$\phn          \\

\sidehead{Limb-darkening coefficients \tablenotemark{c}}
~~~$c_1,i$ (linear term)  \dotfill    & $\hatcurLBii{50}$ & $\hatcurLBii{51}$ & $\hatcurLBii{52}$ & $\hatcurLBii{53}$             \\
~~~$c_2,i$ (quadratic term) \dotfill  & $\hatcurLBiii{50}$ & $\hatcurLBiii{51}$ & $\hatcurLBiii{52}$ & $\hatcurLBiii{53}$            \\

\sidehead{RV parameters}
~~~$K$ (\ms)              \dotfill    & $\hatcurRVK{50}$\phn\phn & $\hatcurRVK{51}$\phn\phn & $\hatcurRVK{52}$\phn\phn & $\hatcurRVK{53}$\phn\phn      \\
~~~$e$ \tablenotemark{d}               \dotfill    & $\hatcurRVeccentwosiglimeccen{50}$ & $\hatcurRVeccentwosiglimeccen{51}$ & $\hatcurRVeccentwosiglimeccen{52}$ & $\hatcurRVeccentwosiglimeccen{53}$          \\
~~~RV jitter HIRES (\ms) \tablenotemark{e}       \dotfill    & $\cdots$ & \hatcurRVjitterA{51} & \hatcurRVjitter{52} & \hatcurRVjitter{53}           \\
~~~RV jitter HDS (\ms)        \dotfill    & \hatcurRVjitterC{50} & \hatcurRVjitterB{51} & $\cdots$ & $\cdots$           \\
~~~RV jitter TRES (\ms)        \dotfill    & \hatcurRVjitterA{50} & $\cdots$ & $\cdots$ & $\cdots$           \\
~~~RV jitter FIES (\ms)        \dotfill    & \hatcurRVjitterB{50} & $\cdots$ & $\cdots$ & $\cdots$           \\

\sidehead{Secondary eclipse parameters}
~~~$T_s$ (BJD)            \dotfill    & $\hatcurXsecondary{50}$ & $\hatcurXsecondary{51}$ & $\hatcurXsecondary{52}$ & $\hatcurXsecondary{53}$       \\
~~~$T_{s,14}$ (days)             \dotfill   & $\hatcurXsecdur{50}$ & $\hatcurXsecdur{51}$ & $\hatcurXsecdur{52}$ & $\hatcurXsecdur{53}$          \\
~~~$T_{s,12}$ (days)             \dotfill   & $\hatcurXsecingdur{50}$ & $\hatcurXsecingdur{51}$ & $\hatcurXsecingdur{52}$ & $\hatcurXsecingdur{53}$       \\

\sidehead{Planetary parameters}
~~~$\mpl$ ($\mjup$)       \dotfill    & $\hatcurPPmlong{50}$ & $\hatcurPPmlong{51}$ & $\hatcurPPmlong{52}$ & $\hatcurPPmlong{53}$          \\
~~~$\rpl$ ($\rjup$)       \dotfill    & $\hatcurPPrlong{50}$ & $\hatcurPPrlong{51}$ & $\hatcurPPrlong{52}$ & $\hatcurPPrlong{53}$          \\
~~~$C(\mpl,\rpl)$
    \tablenotemark{f}     \dotfill    & $\hatcurPPmrcorr{50}$ & $\hatcurPPmrcorr{51}$ & $\hatcurPPmrcorr{52}$ & $\hatcurPPmrcorr{53}$         \\
~~~$\rhopl$ (\gcmc)       \dotfill    & $\hatcurPPrho{50}$ & $\hatcurPPrho{51}$ & $\hatcurPPrho{52}$ & $\hatcurPPrho{53}$            \\
~~~$\log g_p$ (cgs)       \dotfill    & $\hatcurPPlogg{50}$ & $\hatcurPPlogg{51}$ & $\hatcurPPlogg{52}$ & $\hatcurPPlogg{53}$           \\
~~~$a$ (AU)               \dotfill    & $\hatcurPParel{50}$ & $\hatcurPParel{51}$ & $\hatcurPParel{52}$ & $\hatcurPParel{53}$           \\
~~~$T_{\rm eq}$ (K)        \dotfill   & $\hatcurPPteff{50}$ & $\hatcurPPteff{51}$ & $\hatcurPPteff{52}$ & $\hatcurPPteff{53}$           \\
~~~$\Theta$ \tablenotemark{g} \dotfill & $\hatcurPPtheta{50}$ & $\hatcurPPtheta{51}$ & $\hatcurPPtheta{52}$ & $\hatcurPPtheta{53}$          \\
~~~$\log_{10}\langle F \rangle$ (cgs) \tablenotemark{h}
                          \dotfill    & $\hatcurPPfluxavglog{50}$ & $\hatcurPPfluxavglog{51}$ & $\hatcurPPfluxavglog{52}$ & $\hatcurPPfluxavglog{53}$        \\ [-1.5ex]
\enddata
\tablenotetext{a}{
    Times are in Barycentric Julian Date calculated directly from UTC {\em without} correction for leap seconds.
    \ensuremath{T_c}: Reference epoch of
    mid transit that minimizes the correlation with the orbital
    period.
    \ensuremath{T_{14}}: total transit duration, time
    between first to last contact;
    \ensuremath{T_{12}=T_{34}}: ingress/egress time, time between first
    and second, or third and fourth contact.
}
\tablenotetext{b}{
   Reciprocal of the half duration of the transit used as a jump parameter in our MCMC analysis in place of $\arstar$. It is related to $\arstar$ by the expression $\zrstar = \arstar(2\pi(1+e\sin\omega))/(P\sqrt{1-b^2}\sqrt{1-e^2})$ \citep{bakos:2010:hat11}.
}
\tablenotetext{c}{
    Values for a quadratic law, adopted from the tabulations by
    \cite{claret:2004} according to the spectroscopic (SPC) parameters
    listed in \reftabl{stellar}.
}
\tablenotetext{d}{
    As discussed in \refsecl{globmod} the adopted parameters for all
    four systems are determined assuming circular orbits. We also list
    the 95\% confidence upper limit on the eccentricity determined
    when $\sqrt{e}\cos\omega$ and $\sqrt{e}\sin\omega$ are allowed to
    vary in the fit.
}
\tablenotetext{e}{
    Term added in quadrature to the formal RV uncertainties for each
    instrument. This is treated as a free parameter in the fitting
    routine. For HIRES we include an empirical prior constraint
    following \citet{hartman:2014:hat44hat46}.
}
\tablenotetext{f}{
    Correlation coefficient between the planetary mass \mpl\ and radius
    \rpl.
}
\tablenotetext{g}{
    The Safronov number is given by $\Theta = \frac{1}{2}(V_{\rm
    esc}/V_{\rm orb})^2 = (a/\rpl)(\mpl / \mstar )$
    \citep[see][]{hansen:2007}.
}
\tablenotetext{h}{
    Incoming flux per unit surface area, averaged over the orbit.
}
\ifthenelse{\boolean{emulateapj}}{
    \end{deluxetable*}
}{
    \end{deluxetable}
}



\section{Discussion}
\label{sec:discussion}

In this paper we presented the discovery and characterization of four
transiting exoplanets from the HATNet survey, including three hot
Jupiters (\hatcurb{50}, \hatcurb{52} and \hatcurb{53}) and a hot
Saturn (\hatcurb{51}). All four planets have masses and radii
determined to better than 10\% precision. The mass uncertainties are
5.4\%, 5.8\%, 3.5\%, and 3.8\% for \hatcurb{50} through \hatcurb{53},
respectively, while the respective radius uncertainties are 5.0\%,
4.2\%, 7.1\%, and 6.9\%. The stars \hatcur{50}, -51, and -53 also have
fairly precise isochrone-based age determinations (uncertainty less
than 2\,Gyr) thanks to their favorable position within the $T_{\rm
  eff}$--\rhostar\ plane (Fig.~\ref{fig:iso}).

\ifthenelse{\boolean{emulateapj}}{
    \begin{figure*}[!ht]
}{
    \begin{figure}[!ht]
}
\epsscale{1.2}
\plotone{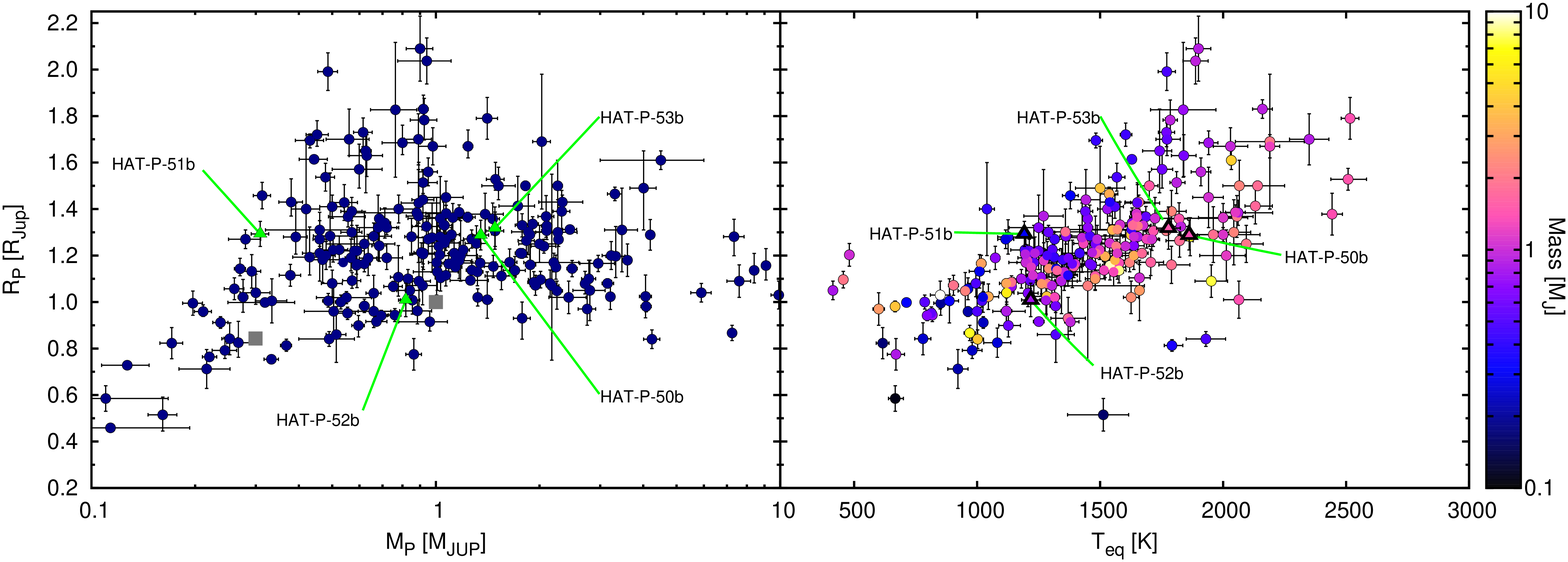}
\epsscale{1.0}
\caption{ Left: Mass--radius diagram for transiting planets with
  $0.1\,\mjup < M < 10\,\mjup$. The four planets discovered here are
  indicated. The two filled gray squares show Saturn and Jupiter. The
  parameters for other transiting planets are compiled from the
  literature (c.f.~\url{http://www.exoplanets.org}). Right: planet
  radius vs.~estimated equilibrium temperature (assuming zero albedo
  and complete redistribution of heat) for the same sample of planets
  as shown at left. Symbols are assigned colors based on the planetary
  masses.
}
\label{fig:planetmr}
\ifthenelse{\boolean{emulateapj}}{
    \end{figure*}
}{
    \end{figure}
}

In \reffigl{planetmr} we show the location of these planets on a
mass--radius diagram, comparing them to the full sample of confirmed
TEPs with $0.1 < \mpl < 10$\,\mjup. The new planets all fall within
the range of values already seen by other planets, with \hatcurb{51}
falling near the upper envelope of the distribution of points in the
mass--radius diagram, and \hatcurb{52} falling near the lower
envelope. We also show the location of each planet on a $T_{\rm
  eq}$--radius diagram. Again we find that the planets all follow the
well-established trends. While not atypical compared to other known
exoplanets, these objects contribute to the growing sample of
well-characterized planets which may be used to explore the population
of planets in the Galaxy through statistical methods.

In terms of potential for additional follow-up observations, we
conclude that it should be feasible to measure the Rossiter-McLaughlin
effect for \hatcurb{50}, \hatcurb{51} and \hatcurb{53} using
Subaru/HDS or Keck/HIRES. For \hatcurb{50} the expected amplitude of
the R-M effect is $42$\,\ms\ for an aligned orbit \citep[using eq.~40
  in][]{winn:2010:teps}. For \hatcurb{51} the expected amplitude is
$27$\,\ms, and for \hatcurb{53} it is $48$\,\ms. The Subaru/HDS
velocity residuals for \hatcur{50} have an RMS of 23\,\ms, with a
median exposure time of 10\,minutes. Assuming $20$ such exposures are
obtained over the course of a single transit, it should be possible to
measure the R-M amplitude to a precision of $5\sigma$ (based on
fitting models to simulated observations). For \hatcur{51}, the
Keck/HIRES RVs have a residual RMS of $5.4$\,\ms, and a median
exposure time of $25$\,minutes. Seven of these exposures could be
collected over a single transit, allowing a $9\sigma$ detection of the
R-M amplitude. For \hatcur{53}, the Keck/HIRES RVs have a residual of
$11$\,\ms\ and an exposure time of 25\,minutes. For this system it
should be possible to collect 6 similar exposures during a transit,
and measure the R-M amplitude with $8\sigma$ confidence. For
\hatcurb{52} the R-M amplitude is only $\la 7$\,\ms\ (limited by the
very slow rotation), and we would not expect to detect it in a single
transit with better than 2$\sigma$ confidence.

The conclusion that the R-M effect should be easier to detect for both
\hatcur{51} and \hatcur{53} than for \hatcur{50}, despite both stars
being significantly fainter than \hatcur{50}, and despite both stars
having a lower $\vsini$, may be counter intuitive. The RV observations
for \hatcur{51} and \hatcur{53} are both significantly higher
precision than those for \hatcur{50}, even though the \hatcur{50}
observations have higher S/N. Some of the difference may be due to the
different instruments (Subaru/HDS for \hatcur{50} vs.~Keck/HIRES for
\hatcur{51} and \hatcur{53}). However, slower rotation and cooler
surface temperatures are also factors which tend to improve the RV
precision. In this respect we expect \hatcur{51} to have higher
precision than \hatcur{53} at fixed S/N, and \hatcur{53} to have
higher precision than \hatcur{50} at fixed S/N, which is what
we see.

Measuring the R-M effect for \hatcurb{51} may be of particular
interest due to its small mass. HAT-P-11b and Kepler-63b are the only
planets smaller than \hatcurb{51} for which this effect has been
measured to date (\citealp{winn:2010:hat11} and
\citealp{sanchisojeda:2013}; the obliquity has also been measured for
the Kepler-30 system by star-spot crossings, see
\citealp{sanchisojeda:2012}).

While the R-M effect should be detectable for \hatcurb{50}, \hatcurb{51}
and \hatcurb{53}, due to the relatively small value of $\rpl/\rstar$
for \hatcurb{50}, and the faintness of the other targets, none of the
new planets are particularly well-suited for atmospheric
characterization.



\acknowledgements 

HATNet operations have been funded by NASA grants NNG04GN74G,
NNX08AF23G, and NNX13AJ15G. Follow-up of HATNet targets has been
partially supported through NSF grant AST-1108686. G.\'A.B, Z.C. and
K.P. acknowledge partial support from NASA grant
NNX09AB29G. J.H. acknowledges support from NASA grant
NNX14AF87G. K.P. acknowledges support from NASA grant
NNX13AQ62G. G.T. acknowledges partial support from NASA grant
NNX14AB83G. We acknowledge partial support also from the Kepler
Mission under NASA Cooperative Agreement NCC2-1390 (D.W.L., PI). Data
presented in this paper are based on observations obtained at the HAT
station at the Submillimeter Array of SAO, and the HAT station at the
Fred Lawrence Whipple Observatory of SAO. The authors wish to
acknowledge the very significant cultural role and reverence that the
summit of Mauna Kea has always had within the indigenous Hawaiian
community. We are most fortunate to have the opportunity to conduct
observations from this mountain. This research has made use of Keck
telescope time granted through NOAO (program A245Hr) and NASA (N154Hr,
N130Hr).
This research was made possible through the use of the AAVSO
Photometric All-Sky Survey (APASS), funded by the Robert Martin Ayers
Sciences Fund.


\clearpage

\begin{thebibliography}{55}
\expandafter\ifx\csname natexlab\endcsname\relax\def\natexlab#1{#1}\fi

\bibitem[{{Alonso} {et~al.}(2004){Alonso}, {Brown}, {Torres}, {Latham},
  {Sozzetti}, {Mandushev}, {Belmonte}, {Charbonneau}, {Deeg}, {Dunham},
  {O'Donovan}, \& {Stefanik}}]{alonso:2004}
{Alonso}, R., {Brown}, T.~M., {Torres}, G., {et~al.} 2004, \apjl, 613, L153

\bibitem[{{Bakos} {et~al.}(2004){Bakos}, {Noyes}, {Kov{\'a}cs}, {Stanek},
  {Sasselov}, \& {Domsa}}]{bakos:2004:hatnet}
{Bakos}, G., {Noyes}, R.~W., {Kov{\'a}cs}, G., {et~al.} 2004, \pasp, 116, 266

\bibitem[{{Bakos} {et~al.}(2010){Bakos}, {Torres}, {P{\'a}l}, {Hartman},
  {Kov{\'a}cs}, {Noyes}, {Latham}, {Sasselov}, {Sip{\H o}cz}, {Esquerdo},
  {Fischer}, {Johnson}, {Marcy}, {Butler}, {Isaacson}, {Howard}, {Vogt},
  {Kov{\'a}cs}, {Fernandez}, {Mo{\'o}r}, {Stefanik}, {L{\'a}z{\'a}r}, {Papp},
  \& {S{\'a}ri}}]{bakos:2010:hat11}
{Bakos}, G.~{\'A}., {Torres}, G., {P{\'a}l}, A., {et~al.} 2010, \apj, 710, 1724

\bibitem[{{Bakos} {et~al.}(2013){Bakos}, {Csubry}, {Penev}, {Bayliss},
  {Jord{\'a}n}, {Afonso}, {Hartman}, {Henning}, {Kov{\'a}cs}, {Noyes},
  {B{\'e}ky}, {Suc}, {Cs{\'a}k}, {Rabus}, {L{\'a}z{\'a}r}, {Papp}, {S{\'a}ri},
  {Conroy}, {Zhou}, {Sackett}, {Schmidt}, {Mancini}, {Sasselov}, \&
  {Ueltzhoeffer}}]{bakos:2013:hatsouth}
{Bakos}, G.~{\'A}., {Csubry}, Z., {Penev}, K., {et~al.} 2013, \pasp, 125, 154

\bibitem[{{B{\'e}ky} {et~al.}(2011){B{\'e}ky}, {Bakos}, {Hartman}, {Torres},
  {Latham}, {Jord{\'a}n}, {Arriagada}, {Bayliss}, {Kiss}, {Kov{\'a}cs},
  {Quinn}, {Marcy}, {Howard}, {Fischer}, {Johnson}, {Esquerdo}, {Noyes},
  {Buchhave}, {Sasselov}, {Stefanik}, {Perumpilly}, {L{\'a}z{\'a}r}, {Papp}, \&
  {S{\'a}ri}}]{beky:2011:hat27}
{B{\'e}ky}, B., {Bakos}, G.~{\'A}., {Hartman}, J., {et~al.} 2011, \apj, 734,
  109

\bibitem[{{Bieryla} {et~al.}(2014){Bieryla}, {Hartman}, {Bakos}, {Bhatti},
  {Kov{\'a}cs}, {Boisse}, {Latham}, {Buchhave}, {Csubry}, {Penev}, {de
  Val-Borro}, {B{\'e}ky}, {Falco}, {Torres}, {Noyes}, {Berlind}, {Calkins},
  {Esquerdo}, {L{\'a}z{\'a}r}, {Papp}, \& {S{\'a}ri}}]{bieryla:2014:hat49}
{Bieryla}, A., {Hartman}, J.~D., {Bakos}, G.~{\'A}., {et~al.} 2014, \aj, 147,
  84

\bibitem[{{Boisse} {et~al.}(2013){Boisse}, {Hartman}, {Bakos}, {Penev},
  {Csubry}, {B{\'e}ky}, {Latham}, {Bieryla}, {Torres}, {Kov{\'a}cs},
  {Buchhave}, {Hansen}, {Everett}, {Esquerdo}, {Szklen{\'a}r}, {Falco},
  {Shporer}, {Fulton}, {Noyes}, {Stefanik}, {L{\'a}z{\'a}r}, {Papp}, \&
  {S{\'a}ri}}]{boisse:2013:hat42hat43}
{Boisse}, I., {Hartman}, J.~D., {Bakos}, G.~{\'A}., {et~al.} 2013, \aap, 558,
  A86

\bibitem[{{Bouchy} {et~al.}(2009){Bouchy}, {H{\'e}brard}, {Udry}, {Delfosse},
  {Boisse}, {Desort}, {Bonfils}, {Eggenberger}, {Ehrenreich}, {Forveille},
  {Lagrange}, {Le Coroller}, {Lovis}, {Moutou}, {Pepe}, {Perrier}, {Pont},
  {Queloz}, {Santos}, {S{\'e}gransan}, \& {Vidal-Madjar}}]{bouchy:2009}
{Bouchy}, F., {H{\'e}brard}, G., {Udry}, S., {et~al.} 2009, \aap, 505, 853

\bibitem[{{Brown} {et~al.}(2013){Brown}, {Baliber}, {Bianco}, {Bowman},
  {Burleson}, {Conway}, {Crellin}, {Depagne}, {De Vera}, {Dilday}, {Dragomir},
  {Dubberley}, {Eastman}, {Elphick}, {Falarski}, {Foale}, {Ford}, {Fulton},
  {Garza}, {Gomez}, {Graham}, {Greene}, {Haldeman}, {Hawkins}, {Haworth},
  {Haynes}, {Hidas}, {Hjelstrom}, {Howell}, {Hygelund}, {Lister}, {Lobdill},
  {Martinez}, {Mullins}, {Norbury}, {Parrent}, {Paulson}, {Petry}, {Pickles},
  {Posner}, {Rosing}, {Ross}, {Sand}, {Saunders}, {Shobbrook}, {Shporer},
  {Street}, {Thomas}, {Tsapras}, {Tufts}, {Valenti}, {Vander Horst}, {Walker},
  {White}, \& {Willis}}]{brown:2013}
{Brown}, T.~M., {Baliber}, N., {Bianco}, F.~B., {et~al.} 2013, \pasp, 125, 1031

\bibitem[{{Buchhave} {et~al.}(2010){Buchhave}, {Bakos}, {Hartman}, {Torres},
  {Kov{\'a}cs}, {Latham}, {Noyes}, {Esquerdo}, {Everett}, {Howard}, {Marcy},
  {Fischer}, {Johnson}, {Andersen}, {F{\H u}r{\'e}sz}, {Perumpilly},
  {Sasselov}, {Stefanik}, {B{\'e}ky}, {L{\'a}z{\'a}r}, {Papp}, \&
  {S{\'a}ri}}]{buchhave:2010:hat16}
{Buchhave}, L.~A., {Bakos}, G.~{\'A}., {Hartman}, J.~D., {et~al.} 2010, \apj,
  720, 1118

\bibitem[{{Buchhave} {et~al.}(2012){Buchhave}, {Latham}, {Johansen},
  {Bizzarro}, {Torres}, {Rowe}, {Batalha}, {Borucki}, {Brugamyer}, {Caldwell},
  {Bryson}, {Ciardi}, {Cochran}, {Endl}, {Esquerdo}, {Ford}, {Geary},
  {Gilliland}, {Hansen}, {Isaacson}, {Laird}, {Lucas}, {Marcy}, {Morse},
  {Robertson}, {Shporer}, {Stefanik}, {Still}, \& {Quinn}}]{buchhave:2012:spc}
{Buchhave}, L.~A., {Latham}, D.~W., {Johansen}, A., {et~al.} 2012, \nat, 486,
  375

\bibitem[{{Burrows} {et~al.}(2007){Burrows}, {Hubeny}, {Budaj}, \&
  {Hubbard}}]{burrows:2007}
{Burrows}, A., {Hubeny}, I., {Budaj}, J., \& {Hubbard}, W.~B. 2007, \apj, 661,
  502

\bibitem[{{Butler} {et~al.}(1996){Butler}, {Marcy}, {Williams}, {McCarthy},
  {Dosanjh}, \& {Vogt}}]{butler:1996}
{Butler}, R.~P., {Marcy}, G.~W., {Williams}, E., {et~al.} 1996, \pasp, 108, 500

\bibitem[{{Cardelli} {et~al.}(1989){Cardelli}, {Clayton}, \&
  {Mathis}}]{cardelli:1989}
{Cardelli}, J.~A., {Clayton}, G.~C., \& {Mathis}, J.~S. 1989, \apj, 345, 245

\bibitem[{{Charbonneau} {et~al.}(2002){Charbonneau}, {Brown}, {Noyes}, \&
  {Gilliland}}]{charbonneau:2002}
{Charbonneau}, D., {Brown}, T.~M., {Noyes}, R.~W., \& {Gilliland}, R.~L. 2002,
  \apj, 568, 377

\bibitem[{{Claret}(2004)}]{claret:2004}
{Claret}, A. 2004, \aap, 428, 1001

\bibitem[{{Dawson} \& {Johnson}(2012)}]{dawson:2012}
{Dawson}, R.~I., \& {Johnson}, J.~A. 2012, \apj, 756, 122

\bibitem[{{Deeming}(1975)}]{deeming:1975}
{Deeming}, T.~J. 1975, \apss, 36, 137

\bibitem[{{Djupvik} \& {Andersen}(2010)}]{djupvik:2010}
{Djupvik}, A.~A., \& {Andersen}, J. 2010, in Highlights of Spanish Astrophysics
  V, ed. {J.~M.~Diego, L.~J.~Goicoechea, J.~I.~Gonz{\'a}lez-Serrano, \&
  J.~Gorgas}, 211

\bibitem[{{Droege} {et~al.}(2006){Droege}, {Richmond}, {Sallman}, \&
  {Creager}}]{droege:2006:tass}
{Droege}, T.~F., {Richmond}, M.~W., {Sallman}, M.~P., \& {Creager}, R.~P. 2006,
  \pasp, 118, 1666

\bibitem[{{Eastman} {et~al.}(2013){Eastman}, {Gaudi}, \& {Agol}}]{eastman:2013}
{Eastman}, J., {Gaudi}, B.~S., \& {Agol}, E. 2013, \pasp, 125, 83

\bibitem[{{Enoch} {et~al.}(2012){Enoch}, {Collier Cameron}, \&
  {Horne}}]{enoch:2012}
{Enoch}, B., {Collier Cameron}, A., \& {Horne}, K. 2012, \aap, 540, A99

\bibitem[{{F\H{u}resz}(2008)}]{furesz:2008}
{F\H{u}resz}, G. 2008, PhD thesis, {Univ. of Szeged, Hungary}

\bibitem[{{Guillot} {et~al.}(2006){Guillot}, {Santos}, {Pont}, {Iro}, {Melo},
  \& {Ribas}}]{guillot:2006}
{Guillot}, T., {Santos}, N.~C., {Pont}, F., {et~al.} 2006, \aap, 453, L21

\bibitem[{{Hansen} \& {Barman}(2007)}]{hansen:2007}
{Hansen}, B.~M.~S., \& {Barman}, T. 2007, \apj, 671, 861

\bibitem[{{Hartman} {et~al.}(2012){Hartman}, {Bakos}, {B{\'e}ky}, {Torres},
  {Latham}, {Csubry}, {Penev}, {Shporer}, {Fulton}, {Buchhave}, {Johnson},
  {Howard}, {Marcy}, {Fischer}, {Kov{\'a}cs}, {Noyes}, {Esquerdo}, {Everett},
  {Szklen{\'a}r}, {Quinn}, {Bieryla}, {Knox}, {Hinz}, {Sasselov}, {F{\H
  u}r{\'e}sz}, {Stefanik}, {L{\'a}z{\'a}r}, {Papp}, \&
  {S{\'a}ri}}]{hartman:2012:hat39hat41}
{Hartman}, J.~D., {Bakos}, G.~{\'A}., {B{\'e}ky}, B., {et~al.} 2012, \aj, 144,
  139

\bibitem[{{Hartman} {et~al.}(2014){Hartman}, {Bakos}, {Torres}, {Kov{\'a}cs},
  {Johnson}, {Howard}, {Marcy}, {Latham}, {Bieryla}, {Buchhave}, {Bhatti},
  {B{\'e}ky}, {Csubry}, {Penev}, {de Val-Borro}, {Noyes}, {Fischer},
  {Esquerdo}, {Everett}, {Szklen{\'a}r}, {Zhou}, {Bayliss}, {Shporer},
  {Fulton}, {Sanchis-Ojeda}, {Falco}, {L{\'a}z{\'a}r}, {Papp}, \&
  {S{\'a}ri}}]{hartman:2014:hat44hat46}
{Hartman}, J.~D., {Bakos}, G.~{\'A}., {Torres}, G., {et~al.} 2014, \aj, 147,
  128

\bibitem[{{Isaacson} \& {Fischer}(2010)}]{isaacson:2010}
{Isaacson}, H., \& {Fischer}, D. 2010, \apj, 725, 875

\bibitem[{{Kambe} {et~al.}(2002){Kambe}, {Sato}, {Takeda}, {Ando}, {Noguchi},
  {Aoki}, {Izumiura}, {Wada}, {Masuda}, {Okada}, {Shimizu}, {Watanabe},
  {Yoshida}, {Honda}, \& {Kawanomoto}}]{kambe:2002}
{Kambe}, E., {Sato}, B., {Takeda}, Y., {et~al.} 2002, \pasj, 54, 865

\bibitem[{{Kov{\'a}cs} {et~al.}(2005){Kov{\'a}cs}, {Bakos}, \&
  {Noyes}}]{kovacs:2005:TFA}
{Kov{\'a}cs}, G., {Bakos}, G., \& {Noyes}, R.~W. 2005, \mnras, 356, 557

\bibitem[{{Kov{\'a}cs} {et~al.}(2002){Kov{\'a}cs}, {Zucker}, \&
  {Mazeh}}]{kovacs:2002:BLS}
{Kov{\'a}cs}, G., {Zucker}, S., \& {Mazeh}, T. 2002, \aap, 391, 369

\bibitem[{{Kurtz}(1985)}]{kurtz:1985}
{Kurtz}, D.~W. 1985, \mnras, 213, 773

\bibitem[{{Latham} {et~al.}(2009){Latham}, {Bakos}, {Torres}, {Stefanik},
  {Noyes}, {Kov{\'a}cs}, {P{\'a}l}, {Marcy}, {Fischer}, {Butler}, {Sip{\H
  o}cz}, {Sasselov}, {Esquerdo}, {Vogt}, {Hartman}, {Kov{\'a}cs},
  {L{\'a}z{\'a}r}, {Papp}, \& {S{\'a}ri}}]{latham:2009:hat8}
{Latham}, D.~W., {Bakos}, G.~{\'A}., {Torres}, G., {et~al.} 2009, \apj, 704,
  1107

\bibitem[{{Laughlin} {et~al.}(2011){Laughlin}, {Crismani}, \&
  {Adams}}]{laughlin:2011}
{Laughlin}, G., {Crismani}, M., \& {Adams}, F.~C. 2011, \apjl, 729, L7

\bibitem[{{McCullough} {et~al.}(2005){McCullough}, {Stys}, {Valenti},
  {Fleming}, {Janes}, \& {Heasley}}]{mccullough:2005}
{McCullough}, P.~R., {Stys}, J.~E., {Valenti}, J.~A., {et~al.} 2005, \pasp,
  117, 783

\bibitem[{{Mullally} {et~al.}(2015){Mullally}, {Coughlin}, {Thompson}, {Rowe},
  {Burke}, {Latham}, {Batalha}, {Bryson}, {Christiansen}, {Henze}, {Ofir},
  {Quarles}, {Shporer}, {Van Eylen}, {Van Laerhoven}, {Shah}, {Wolfgang},
  {Chaplin}, {Xie}, {Akeson}, {Argabright}, {Bachtell}, {Borucki}, {Caldwell},
  {Campbell}, {Catanzarite}, {Cochran}, {Duren}, {Fleming}, {Fraquelli},
  {Girouard}, {Haas}, {He{\l}miniak}, {Howell}, {Huber}, {Larson}, {Gautier},
  {Jenkins}, {Li}, {Lissauer}, {McArthur}, {Miller}, {Morris}, {Patil-Sabale},
  {Plavchan}, {Putnam}, {Quintana}, {Ramirez}, {Silva Aguirre}, {Seader},
  {Smith}, {Steffen}, {Stewart}, {Stober}, {Still}, {Tenenbaum}, {Troeltzsch},
  {Twicken}, \& {Zamudio}}]{mullally:2015}
{Mullally}, F., {Coughlin}, J.~L., {Thompson}, S.~E., {et~al.} 2015, ArXiv
  e-prints, 1502.02038

\bibitem[{{Noguchi} {et~al.}(2002){Noguchi}, {Aoki}, {Kawanomoto}, {Ando},
  {Honda}, {Izumiura}, {Kambe}, {Okita}, {Sadakane}, {Sato}, {Tajitsu},
  {Takada-Hidai}, {Tanaka}, {Watanabe}, \& {Yoshida}}]{noguchi:2002}
{Noguchi}, K., {Aoki}, W., {Kawanomoto}, S., {et~al.} 2002, \pasj, 54, 855

\bibitem[{{Noyes} {et~al.}(1984){Noyes}, {Hartmann}, {Baliunas}, {Duncan}, \&
  {Vaughan}}]{noyes:1984}
{Noyes}, R.~W., {Hartmann}, L.~W., {Baliunas}, S.~L., {Duncan}, D.~K., \&
  {Vaughan}, A.~H. 1984, \apj, 279, 763

\bibitem[{{P{\'a}l}(2009)}]{pal:2009:thesis}
{P{\'a}l}, A. 2009, PhD thesis, Department of Astronomy, E{\"o}tv{\"o}s
  Lor{\'a}nd University

\bibitem[{{P{\'a}l} {et~al.}(2008){P{\'a}l}, {Bakos}, {Torres}, {Noyes},
  {Latham}, {Kov{\'a}cs}, {Marcy}, {Fischer}, {Butler}, {Sasselov}, {Sip{\H
  o}cz}, {Esquerdo}, {Kov{\'a}cs}, {Stefanik}, {L{\'a}z{\'a}r}, {Papp}, \&
  {S{\'a}ri}}]{pal:2008:hat7}
{P{\'a}l}, A., {Bakos}, G.~{\'A}., {Torres}, G., {et~al.} 2008, \apj, 680, 1450

\bibitem[{{Pepper} {et~al.}(2007){Pepper}, {Pogge}, {DePoy}, {Marshall},
  {Stanek}, {Stutz}, {Poindexter}, {Siverd}, {O'Brien}, {Trueblood}, \&
  {Trueblood}}]{pepper:2007}
{Pepper}, J., {Pogge}, R.~W., {DePoy}, D.~L., {et~al.} 2007, \pasp, 119, 923

\bibitem[{{Pollacco} {et~al.}(2006){Pollacco}, {Skillen}, {Collier Cameron},
  {Christian}, {Hellier}, {Irwin}, {Lister}, {Street}, {West}, {Anderson},
  {Clarkson}, {Deeg}, {Enoch}, {Evans}, {Fitzsimmons}, {Haswell}, {Hodgkin},
  {Horne}, {Kane}, {Keenan}, {Maxted}, {Norton}, {Osborne}, {Parley}, {Ryans},
  {Smalley}, {Wheatley}, \& {Wilson}}]{pollacco:2006:wasp}
{Pollacco}, D.~L., {Skillen}, I., {Collier Cameron}, A., {et~al.} 2006, \pasp,
  118, 1407

\bibitem[{{Queloz} {et~al.}(2000){Queloz}, {Eggenberger}, {Mayor}, {Perrier},
  {Beuzit}, {Naef}, {Sivan}, \& {Udry}}]{queloz:2000}
{Queloz}, D., {Eggenberger}, A., {Mayor}, M., {et~al.} 2000, \aap, 359, L13

\bibitem[{{Sanchis-Ojeda} {et~al.}(2012){Sanchis-Ojeda}, {Fabrycky}, {Winn},
  {Barclay}, {Clarke}, {Ford}, {Fortney}, {Geary}, {Holman}, {Howard},
  {Jenkins}, {Koch}, {Lissauer}, {Marcy}, {Mullally}, {Ragozzine}, {Seader},
  {Still}, \& {Thompson}}]{sanchisojeda:2012}
{Sanchis-Ojeda}, R., {Fabrycky}, D.~C., {Winn}, J.~N., {et~al.} 2012, \nat,
  487, 449

\bibitem[{{Sanchis-Ojeda} {et~al.}(2013){Sanchis-Ojeda}, {Winn}, {Marcy},
  {Howard}, {Isaacson}, {Johnson}, {Torres}, {Albrecht}, {Campante}, {Chaplin},
  {Davies}, {Lund}, {Carter}, {Dawson}, {Buchhave}, {Everett}, {Fischer},
  {Geary}, {Gilliland}, {Horch}, {Howell}, \& {Latham}}]{sanchisojeda:2013}
{Sanchis-Ojeda}, R., {Winn}, J.~N., {Marcy}, G.~W., {et~al.} 2013, \apj, 775,
  54

\bibitem[{{Sato} {et~al.}(2002){Sato}, {Kambe}, {Takeda}, {Izumiura}, \&
  {Ando}}]{sato:2002}
{Sato}, B., {Kambe}, E., {Takeda}, Y., {Izumiura}, H., \& {Ando}, H. 2002,
  \pasj, 54, 873

\bibitem[{{Sato} {et~al.}(2012){Sato}, {Hartman}, {Bakos}, {B{\'e}ky},
  {Torres}, {Latham}, {Kov{\'a}cs}, {Csubry}, {Penev}, {Noyes}, {Buchhave},
  {Quinn}, {Everett}, {Esquerdo}, {Fischer}, {Howard}, {Johnson}, {Marcy},
  {Sasselov}, {Szklen{\'a}r}, {L{\'a}z{\'a}r}, {Papp}, \&
  {S{\'a}ri}}]{sato:2012:hat38}
{Sato}, B., {Hartman}, J.~D., {Bakos}, G.~{\'A}., {et~al.} 2012, \pasj, 64, 97

\bibitem[{{ter Braak}(2006)}]{terbraak:2006}
{ter Braak}, C.~J.~F. 2006, Statistics and Computing, 16, 239

\bibitem[{{Torres} {et~al.}(2007){Torres}, {Bakos}, {Kov{\'a}cs}, {Latham},
  {Fern{\'a}ndez}, {Noyes}, {Esquerdo}, {Sozzetti}, {Fischer}, {Butler},
  {Marcy}, {Stefanik}, {Sasselov}, {L{\'a}z{\'a}r}, {Papp}, \&
  {S{\'a}ri}}]{torres:2007:hat3}
{Torres}, G., {Bakos}, G.~{\'A}., {Kov{\'a}cs}, G., {et~al.} 2007, \apjl, 666,
  L121

\bibitem[{{Vogt} {et~al.}(1994){Vogt}, {Allen}, {Bigelow}, {Bresee}, {Brown},
  {Cantrall}, {Conrad}, {Couture}, {Delaney}, {Epps}, {Hilyard}, {Hilyard},
  {Horn}, {Jern}, {Kanto}, {Keane}, {Kibrick}, {Lewis}, {Osborne},
  {Pardeilhan}, {Pfister}, {Ricketts}, {Robinson}, {Stover}, {Tucker}, {Ward},
  \& {Wei}}]{vogt:1994}
{Vogt}, S.~S., {Allen}, S.~L., {Bigelow}, B.~C., {et~al.} 1994, in Society of
  Photo-Optical Instrumentation Engineers (SPIE) Conference Series, Vol. 2198,
  Society of Photo-Optical Instrumentation Engineers (SPIE) Conference Series,
  ed. D.~L. {Crawford} \& E.~R. {Craine}, 362

\bibitem[{{Wang} {et~al.}(2003){Wang}, {Hildebrand}, {Hobbs}, {Heimsath},
  {Kelderhouse}, {Loewenstein}, {Lucero}, {Rockosi}, {Sandford}, {Sundwall},
  {Thorburn}, \& {York}}]{wang:2003}
{Wang}, S.-i., {Hildebrand}, R.~H., {Hobbs}, L.~M., {et~al.} 2003, in Society
  of Photo-Optical Instrumentation Engineers (SPIE) Conference Series, Vol.
  4841, Instrument Design and Performance for Optical/Infrared Ground-based
  Telescopes, ed. M.~{Iye} \& A.~F.~M. {Moorwood}, 1145--1156

\bibitem[{{Winn}(2010)}]{winn:2010:teps}
{Winn}, J.~N. 2010, ArXiv e-prints, 1001.2010

\bibitem[{{Winn} {et~al.}(2010){Winn}, {Johnson}, {Howard}, {Marcy},
  {Isaacson}, {Shporer}, {Bakos}, {Hartman}, \& {Albrecht}}]{winn:2010:hat11}
{Winn}, J.~N., {Johnson}, J.~A., {Howard}, A.~W., {et~al.} 2010, \apjl, 723,
  L223

\bibitem[{{Yi} {et~al.}(2001){Yi}, {Demarque}, {Kim}, {Lee}, {Ree}, {Lejeune},
  \& {Barnes}}]{yi:2001}
{Yi}, S., {Demarque}, P., {Kim}, Y.-C., {et~al.} 2001, \apjs, 136, 417

\bibitem[{{Zacharias} {et~al.}(2012){Zacharias}, {Finch}, {Girard}, {Henden},
  {Bartlett}, {Monet}, \& {Zacharias}}]{zacharias:2012:ucac4}
{Zacharias}, N., {Finch}, C.~T., {Girard}, T.~M., {et~al.} 2012, VizieR Online
  Data Catalog, 1322, 0

\end{thebibliography}

\clearpage
\LongTables

\ifthenelse{\boolean{emulateapj}}{
    \begin{deluxetable*}{lrrrrrrl}
}{
    \begin{deluxetable}{lrrrrrrl}
}
\tablewidth{0pc}
\tablecaption{
    Relative radial velocities, bisector spans, and activity index
    measurements for \hatcur{50}--\hatcur{53}.
    \label{tab:rvs}
}
\tablehead{
    \colhead{BJD} &
    \colhead{RV\tablenotemark{a}} &
    \colhead{\ensuremath{\sigma_{\rm RV}}\tablenotemark{b}} &
    \colhead{BS} &
    \colhead{\ensuremath{\sigma_{\rm BS}}} &
    \colhead{S\tablenotemark{c}} &
    \colhead{Phase} &
    \colhead{Instrument}\\
    \colhead{\hbox{(2,454,000$+$)}} &
    \colhead{(\ms)} &
    \colhead{(\ms)} &
    \colhead{(\ms)} &
    \colhead{(\ms)} &
    \colhead{} &
    \colhead{} &
    \colhead{}
}
\startdata
\multicolumn{8}{c}{\bf HAT-P-50} \\
\hline\\
    \input{\hatcurhtr{50}_rvtable.tex}
\cutinhead{\bf HAT-P-51}
    \input{\hatcurhtr{51}_rvtable.tex}
\cutinhead{\bf HAT-P-52}
    \input{\hatcurhtr{52}_rvtable.tex}
\cutinhead{\bf HAT-P-53}
    \input{\hatcurhtr{53}_rvtable.tex}
\enddata
\tablenotetext{a}{
    The zero-point of these velocities is arbitrary. An overall offset
    $\gamma_{\rm rel}$ fitted to these velocities in \refsecl{globmod}
    has {\em not} been subtracted.
}
\tablenotetext{b}{
    Internal errors excluding the component of astrophysical jitter
    considered in \refsecl{globmod}.
}
\tablenotetext{c}{
    Chromospheric activity index calculated following \citet{isaacson:2010}.
}
\ifthenelse{\boolean{rvtablelong}}{
    \tablecomments{
        Note that for the iodine-free template exposures we do not
        measure the RV but do measure the BS and S index.  Such
        template exposures can be distinguished by the missing RV
        value.
    }
}{
    \tablecomments{
        Note that for the iodine-free template exposures we do not
        measure the RV but do measure the BS and S index.  Such
        template exposures can be distinguished by the missing RV
        value.  This table is presented in its entirety in the
        electronic edition of the Astrophysical Journal.  A portion is
        shown here for guidance regarding its form and content.
    }
} 
\ifthenelse{\boolean{emulateapj}}{
    \end{deluxetable*}
}{
    \end{deluxetable}
}

\end{document}